# MAUNAKEA SPECTROSCOPIC EXPLORER 2018

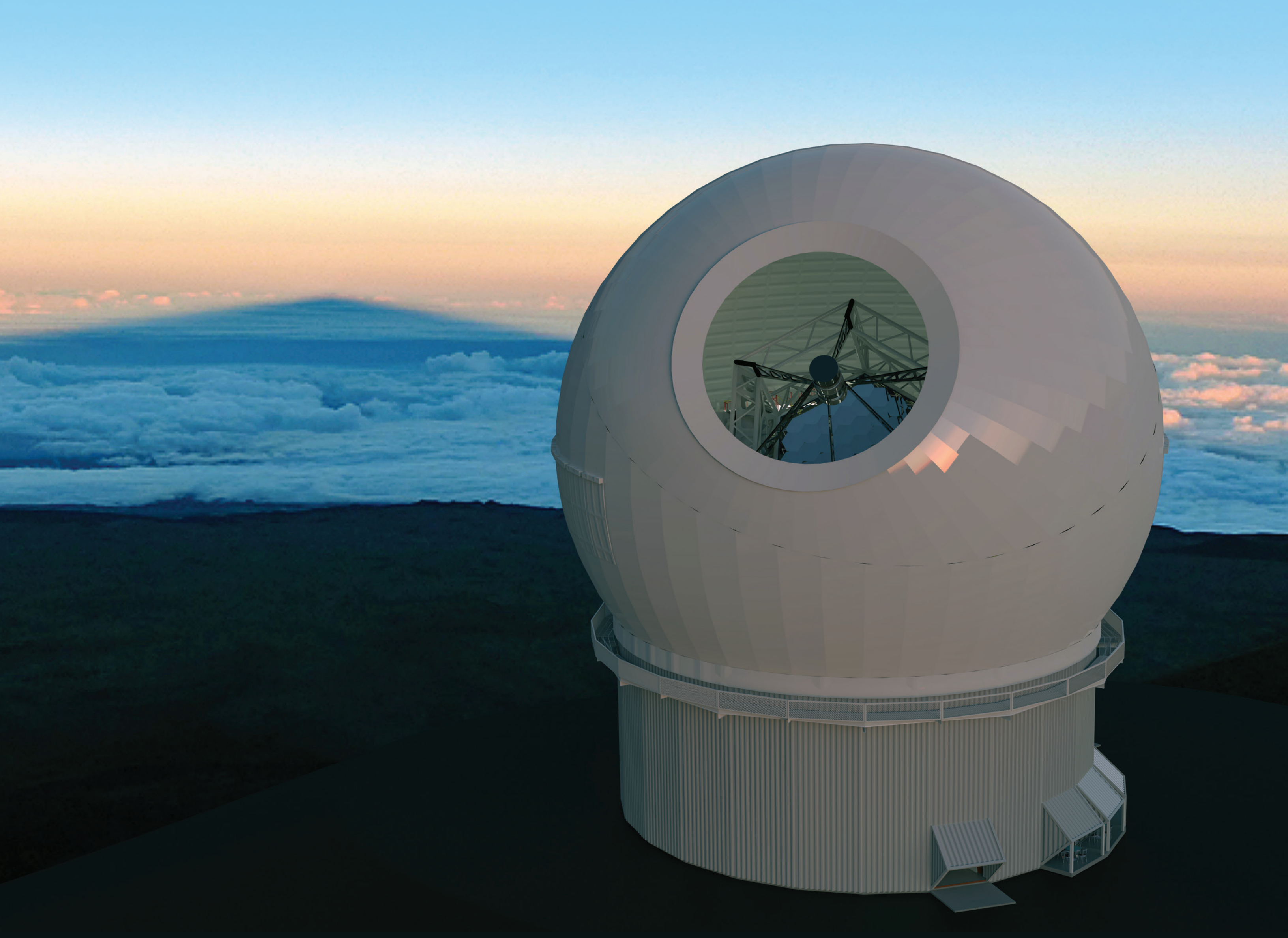

Maunakea Spectroscopic Explorer

# Maunakea Spectroscopic Explorer 2018

Alexis Hill

Nicolas Flagey

Alan McConnachie

Kei Szeto

# Table of Contents

# Table of Figures

# Glossary of Acronyms

| | |
|---|---|
| AAO | Australian Astronomical Observatory |
| ACC | Handling and Accessibility |
| ADC | Atmospheric Dispersion Corrector |
| ADRP | Automatic Data Reduction Pipeline |
| AGC | Acquisition and Guide Cameras |
| AlON | Aluminium OxyNitride |
| API | Application Programming Interface |
| ApJ | Astrophysical Journal |
| APS | Alignment and Phasing System |
| AR | Anti-Reflection |
| CFHT | Canada-France-Hawaii Telescope |
| CMP | Comprehensive Management Plan |
| CNRS | Centre National de la Recherche Scientifique |
| Coat | Mirror Coating System |
| CoDP | Conceptual Design Phase |
| CpS | Computing Services |
| CRAL | Centre de Recherche Astrophysique de Lyon |
| CSW | Common Software |
| DAD | Data Archives and Distribution |
| DIMM | Differential Image Motion Monitor |
| DMS | Data Management System |
| DRP | Data Reduction Pipeline |
| DRS | Design Reference Survey |
| DSC | Detailed Science Case |
| DT-INSU | Division Technique de l'INSU |
| ECMS | Environmental Conditions Monitoring System |
| ECS | Enclosure Control System |
| ELT | Extremely Large Telescope |
| EM | EnVision Maunakea |
| ENCL | Enclosure |
| EPO | Education and Public Outreach |
| ESPaDOnS | Echelle Spectro Polarimetric Device for the Observation of Stars |
| ESS | Enclosure Safety System |
| ETC | Exposure Time Calculator |
| FCSe | Facility Control Sequencer |
| FEA | Finite Element Analysis |
| FIRE | Feedback in Realistic Environments |
| FiTS | Fiber Transmission System |
| FRD | Focal Ratio Degradation |
| FTO | FiberTech Optica |
| GMT | Giant Magellan Telescope |
| GRACES | Gemini Remote Access Echelle Spectrograph |
| GTC | Gran Telescopio Canarias |
| GUIs | Graphical User Interfaces |
| HAA | Herzberg Astronomy and Astrophysics |
| HIRES | High Resolution Echelle Spectrometer |
| HR | High Resolution |
| ICS | Instrument Control System |
| ICSe | Instrument Control Sequencer |
| IFU | Integral Field Unit |
| IIA | Indian Institute of Astrophysics |
| InRo | Instrument Rotator |
| IQ | Image Quality |
| LMR | Low and Moderate Resolution |
| LMT | Large Millimeter Telescope Alfonso Serrano |
| LSST | Large Synoptic Survey Telescope |
| LUT | Lookup Table |
| M1 | Primary Mirror Optics System |
| M1CS | M1 Control System |
| MCS | Mount Control System |
| MG | Management Group |
| MOS | Multi-Object Spectrograph |
| MSA | Mounted Segment Assembly |
| MSE | Maunakea Spectroscopic Explorer |
| MSTR | Telescope Mount Structure |
| NAOC | National Astronomical Observatories of China |
| ngVLA | next generation Very Large Array |
| NIAOT | Nanjing Institute of Astronomical Optics and Technology |
| NIHAO | Numerical Investigation of a Hundred Astrophysical Objects (project) |
| NOAO | National Optical Astronomy Observatory |
| NRC | National Research Council |
| OAD | Observatory Architecture Document |
| OB | Outer Building |
| OBF | Observatory Building and Facilities |
| OCD | Operations Concept Document |
| OCSe | Observatory Control Sequencer |
| ODR | Observatory Data Repository |
| OED | Observing Element Database |
| OESA | Observatory Execution System Architecture |
| OF | Observing Field |
| OM | Observing Matrix |
| OMG | Observing Matrix Generator |
| ORD | Observatory Requirements Document |
| OS | Observing Sequence |
| OSS | Observatory Safety System |
| OTSp | Other Spaces and Facilities |
| PAC | Phasing and Alignment Camera |
| PBS | Product Breakdown Structure |
| PCS | Phasing Camera Systems |
| PDP | Preliminary Design Phase |
| PEAS | Procedure Executive and Analysis Software |
| PESA | Program Execution System Architecture |
| PFHS | Prime Focus Hexapod System |
| PIER | Inner Pier |
| PMA | Polished Mirror Assembly |
| PO | Project Office |
| PosS | Positioner System |
| PSA | Primary Segment Assemblies |
| PSF | Point Spread Function |
| QSO | Queue-scheduled Observing |
| r- | rapid neutron capture |
| RTV | Room Temperature Vulcanizing |
| s- | slow neutron capture |
| SCal | Science Calibration System |
| SCW | Service Cable Wrap |
| SDSS | Sloan Digital Sky Survey |
| sGRB | short Gamma Ray Burst |
| SH | Shack-Hartmann |
| SHO | Shops and Labs |
| SHS | Segment Handling System |
| SIP | Science Instrument Package |
| SKA | Square Kilometer Array |
| SNR | Signal-to-Noise Ratio |
| SoU | Statement of Understanding |
| SPA | Science Products Archive |
| SRD | Science Requirements Document |
| SSA | Segment Support Assembly |
| STEM | Science, Technology, Engineering and Mathematics |
| STR | Structure |
| SWG | Science Working Groups |
| TCSe | Telescope Control Sequencer |
| TEL | Telescope |
| THE | Thermal Management System |
| TMT | Thirty Meter Telescope |
| TOFS | Telescope Optical Feedback System |
| TSS | Telescope Safety System |
| UAM | Universidad Autónoma de Madrid |
| USTC | University of Science and Technology of China |
| UTI | Utilities System |
| UVic | University of Victoria |
| WBS | Work Breakdown Structure |
| WFC | Wide Field Corrector |

# Preface

## Purpose and Scope

This is the *Maunakea Spectroscopic Explorer 2018 book*. It is intended as a concise reference guide to all aspects of the scientific and technical design of MSE, for the international astronomy and engineering communities, and related agencies. The current version is a status report of MSE's science goals and their practical implementation, following the System Conceptual Design Review, held in January 2018.

For a complete description of the science case for MSE, intended for the astronomy community, please see *The Detailed Science Case for the Maunakea Spectroscopic Explorer.*[1]

## A Note on the Structure of the Maunakea Spectroscopic Explorer 2018 book

The book is divided into five chapters:

Chapter 1 contains the Executive Summary.

Chapter 2 focuses on the science case and science capabilities of MSE, including the highlights of the layout of the MSE Observatory and the science operations model. It answers the question: what will MSE do and what will it look like?

Chapter 3 focuses on the overall system design, and highlights important design decisions, which affect the system and major sub-systems. It answers the question: why does MSE look the way it does?

Chapter 4 provides detailed descriptions of the conceptual designs of all MSE's major subsystems. It answers the question: how does MSE work?

Chapter 5 presents a concise overview of the plan for moving MSE forward to the next Preliminary Design Phase and beyond and contains the relevant programmatic information.[2] It answers the question: how do we plan to bring MSE to fruition?

## Credits and Acknowledgements

The *MSE 2018* book represents several years of work on the development of the conceptual design of MSE. This work includes significant engineering contributions from participants based in Canada, France and Hawaii, as well as in Australia, China, India, and Spain, all coordinated from the MSE Project Office, located at the CFHT Headquarters in Waimea, Hawaii. The authors and the MSE collaborators recognize the cultural importance of the summit of Maunakea to a broad cross section of the Native Hawaiian community.

The *MSE 2018* book contains a summary of a large amount of scientific and engineering documentation, produced by the Project Office and the international team, which led to the System Conceptual Design of MSE in 2018.

---

[1] https://arxiv.org/abs/1606.00043

[2] Please refer to the MSE Prospectus, http://mse.cfht.hawaii.edu/docs/mse-science-docs/prospectus/MSE_prospectus_3-6-18_spreads.pdf for the project schedule and cost.

**Contributors**

Anthony, Andre
Ariño, Javier
Babas, Ferdinand
Bagnoud, Gregoire
Baker, Gabriella
Barrick, Gregory
Bauman, Steve
Benedict, Tom
Berthod, Christophe
Bilbao, Armando
Bizkarguenaga, Alberto
Blin, Alexandre
Bradley, Colin
Brousseau, Denis
Brown, Rebecca
Brzeski, Jurek
Brzezik, Walter
Caillier, Patrick
Campo, Ramón
Carton, Pierre-Henri
Chu, Jiaru
Churilov, Vladimir
Crampton, David
Crofoot, Lisa
Dale, Laurie
de Bilbao, Lander
de la Maza, Markel Sainz
Devost, Daniel
Edgar, Michael
Erickson, Darren
Farrell, Tony
Fouque, Pascal
Fournier, Paul
Garrido, Javier
Gedig, Mike
Geyskens, Nicolas
Gilbert, James
Gillingham, Peter
González de Rivera, Guillermo
Green, Greg
Grigel, Eric
Hall, Patrick
Ho, Kevin
Horville, David
Hu, Hongzhuan
Irusta, David
Isani, Sidik
Jahandar, Farbod
Kaplinghat, Manoj
Kielty , Collin
Kulkarni, Neelesh
Lahidalga, Leire
Laurent, Florence
Lawrence, Jon
Laychak, Mary Beth
Lee, Jooyoung
Liu, Zhigang
Loewen, Nathan
López, Fernando
Lorentz, Thomas
Lorgeoux, Guillaume
Mahoney, Billy
Mali, Slavko
Manuel, Eric
Martínez, Sofía
Mazoukh, Celine
Messaddeq, Younès
Migniau, Jean-Emmanuel
Mignot, Shan
Monty, Stephanie
Morency, Steeve
Mouser, Yves
Muller, Ronny
Muller, Rolf
Murga, Gaizka
Murowinski, Rick
Nicolov, Victor
Pai, Naveen
Pawluczyk, Rafal
Pazder, John
Pécontal, Arlette
Petric, Andreea
Prada, Francisco
Rai, Corinne
Ricard, Coba
Roberts, Jennifer
Rodgers, J. Michael
Rodgers, Jane
Ruan, Federico
Russelo, Tamatea
Salmom, Derrick
Sánchez, Justo
Saunders, Will
Scott, Case
Sheinis, Andy
Simons, Douglas
Smedley, Scott
Tang, Zhen
Teran, Jose
Thibault, Simon
Thirupathi, Sivarani
Tresse, Laurence
Troy, Mitchell
Urrutia, Rafael
van Vuuren, Emile
Venkatesan, Sudharshan
Venn, Kim
Vermeulen, Tom
Villaver, Eva
Waller, Lew
Wang, Lei
Wang, Jianping
Williams, Eric
Wilson, Matt
Withington, Kanoa
Yèche, Christophe
Yong, David
Zhai, Chao
Zhang, Kai
Zhelem, Ross
Zhou, Zengxiang

## 1. Executive Summary

Maunakea Spectroscopic Explorer (MSE) is a planned 10-m class, wide-field, optical and near-infrared facility, designed to enable transformative science, while filling a critical missing gap in the emerging international network of large-scale astronomical facilities. MSE is completely dedicated to multi-object spectroscopy of samples of between thousands and millions of astrophysical objects. It will lead the world in this arena, due to its unique design capabilities: it will boast a large (11.25 m) aperture and wide (1.52 sq. degree) field of view; it will have the capabilities to observe at a wide range of spectral resolutions, from R2500 to R40,000, with massive multiplexing (4,332 spectra per exposure, with all spectral resolutions available at all times), and an on-target observing efficiency of more than 80%. With these unrivalled capabilities, MSE will collect a number of spectra equivalent to an entire SDSS Legacy Survey every eight weeks.

MSE will unveil the composition and dynamics of the faint Universe and is designed to excel at precision studies of faint astrophysical phenomena. This will enable unique and transformative science. Among other science cases, MSE will reveal the emergence of the Periodic Table of the elements; it will provide the premier facility for astrophysical exploration of the nature of dark matter; it will chart the growth of supermassive black holes, and discover the connection of galaxies to the large-scale structure of the Universe. MSE will also provide critical follow-up for multi-wavelength imaging surveys, such as those of the Large Synoptic Survey Telescope, Gaia, Euclid, the Wide Field Infrared Survey Telescope, the Square Kilometre Array, and the Next Generation Very Large Array. This new generation of facilities is beginning to collect enormous datasets of billions of new faint objects, datasets which will grow exponentially over the coming decade. MSE provides essential optical and near-infrared spectroscopic follow-up for millions of these faint sources. In this respect, MSE enables a synergy between wide-field imaging surveys and pointed follow-up by the Thirty Meter Telescope, the Giant Magellan Telescope, and the (European) Extremely Large Telescope, by providing the essential filtering of the immensely large survey datasets. MSE has extensive science overlaps and complementary capabilities compared to most of the upcoming generation of astronomical facilities, and the strategic importance of MSE cannot be overstated.

The scientific impact of MSE will be made possible and affordable by upgrading the existing Canada-France-Hawaii Telescope (CFHT) infrastructure on the Maunakea summit, Hawaii. CFHT is located at a world-class astronomical site with excellent free-atmosphere seeing (0.4 arcseconds median seeing at 500 nm). The Mauna Kea Science Reserve Comprehensive Management Plan[i] for the Astronomy Precinct explicitly recognizes CFHT as one of the sites that will be redeveloped. CFHT is an iconic 3.6-m telescope with four decades of operational experience and a legacy of discovery on Maunakea. MSE will build on the experience of CFHT and incorporate the latest technical advancements made by other top astronomical facilities. In order to minimize environmental and cultural impacts to the site, as well as minimizing cost, MSE will replace CFHT with an 11.25 m aperture telescope, while retaining the current summit facility footprint.

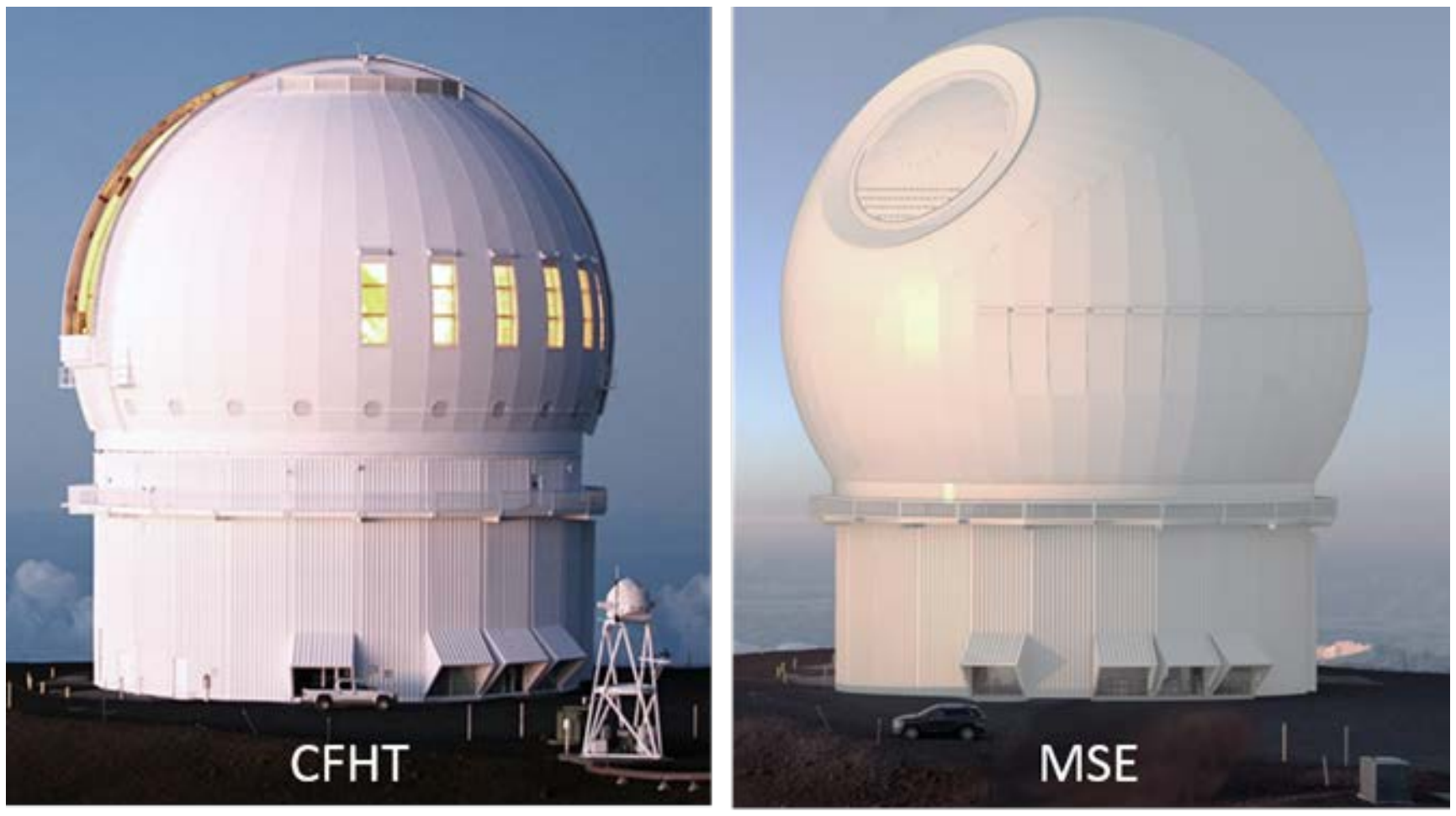

*Figure 1: Exterior views of CFHT (ventilation modules open) and MSE (ventilation modules closed).*

The rotating CFHT enclosure will be replaced by a Calotte enclosure that is only 10% larger than the current size, leaving the foundation and much of the remaining infrastructure intact. Building renovations and structural upgrades will be internal, so the outward appearance of MSE will remain very much unchanged from that of CFHT (Figure 1). Inside, however, a modern observatory will perform cutting-edge science at one of the best astronomical sites in the world, with access to three quarters of the entire night sky.

MSE is designed to take advantage of the proven site on Maunakea, which allows for an extremely sensitive, wide-field, and massively multiplexed facility. These design attributes are the key ingredients that will enable researchers to efficiently conduct large surveys of faint astronomical objects. At the MSE focal surface, 4,332 light-collecting input fibers are packed into a hexagonal array. The fibers are precisely positioned to a submillimeter accuracy, within six microns of error, in order to maximize the amount of light injected from science targets into the input fibers, which collect and transmit light to banks of spectrographs tens of meters away. The exquisite seeing allows the fiber diameters to be kept small,[3] thus keeping the size and cost of the spectrographs attainable. One bank of spectrographs receives light from 3,249 fibers from the focal surface and is switchable between the collection of low resolution spectra at R2500 and moderate resolution spectra at R6000, covering the optical to near-infrared wavelength range of 0.36–1.8 microns. Concurrently, the other bank of spectrographs receives light from 1,083 fibers from the focal surface and is dedicated to collecting the high resolution spectra in targeted optical wavelength windows within the wavelength range of 0.36–0.5 microns at R40,000 and 0.5–0.9 microns at R20,000. All resolution modes have simultaneous full field coverage, and the massive multiplexing results in many thousands of spectra per hour and over a million spectra per month, available to the MSE user community. An Integral Field Unit (IFU) system has been incorporated into the design, as an upgrade path for a second generation capability for MSE.

Aside from the physical infrastructure, MSE's success is enabled by efficiently scheduled and executed surveys, by the quality of the data collected, and by MSE's ability to make the science products available to survey teams in a timely and efficient manner. MSE will devote approximately 80% of available time to executing large, homogeneous legacy surveys, which require several years to complete. More focused strategic programs, which require smaller amounts of observing time and typically lead to more rapid publications, will occupy the remaining 20% of observing time. Proposals for both types of programs will be solicited from the MSE user community at regular intervals, and the selection process will be determined using the MSE partnership governance model. MSE is operated solely in a queue-based mode, requiring sophisticated scheduling software. Data is made available to the survey team immediately, and to the whole MSE community on a short timescale. All data and derived products will have a proprietary period, set by the MSE partnership.

[3] 85 micron diameter, 0.8 arcseconds, for the high resolution spectrographs and 107 micron diameter, 1.0 arcseconds, for the low and moderate resolution spectrographs.

CFHT has been a long-term leader in engaging with the community through innovative public outreach programs for the promotion of science, technology, engineering, and mathematics (STEM), culture, and the environment. As a member of the Maunakea Observatories and international astronomical communities, MSE will continue CFHT's legacy of community engagement within the partnership. Community engagement will be an integral part of MSE's outreach activities throughout the life of the new observatory.

---

[i] http://www.malamamaunakea.org/management/master-plan

## 2. Overview: What is MSE?

MSE is the realization of a long-held ambition: to provide dedicated, large-aperture, wide-field multi-object spectroscopy at optical and near-infrared wavelengths. It is designed to meet the diverse needs of various fields of astronomy, needs which are all centered on a single capability. MSE will be at the hub of the international network of astronomical facilities in the 2020s and beyond. It enables transformative, standalone science and enriches everything from our understanding of the formation of the elements of the periodic table to the nature of dark matter; from the growth of supermassive black holes and the connection of galaxies to the large-scale structure of the Universe. A detailed summary of the science capabilities of MSE is provided in Table 1.

*Table 1: The detailed science capabilities of MSE*

<table>
<tr><th colspan="13">Site characteristics</th></tr>
<tr><td>Observatory latitude</td><td colspan="12">19.9 degrees</td></tr>
<tr><td>Accessible Sky</td><td colspan="12">30,000 square degrees (airmass < 1.55 i.e., δ > -30 degrees)</td></tr>
<tr><td>Median image quality</td><td colspan="12">0.37 arcsec (free atmosphere, zenith, 500 nm)</td></tr>
<tr><td>Average length of night</td><td colspan="12">10.2 hours</td></tr>
<tr><td>Historical weather losses (average)</td><td colspan="12">2.2 hours / night</td></tr>
<tr><td>Observing efficiency (on-sky, on-target)</td><td colspan="12">80%</td></tr>
<tr><td>Expected on-target science observing hours</td><td colspan="12">2336 hours / year</td></tr>
<tr><td>Expected on-target fiber-hours</td><td colspan="12">10,112,544 fiber-hours / year (total): 7,589,664 (LR & MR) / 2,529,888 (HR)</td></tr>
<tr><th colspan="13">Telescope architecture</th></tr>
<tr><td>Structure, focus</td><td colspan="12">Altitude-azimuth, Prime</td></tr>
<tr><td>M1 aperture</td><td colspan="12">80.8 $m^2$</td></tr>
<tr><td>Science field of view</td><td colspan="12">1.52 square degrees</td></tr>
<tr><td>Spectrograph system</td><td colspan="12">6 x LMR spectrographs (4 channels/spectrogrpah, all identical, each channel seperately switchable to provide LR and MR modes<br>2 x HR spectrographs (3 channels/spectrograph), both identical, to provide high resolution mode<br>All spectrographs always available with full multiplexing<br>Deployable IFU system using LR /MR spectrograph system available as second generation capability</td></tr>
<tr><th colspan="13">Fiber positioning system</th></tr>
<tr><td>Multiplexing</td><td colspan="12">4,329 (total): 3,249 (LR & MR) / 1,083 (HR)</td></tr>
<tr><td>Fiber size</td><td colspan="12">1 arcsec (LR & MR) / 0.8 arcsec (HR)</td></tr>
<tr><td>Positioner patrol radius</td><td colspan="12">90.3 arcsecs</td></tr>
<tr><td>Positioner accuracy</td><td colspan="12">0.06 arcsec rms</td></tr>
<tr><td>Positioner closest approach</td><td colspan="12">Two fibers can approach with 7 arcsecs of each other (three fibers can be placed within 9.9 arcsec diameter circle)</td></tr>
<tr><td>Repositioning time</td><td colspan="12">< 120 seconds</td></tr>
<tr><td>Typical allocation efficiency</td><td colspan="12">> 80 % (assuming source density approximately matched to fiber density)</td></tr>
<tr><th colspan="13">Low resolution (LR) spectroscopy</th></tr>
<tr><td>Wavelength range</td><td colspan="3">360 ≦ λ ≦ 560 nm</td><td colspan="3">540 ≦ λ ≦ 740 nm</td><td colspan="3">715 ≦ λ ≦ 985 nm</td><td colspan="3">960 ≦ λ ≦ 1320 nm</td></tr>
<tr><td>Spectral resolution (approx. at center of band)</td><td colspan="3">2,550</td><td colspan="3">3,650</td><td colspan="3">3,600</td><td colspan="3">3,600</td></tr>
<tr><td>Sensitivity requirement<br>(pt. source, 1hr, zenith, median seeing, monochromatic magnitude)</td><td colspan="3">m = 24.0<br>SNR/res. elem. = 2, λ > 400 nm<br>SNR/res. elem. = 1, λ ≦ 400 nm</td><td colspan="3">m = 24.0<br>SNR/resolution element = 2</td><td colspan="3">m = 24.0<br>SNR/resolution element = 2</td><td colspan="3">m = 24.0<br>SNR/resolution element = 2</td></tr>
<tr><th colspan="13">Moderate resolution (MR) spectroscopy</th></tr>
<tr><td>Wavelength range</td><td colspan="3">391 ≦ λ ≦ 510 nm</td><td colspan="3">576 ≦ λ ≦ 700 nm</td><td colspan="3">737 ≦ λ ≦ 900 nm</td><td colspan="3">1457 ≦ λ ≦ 1780 nm</td></tr>
<tr><td>Spectral resolution (approx. at center of band)</td><td colspan="3">4,400</td><td colspan="3">6,200</td><td colspan="3">6,100</td><td colspan="3">6,000</td></tr>
<tr><td>Sensitivity requirement<br>(pt. source, 1hr, 30 degree zenith angle, median seeing, monochromatic magnitude)</td><td colspan="3">m = 23.5<br>SNR/res. elem. = 2, λ > 400 nm<br>SNR/res. elem. = 1, λ ≦ 400 nm</td><td colspan="3">m = 23.5<br>SNR/resolution element = 2</td><td colspan="3">m = 23.5<br>SNR/resolution element = 2</td><td colspan="3">m = 24.0<br>SNR/resolution element = 2</td></tr>
<tr><th colspan="13">High resolution (HR) spectroscopy</th></tr>
<tr><td>Wavelength range</td><td colspan="4">360 ≦ λ ≦ 460 nm</td><td colspan="4">440 ≦ λ ≦ 620 nm</td><td colspan="4">600 ≦ λ ≦ 900 nm</td></tr>
<tr><td>Wavelength band</td><td colspan="4">λ / 30<br>[ baseline: 401 - 415 nm ]</td><td colspan="4">λ / 30<br>[ baseline: 472 - 488.5 nm ]</td><td colspan="4">λ / 15<br>[ baseline: 626.5 - 672 nm ]</td></tr>
<tr><td>Spectral resolution (approx. at center of band)</td><td colspan="4">40,000</td><td colspan="4">40,000</td><td colspan="4">20,000</td></tr>
<tr><td>Sensitivity requirement<br>(pt. source, 1hr, zenith, median seeing, monochromatic magnitude)</td><td colspan="4">m = 20.0<br>SNR/resolution element = 10, λ > 400 nm<br>SNR/resolution element = 5, λ ≦ 400 nm</td><td colspan="4">m = 20.0<br>SNR/resolution element = 10</td><td colspan="4">m = 24.0<br>SNR/resolution element = 10</td></tr>
<tr><th colspan="13">Science calibration</th></tr>
<tr><td>Sky subtraction accuracy</td><td colspan="12">0.5% requirement (0.1% goal)</td></tr>
<tr><td>Velocity precision</td><td colspan="12">100 m/s (HR, SNR/resolution element = 30)</td></tr>
<tr><td>Relative spectrophotometric accuracy</td><td colspan="12">3% (LR, SNR/resolution element = 30)</td></tr>
</table>

This chapter presents an overview of the science that has driven the development and detailed design of MSE (Section 2.1). It describes how MSE complements other front-line facilities (Section 2.2) and provides a summary of the key capabilities of the observatory, particularly with respect to other multi-object spectroscopic instruments (Section 2.3). It concludes with a top-level overview of the design of MSE, which highlights key design elements, and includes a discussion of the anticipated operational model of the facility that will enable partners to conduct vast spectroscopic surveys of the Universe (Section 2.4).

## 2.1. The Science Case: Exploring the Composition and Dynamics of the Faint Universe

MSE will unveil the composition and dynamics of the faint Universe and is designed to excel at precision studies of faint astrophysical phenomena that are beyond the reach of 4-m class spectroscopic instruments. It will be the premier astronomical facility for understanding the nature and evolution of matter in the Universe, whether that mass is baryonic or in the dark sector. It will trace the astrophysical emergence of the periodic table of the elements over cosmic time and determine the allowed properties of the dark matter particle, through a suite of coordinated observations of dynamical probes across all mass scales.

The science case for MSE is described at length in "The Detailed Science for the Maunakea Spectroscopic Explorer" (DSC, May 2015).[4] The impacts of MSE will be felt across astrophysics including characterization of exoplanetary hosts; in-situ chemical tagging of thick disk and halo stars; tomographic mapping of the interstellar and intergalactic media; linking of galaxies to the large-scale structure of the Universe (Figure 2); the growth of supermassive black holes over cosmic time; exploration of the variable Universe in the era of multi-messenger astronomy; and next generation cosmological surveys probing the nature of dark energy. The wealth of science areas impacted by MSE will ensure its lasting legacy and the data produced will provide a rich repository for exploration and discovery. In the following sections, we will highlight some headline science connected with MSE, but the reader is referred to the DSC for a comprehensive description of all the major science topics driving the design of the facility.

[4] An updated version of the DSC will be available in 2019.

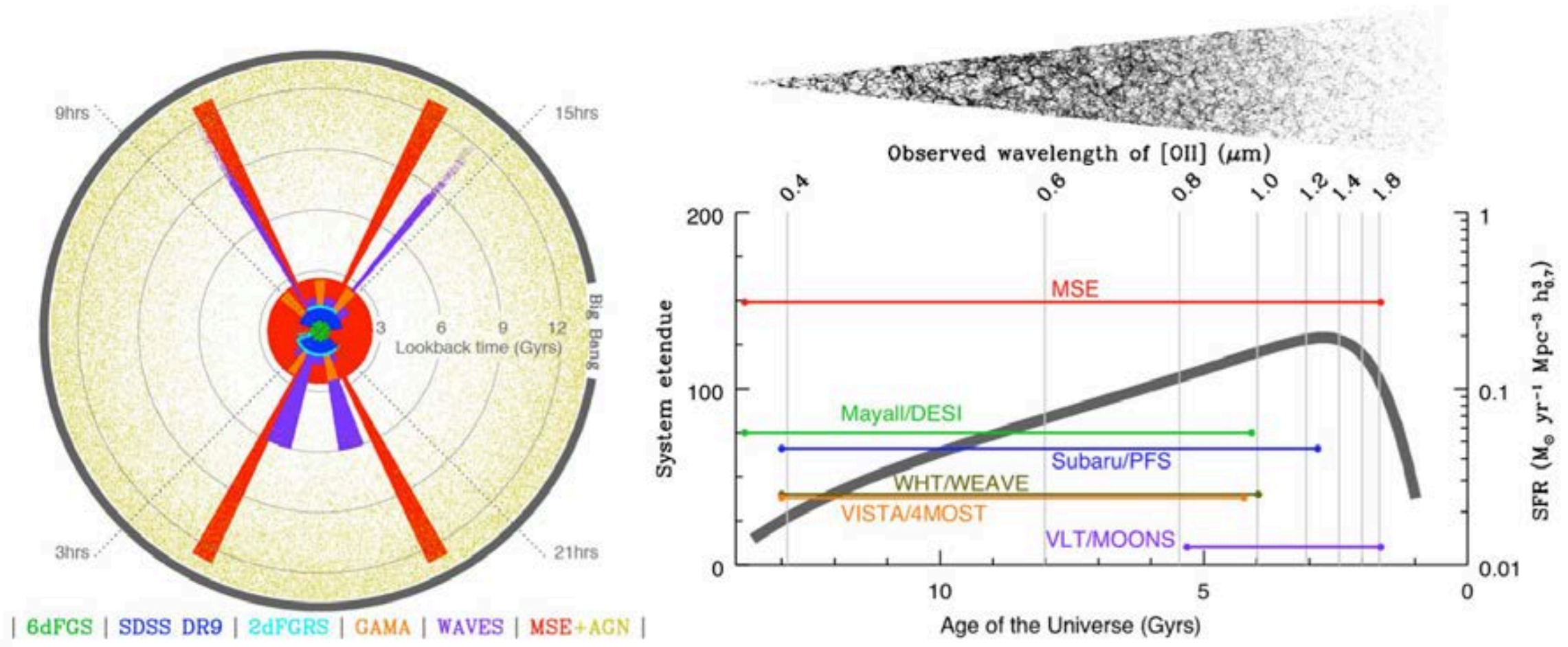

*Figure 2: Left panel: Cone plot showing an illustrative MSE survey, in comparison with other notable benchmark galaxy surveys. Cones are truncated at the redshift at which L* galaxies are no longer visible. Right panels: Lookback time versus cosmic star formation rate (right axis) using the parameterization of Hopkins & Beacom (ApJ 2006: 651, 142). Grey lines indicate the wavelength of OII 3727Å. Also indicated on this wavelength scale is the wavelength coverage of the major highly multiplexed spectrographs in development. These are offset according to their system's étendue (left axis). The light cone demonstrates the range of MSE for extragalactic surveys, using a homogeneous set of tracers at all redshifts.*

## MSE and the emergence of the periodic table of the elements

Understanding when, where, and how the chemical elements were created remains one of the grand unanswered questions in astronomy, and is a primary motivation for building MSE.

Some 13.8 billion years ago, the Big Bang produced all the hydrogen and most of the helium in the Universe, as well as trace amounts of lithium. All the other chemical elements—the oxygen that we breathe, the calcium in our bones, and the iron in our blood—were forged in stars. While the nuclear processes governing the production of the elements and isotopes from helium to uranium are generally understood, the astrophysical locations and details of the stellar nucleosynthesis continue to be debated.

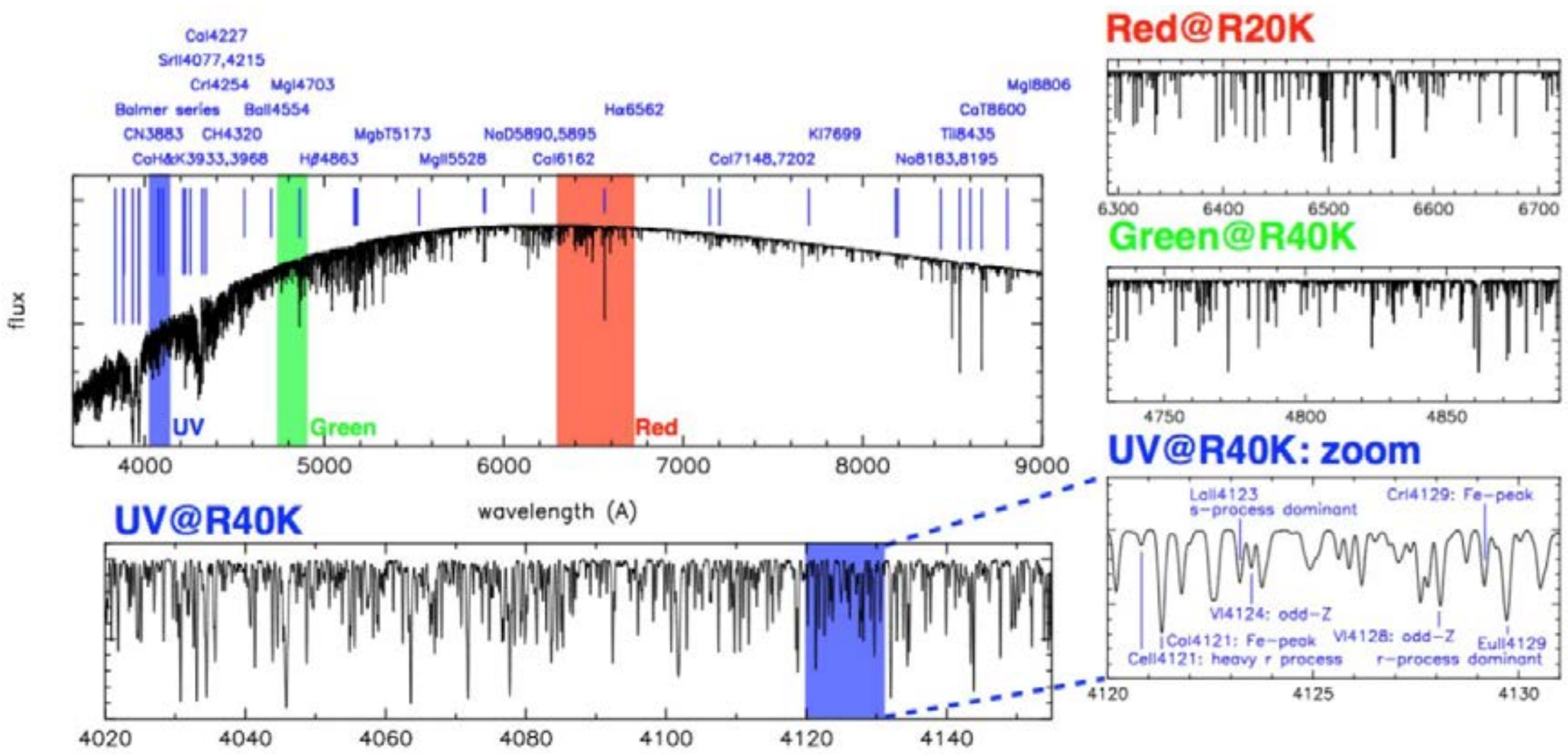

*Figure 3: Main panel shows the relative flux of a synthetic spectrum of a metal-poor red giant star at the moderate MSE spectral resolution of R~6000, along with some of the strong-line stellar diagnostics accessible at this resolution. Highlighted regions show the normalized flux in three windows, observable with the high-resolution mode of MSE. A magnified region of the UV window shows examples of the species that will be identified at high resolution. MSE chemical tagging surveys will identify species by sampling a large and diverse set of nucleosynthetic pathways and processes.*

Chemical elements and their formation sites, along with stellar evolution timescales, provide a useful diagnostic tool which acts as a chronometer, allowing us to study the milestones of galaxy formation and the origin of the solar system’s chemical abundances. For example, the recent detections of a gravitational wave signal from a neutron star merger and an associated short gamma ray burst (sGRB) have provided possible clues to the solution of the pertinent problem of the production sites of the “rapid neutron capture” (r-)process elements. Though neutron star mergers can produce robust r-process synthesis all the way up to the heaviest elements, such as uranium and thorium, they still fail to adequately explain the observed chemical evolution trends. The presence of abundant r-process elements at low metallicities requires a short merging time, which is not easily compatible with sGRB timescales, or with the r-process trends observed at higher metallicities. There are also some indications of an intermediate neutron-capture process, operating at timescales between the r-process and the “slow neutron capture” (s-) process. The production of neutron-capture elements through an intermediate process provides a more coherent explanation of the origin of some stars (CEMP-r+s stars).

MSE will chart the evolution of the chemical elements over cosmic time. The most chemically ancient stars in the Galactic halo and surrounding ultra-faint dwarf galaxies are of particular importance in understanding the formation of the elements in earliest times. These objects are relics that date back to the epoch of reionization and provide remarkable laboratories that permit us to explore the chemical and physical conditions of the earliest star-forming environments in the Universe, and the very first metal-enrichment events. While these ancient objects are incredibly valuable (they are fossils that preserve the signatures of the first generations of stars), they are also incredibly rare.

MSE has an unparalleled competitive advantage in its ability to reveal the origin of the chemical elements. It is imperative to identify chemical species, which trace a variety of nucleosynthetic pathways. Many chemical species—including some unique tracers—have signatures in the blue/near-UV region of the spectrum, and here MSE has very high (R40K) spectral resolution and good throughput (in part due to its large aperture). These capabilities are essential to the detection of potentially very weak lines, in areas of the spectrum that are rich in other features (see Figure 3). Enormous multiplexing is required to build statistical datasets, from which the clustering of stars in chemical space can be mapped. From these datasets, the rarest of all stars—those that are chemically ancient—can be identified and forensically analyzed in significant numbers. Through these studies, MSE will reveal the epochs and timescales associated with the formation of the different nucleosynthetic groups, thus enabling transformative progress in mapping the emergence of the periodic table of the elements.

**MSE and astrophysical probes of dark matter**

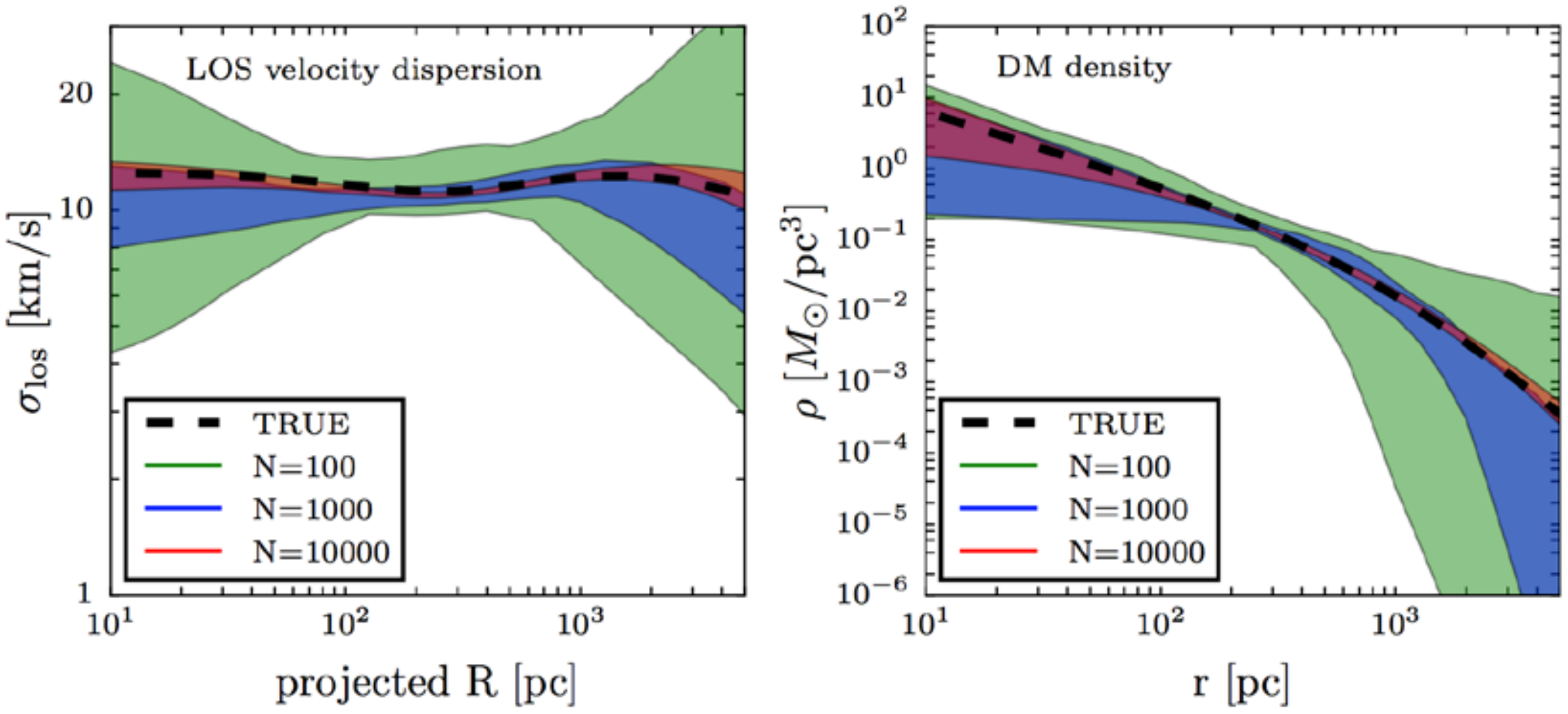

*Figure 4: Recovery of line-of-sight velocity dispersion (left) and dark matter density (right) profiles as a function of stellar spectroscopic sample size for dwarf galaxies, demonstrating the need for extensive datasets with MSE. Shaded regions represent 95% credible intervals from a standard analysis of mock data sets consisting of radial velocities for N=$10^2$, $10^3$, and $10^4$ stars (median velocity errors of 2km/s), generated from an equilibrium dynamical model for which true profiles are known (thick black lines).*

Astronomical observations have provided overwhelming evidence that about 80% of the matter density in the Universe takes the form of “dark matter,” yet dedicated searches for a dark matter particle in the laboratory, in space, and at colliders have yielded no results thus far. MSE will become the premier astronomical facility to advance the decades-old quest for the dark matter particle.

While it is still possible to find viable dark matter candidates in supersymmetric extensions of the Standard Model, the theories have widened, to include models in which dark and normal matter are present in different sectors, which are very weakly coupled to each other. Many possibilities concerning “hidden sector” dark matter, both thermally and non-thermally produced, are currently being investigated. Many of these theories can be tested in the lab: using, for example, models in which the dark matter is produced through its couplings to normal matter. MSE, however, will be essential for the large swathes of parameter space that can only be observed using astrophysical probes.

Analogous to Standard Model particles, dark matter in hidden sectors can be light (e.g., warm or fuzzy dark matter) or have large self-interaction cross sections (such as hydrogen self-scattering or proton-neutron scattering) including energy loss mechanisms (inelastic scattering). These fundamental properties of the dark matter particle can only be verified using astrophysical probes, through their influence on structure formation at small scales. The key astrophysical probes are the densities of dark matter halos in the central regions of halos (roughly 1–10% of the virial radius) and the properties of the satellites of massive halos (such as the satellites of the Milky Way). Collectively, these probes are often labeled "small-scale structures," and can be accessed by MSE via precision measurements of the radial velocities of large numbers of tracer particles: stars, star clusters, and galaxies.

Recent computing advances have led to a better understanding of the impact of stellar feedback on the structure of galaxies, including their dark matter content. Many groups have found that feedback can change dark matter distribution dramatically. For example, two major collaborations (NIHAO and FIRE) with separate codes have found that the dark matter density profile gets shallower with feedback at about one percent of the virial radius. At the same time, there has been progress in pinning down the dark matter content of galaxies in the Local Group and of more massive field galaxies.

There have been concurrent advances in understanding the predictions of self-interacting, warm, and fuzzy dark matter. We now possess modern hydrodynamic codes that track the self-interactions and pressure for fuzzy dark matter in a full cosmological setting. Based on the advances made possible by these simulations, we know that large self-interactions provide an effective way to understand not only the cores of dwarfs and low surface brightness galaxies, but also the lack of evidence for them in high surface brightness galaxies. Taken together, these developments bode well for the settlement of a debate that has been raging for over two decades: what do these small-scale structure puzzles tell us about the nature of dark matter?

A suite of MSE dark matter surveys will provide decisive measurements, vital to pinning down the nature of the dark matter particle. For example, MSE will allow us to obtain a deeper kinematic census of the satellites in the local volume, test various proposed solutions to the "too-big-to-fail problem," and help quantify the prevalence of the planes of satellites, which has important implications for dark matter models. Measurements using MSE will also have important consequences for self-interacting dark matter models. These models transport kinetic energy within halos and alter the dark matter density profiles. In low mass galaxies (such as the Local Group dwarf spheroidal galaxies) this leads to constant density cores, which will be tested using the large data sets of resolved stellar velocities that only MSE can provide (see Figure 4).

MSE will also provide critical data on fuzzy dark matter. Fuzzy dark matter changes the power spectrum and introduces a cut-off in halo formation below a mass scale determined by the dark matter particle mass. This can be tested through substructure detections (resolved and flux-ratio anomalies), using strong lensing or gaps in stellar streams. MSE will be essential in selecting systems (streams and lenses) and in obtaining the requisite kinematic data that can be used to place constraints on the dark matter model spaces. Perhaps most excitingly, we expect to be able to attain new levels of precision, by combining radial velocity surveys using MSE and precision astrometric studies using the Extremely Large Telescope (ELT), the Thirty Meter Telescope (TMT), or the Giant Magellan Telescope (GMT), to provide the ultimate dynamic analyses of dark matter halos. Directly or in conjunction with 30-m class telescopes, MSE will decisively test the idea that dark matter is cold and collisionless.

### 2.2. Science Opportunities: MSE as Part of the Facility Landscape

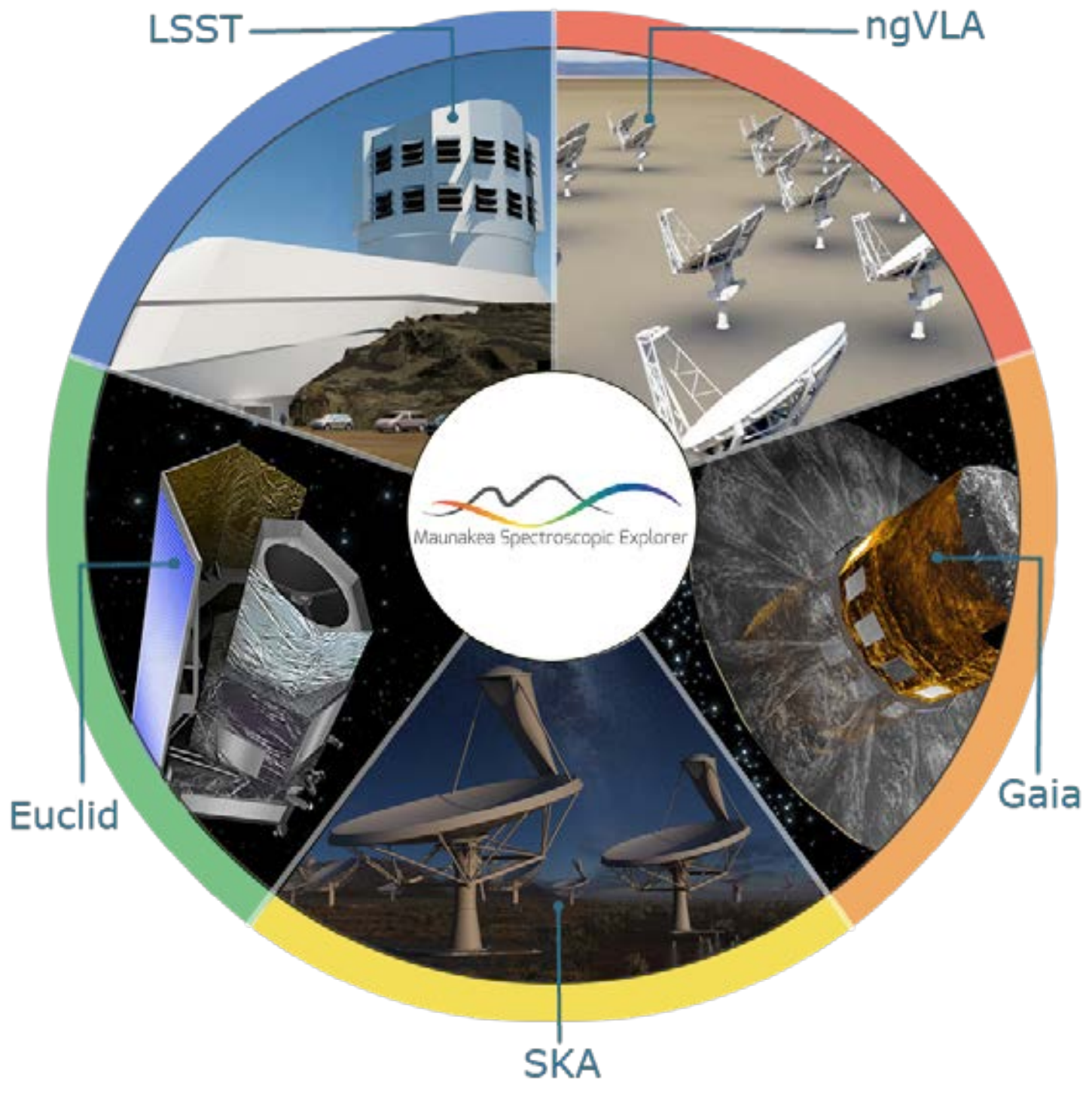

*Figure 5: Some of the most notable next generation facilities that, together with MSE, will help define the international network of astronomical facilities operational beyond the 2020s.*[5]

Astronomy has recently entered the multi-wavelength realm of big facilities. Figure 5 shows some of the most notable new observatories. Collectively, these facilities represent many billions of dollars in investment and many decades of development, by large teams from the international community. Some of these facilities are already hosting scientific exploration: allowing researchers to probe questions as diverse and fundamental as the origin of life and the composition of our universe. The second data release from Gaia is already having an impact on almost all areas of astronomy; the Large Synoptic Survey Telescope (LSST) is being built; and the VLOTs are entering the construction phase. These are no longer next generation scientific ambitions; instead, they are at the cutting edge of what is happening *now*.

Each of these facilities hosts well-defined, standalone scientific projects. Recent history has demonstrated that combining data from multiple facilities will be key to major advances.

---

[5] Artist's conception of the multi-antenna ngVLA, courtesy of NRAO; Gaia image credit: ESA–D (Ducros, 2013); the Square Kilometer Array image credit: SKA Organisation; Euclid artist's impression, based on CAD drawings provided by Thales Alenia Space, Italy; and Airbus (Defence and Space), France © ESA; the Large Synoptic Survey Telescope, image credit: Todd Mason, Mason Productions Inc./LSST Corporation.

Examining this network of astronomical facilities inevitably prompts the question of whether any critical capabilities are still missing?

Multiple international groups have asked this question over the past decade, and all have concluded that the vital piece of equipment needed is a dedicated, large-aperture, multi-object spectroscopic facility. In Canada, the Long Range Plan 2010[i] notes that a 10-m class telescope, equipped with an extremely multiplexed spectrograph, “would be a unique resource for follow-up spectroscopy, both for the European Gaia satellite mission, and also for LSST and Euclid/WFIRST”. The Australian Astronomy Decadal Plan 2016–25[ii] details the ways in which a dedicated, wide-field spectroscopic facility, integrated into a large telescope, would “provide follow-up spectra of objects identified by the SKA and imaging telescopes like the US-led Large Synoptic Survey Telescope.” In Europe, the ESO science program for the 2020s has identified highly multiplexed spectroscopy as a priority, identifying this as an area which merits strong backing from their community, due to its broad importance for a vast range of science projects.[iii] An ESO Working Group[iv] commissioned to study the potential of wide-field MOS has concluded that “such a facility could enable transformational progress in several broad areas of astrophysics, and may constitute an unmatched ESO capability for decades,” and has conducted a science feasibility study. In the US, in response to the National Research Council report, “Optimizing the U.S. Optical and Infrared System in the Era of LSST,” NOAO and LSST convened study groups whose top recommendation was that the US community “develop or obtain access to a highly multiplexed, wide-field, optical, multi-object spectroscopic capability on an 8-m class telescope, preferably in the Southern Hemisphere.”

MSE is a response to the internationally recognized need for highly multiplexed, deep, optical and near-infrared spectroscopy of the faint Universe. The apparent consensus of scientific opinion regarding the strategic importance of the capabilities offered by MSE is a natural conclusion, based on the capabilities and expected science outcomes of the currently anticipated facilities. The strategic importance of MSE within the international network of astronomical facilities cannot be overstated, as a brief survey of these facilities demonstrates:

1. **Ground-based, wide-field imaging:** There has been a long and successful history of exploiting synergies between optical and infrared imaging and spectroscopic surveys. Complete, deep, optical imaging of the sky in a range of filters will soon be available, through a combination of survey programs, including PanSTARRS, SkyMapper, Subaru/HyperSuprimeCam, CFHT/Canada–France Imaging Survey, Blanco/Dark Energy Camera and, most notably, LSST. The capabilities of MSE are highly desirable, as they enhance the discovery potential of both LSST and the other surveys named.
2. **Deep, near-infrared, space-based, wide-field imaging:** Just as with ground-based imaging, researchers have successfully established the benefits of combining space-based imaging and wide-field spectroscopy. Over the next decade, Euclid and WFIRST will provide extremely deep, high-spatial resolution imaging (∼0.2″), mostly in the near-infrared, over vast areas of sky. The combination of Euclid/WFIRST imaging with photometric redshifts (LSST) and MSE can be expected to provide a complete blueprint of galaxy evolution from the present epoch to the peak of the cosmic star-formation era.
3. **Precision astrometry and photometry:** Gaia is a landmark astrometric space mission, which is cataloging all astronomical point sources brighter than V ∼20 mag: around 1 billion objects (1% of the stellar mass of the Milky Way). Gaia is obtaining multi-band photometry for all sources, and observing the spectra of the brightest stars. The need for ground-based spectroscopy to supplement the Gaia data is well established. Using MSE to obtain the high resolution spectroscopy of millions of stars across Gaia's entire luminosity range will provide precise (sub-km/s) radial velocities and chemical abundance information that cannot be obtained by other means.

4. **Multi-wavelength astronomy:** The Square Kilometer Array (SKA) and, potentially, the next generation Very Large Array (ngVLA), will revolutionize astronomy at radio wavelengths. The use of optical spectroscopy is crucial, to maximize the scientific output of these radio facilities. The sensitivity challenge involved is extreme: matching SKA1 detection thresholds requires optical data extending to r ~24[v]. MSE will meet this sensitivity challenge and will have a survey speed that allows for the construction of large statistical datasets, which can be combined with the vast radio datasets that SKA and ngVLA will produce.
5. **Multi-messenger astronomy and the time domain:** Large, dedicated, spectroscopic facilities have a clear role to play as scientific attention increasingly turns to the time domain. MSE will provide a particularly powerful tool for the follow-up of slow and faint transients, and the characterization of the host galaxies of extragalactic explosive transients (both pre- and post-explosion). Perhaps most excitingly, however, the first detection of gravitational waves—by LIGO, in February 2016—promises to provide a fundamentally new way of exploring the Universe, to which MSE will contribute, as the premier large-aperture spectroscopic facility.
6. **Finding the needles and diamonds in the haystack:** The TMT, ELT, and GMT will become the premier optical/infrared facilities for detailed, high spatial resolution views of the faintest astronomical targets. Since they possess fields of view in the order of a few arcminutes, target identification will require a coordinated suite of supporting facilities, to ensure these cutting-edge facilities maximize their scientific impact. MSE will occupy an important role as the essential link—efficiently narrowing the billions of potential targets identified by these wide-field photometric surveys down to the much smaller number of targets that will be followed up by the VLOTs.

## 2.3. Science Capabilities

### 2.3.1. From Science to Requirements

The Detailed Science Case makes extensive reference to Science Reference Observations, presented in the appendices to the DSC. These have been defined by the international Science Team as specific, detailed, scientific programs for MSE, which are transformative of their fields and only possible with MSE. The SROs were selected to span the range of anticipated fields to which MSE is expected to contribute. The Science Requirements for MSE—i.e., the highest-level design requirements for the facility—are defined as the suite of capabilities necessary for MSE to carry out these observations. These capabilities are extensive and cannot be found in combination in any other facility. They will ensure that MSE remains at the cutting edge of astronomical capabilities in the 2020s and beyond, even as the scientific opportunities change in ways that cannot be predicted.

*Table 2: Cross-references between Science Reference Observations (in the columns) and Science Requirements (grouped in rows). Blue borders indicate that the SRO is used in the derivation of the requirement; blue boxes indicate that the requirement is highly relevant; grey boxes indicate that the requirement has some relevance to the SRO.*

| | | Resolved stellar sources | | | | | Extragalactic sources | | | | | | |
|---|---|---|---|---|---|---|---|---|---|---|---|---|---|
| | | *Exoplanet hosts* | *Time domain stellar astrophysics* | *Chemical tagging in the outer galaxy* | *CDM subhalos and stellar streams* | *Local Group Galaxies* | *Nearby galaxies* | *Virgo and Coma* | *Halo occupation* | *Galaxies and AGN* | *The InterGalactic Medium* | *Reverberation mapping* | *Peculiar velocities* |
| Spectral resolution | *Low spectral resolution* | | R~2000 (white dwarfs) | | | | R~3000 | R~3000 | R~2000-3000 | R~3000 | | R~3000 | R~1000-2000 |
| | *Intermediate spectral resolution* | | Any repeat observations | Essential, R~6500 | Essential, R~6500 | Essential, R~6500 | Velocities of low mass galaxies | Velocities of low mass galaxies | | | R~5000 | | |
| | *High spectral resolution* | R>=40000 | R>=50000 | Essential, R~20-40k | Essential | Young stars | | Bright globular clusters | | | | | |
| Focal plane input | *Science field of view* | ~2000 sq. deg | all-sky | 1000s sq. deg | 1000s sq. deg | 100s sq. deg | 3200 (100) sq. deg | ~100 sq. deg | ~1000s sq. deg | ~300 sq. deg | 40 sq. deg | 7 sq. deg | all-sky |
| | *Multiplexing at lower resolution* | | | | | | ~5000 galaxies/sq. deg | 100s target (galaxies/GCs) /sq.deg | >5000 galaxies/sq. deg | 770 galaxies/sq. deg | | 600AGN/ deg | 1000s galaxies/sq. deg |
| | *Multiplexing at moderate resolution* | | | 1000s stars/sq.deg to g~23 | 1000s stars/sq.deg to g~23 | few-thousands stars/sq.deg | ~5000 galaxies/sq. deg | | | | 500 galaxies/sq.deg | | |
| | *Multiplexing at high resolution* | ~100 stars/sq.deg @ g=16 | ~1000 stars/sq.deg to g=20.5 | ~1000 stars/sq.deg to g=20.5 | ~1000s stars/sq.deg to g~23 | | | | | | | | |
| | *Spatially resolved spectra* | | | | | | Goal | Goal | | | | | Yes |
| Sensitivity | *Spectral coverage at low resolution* | | | | | | 0.37 - 1.5um | 0.37 - 1.5um | 0.36 - 1.8um | 0.36 - 1.8um | 0.36 - 1.8um | 0.36 - 1.8um | Optical emission lines |
| | *Spectral coverage at moderate resolution* | | | Strong line diagnostics in optical | CaT essential | CaT essential | Goal: Complete | Goal: Complete | | | Goal: Complete | | |
| | *Spectral coverage at high resolution* | Strong lines for velocities; tagging | Strong lines for velocities | Chemical tagging | Strong lines for velocities | | | | | | | | |
| | *Sensitivity at low resolution* | | | | | | i=24.5 | i=24.5 | i=25.3 | i=25 / H=24 | | i=23.25 | i=24.5 |
| | *Sensitivity at moderate resolution* | | | g>20.5 | g~23 | i=24 | i=24.5 | i=24.5 | | | r~24 | | |
| | *Sensitivity at high resolution* | g=16 @high SNR | g=20.5 | g=20.5 | g~22 | | | | | | | | |
| Calibration | *Velocities at low resolution* | | | | | | v~20km/s | v~20km/s | v~100km/s | v~20km/s | | v~20km/s | v~20km/s |
| | *Velocities at moderate resolution* | | | v~1km/s | v~1km/s | v~5km/s | v~9km/s | v~9km/s | | | v~20km/s (10km/s goal) | | |
| | *Velocities at high resolution* | v<100m/s | v~100m/s | v~100m/s | v<1km/s | | | | | | | | |
| | *Relative spectrophotometry* | | | | | | ~4% | | | | | Critical: 3% | |
| | *Sky subtraction, continuum* | few % | few % | few % | few % | <1% | <1% | <1% | <0.5% | <0.5% | <1% | <1% | <1% |
| | *Sky subtraction, emission lines* | | | | important (CaT region) | important (CaT region) | critical | critical | critical | critical | critical | critical | |
| Operations | *Accessible sky* | Plato footprint (ecliptic) | Gaia footprint (all sky) | Gaia footprint (all sky) | Gaia, PS1, HSC footprint | Northern hemisphere (M31, M33, dec~+40) | LSST overlap useful (10000 sq. Deg) | NGVS footprint dec+12 | LSST overlap useful (10000 sq. Deg); Euclid (all sky) | LSST overlap useful (10000 sq. Deg); Euclid (all sky) | LSST overlap useful (10000 sq. Deg); Euclid (all sky) | all sky target distribution | all sky target distribution |
| | *Observing efficiency* | maximize | maximize | maximize | maximize | maximize | maximize | maximize | maximize | maximize | maximize | maximize | maximize |
| | *Observatory lifetime* | Monitoring >= years | Monitoring >= years | Survey >= 5 years | Survey >= 5 years | Survey ~100 nights | Survey ~few years | Survey ~100 nights | Survey ~7 years | Survey ~100 nights | Survey ~100 nights | Monitoring ~5 years | Survey ~years |

Table 2 shows the cross-references between each SRO and each of the high-level MSE science requirements. The science requirements are outlined and discussed in detail in the MSE Science Requirements Document.[vi]

### 2.3.2. Key Capabilities

Table 1 summarizes the scientific capabilities of MSE. The telescope will have an 11.25 m segmented primary mirror and a 1.52 sq. degree field of view. Three different spectral resolution settings are possible. At the lowest resolution, 3,249 spectra, spanning the entire optical spectrum in the J-band, can be obtained in a single pointing. At moderate resolution, the same number of spectra can be obtained for approximately half the optical waveband and for the full H-band. The low and moderate resolutions are provided by the same spectrograph system (six identical spectrographs, with four channels each), and each channel can switch settings independently. At the highest resolution, 1,083 spectra will be obtained per pointing for three windows in the optical region of the spectrum. At all resolutions, MSE can encompass the faintest targets, and will remain well calibrated and stable over its lifetime. Located at the equatorial site of Maunakea ($\ell = 19.9°$), MSE will access the entire northern hemisphere and half of the southern sky, making it an ideal follow-up and feeder facility for a large number of both existing and planned ground- and space-based facilities.

*(1) Key Capability 1: Survey Speed and Sensitivity*

MSE must have a high survey speed. This is a function of sensitivity, field of view, multiplexing, and observing efficiency. Many of the surveys MSE is expected to undertake cover large areas of sky. MSE therefore has a 1.52 square degree field of view, hexagonal for ease of tessellation, which will provide an observing efficiency of 80%, defined as the fraction of night time spent collecting science photons (excluding losses due to weather). This will make it one of the most efficient optical observatories in the world.

MSE must have a large aperture, to provide sufficient sensitivity to follow up faint sources identified by imaging surveys conducted by large-aperture telescopes, such as LSST. As a filtering facility for the VLOTs, MSE must also provide excellent sensitivity to faint sources, which can then be followed up using higher spatial resolution (e.g., Integral Field Units), higher spectral resolution, and/or higher signal-to-noise (SNR) by these giant facilities. At a low spectral resolution, MSE will obtain an SNR per resolution element of two for magnitude 24 sources at all wavelengths (point sources, monochromatic AB magnitude; i.e., the approximate depth of a single LSST visit) over the course of an hour-long observation. At high spectral resolution, MSE will obtain an SNR per resolution element of 10 for magnitude 20 sources at all wavelengths (point sources, monochromatic magnitude; i.e., covering the full luminosity range of Gaia sources) over the course of an hour-long observation.

*(2) Key Capability 2: Spectral Performance and Multiplexing*

A range of spectral resolutions and wavelengths, from UV, through optical, to NIR, are needed to enable a diverse range of scientific investigations. The multiplexing requirements of each mode are determined by a consideration of the expected target densities. The source density of galaxies at $z < 0.2$ brighter than $i = 23$ is 2,100/sq. degree (or ~3,200/1.5 sq. degree), which determines the minimum multiplexing for the low and moderate spectral resolution modes (the density of even fainter sources makes it desirable to achieve a value as close to 1 fiber per square arcminute as possible, but technical limitations make it difficult to reach that goal). The source density of thick disk and halo stars at high Galactic latitudes in the critical magnitude range for MSE of $17 < g < 21$ is ~700 per sq. degree or ~1000 per 1.5 sq. degree, which determines the minimum multiplexing for the high spectral resolution mode.

MSE is also being designed to incorporate multi-object Integral Field Units after first light. We anticipate that these will feed the low/moderate spectrograph suite, which will be achieved by switching the fiber positioning and fiber transmission systems.

The wavelength coverage in each mode is determined by the consideration of primary scientific goals. As part of its low/moderate spectral resolution studies, MSE will probe aspects of galaxy evolution in the distant Universe. Sensitivity out to and including H-band ensures that galaxies and AGN can be studied, using the same set of tracers from $z = 0$, to beyond the peak of the star formation history of the Universe. As part of its high spectral resolution studies, a large number of critical nucleosynthetic tracers need to be accessed at blue/UV wavelengths. Many of these features are weak and in crowded spectral regions, which requires a spectral resolution of R>20K.

### *(3) Key Capability 3: Dedicated and Specialized Operations*

MSE is specialized for one task: the efficient acquisition of multi-object spectra. This basic operation philosophy enables the production of stable, high quality data, which is much more difficult to achieve for observatories, in which instruments move on and off the telescope at regular intervals. This allows MSE to address science cases that are very difficult to address using other MOS instruments (for example, time-resolved, high-resolution spectroscopy and quasar reverberation mapping).

The impressive power that a large, homogeneous and well characterized dataset can offer has been most successfully demonstrated by the Sloan Digital Sky Survey (SDSS). A search of NASA ADS[vii] for refereed publications mentioning SDSS in their abstracts returns over 8,600 publications, with more than 436,000 citations.[6] The success of SDSS comes despite the fact that it is a relatively small-aperture facility by modern standards, located at a site that cannot compete with Maunakea in terms of median image quality. However, a large part of its success can be traced to the extremely well calibrated and well characterized nature of the data. MSE can be viewed as an evolution of the SDSS concept, using a telescope with a collecting area around twenty times larger, situated at arguably the best optical astronomical site on the planet.

The fiber-positioning technology chosen by MSE ensures that all spectrographs are available at all times, so every MSE observation will use all 4,332 fibers. High spectral resolution observations of (relatively) bright targets will be prioritized during bright times, and low-resolution observations of fainter, extragalactic targets will be prioritized during dark times. Assuming a (conservative) baseline exposure of one hour per field, with eight hours per night available for observations (10.2 hours before weather losses), then MSE will observe around one million astronomical spectra every month: the equivalent of a SDSS Legacy Survey—1,640,960 spectra—every eight weeks.

---

[6] Search conducted on September 18 2018.

### 2.3.3. MSE Compared with Other Multi-object Spectrographs

*Table 3: Summary of new optical and infrared multi-object spectroscopic instruments and facilities.*

| *Class* | *Facility / Instrument* | *First light (anticipated)* | *Aperture (M1 in m)* | *Field of View (sq. deg)* | *Etendue* | *Multiplexing* | *Wavelength coverage (um)* | *Spectral resolution (approx)* | *IFU* | Dedicated facility |
|---|---|---|---|---|---|---|---|---|---|---|
| 2-m | **SDSS I - IV** | Existing | 2.5 | 1.54 | 7.6 | 640 | 0.38 - 0.92 | 1800 | Yes | Yes |
| 4-m | **Guo Shoujing / LAMOST** | Existing | 4 | 19.6 | 246 | 4000 | 0.37 - 0.90 | 1000 | No | Yes |
| | **AAT / HERMES** | Existing | 3.9 | 3.14 | 37.5 | 392 | windows | 28000, 50000 | No | No |
| | **WHT / WEAVE** | 2018[a] | 4.2 | 3.14 | 43.5 | 960 | 0.37 - 0.96 | 5000 | Yes | Yes |
| | | | | | | | windows | 20000 | | |
| | **Mayall / DESI** | 2019[b] | 4 | 7.1 | 89.2 | 5000 | 0.36 - 0.98 | 4000 | No | Yes |
| | **VISTA / 4MOST** | 2022[c] | 4 | 4.1 | 51.5 | 2436 | 0.39 - 0.95 | 5000 | No | Yes |
| | | | | | | | windows | 18500 | | |
| 8-m | **VLT / MOONS** | 2020[d] | 8.2 | 0.14 | 7.4 | 1000 | 0.65 - 1.80 | 4000 | No | No |
| | | | | | | | windows | 18000 | | |
| | **Subaru / PFS** | 2021[e] | 8.2 | 1.25 | 66 | 2394 | 0.38 - 1.26 | 3000 | No | No |
| | | | | | | | 0.71 - 0.89 | 5000 | | |
| 10-m | **MSE** | 2027 | 11.25 | 1.52 | 151 | 4329 | 0.36 - 1.3 | 3000 | Second generation | Yes |
| | | | | | | | 0.36 - 0.95 (50%) 1.5 - 1.85 | 6500 | | |
| | | | | | | | windows | 40000 | | |

*a* http://www.ing.iac.es/Astronomy/telescopes/wht/weavepars.html#dflow
*b* https://www.desi.lbl.gov
*c* https://www.eso.org/sci/facilities/develop/instruments/4MOST.html#status
*d* https://www.eso.org/sci/facilities/develop/instruments/MOONS.html#status
*e* http://pfs.ipmu.jp/schedule.html

Table 3 compares the core capabilities of a large number of MOS instruments and facilities, including all those that are at advanced stages of design, many of which will be operating on timescales that overlap with MSE. The ESO spectroscopic telescope concept[viii] is not listed in these tables, since only a feasibility study has been conducted and detailed technical specifications are not yet available.

Table 4 lists only those facilities from Table 3 that have comparable sensitivity to MSE (i.e., 8–10 m class facilities).

*Table 4: MSE in comparison to other planned MOS instruments on 8-m class telescopes.*

| | **8 - 12 m class facilities** | | | | | | |
|---|---|---|---|---|---|---|---|
| | **VLT / MOONS** | | **Subaru / PFS** | | **MSE** | | |
| *Dedicated facility* | No | | No | | Yes | | |
| *Aperture (M1 in m)* | 8.2 | | 8.2 | | 11.25 | | |
| *Field of View (sq. deg)* | 0.14 | | 1.25 | | 1.52 | | |
| *Etendue* | 7.4 | | 66 | | 151 | | |
| *Multiplexing* | 1000 | | 2394 | | 4329 | | |
| *Etendue x Multiplexing* | 7400 *( = 0.01 )* | | 158004 *( = 0.24 )* | | 653679 *( = 1.00 )* | | |
| *Observing fraction* | < 1 ? | | 0.2 (first 5 years) 0.2 - 0.5 afterwards ? | | 1 | | |
| *Spectral resolution (approx)* | 4000 | 18000 | 3000 | 5000 | 3000 | 6500 | 40000 |
| *Wavelength coverage (um)* | 0.65 - 1.80 | windows | 0.38 - 1.26 | 0.71 - 0.89 | 0.36 - 1.8 | 0.36 - 0.95 50% | windows |
| *IFU* | No | | No | | Second generation | | |

Table 4 presents a graph showing the wavelength coverage and étendue of the upcoming generation of facilities, listed in Table 3. In addition to its étendue—larger than that of any other upcoming facility by almost a factor of two—the wavelength coverage of MSE is unmatched by any other wide-field, spectroscopic facility in any aperture class. Figure 6 shows MSE in comparison to the facilities listed in Table 3 that are closest to MSE in terms of sensitivity, but still have collecting areas which are smaller by almost a factor of two. Here, étendue is combined with the multiplexing and the observing fraction, to define a quantity that is essentially the survey speed. With VLT/MOONS, we are assuming that the MSE will remain on the telescope at all times; for Subaru/PFS, the solid blue rectangle reflects our assumption that PFS will account for 20% of telescope time, in line with other Subaru strategic survey programs.[ix] The hatched rectangle shows its survey speed were PFS to occupy 100% of telescope time. These considerations make MSE the ultimate spectroscopic facility—more than an order of magnitude more efficient at surveys than its closest competitor.

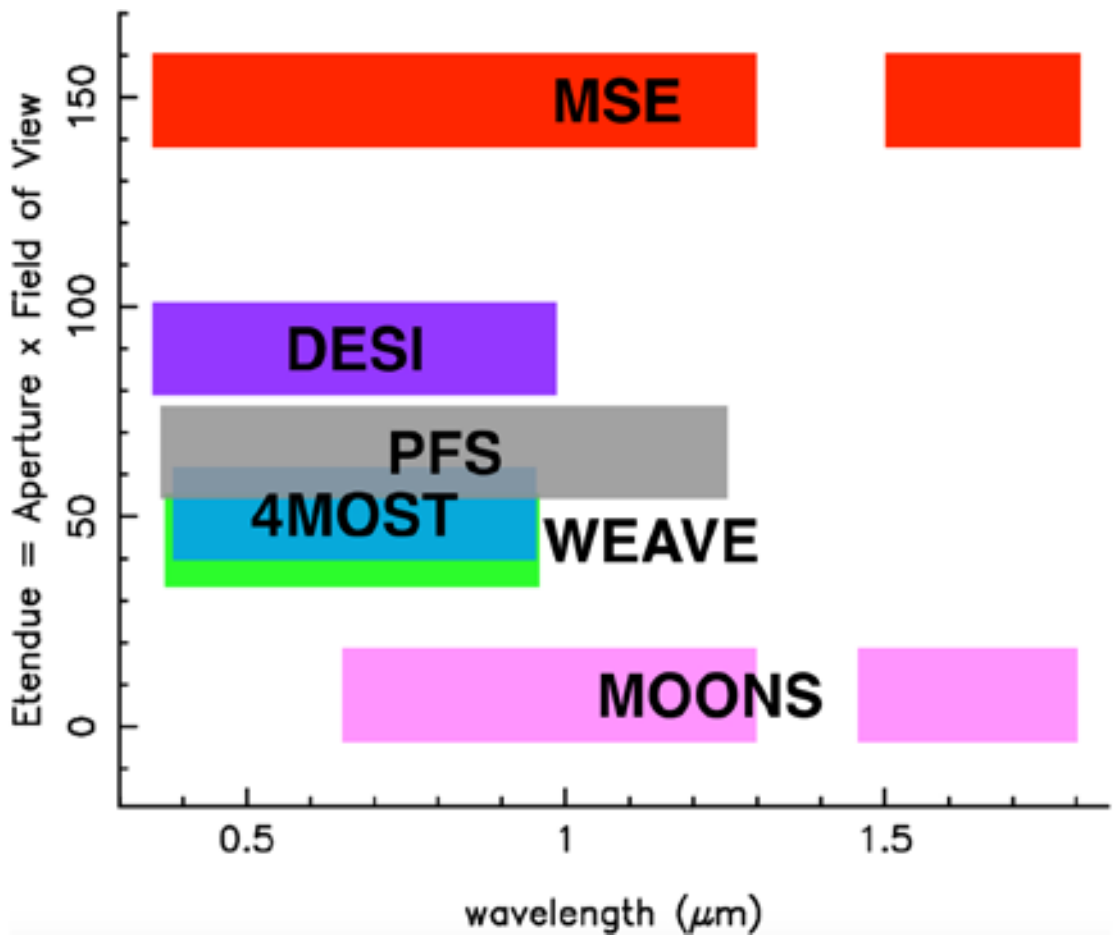

*Figure 6: Etendue vs. wavelength coverage for the upcoming wide-field MOS facilities listed in Table 3.*

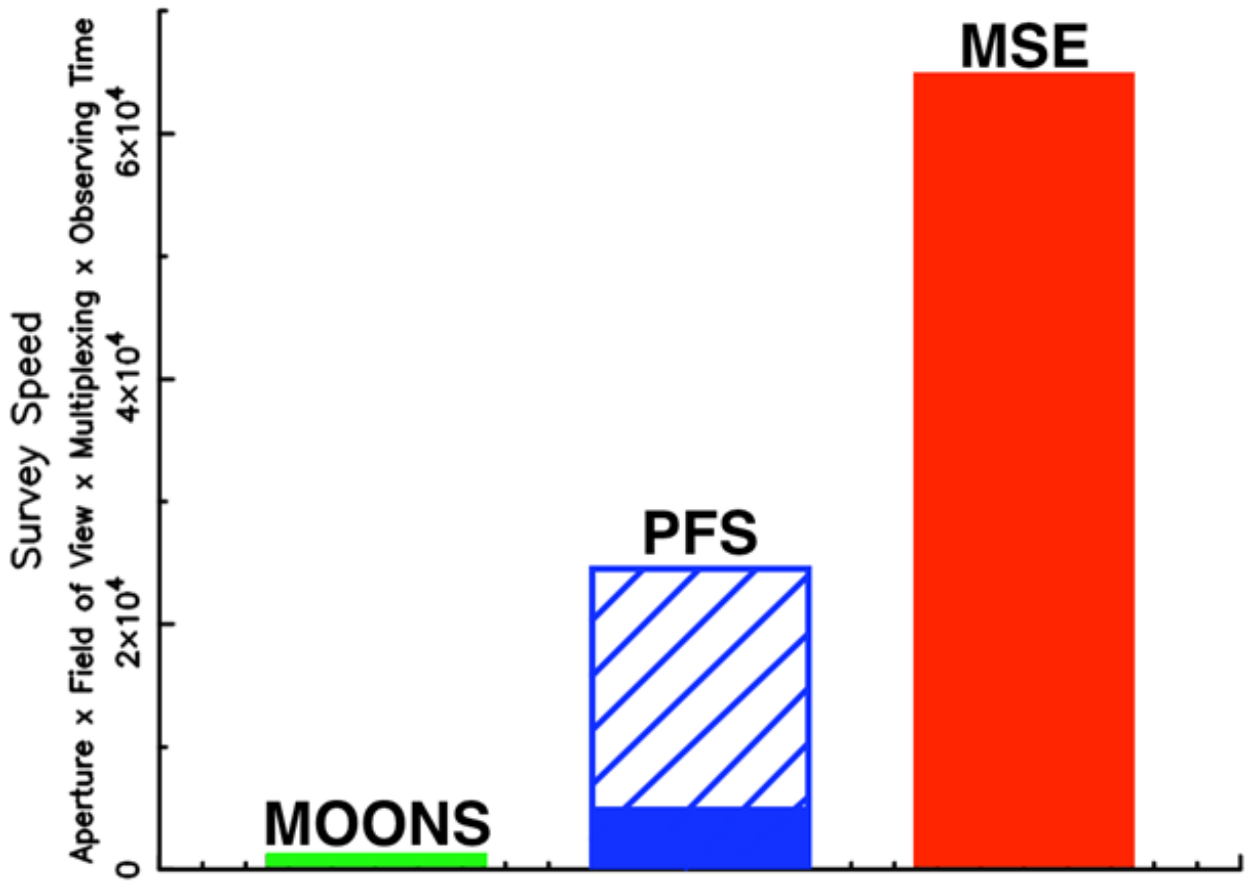

*Figure 7: Comparison of the survey speeds[7] of the three 8–10 m class wide-field MOS capabilities in design or under construction.*

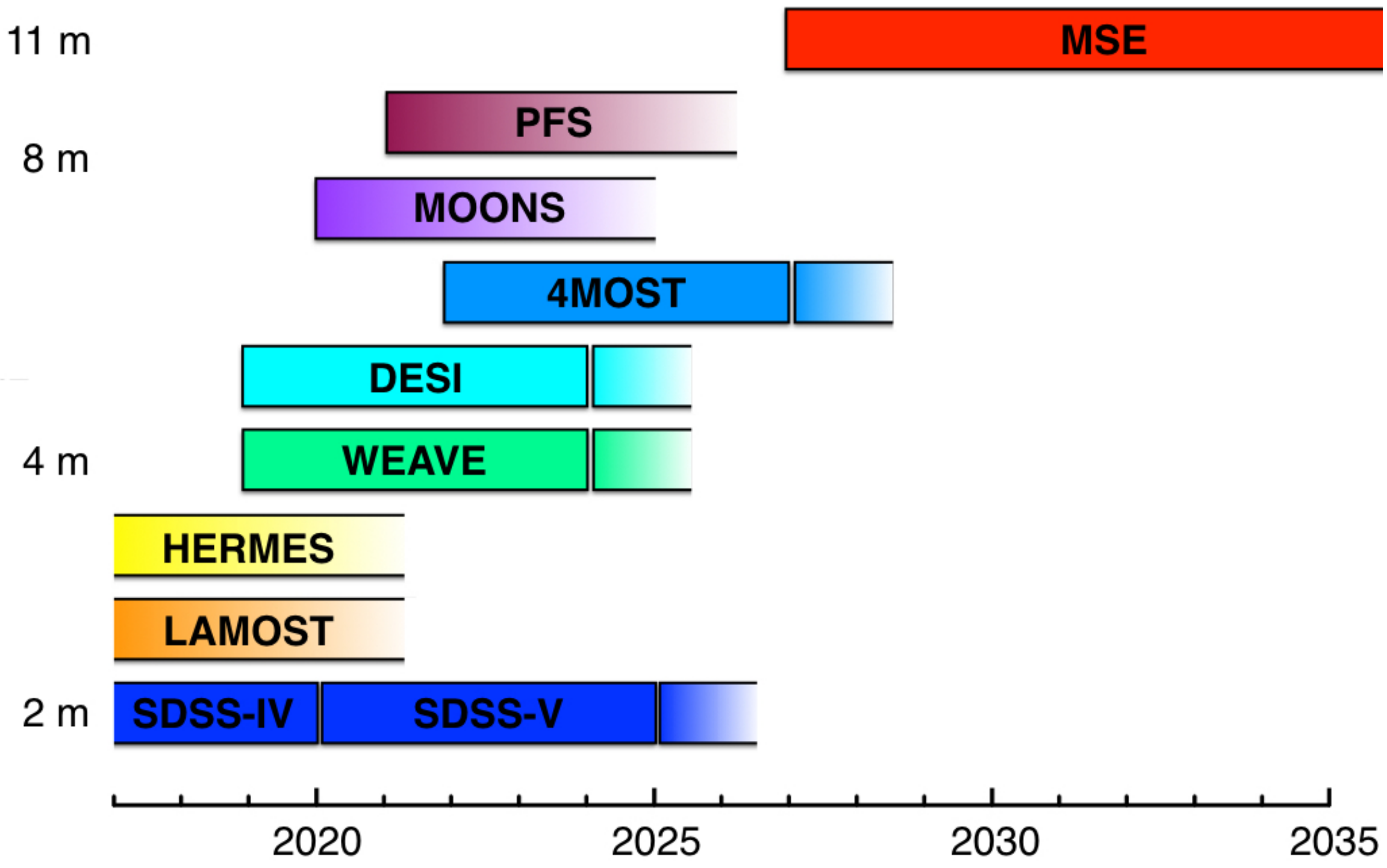

*Figure 8: Current anticipated timelines of the wide-field MOS facilities listed in Table 3. Bounded boxes indicate the duration or lifetime of the survey or facility, and the absence of a vertical solid line indicates that the facility has no clear end date. Several of the facilities will operate initially for a set period of*

[7]Survey speed is defined as the product of aperture, field of view, multiplexing, and observing time (i.e., the fraction of time the instrument is on the telescope). With MOONS, MSE is assumed to be on the telescope 100% of the time. For PFS, the solid blue rectangle shows the results if it is on the telescope 20% of the time, and the hatched rectangle displays the results if it is on the telescope 100% of the time.

*years, with the expectation that their lifespan will be extended beyond the nominal end date. This figure was inspired by a similar figure created by J. Newman.*

Figure 8 shows the anticipated timelines of the wide-field, spectroscopic facilities listed in Table 3, sorted by aperture (a reasonable proxy for sensitivity). Bounded boxes indicate the duration of the principal surveys or the lifetime of the instrument, if known (e.g., DESI will undertake a five-year survey). In many cases, we anticipate that the facilities will be extended beyond their nominal end dates. The absence of a vertical solid line indicates that the survey or facility does not have a clear end date. MSE is the largest of these facilities and the only dedicated facility on a large aperture telescope that could be operational in the 2020s, to overlap with facilities such as LSST, Euclid, and WFIRST.

### 2.4. The MSE Observatory

In consideration of the scientific cases driving MSE and the synergies with other leading astronomical facilities produced the science requirement described in Section 2.3. What follows is a high-level overview of the engineering and operational design of MSE, a design that will realize the scientific ambitions of the international community for wide-field MOS over the coming decades.

#### 2.4.1. Observatory Layout

The overall layout of the observatory is shown in Figure 9. Light from science targets travels through the atmosphere and enters the calotte-style enclosure. It is reflected off the 11.25 m primary mirror (its size is defined by the circumscribing circle of the same diameter), consisting of sixty 1.44 m hexagonal segments, and is directed towards the prime focus of the telescope. Each primary mirror segment is similar in size to those used by TMT and ELT, in order to leverage state-of-the-art, VLOT-class, segmented, mirror-system technologies into the MSE design. MSE is an altitude–azimuth mount, prime-focus telescope, which points at the appropriate fields in the sky, using two independent structures. Azimuth rotation is provided by the azimuth structure, while altitude motion is provided by the elevation structure. The structures are supported by hydrostatic bearings, which provide the rotational motions required. The primary mirror and prime focus components are supported by the elevation structure. The primary mirror is located below the elevation axis, while the prime focus components are supported at the top of spiders, on the elevation structure.

At the top end, light passes through a five-element Wide Field Corrector (WFC), which delivers a convex focal surface. The WFC incorporates an Atmospheric Dispersion Corrector (ADC). The novel design of this system also allows the effect of differential atmospheric refraction to be reduced by a factor of around two, allowing it to maintain an excellent delivered image quality, at the focal surface that enables small fiber sizes.

The WFC/ADC feeds light to the “Sphinx” fiber positioning system, which consists of 4,332 positioners, each integrated with an individual fiber. The Sphinx positioner system covers a hexagonal 1.52 sq. degree science field of view, so that the end of each fiber acquires a target in the sky within its circular patrol area of 7.12 sq. arcminutes. The Sphinx positioners are piezo-actuated tilting spines, modeled on the FMOS/Echidna[8] and 4MOST/AESOP designs.

Light enters the fibers at the focal surface at f/1.97. Out of a total of 4,332 fiber positioners, 3,249 carry 1 arcsecond diameter fibers, which feed light to the low and moderate resolution spectrograph suite; and 1,083 positioners carry 0.8 arcsecond fibers, which feed the high resolution spectrograph suite. Each positioner has a large circular patrol region (180.6 arcseconds in diameter), thus ensuring that full field coverage is simultaneously provided for both the high resolution and low/moderate resolution positioners, despite the difference in multiplexing. In addition, the low and moderate resolution positioners provide high redundancy, since 97% of the field of view can be reached by three or more positioners simultaneously. The acquisition and guide cameras (AGC) occupy the edges, between the hexagonal science field of view and the circular field of view delivered by the WFC. The field of view of the three AGCs ensures that there are always guide stars of sufficient magnitude available for any given observing field.

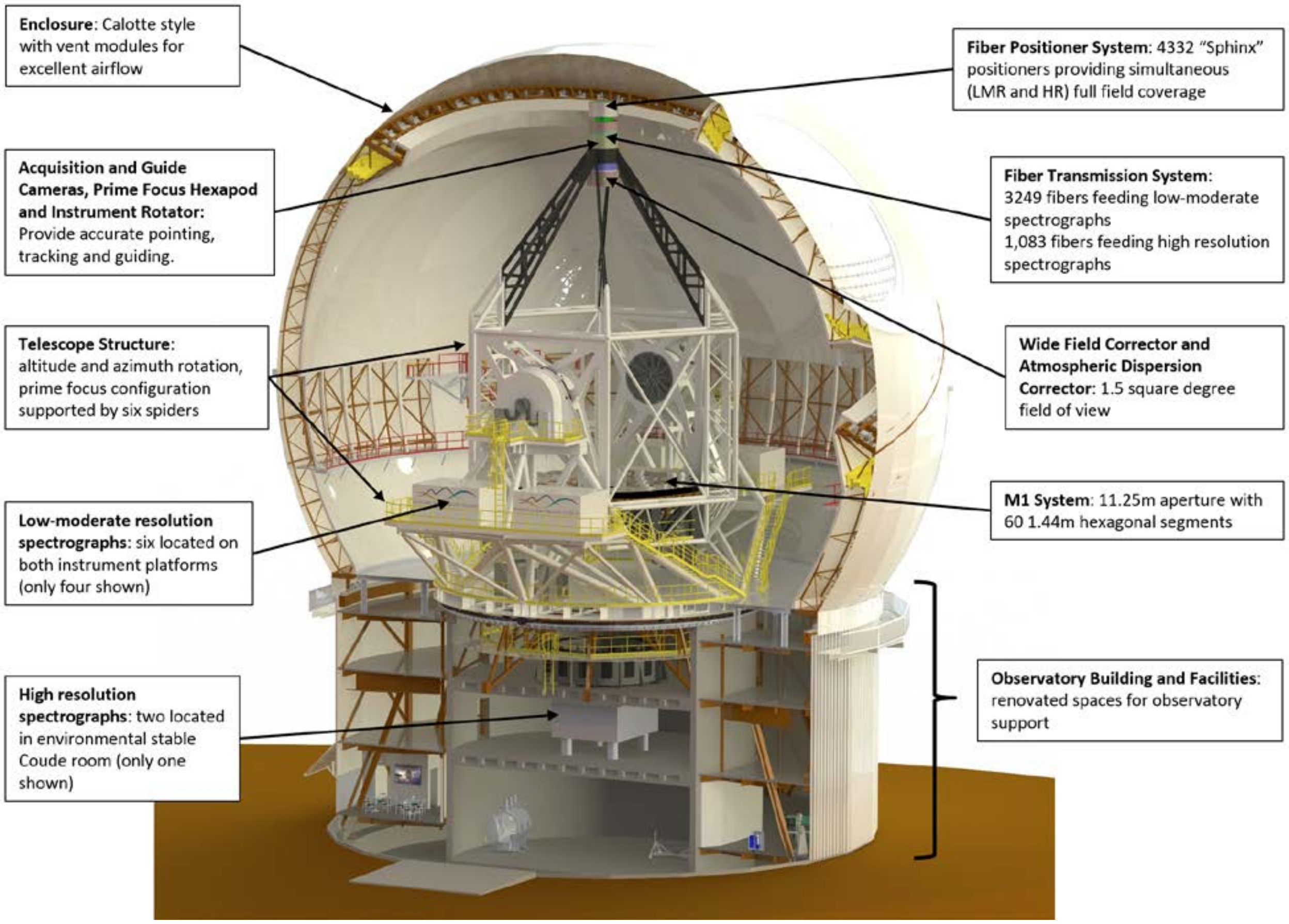

*Figure 9: MSE Observatory architecture.*

Photons bound for the low/moderate resolution spectrographs travel along ~30 m fiber cables to the spectrographs, which are located on two platforms supported by the telescope elevation structure. Photons bound for the high-resolution spectrographs travel along ~50 m fiber cables to the spectrographs located in the very stable environment of the pier lab.

---

[8] In Greek mythology, Sphinx is one of the offspring of Echidna (possibly her child and/or grandchild by her son, Orthrus).

Spectrograph designs respond to the spectral resolution requirements, which define the type, number, and complexity of the spectrographs. MSE's excellent image quality is critical here, since it permits feasible optical designs that do not require image slicing to reach the required spectral resolution, while also minimizing the total detector area required, despite the large number of targets per exposure. Both high resolution and low/moderate resolution spectrographs utilize off-axis f/2 collimators, which complement the high-numerical aperture fibers. These design choices eliminate the need for micro-lenses at either the fiber inputs or outputs and ensure high system throughput. There are six identical, low/moderate spectral resolution spectrographs, each of which takes a sixth of the relevant fibers as an input. Each of these spectrographs is independently switchable between low and moderate resolutions. There are two identical, high spectral resolution spectrographs, each of which takes half of the relevant fibers as an input.

#### 2.4.2. Site Considerations

A unique aspect of the MSE development is its repurposing of a premier astronomical site. The original CFHT building and telescope pier have been repurposed as the support facility for MSE, which includes structural supports for both the enclosure and the telescope, as well as space for labs and shops, and other operational activities. The current building and pier will be retrofitted to support the new telescope, instrumentation suite, and enclosure, while preserving the existing outer building and inner pier and their foundations, to avoid disturbing the soil on site. Engineering studies have shown this to be feasible, taking into account the known mass, structural, and seismic limitations that were established during the conceptual design phase.

The natural seeing at the Maunakea site is widely recognized as excellent. According to statistical data on the image quality at CFHT and neighboring observatories, CFHT's summit ridge site has a stable median atmospheric seeing of 0.37″ FWHM at 500 nm at zenith. The phenomenal image quality provided by the site is an essential enabler of MSE's scientific efficacy and technical development. At any given fiber diameter, better image quality increases the fraction of light from an astronomical target that can enter the science fibers, increasing the overall sensitivity of the system. The size—and therefore cost—of spectrographs is directly proportional to the size of the fibers used. MSE's ability to use small fibers is a direct consequence of its excellent image quality. Significantly poorer image quality would likely either increase the cost of the facility and/or decrease its science efficiency (see discussion in Section 3.5).

#### 2.4.3. Science Operations

In addition to its hardware, the success of MSE as a scientific survey facility hinges on the observatory operations concept. The science operation phases are summarized in Figure 10.

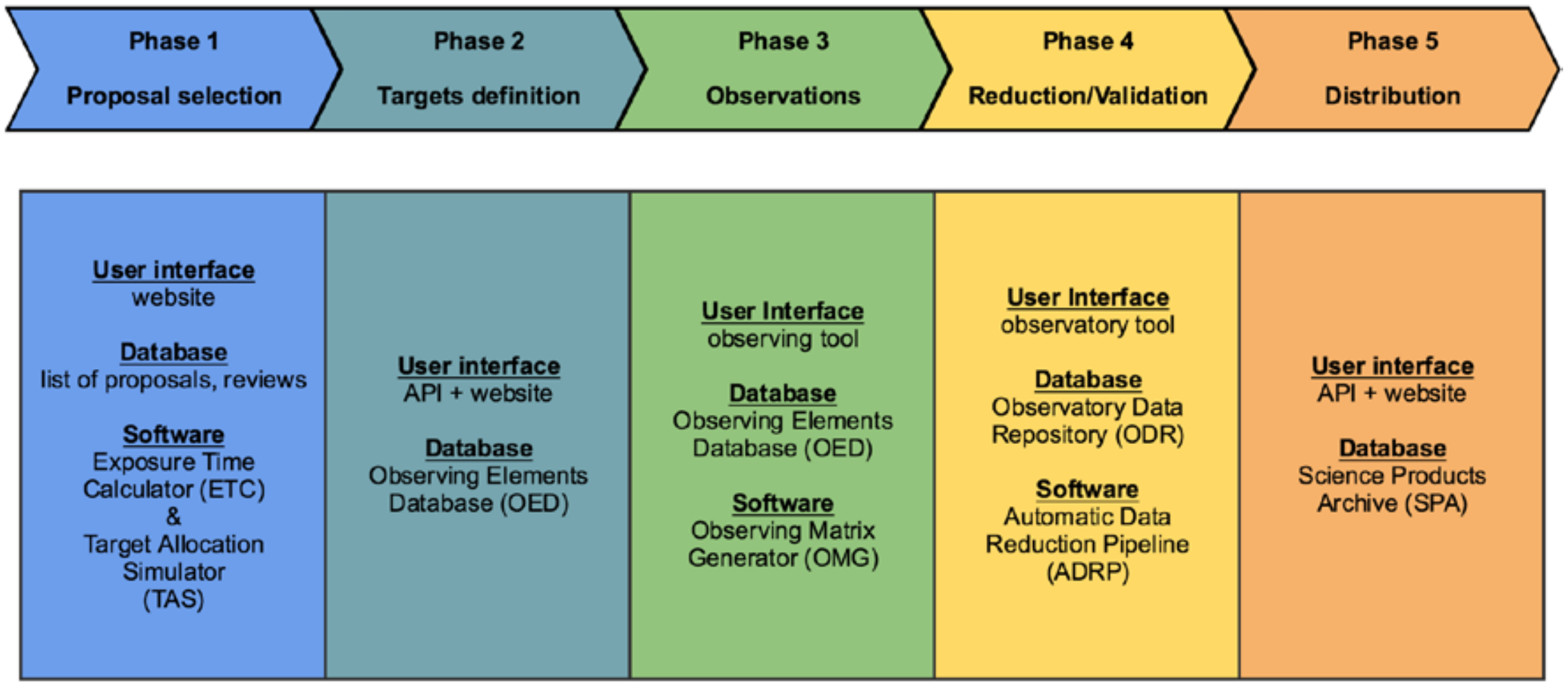

*Figure 10: Phases of the science operations.*

In Phase 1, survey teams propose observing programs at regular intervals. MSE will streamline the selection process and automatically transfer approved programs to Phase 2.

In Phase 2, survey teams use a more sophisticated tool to populate a common Observing Element Database (OED), which consolidates the science targets for all surveys. MSE will provide the capability to automatically upload from thousands to millions of science targets and related information, using an Application Programming Interface (API) to validate targets and allow for additions and modifications during the surveys, which may potentially have multi-year durations.

In Phase 3, observations will be executed based on the content of the OED. The target information from Phase 2 will be used by an Observing Matrix Generator (OMG), to generate optimized schedules for that night, week, year, etc. The OMG is a suite of scheduling software, which performs a complex optimization, in order to dynamically schedule the observations of millions of targets within a range of magnitude, color, coordinate, and validation requirements, while accounting for calibration targets and guide star availability. Observing conditions, including the weather, seeing, and system status, will be monitored. The OMG responds dynamically and updates schedules based on these parameters, in order to minimize overheads and ensure a high level of overall observing efficiency. Within this infrastructure, MSE will be able to handle targets of opportunity, or other transient targets defined on a nightly basis, following observations carried out at other observatories (e.g., LSST). All raw data relevant to science observations will be stored in an Observatory Data Repository (ODR) database.

In Phase 4, data from the ODR is reduced and quality analysis performed. Raw data, including a substantial amount of calibration data, will be processed by an Automatic Data Reduction Pipeline, and can be used by observatory staff for quality assessment and troubleshooting.

Phase 5 is the distribution and archiving of data and data products. There are several levels of data products planned. These range from quick-look to high-level data products, developed in collaboration with the Survey Team and released on annual or semi-annual timescales.

### 2.4.4. Survey Selection

MSE is a dedicated facility, designed to have an operational lifetime of many decades, during which time its user community can apply to conduct spectroscopic surveys with a broad range of science goals. In this respect, it differs significantly from facilities such as SDSS, 4MOST, or DESI, which have been designed to conduct dedicated surveys (or suites of surveys) with a fixed duration (usually five years). Participants/partners in these facilities usually join for the duration of the surveys: in order to obtain potential access to the data, or to have a role in defining the surveys to be conducted.

MSE's anticipated survey model is considerably different from that of the other survey facilities listed above. We anticipate that, during regular operations, MSE will devote approximately 80% of its observing time to large scale "legacy" surveys. These surveys will be multi-year in duration. Only a small number of legacy programs will be scheduled at any one time. The remaining 20% of observing time will be dedicated to shorter, "strategic" surveys. Typically, these will lead to publications on more rapid timescales, and will not require the multi-year allocations of legacy programs.

There will be frequent calls for strategic programs, and less frequent calls (possibly annual or bi-annual) for legacy programs. The procedures for these calls will bear some similarities to the procedures used in ESO Public Surveys.[x] That is, letters of intent will be solicited and large community-based teams will respond. Teams with similar science goals or complementary observational requirements will be asked to merge. All surveys are expected to have compelling, transformational science goals and significant legacy value, and will have broad community support and involvement. After the main proposals have been submitted and ranked by a review panel, the selected survey teams will be responsible for interfacing with the MSE Observatory, to provide the necessary target catalogs for scheduling.

MSE will work closely with the survey team to ensure high quality data and basic data products are rapidly made available to the survey team. These products will be made available to the entire MSE community on short timescales. Standard, homogeneous, derived-data products will be released by the MSE Observatory at regular intervals. Specialized (i.e., survey-specific) derived-data products will generally be the responsibility of the survey team, working in collaboration with the MSE Observatory, where necessary. All raw data, homogeneously derived basic data products, and possibly advanced data products, will eventually be made available to the international community, after a proprietary period set by the MSE partnership.

---

[i] http://www.casca.ca/lrp2010/
[ii] https://www.science.org.au/supporting-science/science-sector-analysis/reports-and-publications/decadal-plan-australian
[iii] http://www.eso.org/public/about-eso/committees/stc/stc-85th/public/STC-551_Science_Priorities_at_ESO_85th_STC_Mtg_Public.pdf
[iv] www.iap.fr/pnc/pncg/Actions_files/eso_moswg_report.pdf
[v] Meyer, M., A. Robotham, D. Obreschkow, et al. "Connecting the Baryons: Multiwavelength Data for SKA HI Surveys." *Advancing Astrophysics with the Square Kilometre Array (AASKA14)* (2015): 131.
[vi] http://mse.cfht.hawaii.edu/docs/
[vii] http://adsabs.harvard.edu/abstract_service.html
[viii] www.iap.fr/pnc/pncg/Actions_files/eso_moswg_report.pdf

[ix] https://www.naoj.org/Science/SACM/Senryaku/senryaku_E.html
[x] https://www.eso.org/sci/observing/PublicSurveys/policies.html

# 3. MSE System Design

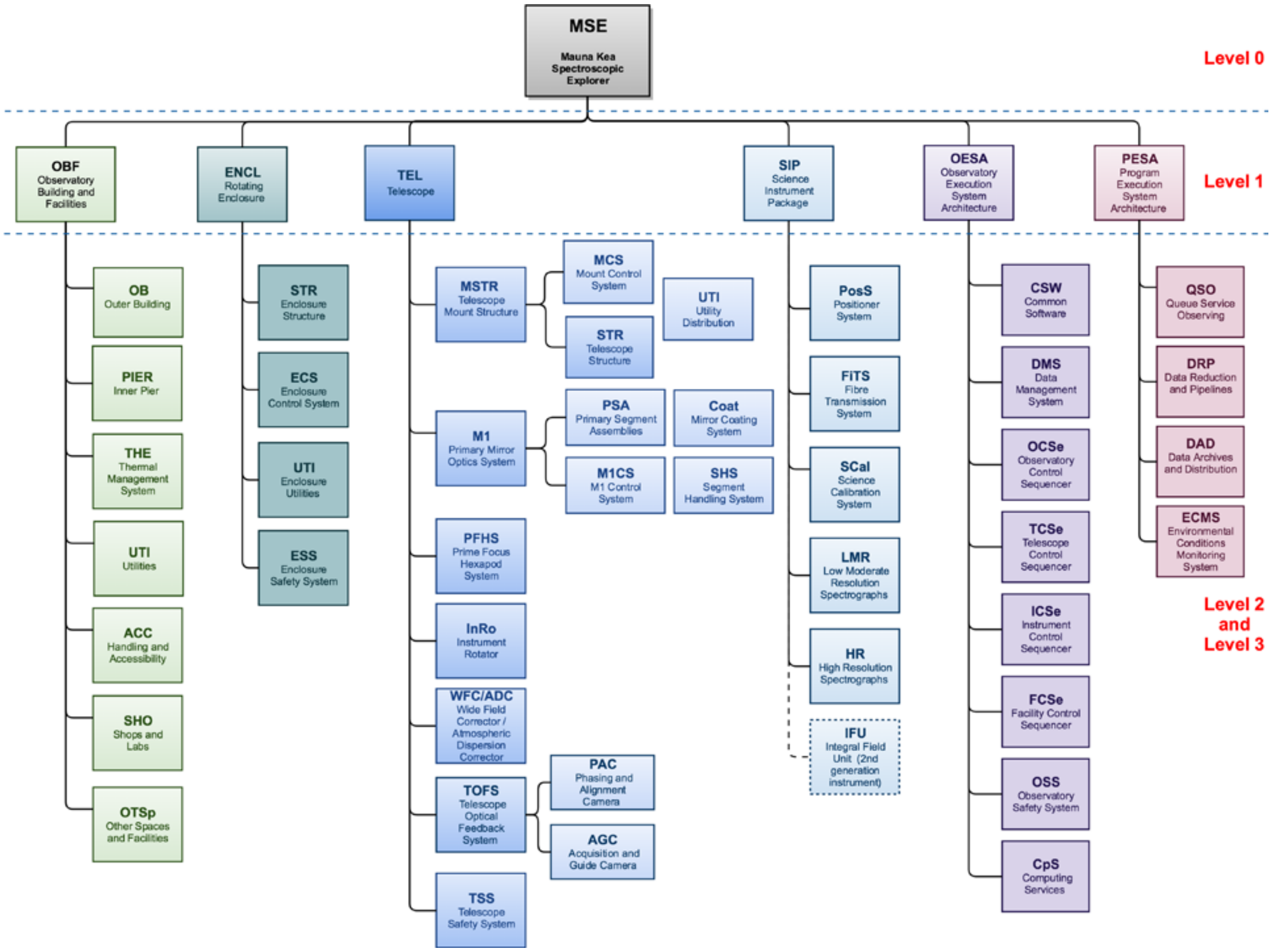

*Figure 11: The Product Breakdown Structure for MSE. This chart shows all the major physical elements of MSE and their relation to each other.*

The science-based requirements and specifications of the field have led MSE to develop an architecture and operations concept that take advantage of one of the best astronomical sites in the world, to provide the high sensitivity expected of a large telescope aperture, massive multiplexing, and dedicated survey operations. The MSE architecture is described in a Product Breakdown Structure (PBS), which identifies MSE's subsystems, and which is the result of a functional analysis of MSE (Figure 11).

This section of the document describes MSE's overarching system architecture and performance, and describes the critical design choices that were made during the development of the system. Specifically, Section 3.1 describes the overall design philosophy adopted by MSE, which guides the decision-making process. Section 3.2 describes the specific design constraints imposed on MSE, in addition to the explicit constraints imposed by the science requirements. Section 3.3 presents an overview of the system architecture adopted, and includes a Product Breakdown Structure and functional analysis. Section 3.4 describes the major trade studies and critical design choices made by MSE. Section 3.5 summarizes the overall system-level performance of MSE, and highlights major performance budgets.

### 3.1. Design Philosophy

MSE incorporates state-of-the-art engineering for large facilities and is inspired by the latest technical advancements, made by other top astronomical facilities around the world. Their general design philosophy involves reusing reliable existing designs wherever possible, building on the knowledge and experience gained from astronomical facilities and instruments currently in use or under development. Where necessary, MSE stretches the limits of current manufacturing and engineering capabilities, in order to achieve the sensitivity and resolution demanded by science. This development philosophy ensures the best possible performance and minimizes development costs and risks, while ensuring an efficient, practical, and cohesive design. The MSE's facility design builds on CFHT's four decades of successful operation atop Maunakea—a strategy reflected in many facility-wide decisions, which adopt the proven achievements of a reliable and efficient facility.

### 3.2. Design Constraints

The overall constraints on MSE originate from the Office of Mauna Kea Management Comprehensive Management Plan[i] (CMP), which has given us permission to redevelop CFHT. MSE is committed to minimizing disturbance to the soil of Maunakea, and maintaining a very similar external appearance to that of the current CFHT facility.

These constraints impose important considerations regarding the overall design of the facility. The requirement to maintain a similar external appearance constrains the enclosure radius and imposes limitations on the space available inside the enclosure for construction, operation, and maintenance. The mass of the entire facility will necessarily be closely tracked to ensure the existing foundations and the soil-bearing capacities are not exceeded. Reusing the existing building also means that it must be retrofitted to meet modern building codes and upgraded, using modern techniques, to support the enclosure and the telescope and to ensure safety and survivability, while operating in the active seismic environment on Maunakea.

In addition to site constraints, many overall architectural decisions are inherent in the science requirements, as MSE would be unlikely to meet these requirements optimally, under a different configuration. This also influenced the decision to install a highly multiplexed, fiber-fed spectrograph system, with a 10-m class segmented mirror and an altitude–azimuth telescope system.

Given these constraints, the MSE conceptual design is the result of several design and trade study efforts.

### 3.3. System Design

MSE collects light from targets in the sky and converts it into science data, which is distributed to scientists. These basic activities of MSE were broken down into the functions that MSE needs to perform, which were then systematically mapped onto MSE's architecture.

MSE's functions are derived from a formal functional analysis, based on envisaged observatory activities (Figure 12), necessary to accomplish the required science observations. The five activity groups in Figure 12 form the first-level functional blocks in the Functional Architecture Diagram (Figure 13), and have the additional on-going function of providing maintenance and support for survey operations. The second-level functions in Figure 13 are informed by the physical products[ii] that MSE needs to provide (Figure 11).

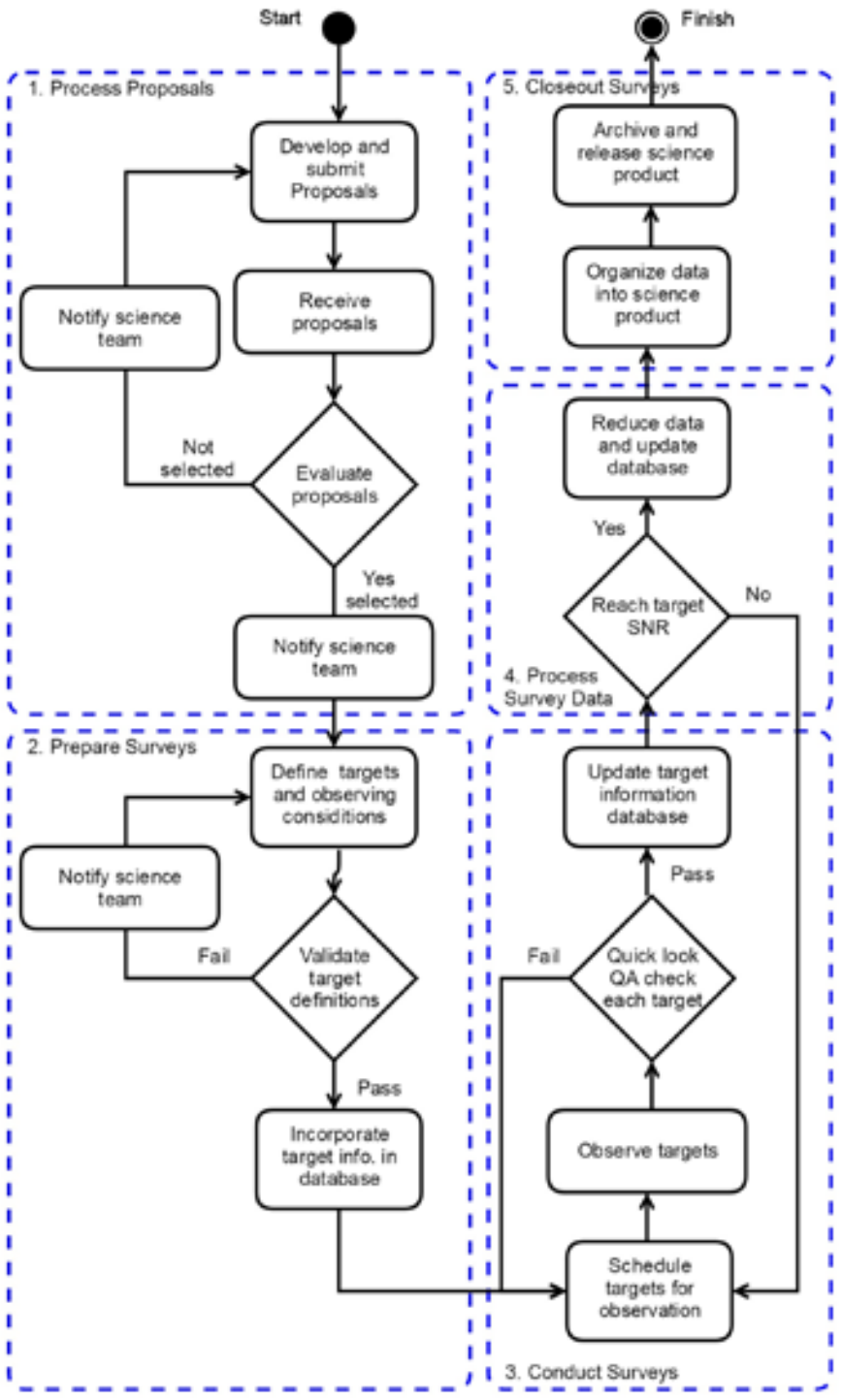

*Figure 12: Observatory activities organized into five activity groups.*

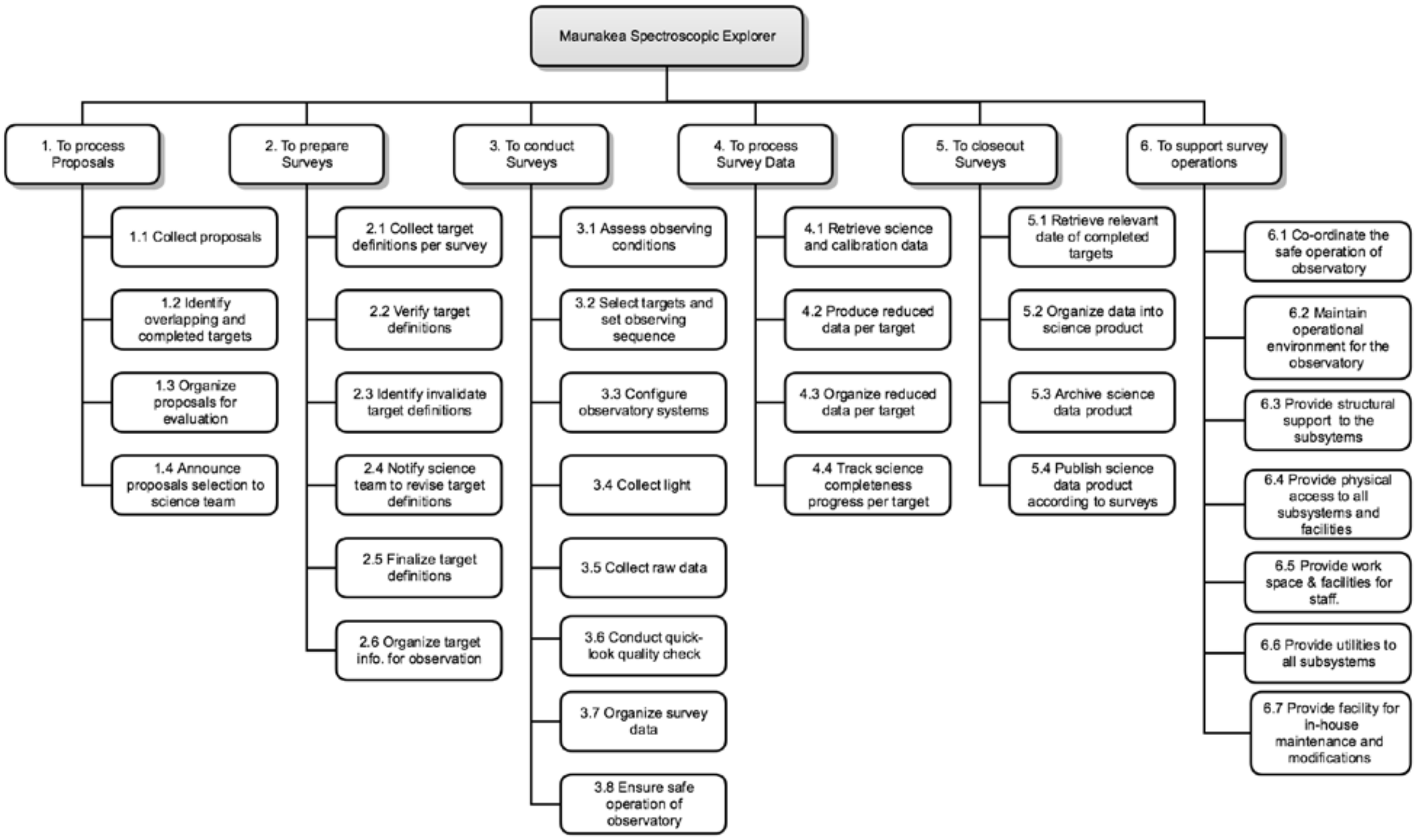

*Figure 13: Functional Architecture Diagram with first- and second-level functional blocks.*

The physical products identified by the functional analysis have been incorporated into the Product Breakdown Structure (PBS). Figure 11 shows the physical organization of the MSE Observatory and groups products by their functions. This was done with the goal of minimizing the number of interfaces and simplifying the verification and test processes during the observatory's future integration phase. This way of organizing the project's products also facilitates workshare among the MSE participants, based on their respective interests, expertise, and resources.

The six Level 1 PBS elements are organized according to MSE's overall functions:

- The Observatory Building and Facilities (OBF) protect the internal components, support the Telescope and Enclosure, and provide thermal management, utilities, safety, access and handling, and other support capabilities for the observatory.
- The Enclosure (ENCL) similarly protects the dome environment from the elements, helps maintain the thermal environment—minimizing air conditioning heat load passively during the day and actively, using ventilation modules, at night—and provides handling and lifting capabilities, along with the important additional function of an open aperture, to allow light to enter during observations.
- The Telescope (TEL) is a complex set of subsystems at the heart of MSE, whose primary function is to collect light and transmit it to the prime focus, to achieve the prescribed image quality. This involves the participation of all the telescope's subsystems, working together to ensure night time observations are enabled.
- The Science Instrument Package (SIP) collects and transmits light from the prime focus to the science detectors, where photons are converted into raw data.
- The Observatory Execution System Architecture (OESA) configures all the subsystems involved in survey operations. It coordinates, controls, and monitors observations, logs system and environmental data, and enforces safety in the observatory.
- The Program Execution System Architecture (PESA) executes all the surveys, as dictated by the science teams, by processing proposals, preparing survey data, executing observations (in concert with OESA), processing science data, and closing out surveys by archiving and distributing science products. PESA also monitors the external environment of the observatory on an ongoing basis.

The developed performance and functional requirements that govern the design, manufacture, assembly, and integration of MSE as a whole have been formalized in a set of documents: the Observatory Architecture Document (OAD), the Operations Concept Document (OCD), and the Observatory Requirements Document (ORD). These three documents represent the Level 1 documents of MSE. Together, they describe the design of the system in response to the Science Requirements Document (SRD, the sole Level 0 document).

## 3.4. Major Trade Studies and Design Choices

This section outlines the major trade studies that were conducted during the Conceptual Design Phase, which resulted in significant design choices for MSE.

### 3.4.1. Enclosure Design Selection

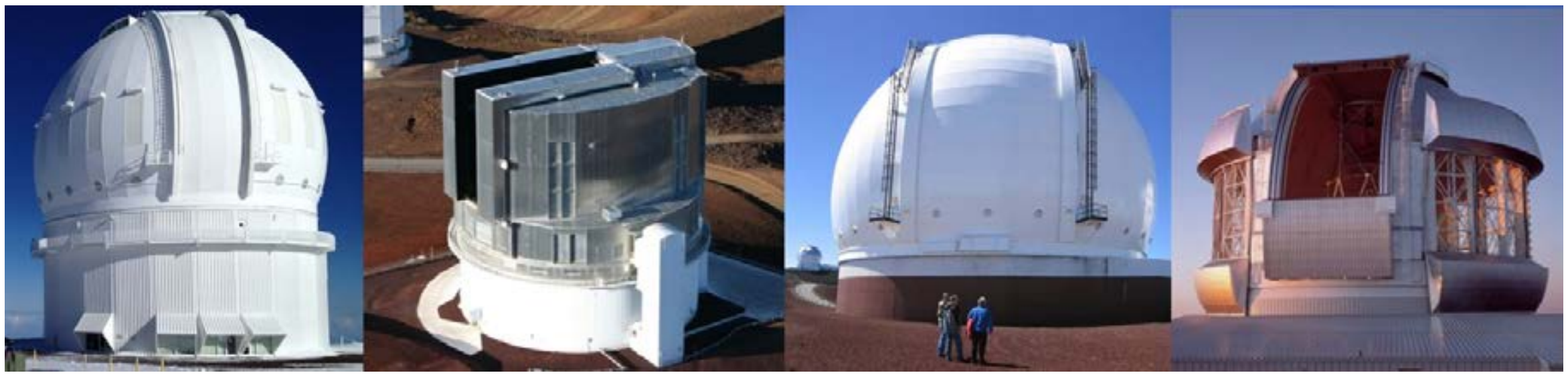

*Figure 14: Enclosure styles. From left to right: CFHT, Subaru, Keck, Gemini.*

There are multiple enclosure styles available for 10-m class telescopes. A selection of these are shown in Figure 14, alongside the current CFHT enclosure. After reviewing the extensive enclosure trade study of the Thirty Meter Telescope (TMT) project, a calotte-style enclosure (similar in design concept to the one selected for TMT) was selected and developed for MSE (Figure 15). The calotte's spherical shape and circular aperture opening are more structurally efficient than conventional enclosure configurations for the same size of telescope, thus allowing for lower mass and smaller dome size. The relative compactness of the calotte enclosure compared to other styles will result in lower construction and operation costs.

Other advantages of the calotte design include intrinsic wind protection, due to its circular aperture opening, and minimal drive and power demands, thanks to the counterbalanced rotating components. The MSE enclosure will incorporate CFHT-style vent doors: these have been demonstrated to be good for ventilation and to minimize thermally induced seeing effects.

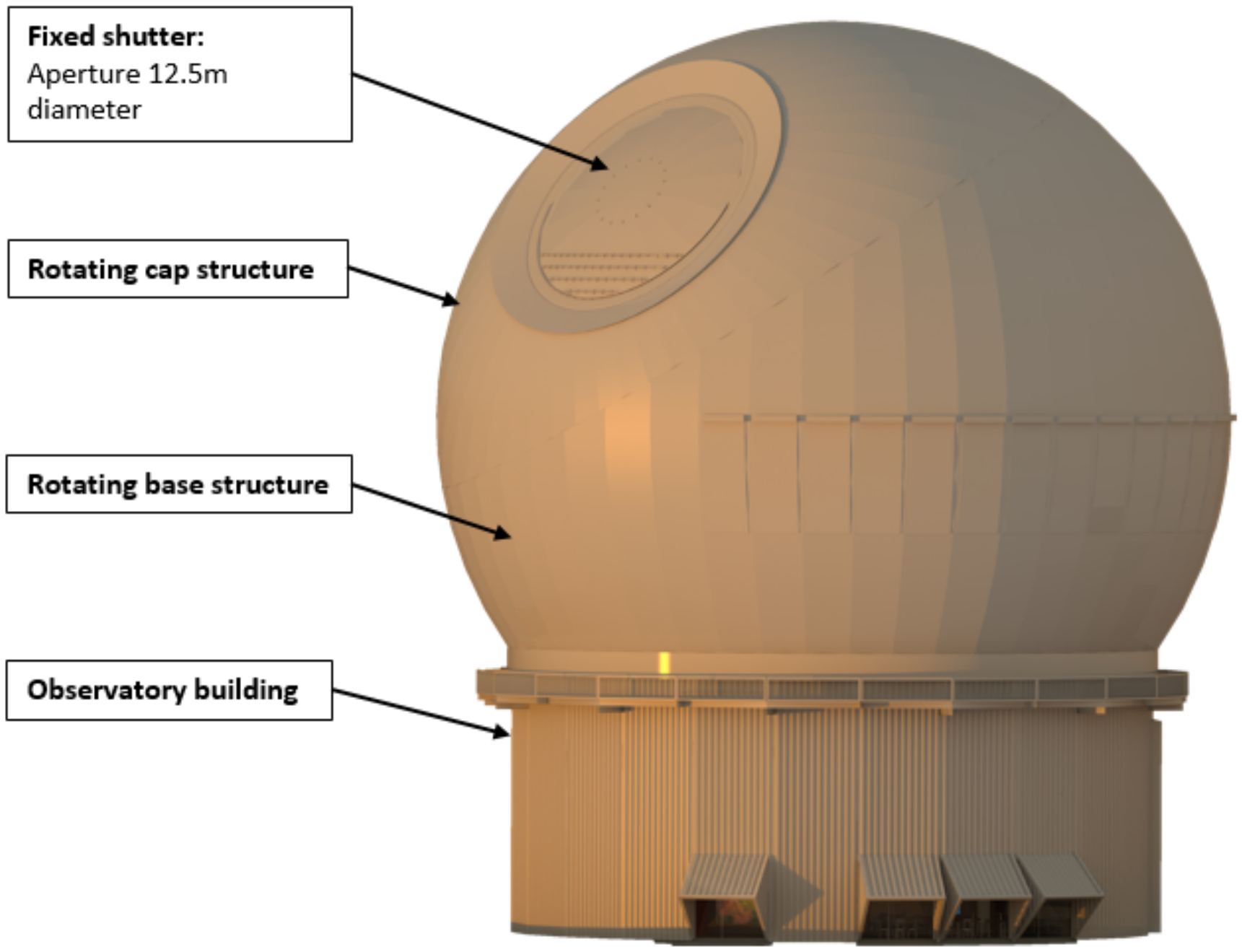

*Figure 15: MSE calotte enclosure concept.*

The adopted style of enclosure includes independently rotating base and cap structures. The biggest difference between the design of MSE's enclosure and that of TMT—other than the fact that MSE's enclosure is significantly smaller—is that MSE's rotating base supports a fixed shutter plug. This eliminates the complexity and the cost of a separate shutter track and drive system. When in operation, the cap rotates over the shutter to close the enclosure aperture. The enclosure incorporates ventilation modules, cranes, and platforms, to support daytime and night time observatory operations.

The aforementioned features mean that the calotte design is the best candidate in meeting MSE's fundamental requirements, while closely matching the existing CFHT enclosure size, except with a larger aperture opening, to accommodate MSE's larger primary mirror.

### 3.4.2. Telescope Optical Configuration Trade Study

A large aperture is critical, to provide a facility with the capacity to follow up on faint sources detected by LSST and other imaging surveys, and to ensure that there is sufficient signal-to-noise ratio in the data collected by MSE to generate clear target selections for follow-up studies with the upcoming VLOTs. The MSE Project Office conducted studies[iii] to assess the optimal optical arrangement, with the goal of finding the best telescope configuration and mirror size, to meet the required sensitivity and operational needs.

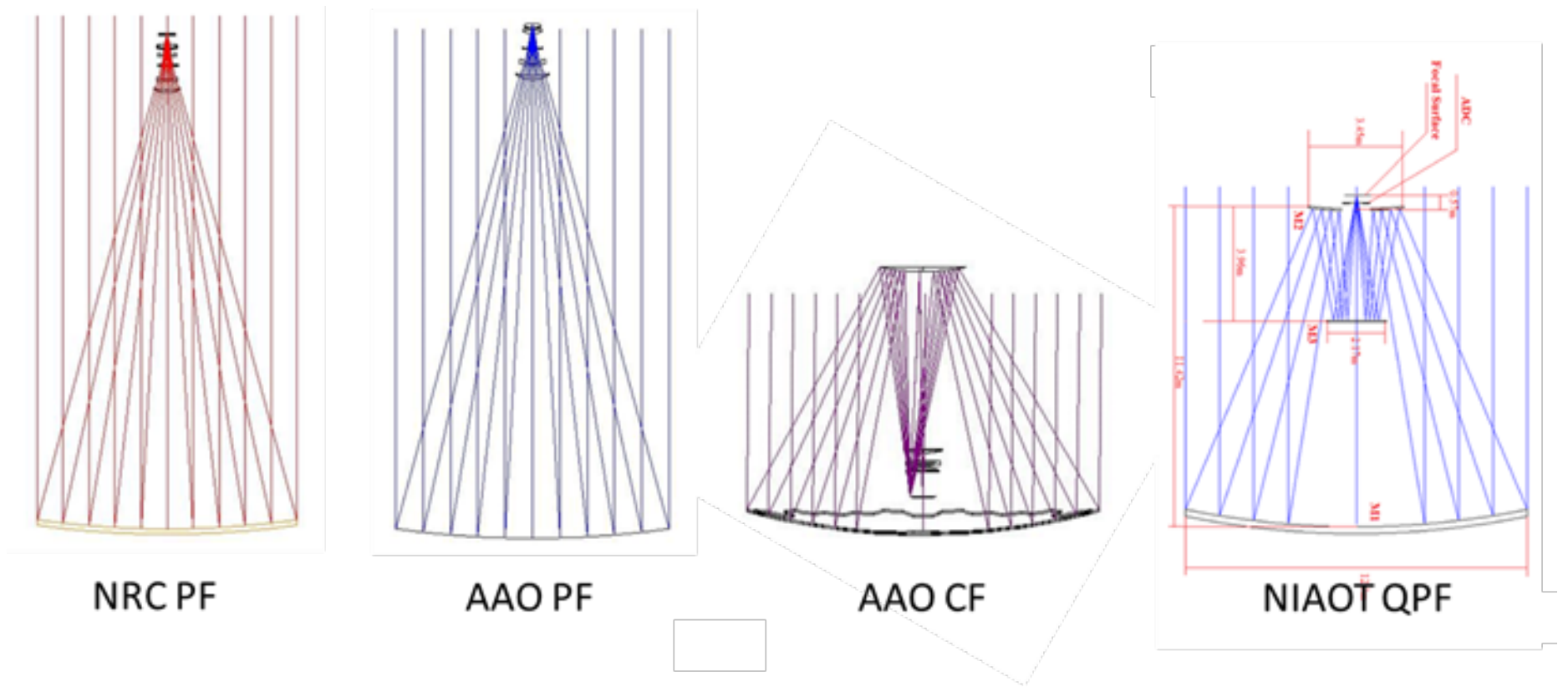

*Figure 16: Primary mirror and optical configuration trade study.*

Four optical configurations (Figure 16) representative of the design space—including a Wide Field Corrector (WFC) and Atmospheric Dispersion Corrector (ADC)—were developed. The National Research Council of Canada (NRC) and the Australian Astronomical Observatory (AAO) provided prime focus designs with a sixty-segment primary mirror and small central obscuration; while AAO and the Nanjing Institute of Astronomical Optics and Technology (NIAOT) provided non-prime focus designs, incorporating a larger primary mirror and significantly higher central obscuration. The overall mirror diameters were 11.25 m for the prime focus and 12.3 m for the non-prime focus options (Figure 17).

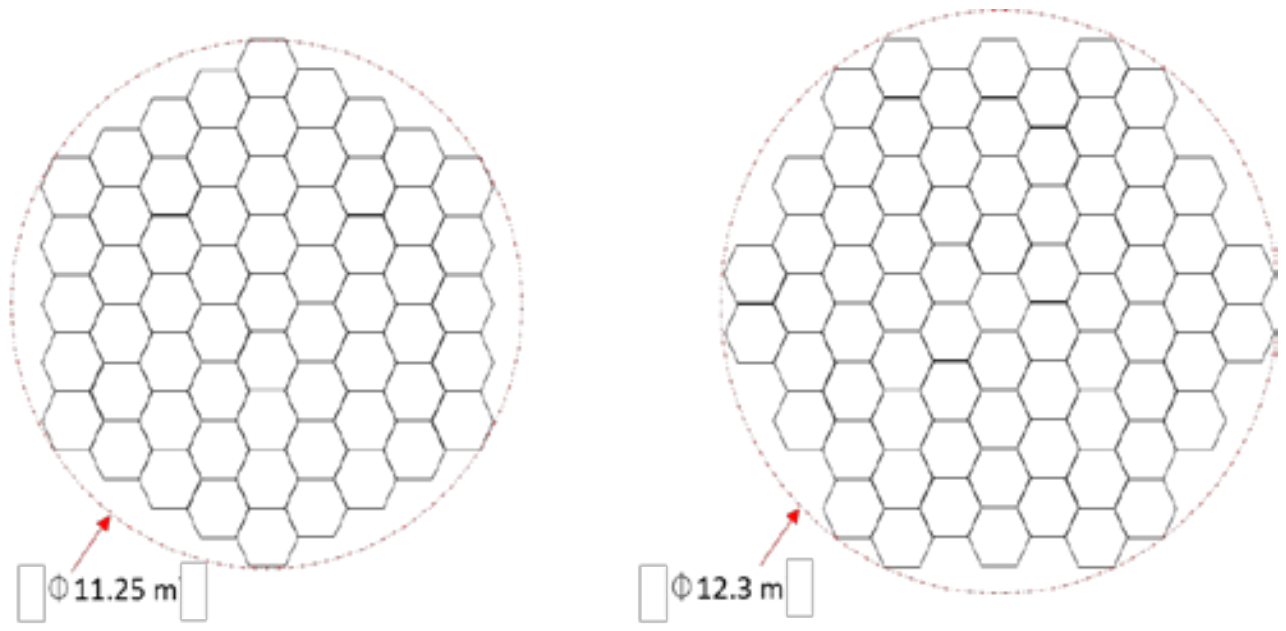

*Figure 17: Prime focus and non-prime focus primary mirror options.*

A total of forty-four items in eight categories, affecting the design of interfacing systems, operations, and programmatic cost and risks, were quantitatively evaluated, from a system perspective in a trade matrix. CAD models of the observatory (Figure 18) were created in order to assess impacts on operations, given the relative sizes of the telescopes with respect to the calotte enclosure.

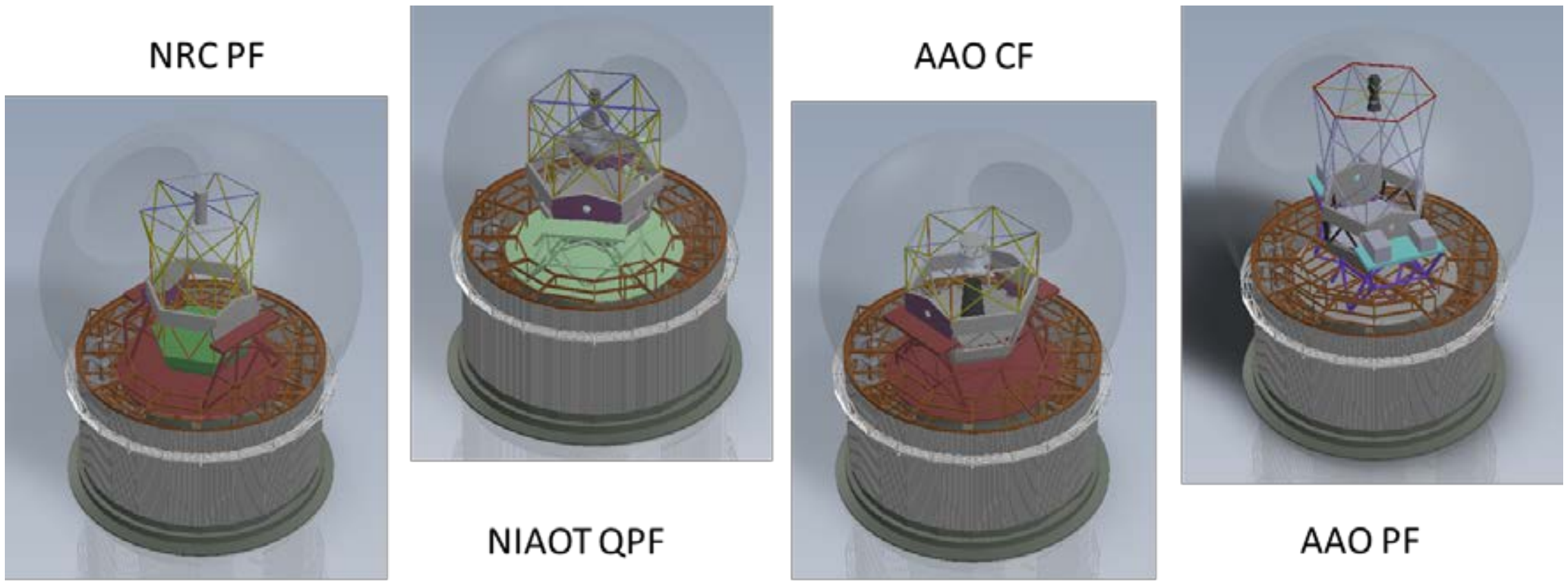

| | | | |
|---|---|---|---|
| $D_{NRC\ PF}$ = 10.0 m | $D_{NIAOT\ QPF}$ = 10.3 m | $D_{AAO\ CF}$ = 10.4 m | $D_{AAO\ PF}$ = 10.1 m |

*Figure 18: CAD models and effective diameters, D, of the optical configurations in the trade study.*

One important finding is that non-prime focus designs do not result in a significant increase in effective collecting area and sensitivity. Once we take the central obscurations into account, the effective diameters are very similar in all the various configurations. The non-prime focus telescopes, by contrast, are more compact, but present more challenges in terms of operation, structural designs, and optics. For example, the non-prime focus versions would require faster f/# for mirror segments, additional large secondary (and, in one case, tertiary) mirrors and bigger spectrographs. A fast primary mirror would require tighter optical tolerances and challenging asphericity.

Following an analysis of the four optical designs and their impacts on the overall observatory system, an 11.25 m prime-focus optical design, with a 10 m effective M1 diameter, was adopted. This represents a “sweet spot” in the design space, in terms of optical fabrication feasibility, performance, and cost.

### 3.4.3. Development of Multiplexing Strategies

The science requirements for the multiplexing levels at the three designated spectral resolutions are:

- ≥3,200 spectra at ~R3000 (low) with full field coverage,
- >3,200 spectra at ~R6000 (moderate) with full field coverage, and
- ≥1,000 spectra at ~R40K (high) with full field coverage

The systems that directly affect the multiplexing capabilities are the Positioner System (PosS), the Fiber Transmission System (FiTS), and the spectrographs. Due to the multiplexing requirements, the MSE system architecture divides the spectrographs into two groups: low and moderate resolution (LMR) and high resolution (HR), and has opted for a ratio of 3:1 LMR to HR fibers “feeding” the two groups.

Given the adopted architecture, the corresponding multiplexing options considered were:

1. Positioner option

a. The FiTS fiber inputs "positioned" by one of the two alternate positioner technologies: tilting spines or phi-theta.

2. FiTS fiber bundles option

   a. Two sets of dedicated fiber bundles—one for each spectrograph group—with fixed slit inputs.

   b. Alternatively, a single set of fiber bundles, shared among the LMR and HR spectrographs, with reconfigurable slit inputs between the two groups.

3. Full field coverage at all resolutions, based on the 3:1 fiber ratio

   a. Three out of four tilting spines positioners carry LMR fibers and one spines positioner in four carries separate HR fibers.

   b. Every phi-theta positioner carries both an LMR and an HR fiber and relies on fiber switching technology to select one HR fiber input out of every four positioners, to feed the HR spectrographs.

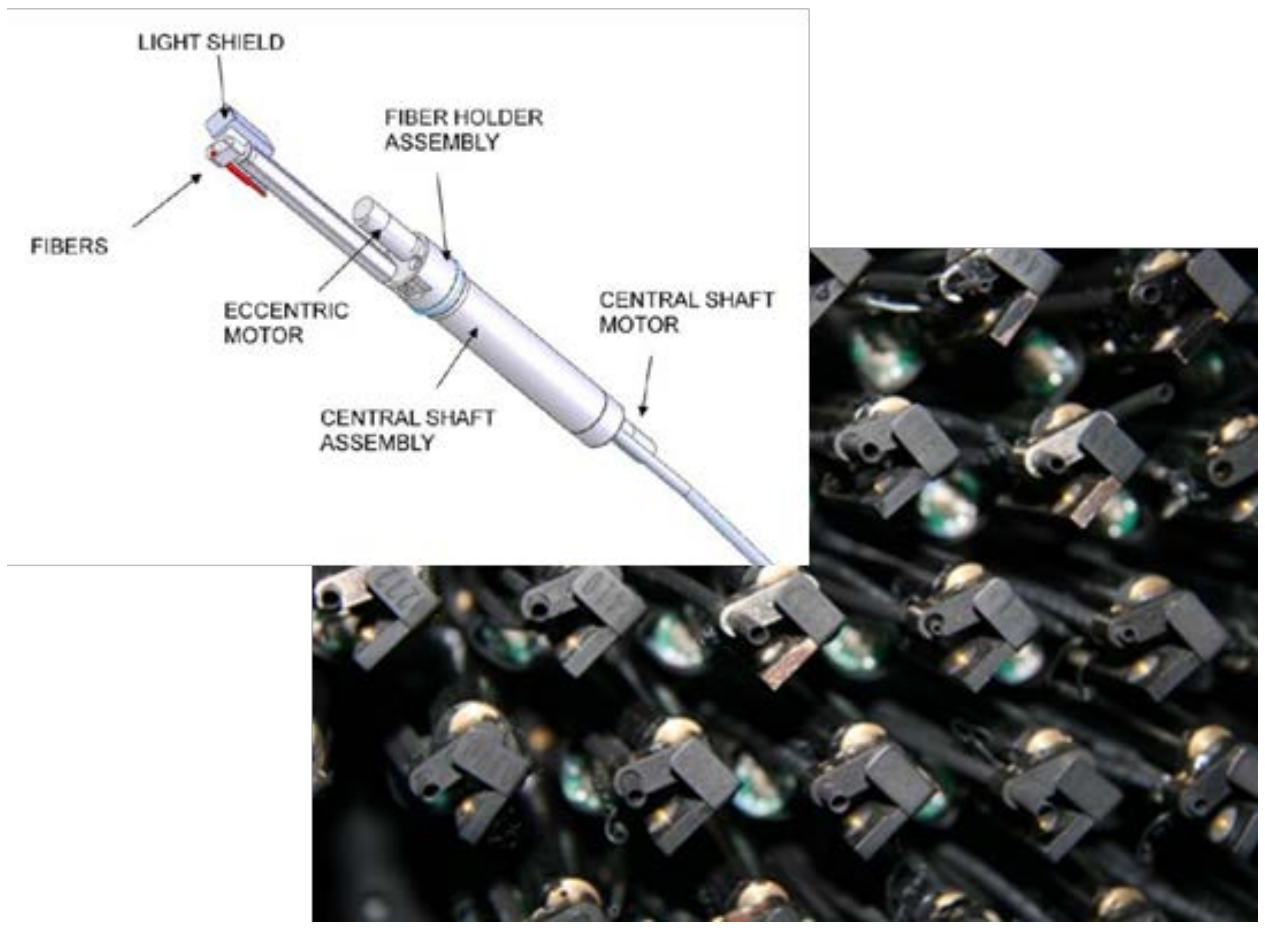

*Figure 19: Phi-theta positioner system—individual unit and positioner array.*

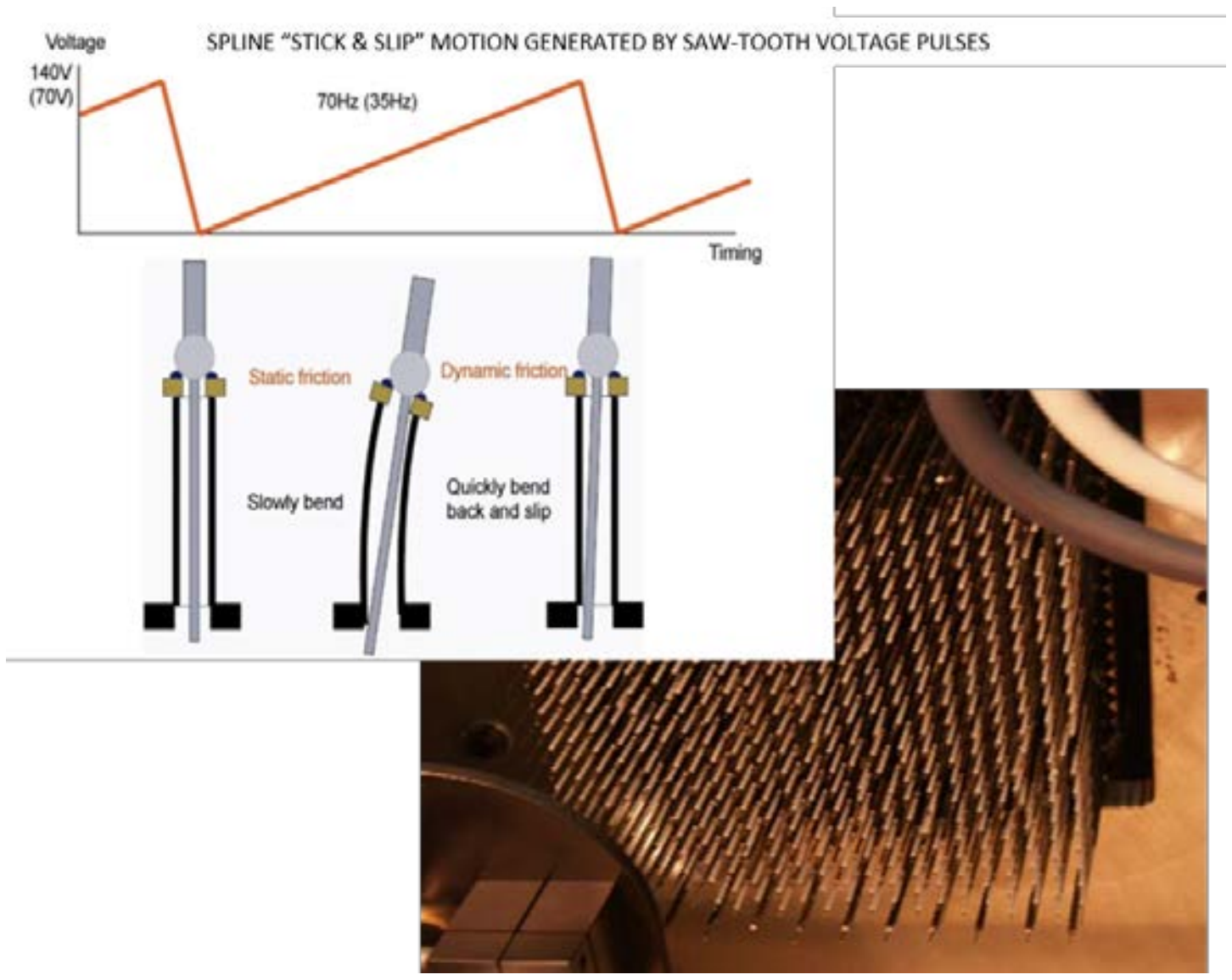

*Figure 20: Tilting spine positioner system—operating principle and positioner array.*

Combinations of the aforementioned options were included in a trade study, to determine the viable system multiplexing configurations for the conceptual designs of the PosS, FiTS, and spectrographs. As in the telescope configuration trade study, a trade matrix was constructed with the same level of quantitative detail, divided into ten categories and twenty-five items. It was based on anticipated performance in areas such as throughput, positioner accuracy, and configuration time; operational considerations, such as reliability, interface requirements, and versatility; and programmatic appraisals, such as technical readiness level, schedule, and costs.

Guided by the multiplexing configuration trade study findings, the following specifications were carried forward, to be confirmed in the conceptual designs:

- Phi-theta positioner system (Figure 19): 3,468 in total, each containing two fibers
- Tilting spines positioner system (Figure 20): 4,624 in total, ¾ of which contain LMR fibers and ¼ HR fibers
- LMR spectrograph with a modular design, for a total capacity of 3,468 spectra
- HR spectrograph with a modular design, for a total capacity of 1,156 spectra
- Fiber transmission system with two sets of dedicated fiber bundles and fixed slit inputs, with 3,468 LMR and 1,156 HR fibers
- Optical switch feasibility study with the ability to switch from 3,468 fiber inputs to 1,156 fiber outputs.

After the completion of the conceptual designs, the final numbers were 1,083 fibers and positioners for the HR and 3,249 for the LMR, respectively.

### 3.4.4. Fiber Positioning System Down-select

Three conceptual positioner designs were developed in parallel by three experienced design teams: the Sphinx tilting spines system, developed by the Australian Astronomical Observatory (AAO); and two phi-theta systems, developed by the University of Science and Technology of China (USTC) and the Universidad Autónoma de Madrid (UAM), Spain. An external committee and the MSE Project Office (PO) independently reviewed the three designs, during the down-select process.

The Sphinx system was selected, based on the recommendations of the external review committee and on the trade matrix, which ranked the designs according to system attributes such as performance, interface requirements, reliability, and operational considerations, dividing these considerations into more than forty areas, organized under nine categories. The PO also analyzed and compared each design’s survey efficiency: i.e., its injection and (target) allocation efficiencies, and found the Sphinx system to be at least 13% more efficient than its rivals, due to its superior allocation efficiency and negligible amount of injection efficiency losses from defocus caused by the spine tilts.

The tilt of the spines also causes the point-spread function (PSF) to vary as a function of tilt, which is not a desirable effect, but has been judged manageable (i.e., it can be compensated by calibration). Moreover, other factors contributed to the PO’s preference of the Sphinx system:

- Flexibility and multiplexing are enabled by conducting simultaneous observations using HR and LMR spectrographs at full field coverage: this increases the system’s survey efficiency advantage to at least 49% higher than those of the phi-theta systems.
- There were concerns that phi-theta systems require optical switches to enable HR observation with full field coverage: the feasibility study identified a potential commercial candidate switch, used in the telecoms industry, but it would take additional research and development to assess its suitability for an astronomical application.
- Another benefit of the Sphinx tilting spine is that it produces less motion-induced fiber stress, which minimizes focal ratio degradation (FRD) and throughput variations that affect wavelength-based solutions.

### 3.4.5. Fiber Transmission System Trade Study

The projected size of the MSE fibers on the sky impacts several aspects of the MSE design and science performance—in all these aspects, the design benefits from the exceptional image quality provided by the MSE site.

Sensitivity is critical to MSE’s science performance. For astrophysical objects that are faint relative to the sky, there is an optimal fiber diameter, which maximizes the signal-to-noise obtained from a source in a given time. Fibers that are too small do not receive enough light from the target, whereas fibers that are too large do not receive enough additional source light relative to the increase in sky flux.

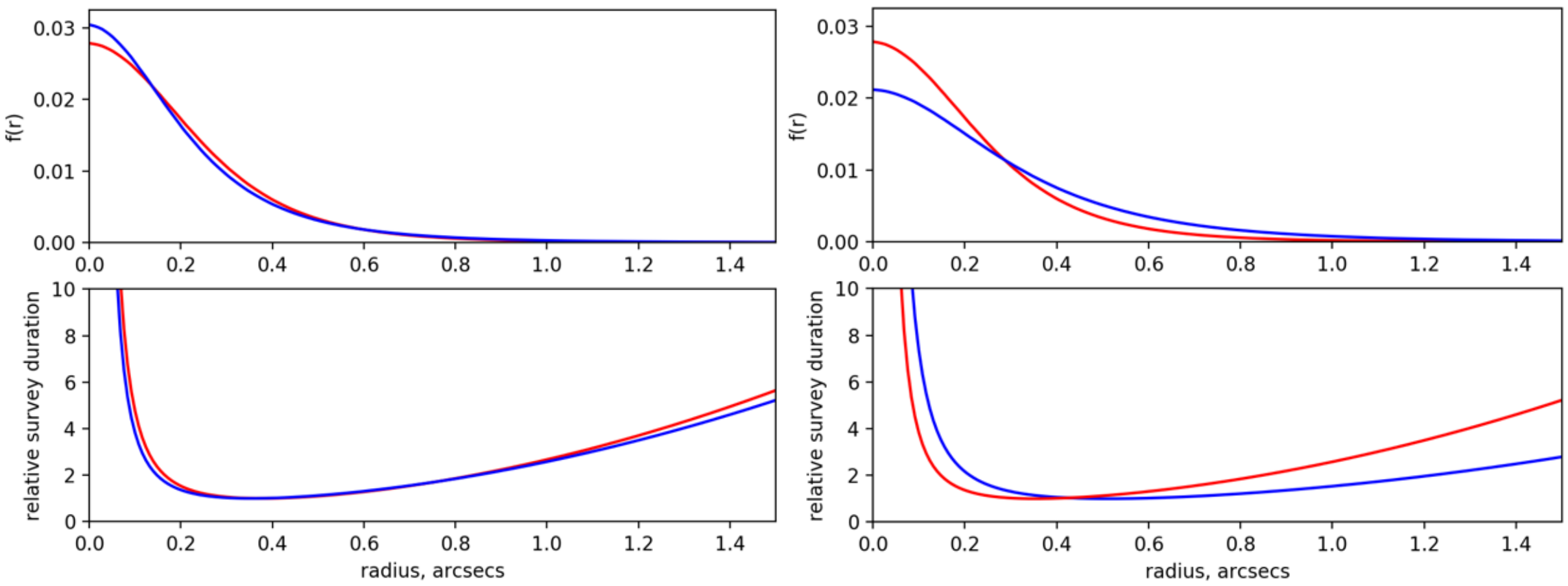

*Figure 21: Results of the fiber diameter sensitivity analysis for high resolution (left) and low resolution (right). Top panels—The red curve shows the Moffat function corresponding to the median delivered IQ for MSE. Blue curves represent the time-averaged source profiles. For high resolution, only point sources have been considered. For low resolution, average galaxy profiles have been considered, in addition to the PSF variations. Bottom panels—The relative time taken to reach a given SNR as a function of fiber radius, for the flux distributions in the top panel and assuming sky-limited targets.*

MSE has conducted a trade study analysis of the optimal fiber size for both the HR setting (whose targets are expected to be mostly stars, i.e., point sources) and the LMR setting (whose targets may be partially resolved). This analysis (Figure 21) takes into account the full range of image quality (IQ) distribution, given the atmospheric seeing distribution observed at the current CFHT site, as well all other factors that contribute to the image quality at the focal plane (Section 3.5.2, budget 4). It also considers all issues relating to the injection efficiency of the fibers (Section 3.5.2, budget 3). In the low-resolution analysis, the profile of the sources was estimated from analysis of galaxy profiles,[iv] in particular those in the AEGIS survey,[v] combined with the image quality distribution. Full details of this analysis will be provided in a forthcoming publication.

The left-hand panels in Figure 21 provide a summary of the results of the HR analysis; the right-hand panel shows the results of the LMR analysis. The top panels show the input source flux distribution. The red curves show the 1D Moffat profile corresponding to the median IQ, for reference. For HR, the blue curve shows the time-averaged PSF, after taking the full IQ distribution into account. For LMR, the blue curve shows the time-averaged source flux distribution, after taking the extended nature of the sources and the IQ distribution into account. The lower panels display the corresponding curves for the relative time it takes to observe sources up to a certain SNR threshold, using fibers of a specific radius, as indicated on the x-axis.

Based on these analyses, it is possible to determine the optimal fiber size to maximize the sensitivity of MSE based on the expected image quality of the telescope and site. These are ~0.8 arcsecond diameter fibers for HR (point sources) and ~1.0 arcsecond diameter fibers for LMR.

In addition to sensitivity considerations, the spectrograph designs are more attainable if the fibers are smaller. For example, for an unsliced spectrograph design:

$$R \propto \frac{\sin\alpha + \sin\beta}{\cos\alpha} \frac{D_c}{D_T} \frac{1}{\varphi}$$

Here, $R$ is the spectral resolution; $\alpha$ and $\beta$ are the angles of incidence and diffraction; $D_c$ is the size of the collimated beam in the spectrograph; $D_T$ is the size of the entrance pupil (equivalent to the size of the primary mirror); and $\varphi$ is the diameter of the fiber. In the case of MSE, $D_T$ is very large. However, a large $D_T$ value cannot be fully compensated for by using a large value for $D_c$, since, in practice, the size of the collimated beam is limited by the availability of large optics. Similarly, it is not desirable to compensate for the large $D_T$ value by using extremely large angles on the grating, since this will increase the physical size of the grating required and decrease its efficiency. In this context, it is clearly essential that the diameter of the fiber is restricted to as small a value as is practical.

Current designs for the two spectrograph systems use 0.8″ diameter fibers for HR and 1.0″ diameter fibers for LMR. The possibility of using even smaller fibers can be reanalyzed, if desired, using the clear cost-science-technical framework that was developed through the fiber diameter sensitivity analysis.

### 3.4.6. Multi-object Integral Field Unit Upgrade Path

A multi-object Integral Field Unit (IFU) system has been identified as a possible second-generation instrument upgrade. The current IFU operation assumptions are:

- The IFU system will feed the same set of LMR spectrographs as now;
- The IFU and (positioner-based) multi-object spectroscopic (MOS) observations will not be concurrent.

The conceptual design of the prime focus components is modular, such that the existing PosS system can be replaced by an IFU system, and vice versa. Similarly, the spectrograph slit inputs are designed to be switchable between IFU and MOS fiber bundle cables, while maintaining optical alignment.

Once the specific science cases for the IFU have been further defined, engineering work will determine the optimal design configuration (e.g., number of spaxels per IFU versus number of IFUs). Additional dedicated FiTS cables may be provided for the IFU and designed and built as an integral part of the upgrade. Switching from IFU to MOS would require switching the corresponding top-end components, along with their FiTS cables. To maintain MSE’s high stability and repeatability, this switch could only happen infrequently. One alternative would be to produce a set of shared FiTS cables, with a system of connectors compatible with both MOS and IFU at the top end. This alternative would allow the telescope structure to accommodate the mounting and routing of only one set of fiber cables between prime focus and the LMR spectrographs, which would minimize handling and ensure system stability at the cost of lower throughput.

The science and design teams will further develop the IFU upgrade path as the project moves into its Preliminary Design Phase. In the meantime, the conceptual design of MSE has been developed to allow for this upgrade.

## 3.5. System Performance and Performance Budgets

The scientific success of MSE depends on its ability to obtain vast numbers of high quality, science-ready spectra from the plethora of faint astrophysical objects identified by current and future photometric and astrometric surveys, such as Gaia, LSST, SKA, Euclid, and WFIRST, at a range of wavelengths. Survey speed—which depends on both system sensitivity and observing efficiency, among other factors—is critical, as is the ability to scientifically calibrate the raw data.

There are Level 0 science requirements that define the sensitivity, observing efficiency, and calibration standards necessary for MSE science. Each of these performance requirements impacts the system as a whole, and multiple subsystems must interact in complex ways to ensure that these requirements are met.

To achieve these system performance requirements, the Level 0 requirements were translated into requirements for all of the relevant subsystems and operational procedures. These system budgets describe how the high-level requirements flow down to each of the subsystems of the MSE design. Estimates of the actual performance of these subsystems can determine whether the system-level specification has been met, and modifications can be made across the system, to reconcile discrepancies, if required.

Here, we describe the principal science performance requirements of MSE and highlight important system-level considerations as described in the relevant system budgets.

### 3.5.1. MSE Sensitivity

There are three high-level science requirements relating to sensitivity—one each for the low, moderate and high spectra resolution modes of MSE. They describe the (monochromatic) magnitude limit that MSE must observe at a certain signal-to-noise ratio per resolution element under certain sky conditions (typical dark time for low/moderate resolution; typical bright time for high resolution) over the course of a one-hour observation, at an air mass of 1.2 and in median seeing. Of course, MSE will undertake observations of many different sources with different luminosities and exposure times, and under different observing conditions. As such, the observation described in the science requirements is a “reference” only, whose sole purpose is to define the sensitivity of the MSE.

Figure 22 shows the required limiting sensitivity of MSE at the three different spectral resolution settings (red lines). The left panels show the SNR obtained for the reference targets, and the right panels show the limiting magnitude at the reference SNR. Specifically:

- At low resolution, MSE is required to reach magnitude 24 targets at all wavelengths longer than 400 nm (monochromatic AB magnitudes) at an SNR per resolution element of two per hour. At wavelengths shorter than 400 nm, MSE is required to obtain an SNR per resolution element of one on these targets.
- At moderate resolution, MSE is required to reach magnitude 23.5 targets at all wavelengths longer than 400 nm (monochromatic AB magnitudes) at a SNR per resolution element of two per hour. At wavelengths shorter than 400 nm, MSE is required to obtain an SNR per resolution element of one on these targets.

- At high resolution, MSE is required to reach magnitude 20 targets at all wavelengths longer than 400 nm (monochromatic AB magnitudes) at an SNR per resolution element of 10 per hour. At wavelengths shorter than 400 nm, MSE is required to obtain an SNR per resolution element of five on these targets.

The blue lines in Figure 22 show the estimated performance for each setting, at the end of the Conceptual Design Phase, and anticipate the changes to the relevant subsystem designs discussed below. Currently, MSE meets or exceeds requirements everywhere, except in the H-band and at the bluest wavelength in the HR mode, and possible design modifications in these regimes are being investigated. However, it is clear that MSE will provide an exceptionally sensitive facility, suitable for the efficient observation of faint targets across the optical–NIR waveband.

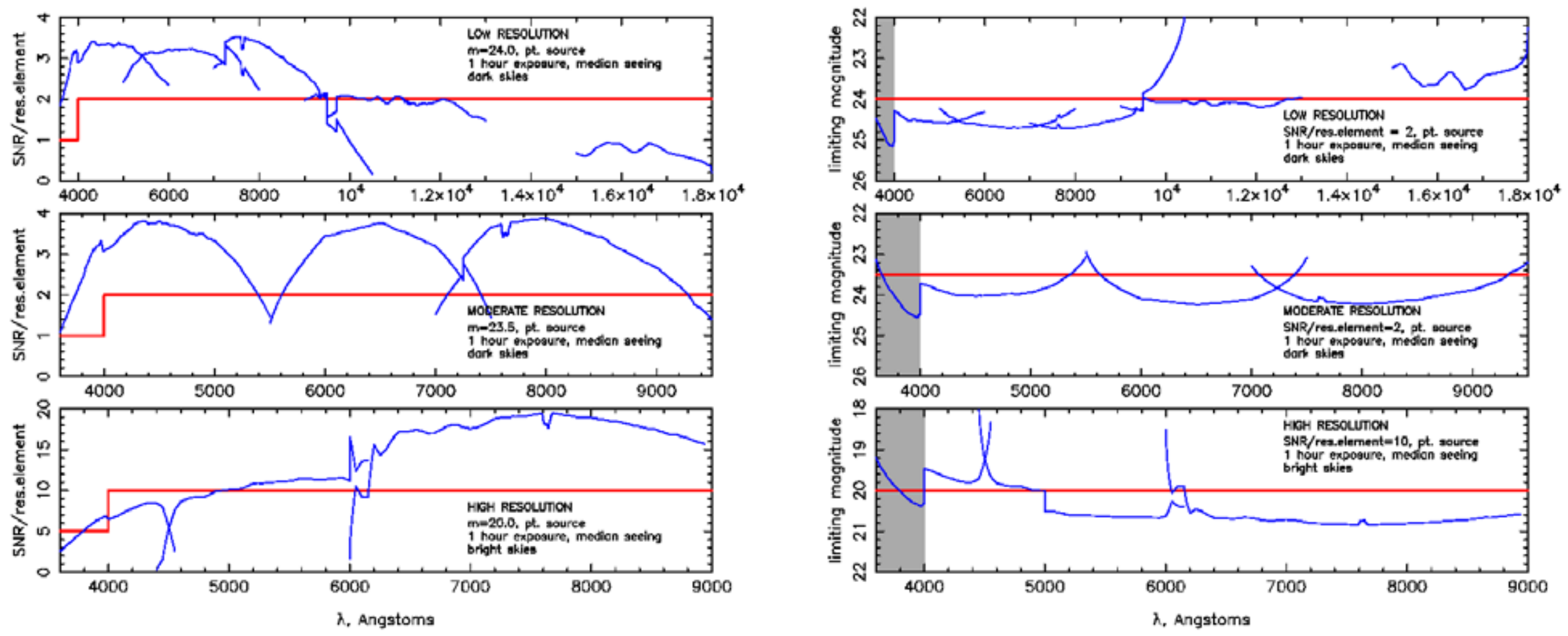

*Figure 22: The required sensitivity of MSE during a one-hour exposure, under median seeing conditions expressed in SNR per resolution elements (left) and limiting magnitude (right). On each side, the top, middle, and bottom panels show the low, moderate, and high spectral resolution settings, respectively. The red lines show the requirements, and the blue lines show estimates of performance, as of the end of the Conceptual Design phase. Line segments correspond to the different arms of the spectrographs.*

### 3.5.2. Sensitivity Budgets and Upcoming Trade Studies

Three principal factors affect MSE sensitivity: system throughput; system noise; and injection efficiency (the fraction of light from an astronomical source incident on the focal plane that enters a fiber, which is considered separately from all other throughput factors). Injection efficiency is also dependent on delivered image quality at the focal plane.

The three principal sensitivity budgets interact with each other through the usual SNR formalism.[vi] This formalism demonstrates that throughput, noise, and injection efficiency are not independent quantities, but budgets that can be traded against each other as necessary, to ensure that high-level sensitivity requirements are met.

### *(1) Performance Budget 1: Throughput*

At each spectral resolution setting for MSE, the dominant source of light losses is the spectrographs. Within the high spectral resolution spectrograph, there is a large contrast between the efficiency of the grating at its peak (~80%) and at the edges of the wavelength window (~50%), due to the fact that the grating operates at such large incidence angles.

Figure 22 demonstrates that sensitivity at the bluest wavelengths, particularly for high spectra resolution science, is particularly challenging. It is anticipated that MSE will adopt blue-enhanced coatings, such as the ZeCoat Corporation[vii] UV-enhanced protective silver coating on M1, for example. The efficiency of broadband optics degrades quickly at UV wavelengths, and any gains here are significant, especially for high spectral resolution chemical abundance studies and studies of the intergalactic medium.

Light losses in the fibers are another significant contributor to the sensitivity of MSE, especially at the critical high spectral resolution windows at < 500 nm. Given the preciousness of blue photons in the high-resolution mode and the need to increase overall sensitivity at this wavelength interval, MSE will conduct a trade study, to examine switching the baseline locations of the two suites of spectrographs, so that the high resolution spectrographs can use shorter fibers. Early studies suggest that this switch will provide a significant gain in sensitivity at high resolution, without impacting MSE’s ability to meet sensitivity requirements at the other spectral resolutions.

### *(2) Performance Budgets 2: Noise*

A dominant and inevitable source of noise for MSE is emission from the sky. All other sources of noise therefore need to be minimal by comparison, so as not to degrade the sensitivity of MSE. We therefore anticipate that the dark current in the NIR detectors will be reduced by operating them at temperatures close to 75K, instead of at the current baseline of 95K (this will reduce the dark current by more than a factor of two). We also anticipate needing to revisit the sensitivity requirement for MSE in this range, given the brightness of the H-band sky, while still ensuring it is significantly more sensitive than that of any other wide-field spectroscopic instrument. To this end, MSE is undertaking a detailed investigation of the NIR background, based on empirical data from the Maunakea observatories.

### *(3) Performance Budget 3: Injection Efficiency*

The injection efficiency budget[viii] is calculated by modeling all known factors that impact the light distribution of a point source at the focal plane, and MSE’s ability to position a fiber accurately at a specific position on the focal plane. Table 5 shows the various factors which have been taken into consideration and the project cycle to which components of the budget have been assigned. These allocations are based on the deviations from the idealized fiber location in two directions, longitudinal (z) and lateral (xy), in units of microns (the maximum amount of attainable flux enters the fiber at z=0, x=0, and y=0).

*Table 5: Injection efficiency budget-allocated longitudinal and lateral deviations.*

| | Total allocation, um | | | | Partition | | |
|---|---|---|---|---|---|---|---|
| **Budget Group** | **Lateral** | | **Longitudinal** | | | | **Discussion** |
| **Theoretical Model** | 45 | max | | | | | Lateral chromatic aberrations |
| | | | | | 45 | max | The residual chromatic aberrations after ADC correction are estimated to lead to a lateral chromatic displacement of 41 microns maximum, i.e. separation, between the foci of any two wavelengths. The separation is derived from optical design and defined by the delivered PSF computed in Zemax. |
| **As-Delivered** | | | 30 | max | | | Longitudinal installation errors of the combined PosS+FiTS |
| | | | | | 25 | max | Scattering of PosS in Z position relative to theoretical focal surface, based on AAO-Sphinx CoDR |
| | | | | | 5 | max | Scattering of fibre tip distances from ferrules, based on FiTS information |
| **Assembly Integration Verification** | | | 50 | max | | | Longitudinal alignment errors of the top end assembly |
| | | | | | 50 | max | Residual alignment errors of the PosS+FiTS focal surface in tip/tilt and/or Z position after Assembly, Integration and Verification (AIV) |
| **Operation - Focus model residual error (after setup)** | | | 10 | max | | | Longitudinal flexure of telescope structure due to zenith angle (gravity) and thermal changes affecting location of the fibre tips |
| | | | | | 10 | max | Residual modeling error in lookup table correction for best-focus setting |
| **Operation - Relative lateral fibre positioning errors (after acquisition)** | 10 | max | | | | | Target coordinates error |
| | | | | | 10 | max | Error in target coordinates due to astrometry inaccuracy and coordinate conversion error |
| | 5 | max | | | | | Sky coordinates to focal surface mapping |
| | | | | | 1 | max | Acquisition/guide cameras registration error with respect to the sky |
| | | | | | 4 | max | Metrology system residual calibration error with respect to the focal surface |
| | 6 | rms | | | | | Positioner closed-loop accuracy |
| | | | | | 4 | rms | Positioner contribution based on AAO-Sphinx CoDR |
| | | | | | 2 | rms | Metrology system contribution based on AAO-Sphinx CoDR |
| **Operation - Fibre defocus (spine tilt after acquisition)** | | | 80 | max | | | |
| | | | | | 80 | max | Defocus at maximum tilt for 300 mm spine with AAO Sphinx positioner (Note: defocus due to lateral positioning error is negligible, <1 um.) |
| **Operation - Telescope motion errors (guiding during exposure)** | 5 | rms | | | | | Instrument rotator rotation error |
| | | | | | 5 | rms | Position error resulting from imperfect control of the rotation rate; allocation of 3.5" rms rotation rate error, which corresponds to 5 um at the edge of the field of view |
| | 10 | rms | | | | | Based on CFHT guiding accuracy of 0.01" which corresponds to 1 um with MSE plat scale; increase allocation by 100% to 2 um; allocations add in quadrature |
| | | | | | 7 | rms | Allocation for mount control |
| | | | | | 7 | rms | Allocation for guide camera |
| **Operation - Differential atmospheric refraction (during exposure)** | 15 | max | | | | | Residual DAR drift after ADC correction |
| | | | | | 15 | max | Based on optical design information and assuming observations between zenith distance 0° to 40°, where most observations are executed. |
| **Operation - Plate scale variation (during exposure)** | 10 | max | | | | | Plate scale variations due to unassigned optics affects |
| | | | | | 10 | max | Margin to be managed by the MSE Project Office |
| **Operation - Motion of fibre tips during exposure** | | | 1 | max | | | Gravity effect |
| | | | | | 1 | max | Due to gravity sag of the positioner support structure; AAO-Sphinx CoDR reports predicted 4 um at zenith |
| | 14 | max | | | | | Thermal effect |
| | | | | | 14 | max | Based on thermal expansion of steel focal plate, 0.59 m dia x 60 mm thk, with 2° temperature increase in one hour |
| | | | 2 | max | | | Thermal effect |
| | | | | | 2 | max | Based on thermal expansion of steel focal plate, 0.59 m dia x 60 mm thk, with 2° temperature increase in one hour |
| | | | 30 | max | | | Instrument rotator axis tilt leading to longitudinal displacement |
| | | | | | 30 | max | Based on 50 urad of tilt over 180° rotation in one hour |
| | | | 5 | max | | | Hexapod focus adjustment error |
| | | | | | 5 | max | Hexapod positional errors; PFHS is assumed to adjust the focal surface position every 15 min. |
| | 1 | max | | | | | Lateral fibre tip displacement due to vibration |
| | | | | | 1 | max | |

### *(4) Performance Budgets 4: Image Quality*

The image quality of MSE is critical for the Injection Efficiency—and hence the overall sensitivity—of the observatory. MSE benefits from an extraordinary site with exceptional free-atmosphere seeing. MSE will maintain this advantage by ensuring that all other factors affecting the delivered image quality (IQ) at the focal plane are small by comparison. The detailed budget is shown in Table 6, and the breakdown of the total IQ (under median seeing conditions) is shown in Figure 23.

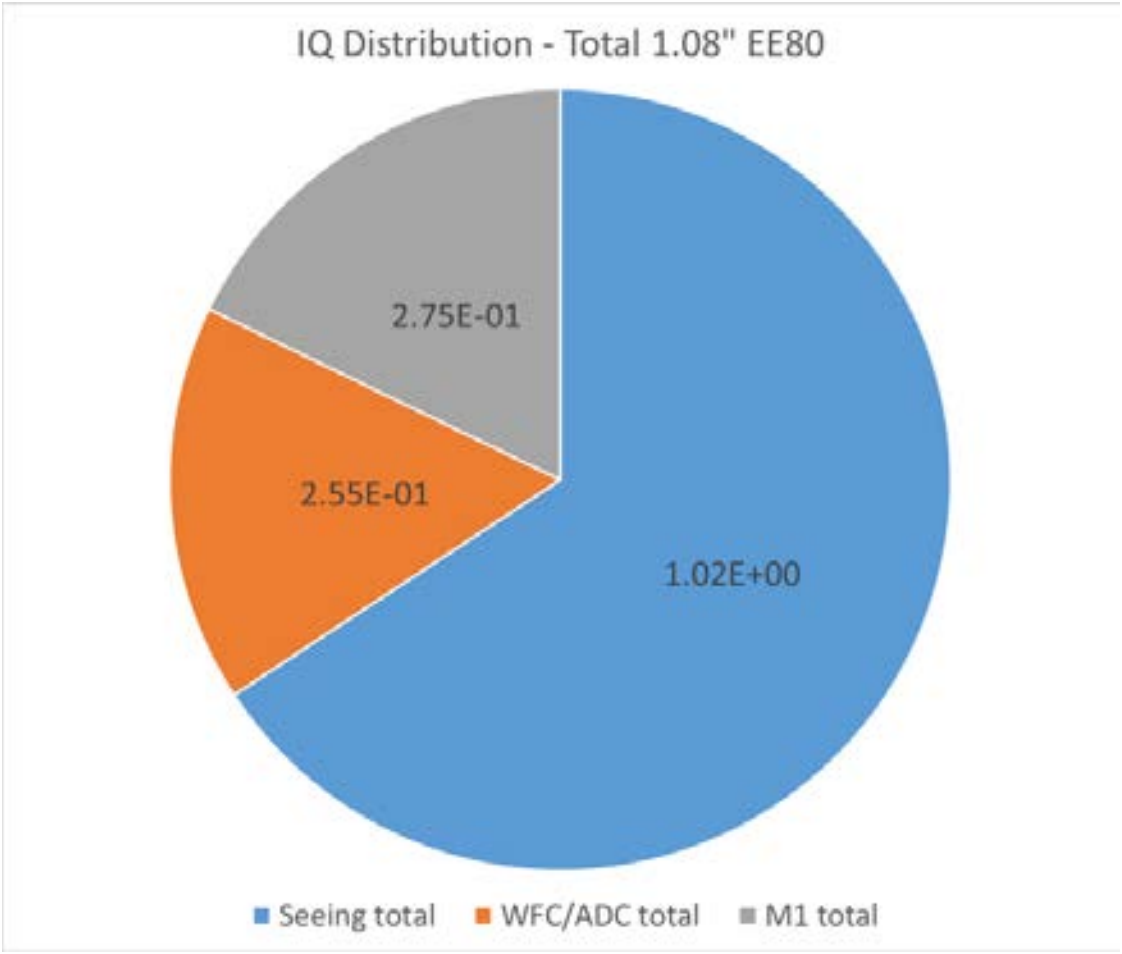

*Figure 23: IQ contributors—Seeing (66%), WFC/ADC (16%), and M1 (18%)*

*Table 6: IQ budget allocations—1.08″ in EE80 or 0.50″ FWHM.*

| Item ID | | | | | | Value | Unit |
|---|---|---|---|---|---|---|---|
| Seeing | Atmospheric | | Natural Site Seeing | | | 0.81 | arcsec ee80 (5/3 cumulative sum) |
| | Uplift | | Due to building affecting ground layer | | Observatory and Facilities | 0.44 | arcsec ee80 (5/3 cumulative sum) |
| | Thermal | Enclosure | Due to temperature gradients in dome | | Enclosure | 0.22 | arcsec ee80 (5/3 cumulative sum) |
| | | M1 | Due to temperature difference between M1 and dome | | Primary Mirror | | |
| | | Top End | Due to thermal dissipation by al lcomponents in top end | | Instrument Rotator | | |
| | | | | | Prime Focus Hexapod | | |
| | | | | | Telescope Optical Feedback System | | |
| | | | | | **Seeing total** | **1.02** | **arcsec ee80 (5/3 cumulative sum)** |
| As- delivered | **Monochromatic aberrations** | | | | | | |
| | WFC/ADC + idealized M1 | | | | Primary Mirror | 0.25 | arcsec ee80 |
| | | | | | **Total design** | **0.25** | **arcsec ee80** |
| | **M1 segments figuring abberations** | | | | | | |
| | Residual error after IBF | M1 | Mirror polishing & SSA mounting | | Primary Mirror | 0.24 | arcsec ee80 |
| | | | | | **Total M1 segments** | **0.24** | **arcsec ee80** |
| | **WFC/ADC aberrations due to mounting and internal alignment errors** | | | | | | |
| | WFC/ADC lens figure error | Radius of curvature | | | WFC/ADC | 1.98E-02 | arcsec ee80 |
| | | Aspheric & conic constant | | | WFC/ADC | 2.17E-02 | arcsec ee80 |
| | | Slope errors | | | WFC/ADC | 1.98E-02 | arcsec ee80 |
| | | Thickness errors | | | WFC/ADC | 6.25E-03 | arcsec ee80 |
| | | Tilt and decentre between lens' surfaces | | Relative surface-wise | WFC/ADC | 2.42E-02 | arcsec ee80 |
| | WFC/ADC materials | Homogeneity | | | WFC/ADC | 1.25E-02 | arcsec ee80 |
| | Alignment errors | Tilt and decentre between lenses | | Relative lens-wise | WFC/ADC | 1.65E-02 | arcsec ee80 |
| | | Axial separation between lenses | | Barrel assembly | WFC/ADC | 6.25E-03 | arcsec ee80 |
| | | | | | **Total WFC/ADV abberations** | **4.84E-02** | **arcsec ee80** |
| | | | | | **Total as-deliverd** | **2.45E-01** | **arcsec ee80** |
| AIV | Installed M1 segments residual error in the mirror cell | | | Mirror cell | Primary Mirror | 0.12 | arcsec ee80 |
| | | | | | **Total M1** | **1.20E-01** | **arcsec ee80** |
| | Alignment errors - WFC/ADC barrel alignment with respect to M1 | | | | TOFS | 0 | arcsec ee80 |
| | | | | | **Total WFC/ADC alignment** | **0** | **arcsec ee80** |
| | | | | | **Total AIV** | **1.20E-01** | **arcsec ee80** |
| Operations | Dynamic segment alignment residuals | | | | Primary Mirror | 6.20E-03 | arcsec ee80 |
| | | | | | **Total M1 dynamic** | **6.20E-03** | **arcsec ee80** |
| | Alignment Errors | Precision of TOFS metrology errors | | | TOFS | 8.84E-03 | arcsec ee80 |
| | | LUT modelling errors | | | PFHS | 8.84E-03 | arcsec ee80 |
| | | | | | **Total WFC/ADC alignment** | **1.25E-02** | **arcsec ee80** |
| | | | | | **Total operations** | **6.32E-02** | **arcsec ee80** |
| | | | | | **Total IQ allocation** | **1.08E+00** | **arcsec ee80** |

### 3.5.3. Observing Efficiency

The observing efficiency of MSE is defined as the fraction of the night time during which the observatory is able to collect on-target science photons, excluding only time lost to weather. Defined in this way, MSE is required to have an average observing efficiency of 80%.

The average duration of a night on Maunakea is 10.2 hours (with a 12 degree twilight limit), and historical weather losses average out to approximately 2.2 hours per night. The observing efficiency budget therefore considers all other possible losses for MSE, including on-sky engineering time, M1 phasing time, subsystem failure rates, and observing overheads. To calculate the observing overheads involves defining the observing sequences and understanding which processes are parallel and which sequential. The baseline night time observing sequence for MSE comprises a set of calibration exposures before and after each science exposure. Overall, current estimates indicate that for nominal exposure sequences lasting longer than ~40 minutes the overall observing efficiency of MSE will exceed 80%.

### 3.5.4. Calibration budgets

Accurate scientific calibration of science spectra obtained using MSE is an absolute necessity. Six high-level MSE Science Requirements explicitly focus on calibration, specifically velocity accuracy, spectrophotometric accuracy, and sky subtraction. The guiding principles of MSE’s calibration strategies are:

- Science calibration should not introduce any significant source of noise into the observation (either directly, such as a spectral flat, or indirectly, such as through a model of the flux response across the focal plane).
- Calibration exposures should not add significant overheads (in terms of time) to science observations.
- Calibration exposures should be obtained in a configuration that is as close as possible to that of the science observations, in order to reproduce as closely as possible the system-wide behavior at the time of the science observation.

The practical ramifications of the calibration requirements and this guiding philosophy are described in the MSE calibration budgets, which detail the operational strategies, subsystem requirements, and proposed observations that together provide the scientific and ancillary data necessary for precision calibration of the MSE science observations.

Science calibration for MSE[ix] essentially requires the accurate estimation of the differential throughput, PSF behavior, and wavelength solutions for all the objects observed during each observation. Complexity arises, given the simultaneous desire to minimize night time overheads and maximize observing efficiency. MSE uses a combination of night time, daytime, and twilight calibration observations. During the night, calibration exposures are obtained using dedicated lamps, with the system in the exact same configuration as during the science exposures, which are very short in duration. At daytime, MSE uses calibration sources that best reproduce the far-field illumination of fibers as obtained during science exposures (i.e., that best reproduce the PSF), which are more time-consuming to obtain.

---

[i] http://www.malamamaunakea.org/uploads/management/plans/CMP_2009.PDF
[ii] Szeto, K., A. Hill, N. Flagey, et al. "Maunakea spectroscopic explorer (MSE): Implementing systems engineering methodology for the development of a new facility." *Modeling, Systems Engineering, and Project Management for Astronomy VIII* (2018): 10705, 107050H.
[iii] Szeto, K., H. Bai, S. Bauman, et al. "Maunakea spectroscopic explorer design development from feasibility concept to baseline design." *Ground-based and Airborne Telescopes VI* (2016): 9906, 99062J.
[iv] Griffith, R. L., M. C. Cooper, J. A. Newman, et al. "The advanced camera for surveys general catalog: Structural parameters for approximately half a million galaxies." *ApJS* (2012): 200, 9.
[v] Davis, M., P. Guhathakurta, N. P. Konidaris, et al. "The All-Wavelength Extended Groth Strip International Survey (AEGIS) data sets." *ApJL* (2007): 660, L1–L6.
[vi] McConnachie, A. W., N. Flagey, K. Szeto, et al. "Maximising the sensitivity of next generation multi-object spectroscopy: System budget development and design optimizations for the Maunakea Spectroscopic Explorer." *Modeling, Systems Engineering, and Project Management for Astronomy VIII* (2018): 10705, 1070522.
[vii] http://www.zecoat.com/
[viii] Flagey, N., K. Szeto, S. Mignot, et al. "Modeling and budgeting fiber injection efficiency for the Maunakea Spectroscopic Explorer (MSE)." *Modeling, Systems Engineering, and Project Management for Astronomy VIII* (2018): 10705, 107051O.
[ix] McConnachie, A. W., N. Flagey, P. Hall, et al. "The science calibration challenges of next generation highly multiplexed optical spectroscopy: The case of the Maunakea Spectroscopic Explorer." *Observatory Operations: Strategies, Processes, and Systems VII* (2018): 10704, 107041O.

# 4. Observatory Subsystem Description

This section provides detailed descriptions of all MSE Observatory subsystems, arranged under the Level 1 elements outlined in the Production Breakdown Structure (PBS):

- Section 4.1 Observatory Building and Facilities
- Section 4.2 Enclosure
- Section 4.3 Telescope
- Section 4.4 Science Instrument Package
- Section 4.5 Observatory Execution System Architecture
- Section 4.6 Program Execution System Architecture

The remaining chapter is divided into six sections, according to the "parent" Level 1 PBS elements. Each section generally contains three parts: (1) An overview of the Level 1 PBS element and its relationship to the other Level 1 PBS elements; (2) A formal description of the functions provided by the Level 1 PBS element and a delineation of the functionalities of the "child" Level 2 and, if applicable, Level 3 PBS elements; (3) Detailed design descriptions of all Level 2 and, if applicable, Level 3 PBS elements.

Readers who would like to gain a general understanding of the overall design of the Observatory should read all six overviews. Readers who wish to systematically link the functionalities of the Observatory with the actual products involved should read all six overviews and product function descriptions. Readers who wish to obtain a detailed understanding of the Observatory's design should read this chapter in its entirety. Readers who wish to obtain a detailed understanding of a particular Level 1 PBS element should read the corresponding section in its entirety.

This chapter is organized in a way intended to provide Observatory design information at a level appropriate and relevant to readers' interests. Each section is designed to stand alone and can be read independently. In the three parts of each section, information is presented progressively, moving from the general to the specific, in accordance with the level of detail required. However, due to the formalism of the functional descriptions in part two, some descriptions have been repeated between parts within each section.

## 4.1. Observatory Building and Facilities (OBF)

### 4.1.1. Observatory Building and Facilities Overview

The Observatory Building and Facilities (OBF) is the infrastructure that supports MSE's daytime and nighttime operations. The OBF covers the upgraded CFHT summit and Waimea headquarters facilities. This includes the structurally upgraded Outer Building and Inner Pier, as well as shops, labs, a control room, offices, and staff space, which have been renovated and reorganized, to accommodate modern building codes and a new operations concept. The OBF also includes active and passive thermal management, updated utilities infrastructure, handling and access equipment, and other infrastructure provisions. The planned overall layout of the OBF is shown in Figure 24.

The OBF renovations have been planned and designed to a conceptual level[i] by the technical operations staff at Canada France Hawaii Telescope (CFHT) in Hawaii, USA.

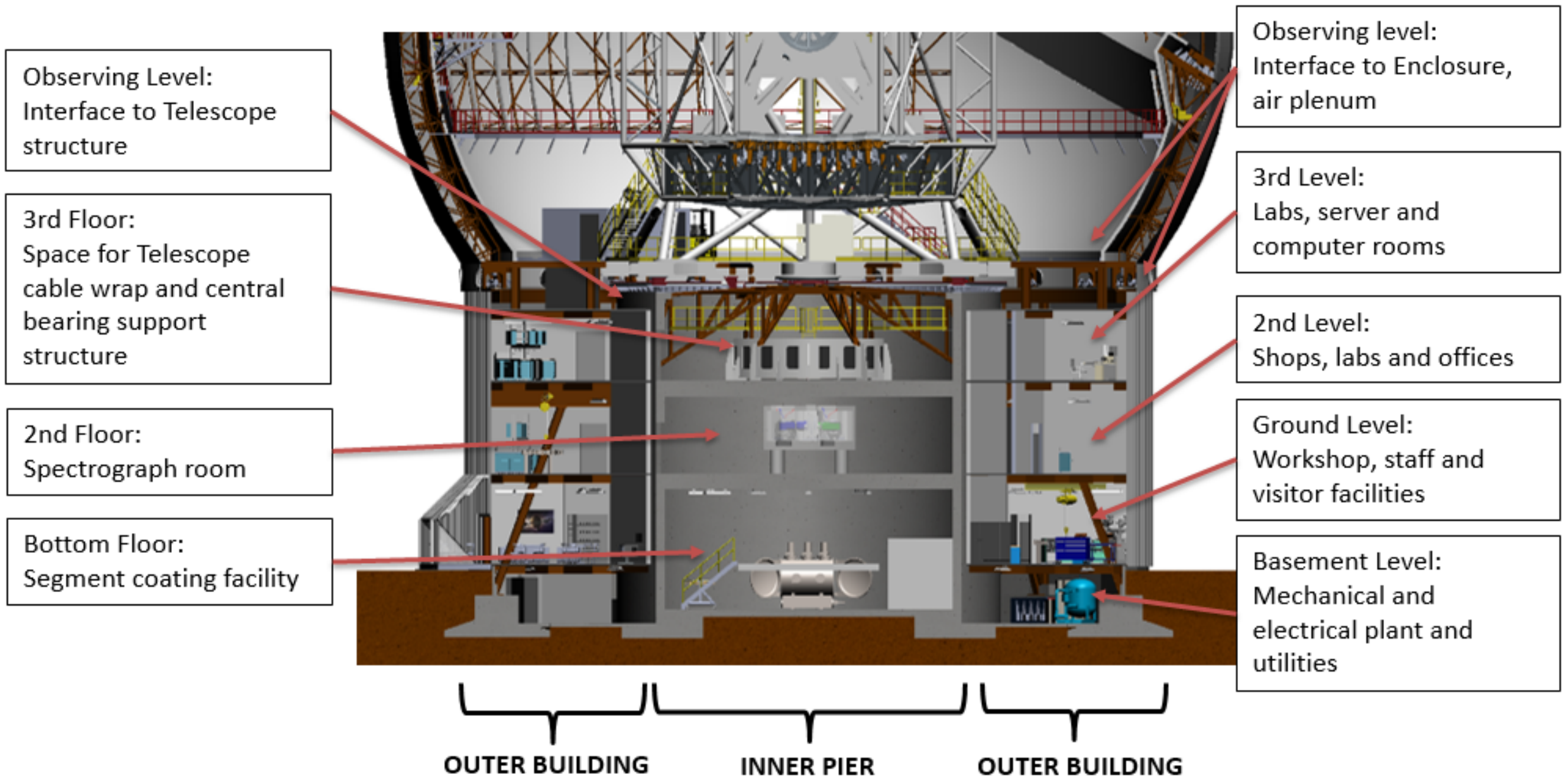

*Figure 24: MSE Observatory Building and Facilities, layout*

The basement level will continue to house machinery used to supply the Observatory with electrical power, cooling, and other utilities. As far as possible, we are reusing existing systems: however, we anticipate upgrades to those systems, as the Observatory is modernized. The current mirror coating facility for CFHT's 3.6 m mirror, housed on the bottom floor, will be completely converted into a mirror segment handling, storage, and coating facility. At least two spectrographs will be housed in the second floor Spectrograph Room, which currently is one of the two Coudé rooms at CFHT. The third floor Inner Pier Coudé room will be converted into access space for the azimuth cable wrap and central bearing support structures for the Telescope System. In the Outer Building, the third level will house the observatory control room and main server computer room.

The OBF Production Breakdown is shown in Figure 25. The OBF products, together with planned renovations and organization, are described in the following sections.

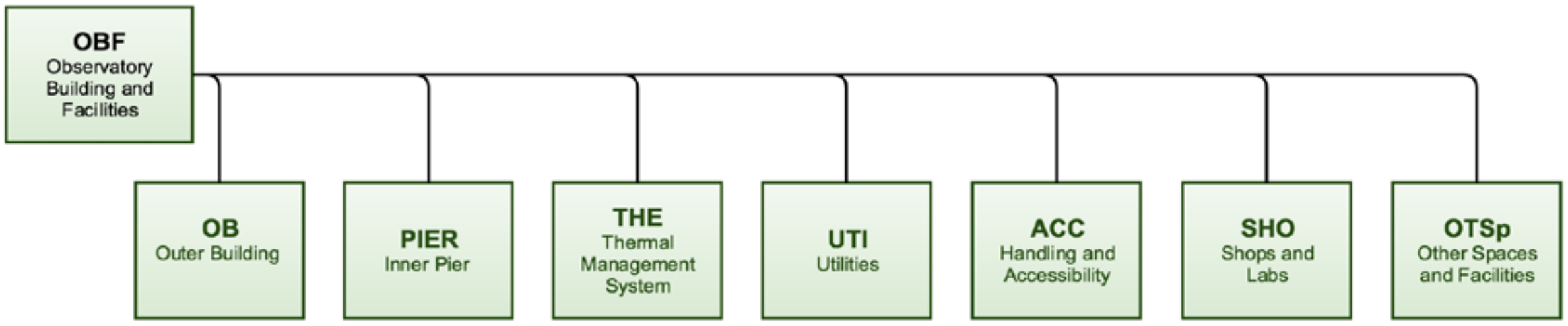

*Figure 25: OBF Product Breakdown Structure*

### 4.1.2. Outer Building (OB) and Inner Pier (PIER)

*(1) Structural Upgrades*

The OB and PIER are separate structures, supporting the Enclosure and the Telescope, respectively, as shown in Figure 26. The OB is a structural steel framework and the PIER is a concrete cylindrical structure, with rebar reinforcement. The foundation of the PIER is structurally separated from the foundation of the OB, to limit the transmission of vibrations due to enclosure rotation and wind buffeting, and from other sources, such as machinery.

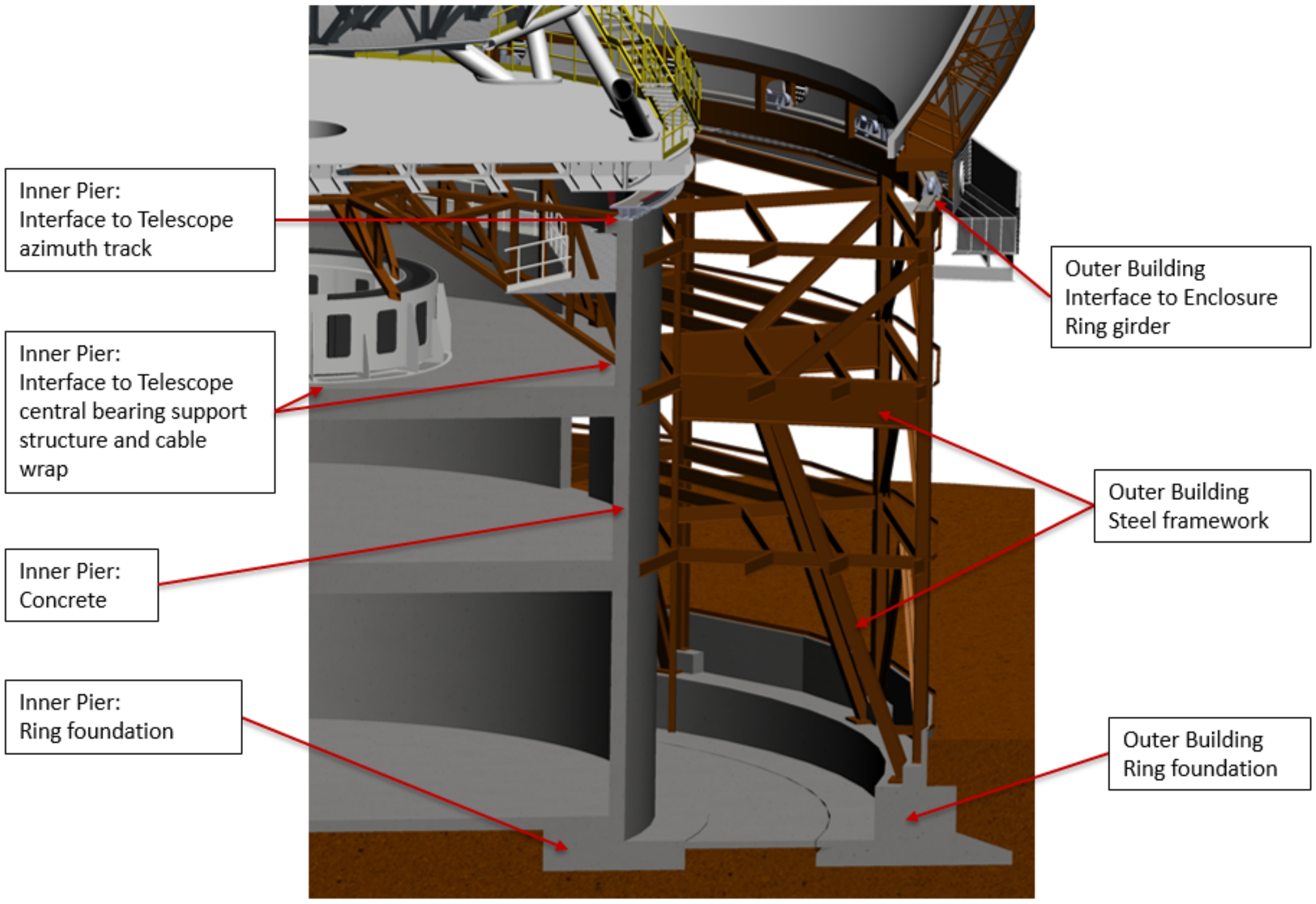

*Figure 26: OBF structural components—Outer Building and Inner Pier*

The new Enclosure has an as-designed mass of 583 tonnes and the new Telescope has a mass of 338 tonnes. The mass of the current CFHT enclosure is 386 tonnes and the current telescope has a mass of 255 tonnes, according to historical documents.

To accommodate the MSE Enclosure, the fifth level observing floor and the existing OB's mezzanine structure will be removed and replaced by a new steel framework and ring girder to support the azimuth track just above the third level. The gap between the new observing floor and the third level ceiling will become an air plenum, to facilitate heat removal by flushing. The third floor ceiling will be insulated, to further limit heat dissipation into the Enclosure space. A mass reduction of 107 tonnes is estimated, once the fifth level observing floor and mezzanine structure have been removed.

To accommodate the MSE Telescope and facilitate the installation of the azimuth track, the top of the PIER will be removed. A new 2' x 2' circular concrete haunch will be added, to secure the anchor bolts for mounting the new azimuth track (Figure 27). The concrete at the top of the existing wall will be removed and replaced by a new haunch section, to which eight circumferential concrete pilasters will be added. A mass reduction of 142 tonnes is estimated once the top of the PIER has been removed.

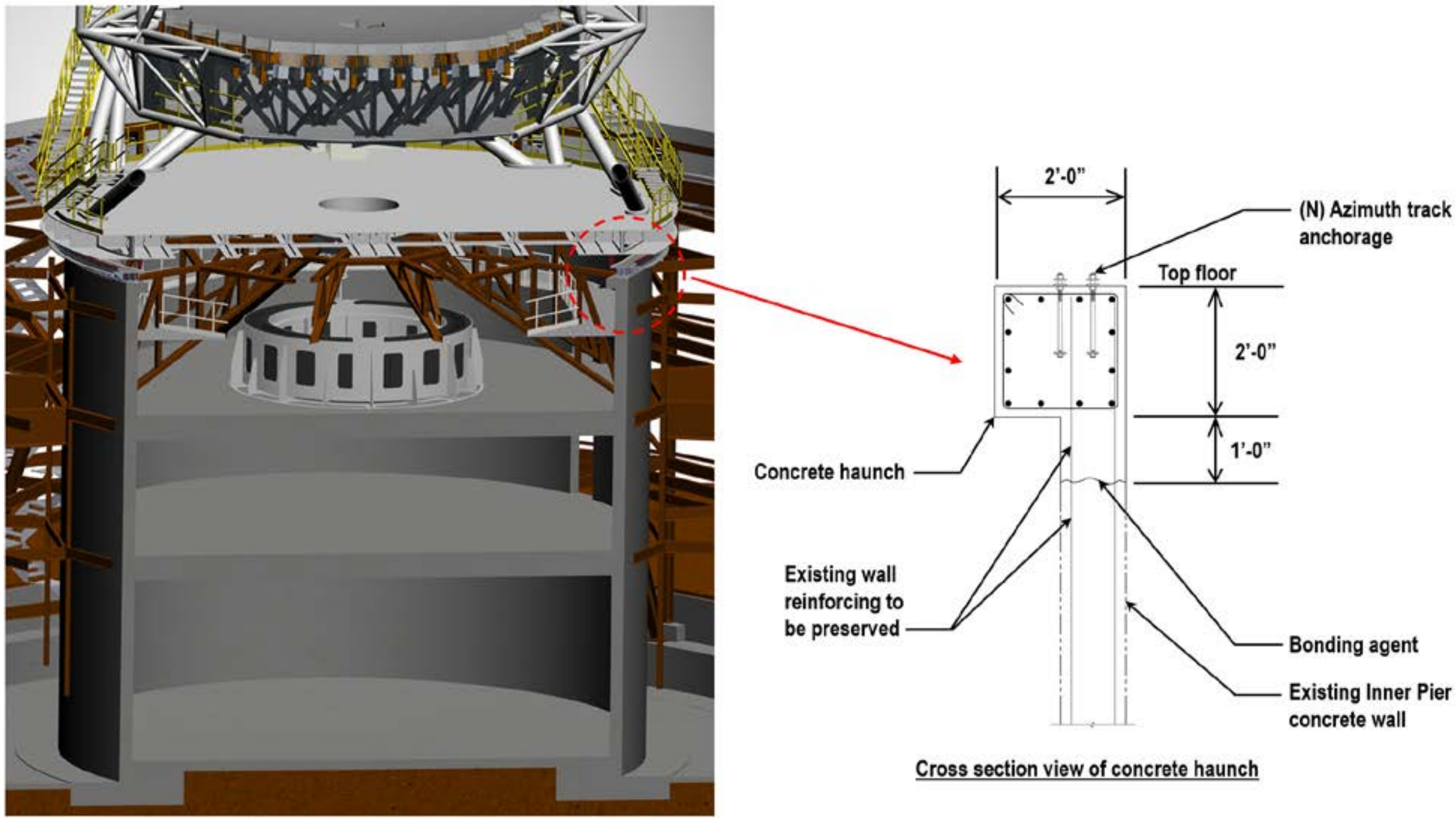

*Figure 27: Inner Pier modification—New concrete haunch*

The structural capacities of the OB and PIER were evaluated under the current design codes (ASCE 7–10), based on the new mass, geometry and mass properties of the Enclosure and Telescope.

The evaluation findings showed that extensive structural upgrades to the OB are needed, in order to meet modern building code requirements, specifically with regard to wind and seismic loads (Figure 28). New diagonal braces will be added, linking the third level to the new enclosure azimuth ring girder, to reinforce the framework supporting the Enclosure. All I-beam sections of exterior vertical columns and horizontal beams will be reinforced by welding doubler-plates (3/8" thick) onto the flanges.

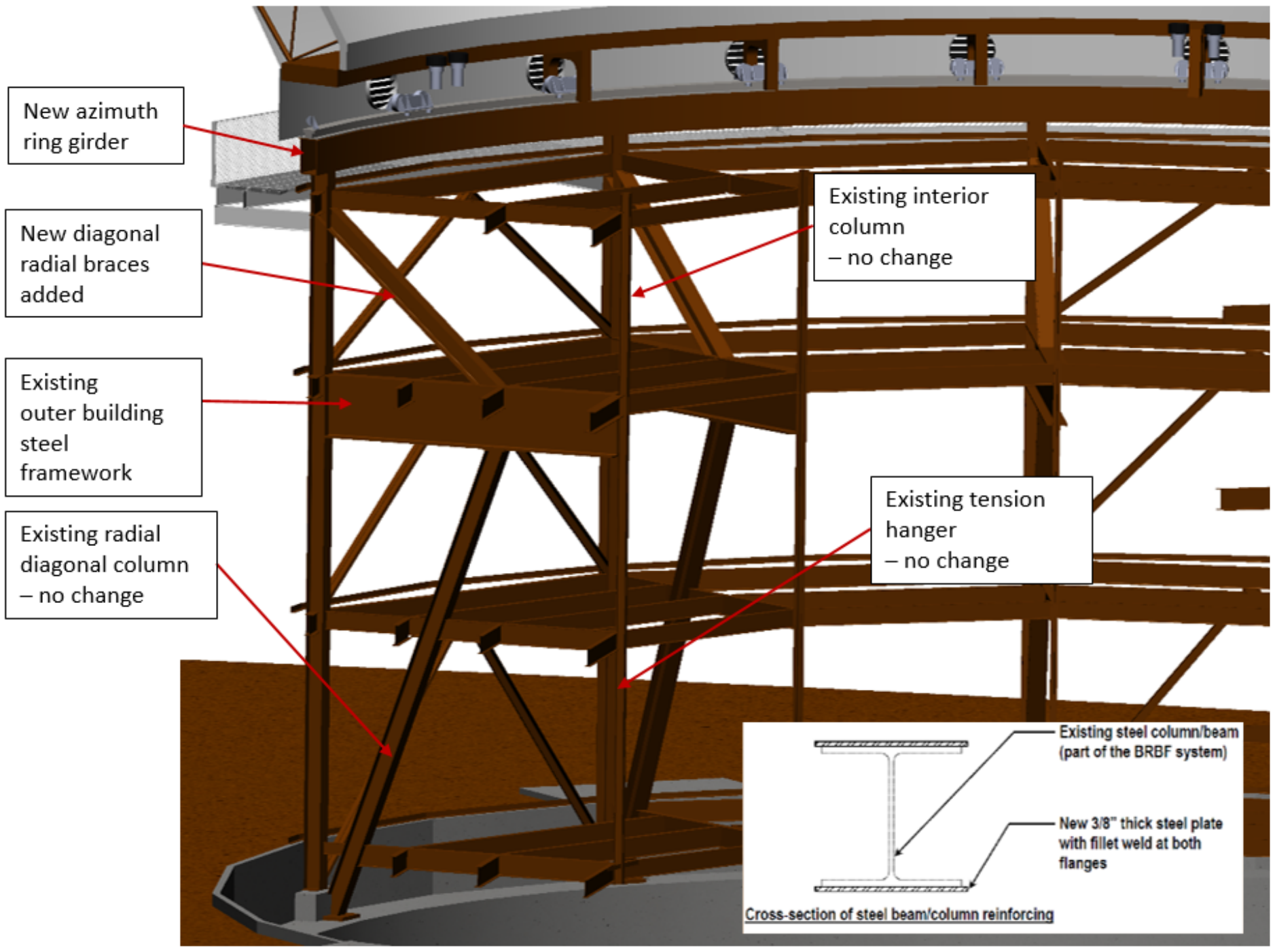

*Figure 28: Outer Building modifications and retrofits.*

Other code-dictated structural changes include replacing the existing chevron bracings with new buckling-restrained braces in every other bay, to increase seismic shear capacity (Figure 29), and reinforcing the anchorage of the vertical columns to the existing concrete ring foundation, to increase their "pullout" capacity, in response to uplift from wind.

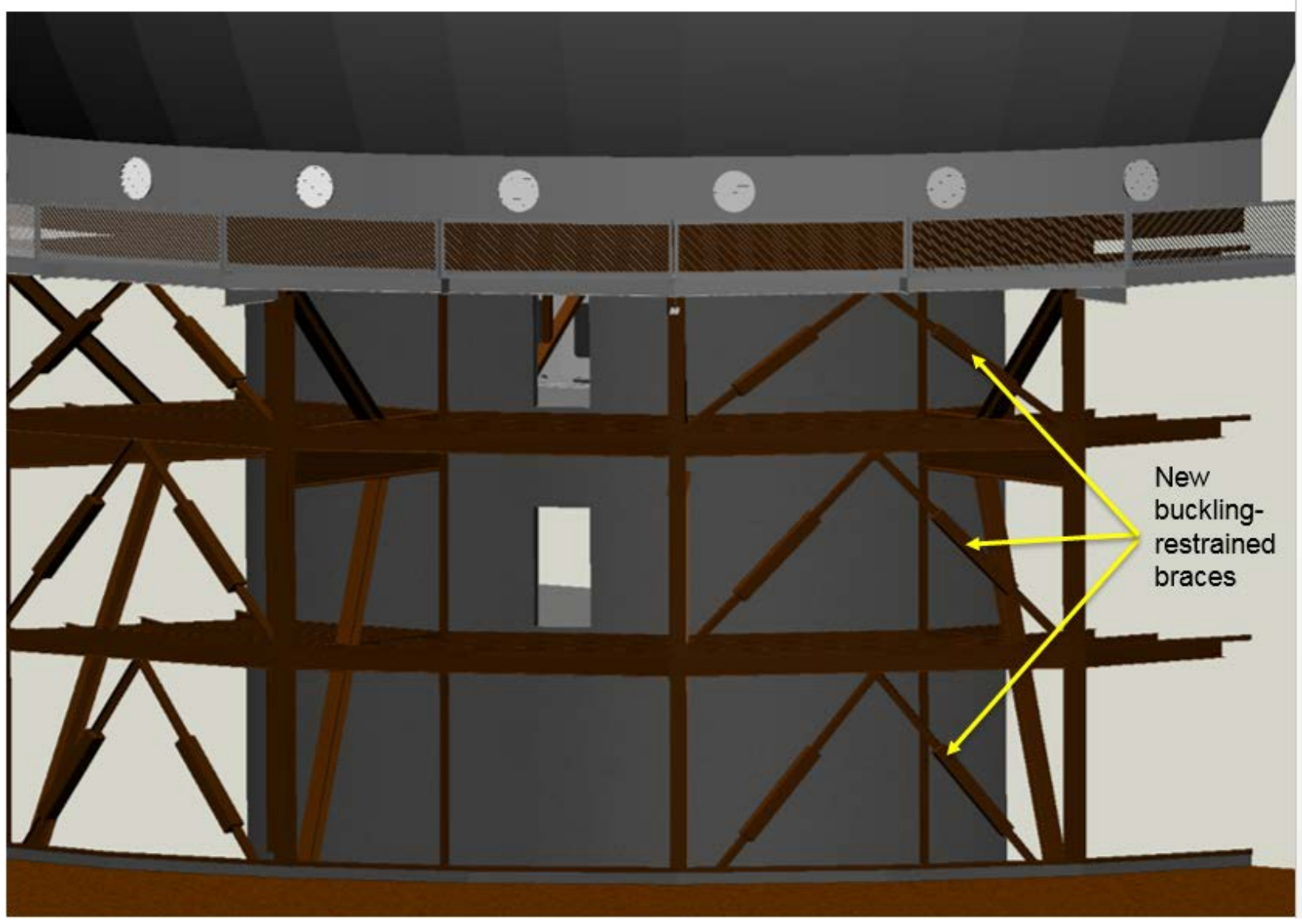

*Figure 29: Outer Building modification—buckling-restrained braces.*

The evaluation findings also showed that the inner concrete pier presents no structural issues when supporting the Telescope. However, localized reinforcement, using carbon fiber wrap at door openings, is recommended. The PIER and its ring foundation are fundamentally stable and able to resist sliding forces and overturning-moment-generated seismic events, while supporting the MSE telescope mass.

*(2) Soil Capacity Testing*

A geotechnical site soil study was performed to determine the maximum allowable weight for both the new Enclosure and OB, and Telescope and PIER, following the structural renovations, under the site's soil conditions. After obtaining permission from the Office of Maunakea Management, site drilling and core samples were taken to evaluate the soil's bearing capacity. The study findings show that foundation contact pressure should be limited to 4,500 psf for the ring foundations of both the OB and PIER.

For the MSE Enclosure and Telescope configuration, the applied foundation pressures on the soil from the OB and PIER do not exceed the allowable limit. The applied pressures are 4,100 psf and 2,670 psf respectively, for a fully loaded facility, including dead, live, and environmental loads.

#### 4.1.3. Thermal Management (THE)

The OBF provides active and passive thermal management strategies, as part of a coordinated system strategy (Figure 30). This includes minimizing air conditioning requirements through insulation and interstitial space venting of the Enclosure, minimizing daytime solar loads by applying low emissivity paint, and limiting heat dissipation from subsystems into the enclosure space.

To that end, equipment that generates heat will be housed in the basement and at ground level. Waste heat will be extracted through the existing exhaust tunnels and released at a distance from the observatory. The ceilings of the lower floors, which contain heat generators, will be insulated, to prevent upward heat transfer. The air plenum above the third level will also prevent the escaped heat from reaching the enclosure space, by flushing.

As part of the Thermal Management System (THE), one of the OBF's active strategies is to provide daytime air conditioning inside the Enclosure, to cool the Telescope, with the goal that the Telescope's temperature should match the forecasted ambient temperature to within ±1.0°C at evening twilight, i.e., at the start of observation. Basically, THE is taking proactive steps to minimize thermally-induced seeing and ensure as-delivered image quality is as close to the natural site seeing as possible. The other active strategy is controlled dome flush by operating the enclosure ventilation modules during nighttime observations.

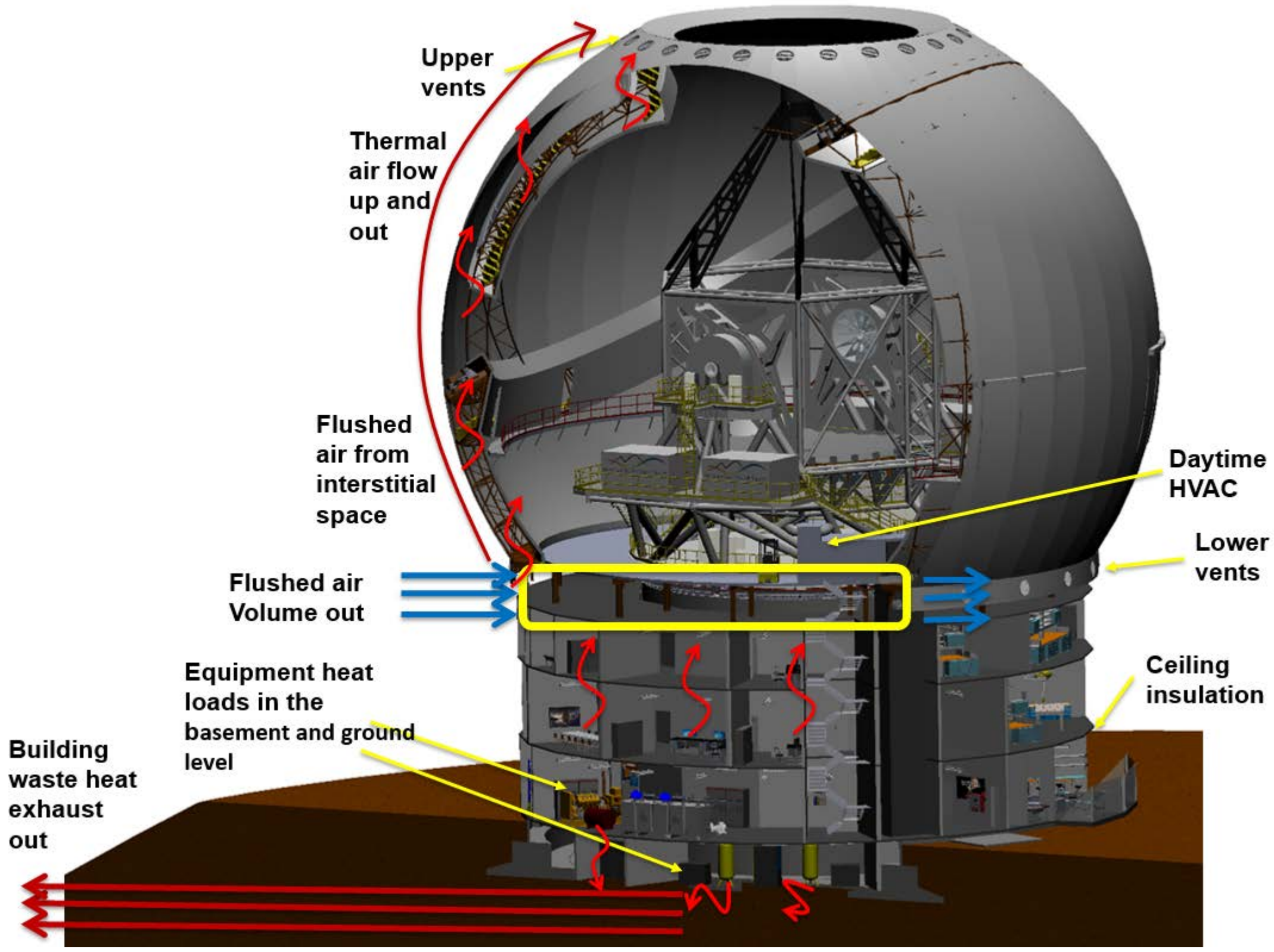

*Figure 30: Thermal management strategies implemented in OBF and Enclosure*

Computational fluid dynamics, aerothermal, and water-tunnel tests are planned for the next design phase, in order to refine the thermal management strategy and consolidate overall thermal management design at the system level.

### 4.1.4. Utilities (UTI)

The Utilities System (UTI) provides all the electrical power, lighting, water and plumbing, fire detection and alarms, and other services associated with supporting MSE operations, whether daytime or nighttime.

The capacity requirements of the overall system utilities will be consolidated during the next design phase, and managed centrally in the form of utility budgets, to facilitate the design of the mechanical and electrical plants. In general, the OBF utility network distributes to eight connection panel locations: one main panel is located in the mirror-coating lab; one in the spectrograph room; one on the observatory floor for the Enclosure; one at the base of the Telescope azimuth structure; one on each of the instrument platforms; one at the mirror cell; and one at the elevation structure hexagon ring, as illustrated in Figure 31.

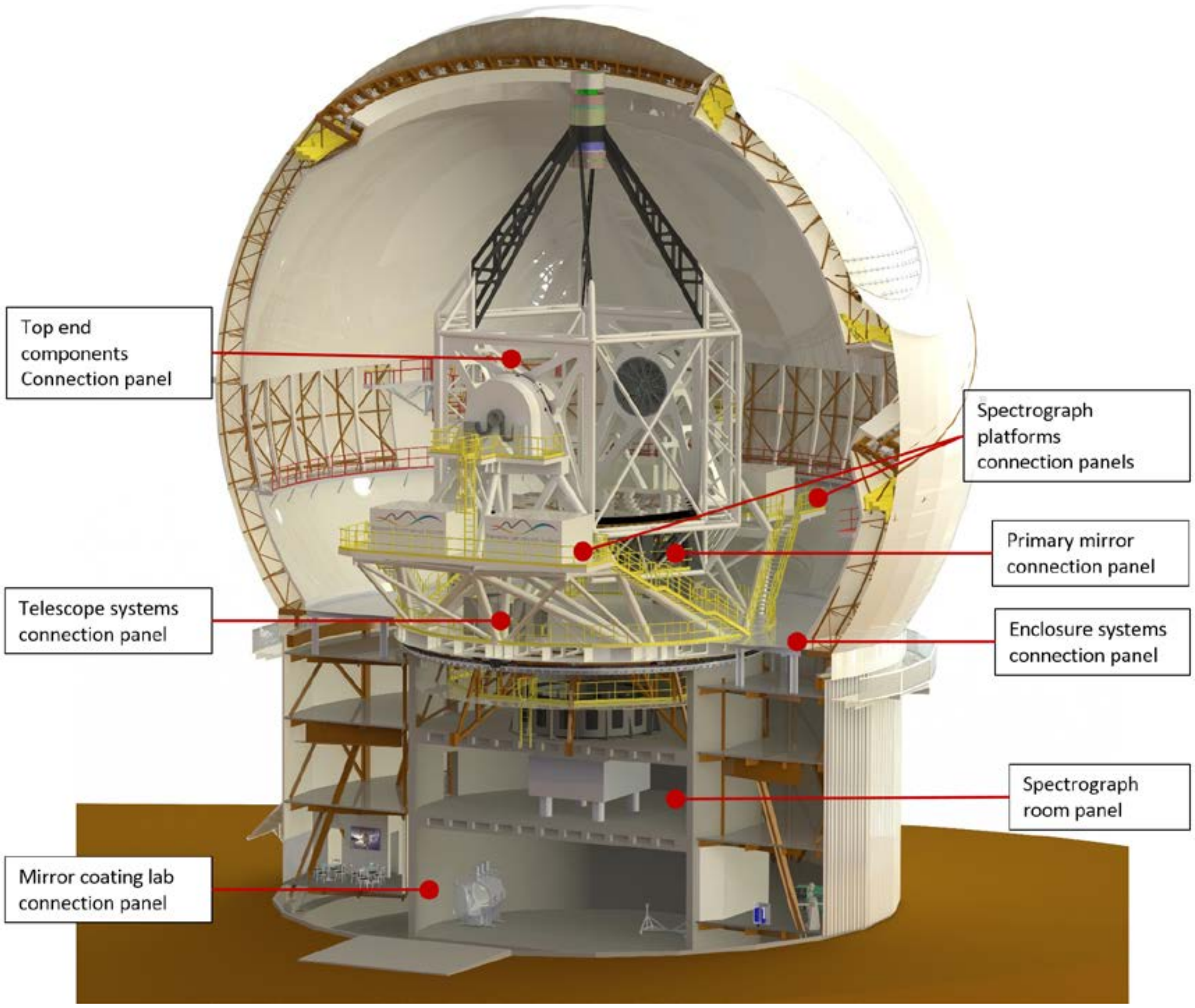

*Figure 31: Utilities distribution on the Telescope—Red dots show provisional main connecting panel locations*

#### 4.1.5. Handling and Accessibility (ACC)

Handling and Accessibility (ACC) provides all the infrastructure needed to allow safe and effective daily servicing and maintenance.

The ACC includes stairs and walkways; elevators and lift platforms; a mobile personnel lift and an observatory-floor-mounted, articulated personnel lift; general cranes in the shop and lab areas; and general rigging and lifting equipment.

Personnel access to the Telescope and Enclosure is from the observing floor level. The Telescope and Enclosure also provide internal walkways, ladders, and platforms, for access to different levels of the Enclosure and different locations on the Telescope.

#### 4.1.6. Shops and Labs (SHO)

Shops and Labs (SHO) provides all the shop and lab facilities needed for MSE operations, within the OB and PIER. SHO includes facilities, such as mechanical workshops; a shipping and receiving area; a refuse and transfer area; an engineering lab facility; a spectrograph room; a summit control room; a computer server room; a mirror-coating lab; and storage space for general supplies, specialized equipment, handling fixtures, etc. Laboratories and shops will be supplied with basic furnishings and equipment for handling and servicing electronics, mechanics, and optics, and storage cabinets to house lab supplies and consumables.

The summit control room has all the capabilities needed to operate the observatory. However, it is only intended as a daytime control room for engineering development, maintenance, and servicing activities.

#### 4.1.7. Other Spaces and Facilities (OTSp)

The Other Spaces and Facilities (OTSp) subsystem provides other spaces that indirectly support observatory operations at the summit. These include conference rooms, staff areas, washrooms, and offices. The OTSp includes all remaining rooms and spaces that are not classified as shops or labs, including a staff lounge and kitchen, offices, a conference room, washrooms, janitorial space, a building security system, an intercom, the summit entrance, the parking area, and general furnishings and equipment.

The OTSp also includes the reorganized Waimea headquarters, which supports and manages the MSE operation. The observatory control room at headquarters has all the capabilities needed to operate the observatory remotely.

## 4.2. Enclosure (ENCL)

### 4.2.1. Enclosure Overview

The Enclosure (ENCL) was designed to a conceptual level by Dynamic Structures Ltd.[ii] in Canada, based on the project's design choice, discussed in Section 3.4.1.

The ENCL is a structurally efficient, Calotte-style dome, composed of two major structural assemblies: base and cap structures, which rotate independently. The Calotte enclosure is shown in Figure 32, with its major components highlighted. The base structure forms the lower portion of the enclosure, rotates about the azimuth (vertical) axis, and contains the ventilation modules. The shutter is a fixed circular structure, attached to the top ring girder of the base structure and recessed such that it does not interfere with cap rotation.

The cap has an aperture opening, which rotates about an inclined cap axis (Figure 32, right-hand panel). The cap rotates with respect to the base, to open the shutter. When the shutter and aperture opening are aligned, the aperture is in the closed position and (together with the closed vents) it forms a weathertight "building envelope," which protects the internal components.

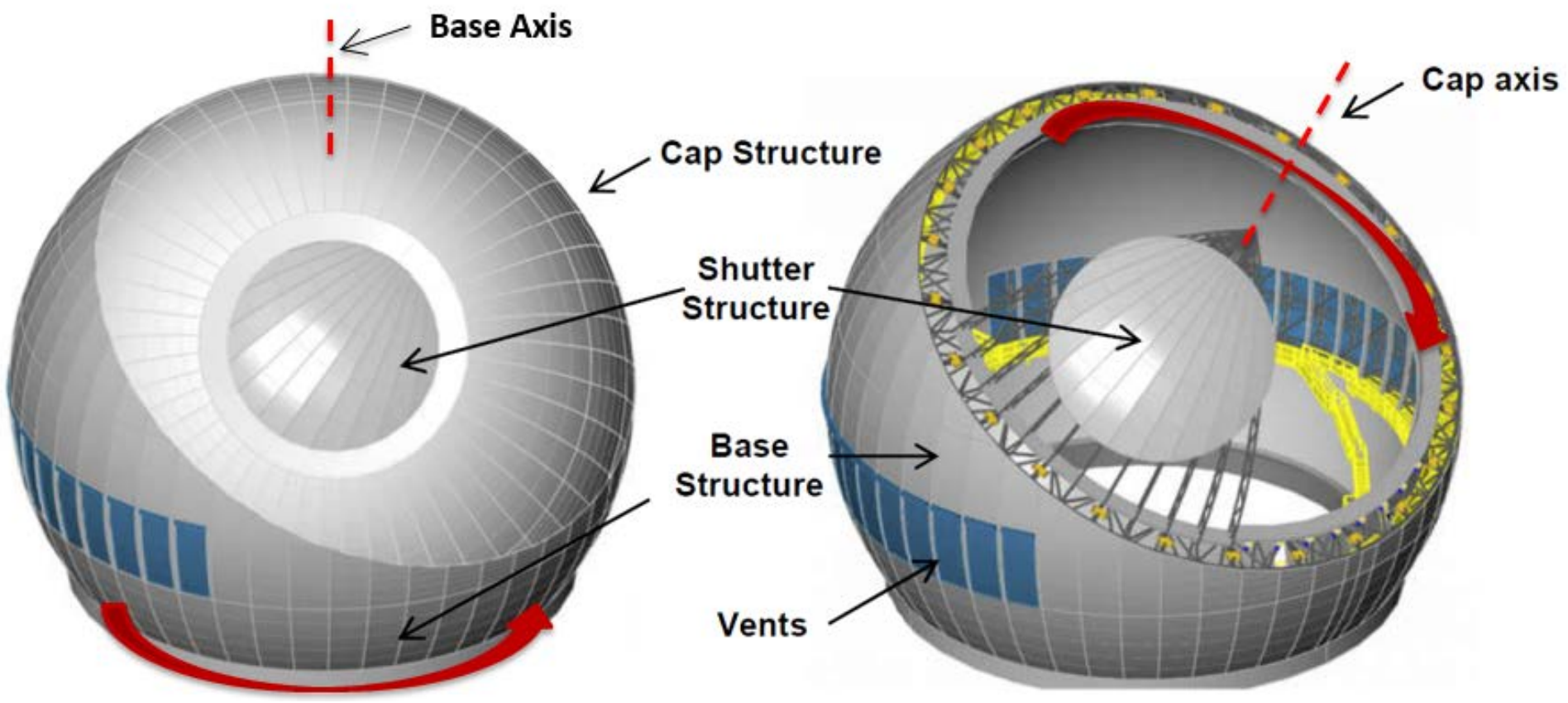

*Figure 32: Enclosure with major components shown, with cap structure (left) and without (right).*

The ENCL is supported by the Outer Building, with the azimuth ring girder as the interface plane. The combined "building" height of the Enclosure and Outer Building, and the Enclosure diameter closely conform to the project's self-imposed limit of within 10% of the exterior dimensions of CFHT. This translates into an enclosure radius of 18.4 m and an overall height above the ground of 42 m. The diameter of the Outer Building will remain unchanged.

The ENCL also includes handling and access infrastructure, including multi-purpose cranes; a dedicated crane for the lifting primary mirror segments; an access platform for the top end of the telescope; and walkways and stairways, which permit access to major enclosure subsystems, such as the vents, bogies, and shutter.

### 4.2.2. Enclosure Functions

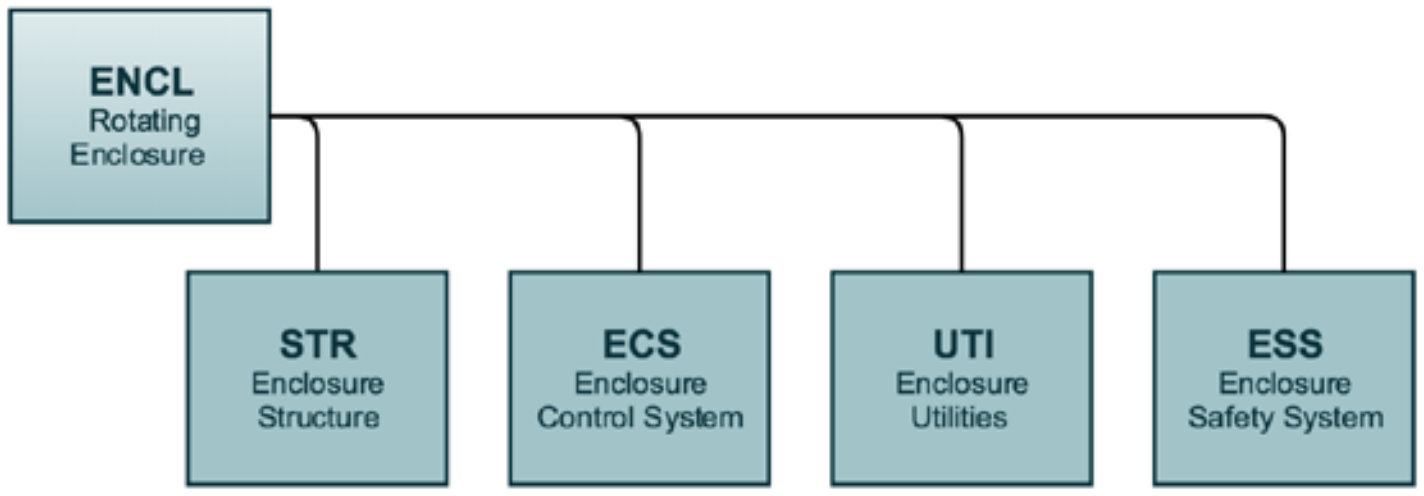

*Figure 33: Enclosure PBS*

The ENCL performs a multitude of functions, including protecting the facility; facilitating observations through the aperture; minimizing dome and mirror seeing; protecting the telescope from wind buffeting; facilitating observatory maintenance and operations; and ensuring a safe environment for all observatory personnel. These functions are realized through the following four products of the Enclosure subsystem (Figure 33):

- Enclosure Structure (STR) contains the structural, mechanical, and electrical "hardware" to enable the aforementioned functions, and a building envelope, which minimizes heat influx and air infiltration.
- Enclosure Control System (ECS) provides command and monitoring capabilities for the drive system, to enable the opening and closing of the aperture, and to follow Telescope tracking during observations and slewing to new astronomical fields.
- Enclosure Utilities (UTI) provides distribution of electricity and data communication; ambient lighting; and lightning protection and grounding.
- Enclosure Safety System (ESS) supervises the Enclosure subsystems, to ensure personnel and equipment safety both locally and globally, in concert with the Observatory Safety System.

In the following sections, the subsystems of ENCL will be described in more detail.

### 4.2.3. Enclosure Structure (STR)

*(1) Structure*

The Enclosure's structural concept uses steel rib-and-tie type truss elements, combined with a welded shell plate (Figure 34). This configuration has been successfully used on many existing enclosures, including CFHT, Keck, and Gemini. The design includes ribs (longitudinal truss elements), ties (truss elements that connect the ribs), ring girders (circular trusses connecting the rib-and-tie trusses at each end, to form partial spherical structures such as the base and cap), and skin plates (welded exterior shell profiles placed between adjacent rib-and-tie trusses, to create a continuous weathertight surface). Insulation is provided by cladded insulated panels, supported by the inner chords of the ribs. This creates a vented interstitial space between the outer skin plates and insulation panels (Figure 30).

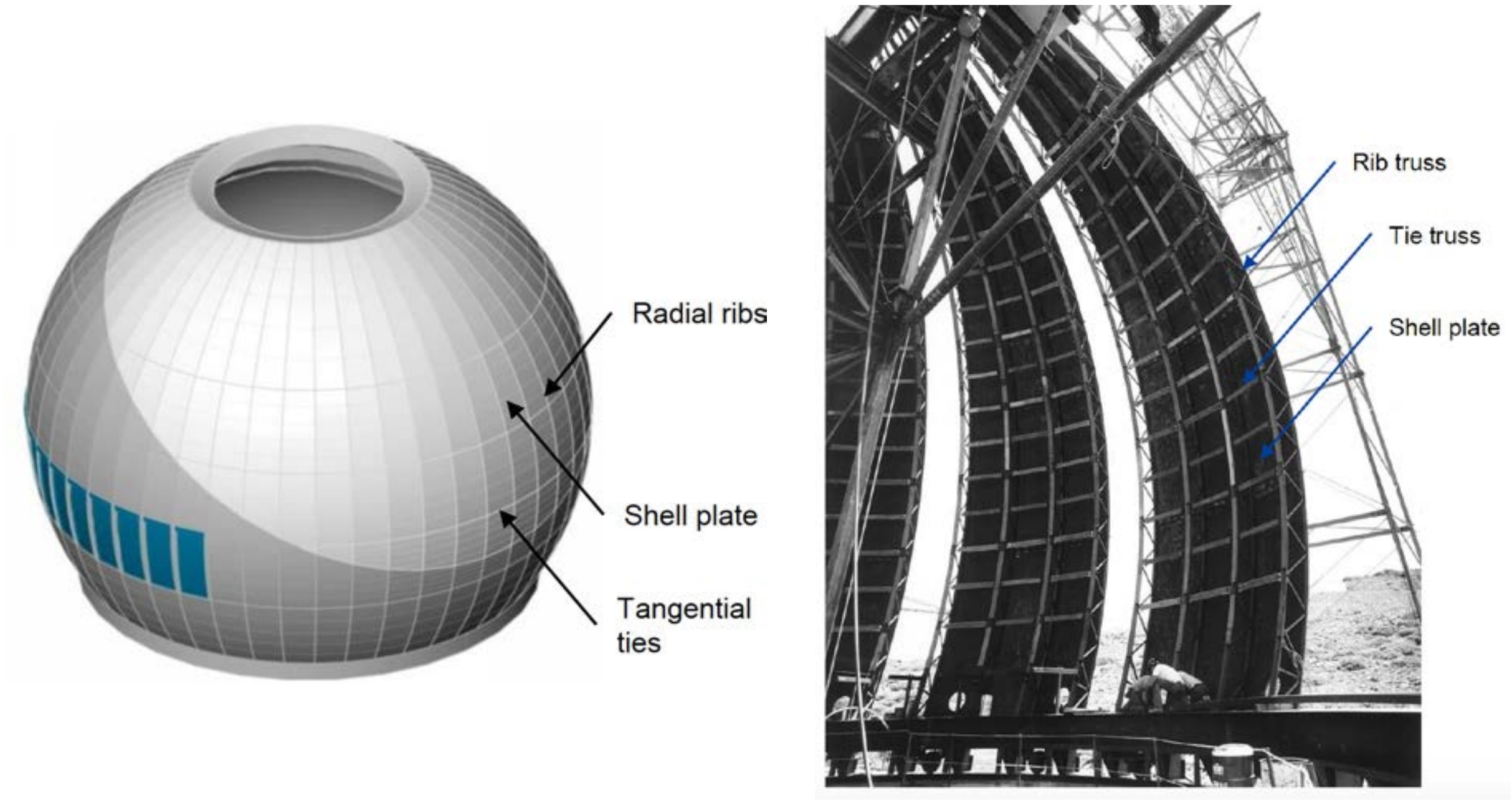

*Figure 34: Enclosure rib and tie structure—MSE (left) and CFHT (right), under construction.*

The structural design has been validated by finite element analysis (FEA) under load case combinations of dead load, environmental loads (wind, snow, and ice), and seismic load for, the Maunakea location, according to the ASCE 7–10 methodology. Under maximum seismic load, major structural members are not allowed to plastically deform in order to preserve the weathertight building envelope, structurally and mechanically. This requirement is a crucial constraint on ENCL mass, since local soil bearing capacity must be taken into consideration.

The results of the FEA have also informed the design of the base and cap drive systems and their bogie loads, drive motor capacities, suspension dynamics characteristics, etc.

*(2) Shutter*

The aperture is positioned in azimuth and zenith by the combined rotation of the base and cap structures about their respective axes. Figure 35 shows the aperture zenith positions.

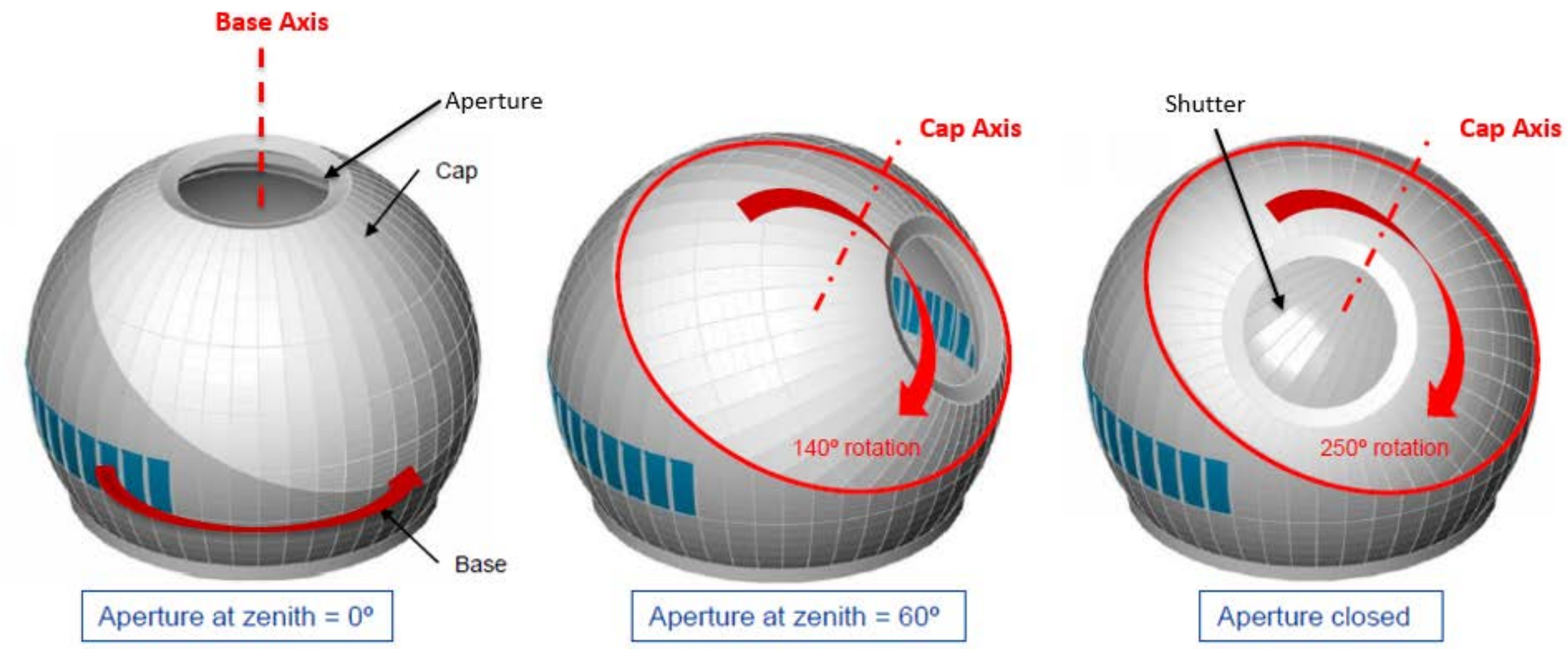

*Figure 35: Enclosure aperture positions by cap rotation.*

The aperture is an unobstructed 12.5 m diameter opening, which can reach any zenith angle between 0° and 60° and all azimuth directions, by rotating the cap relative to the base. The fixed shutter eliminates the need for a shutter rotation system and reduces the complexity and mass of the ENCL. However, this also means that the cap cannot be rotated independently when the aperture is closed. We therefore cannot use cap rotation to mitigate ice and snow accumulation on top of the dome during snowstorms. The Enclosure faces the same problems as traditional spherical domes, which utilize azimuth rotation to mitigate ice and snow accumulation. During winter storms, residual ice and snow may build up on the shutter and fall into the Enclosure when it is opened. We recognize this as a potential hazard: finding ways to mitigate this problem will be an on-going consideration during the next design phase.

The shutter is sealed using an inflatable seal, located on the shutter perimeter (Figure 36). Its main components are a membrane, bladder, and tensioners. A secondary lip seal and drainage system collect any leakage from the primary seal, to keep it away from the dome. Ice deflectors on the lower portion of the shutter prevent falling ice from directly impacting the inflatable seal membrane. The bladder is inflated using a compressor and reservoir system, controlled by the ECS.

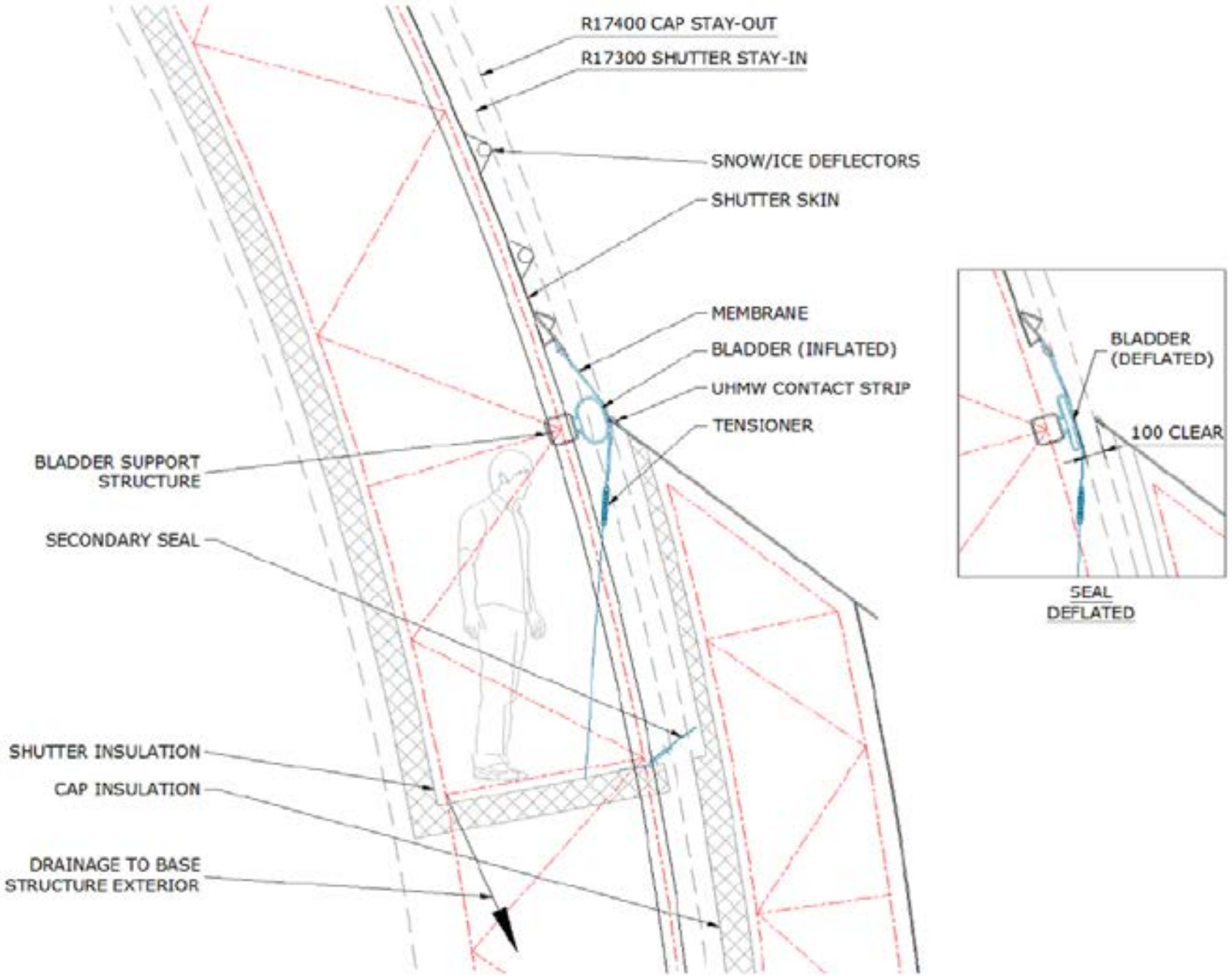

*Figure 36: Shutter seal operation.*

There are concerns about the longevity of inflatable seals in operation at the summit of Maunakea, especially in freezing conditions. A similar seal section has been prototype tested by the TMT project. The MSE Project Office is currently collaborating with the TMT Enclosure Group to evaluate and validate the common seal design.

*(3) Thermal Management*

As stated above, during daytime, the ENCL is air conditioned to the forecasted ambient temperature at evening twilight, using OBF air handling units. It also incorporates low emissivity paint, interstitial space, and insulation, to support the thermal management strategy. In addition, air infiltration is controlled by rotating seals at the interfaces between the Outer Building and the rotating base and between the rotating base and rotating cap. Two types of seals are used: exterior and interior (Figure 37). The exterior seals are designed to provide a weather barrier and the interior seals are designed as thermal and air barriers. Geometrically, they are both labyrinth seals, with overlapping “circumferential walls,” to accommodate axial and radial variations during dome rotation. The sealing surfaces are Teflon coated, to minimize friction and prevent freezing.

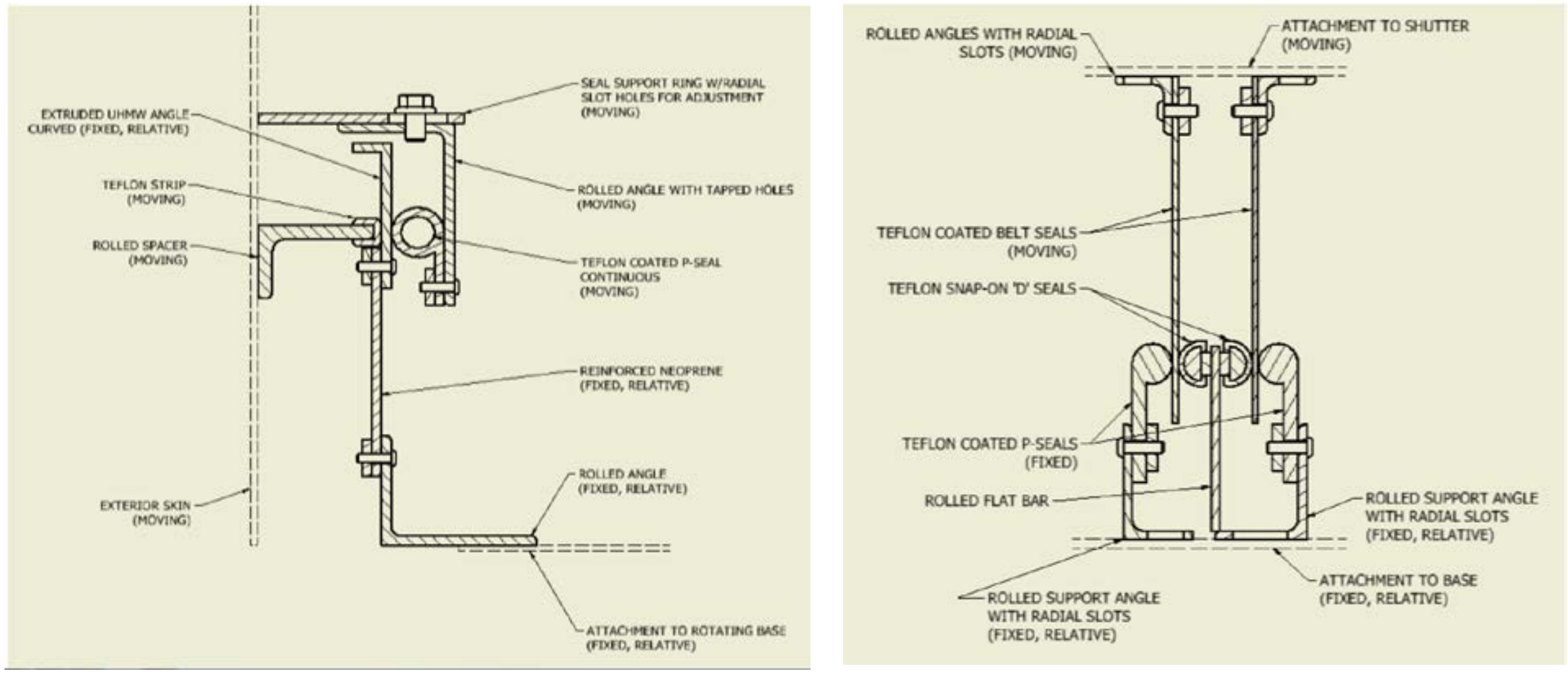

*Figure 37: Enclosure seals—exterior (left) and interior (right)*

A row of 28 vents will be installed on the base structure, with vertically openings centered at the same elevation as the primary mirror. The vents minimize the thermal boundary layer at the optical surface, which causes mirror seeing, and thermal gradient in the optical path, which causes dome seeing. The total projected area of the vents[iii] will be optimized, to provide flushing at low to moderate external wind speeds. The vents have a two-layer door system. The exterior layer is a hurricane roll-up door and the interior layer is a pair of side-hinged, swinging, insulated freezer doors. Informed by the same philosophy as the rotating seal design, the hurricane door provides wind protection up to survival wind speed, and the freezer doors prevent heat and air infiltration. This is also a common design in the TMT project, though the MSE vent openings are smaller. Both sets of doors are commercially available. The roll-up door is the same type currently used for the CFHT vents.

### *(4) Access and Handling*

The ENCL has access walkways in the interstitial space, to allow observatory staff to service the vent modules, cap bogies and drives, cranes, and top-end servicing platform within the interior of the dome (Figure 38). Cap drives are located at the base of the cap walkway, where it joins the vent walkway, for easy access.

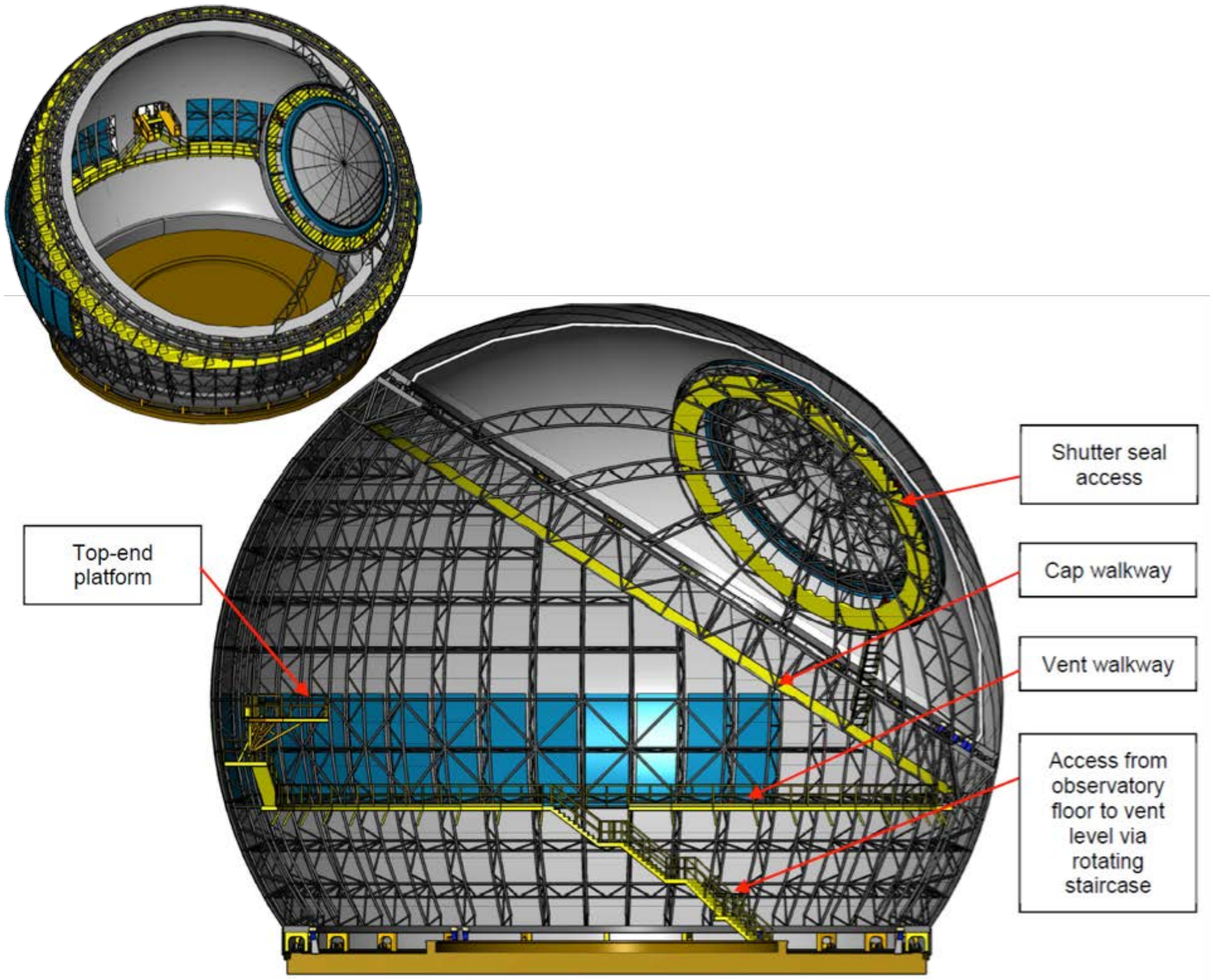

*Figure 38: Enclosure access walkways and platforms*

Access to the perimeter of the shutter is provided by a circular stairway. A top-end service platform provides access to the top-end components of the telescope at horizon pointing, to facilitate maintenance and servicing operations. The geometry of the service platform has yet to be finalized, pending a decision on the interfacing top-end telescope geometry. Figure 39 shows an earlier concept of telescope structure geometry.

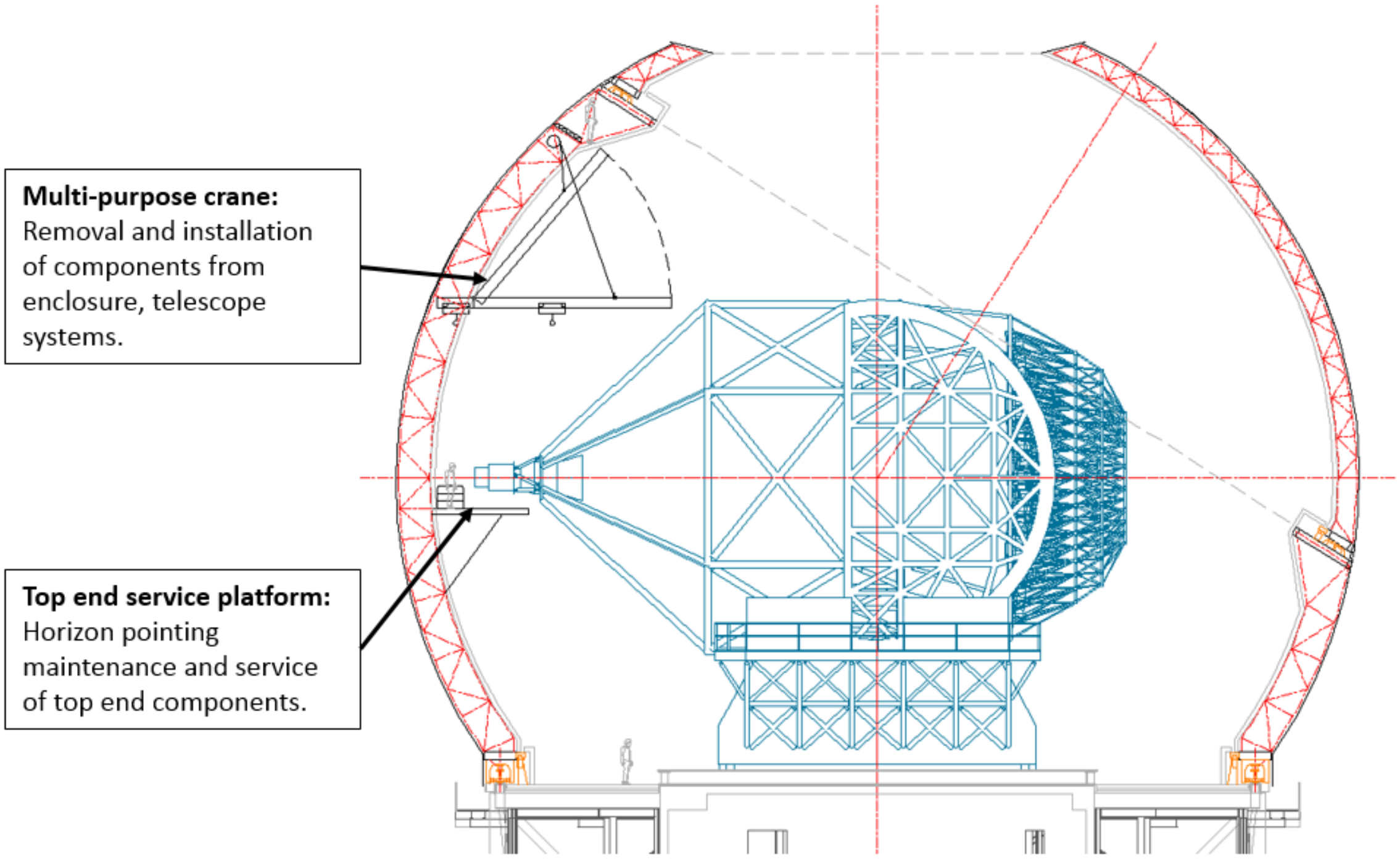

*Figure 39: Top-end service platform enables access to the telescope's top end at horizon pointing*

The enclosure's multi-purpose crane facilitates the removal and installation of components of the enclosure, telescope structure, and telescope-mounted systems, such as the primary mirror assemblies, top-end components, optical test equipment, spectrographs, etc.

### *(5) Mechanisms*

There are two drive and bogie systems: one for base (azimuth) rotation and one for cap rotation.

The azimuth drive and bogie system have been adapted from the current CFHT friction drive system. A total of four drive assemblies are proposed, each equipped with a pair of rubber tire drive wheels, (Figure 40). There is one extra assembly to provide redundancy. The drive assemblies are mounted on the observatory floor, next to the enclosure azimuth track. The CFHT azimuth bogies have been adapted for MSE use. To compensate for the higher enclosure mass, 48 units are evenly mounted on the rotating base azimuth ring girder. Azimuth rotation is designed to operate with one missing bogie.

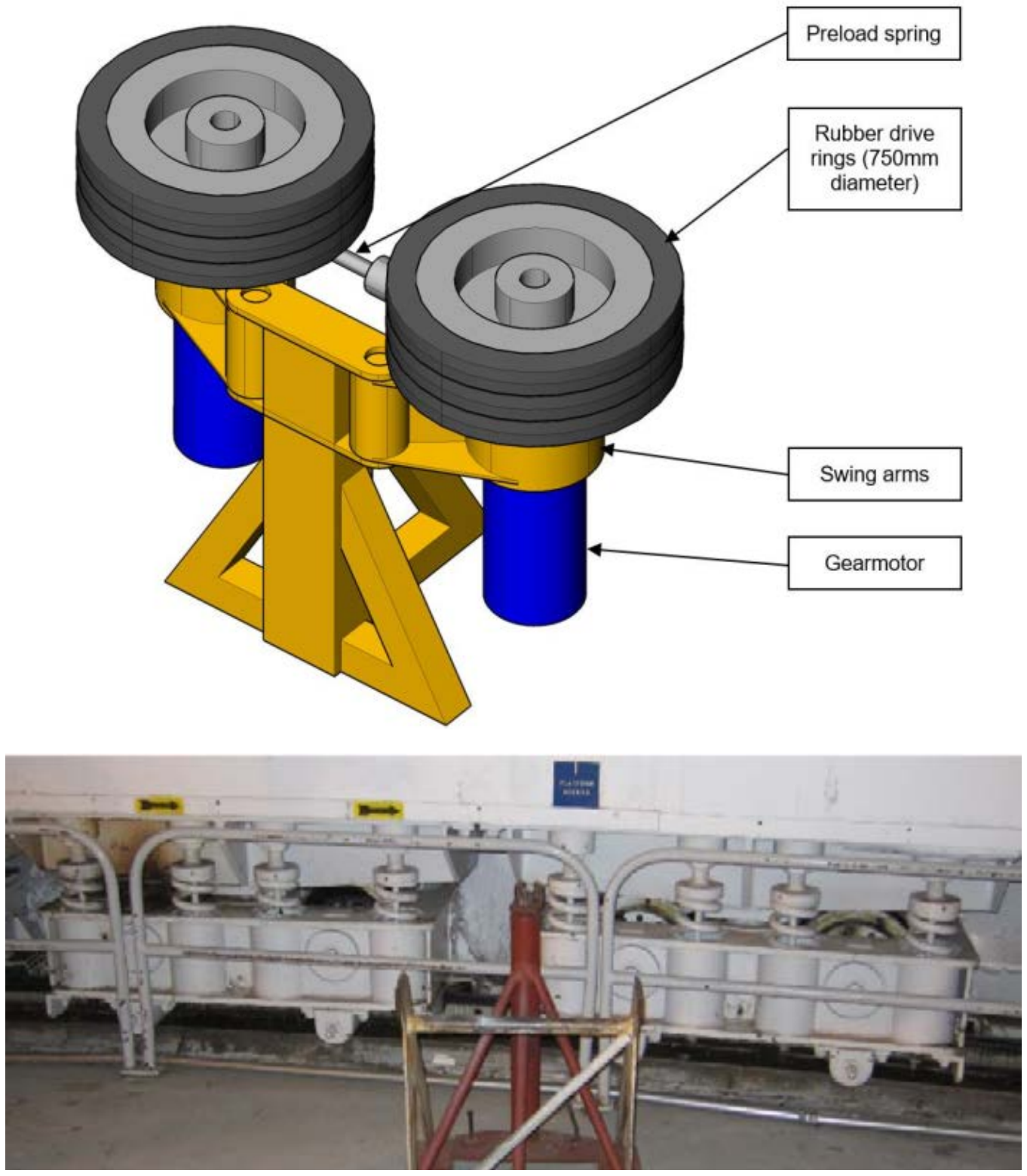

*Figure 40: Azimuth mechanisms—drive wheels (top), CFHT bogies (bottom)*

Similarly, the cap structure is supported by cap bogies, located on the inclined cap–base interface plane. The cap bogies support the weight of the cap in both normal and radial directions with respect to the inclined plane. Bogies are attached to the base structure and the drive bogies have drive units mounted to their frames (Figure 41). Seven drive units have been fitted, including one extra unit, to provide redundancy. Cap rotation is designed to operate with one missing bogie. Brakes and gear reducers have been sized for survival holding torque requirements, which are governed by the unbalanced ice load. In other regards, the cap structure is mass balanced, without ice and snow.

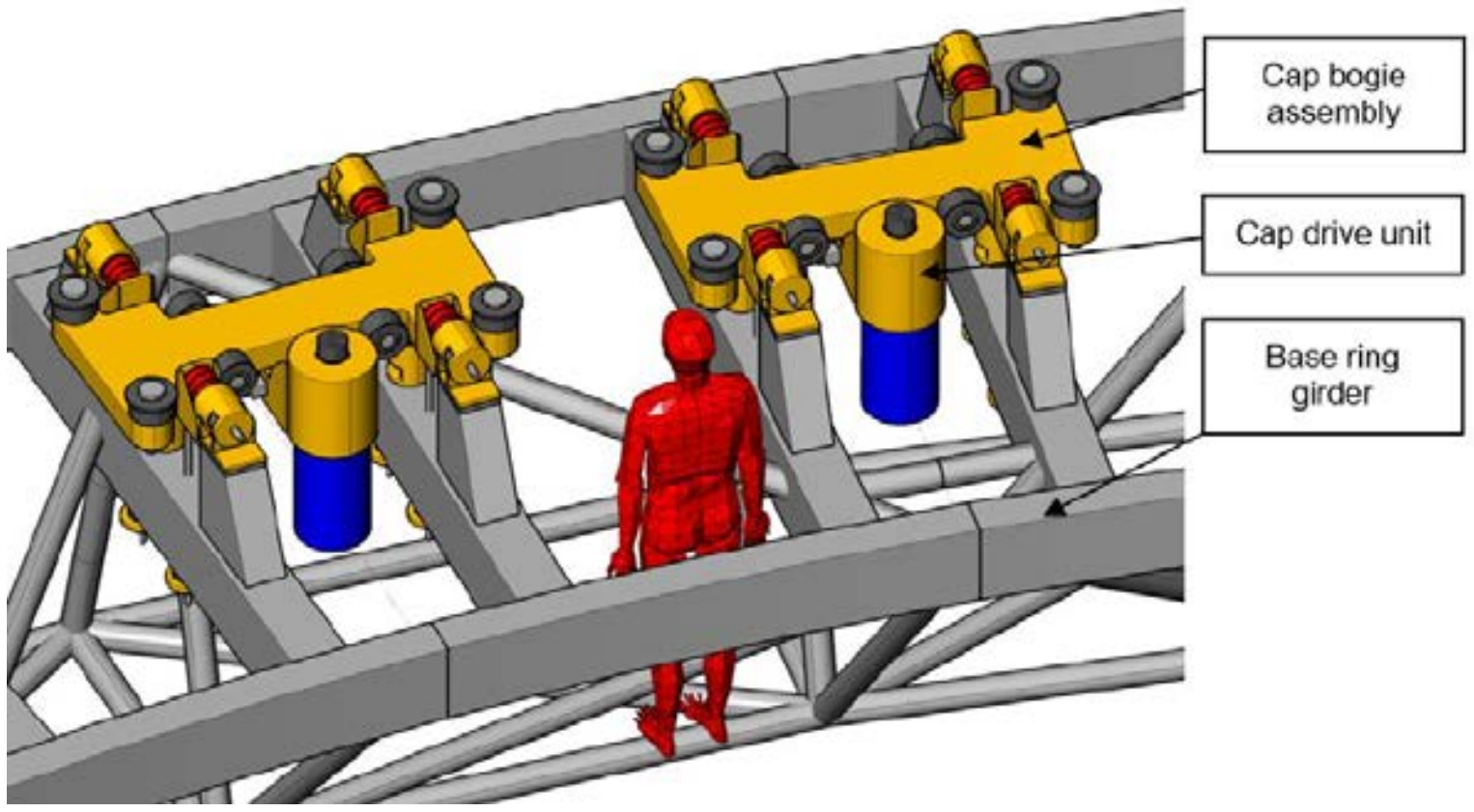

*Figure 41: Cap drive bogies attached on base structure top ring girder.*

Vent doors are operated remotely, using individual drive systems for each door. Exterior doors use a roll-up “reel drive” arrangement supplied by the vendor. Interior doors are commercially available freezer door pairs, with retrofitted drive mechanisms on each module.

The Enclosure is capable of opening or closing (and sealing) its shutter and vents within two minutes and can slew between observing fields anywhere on the sky within three minutes, to an accuracy of 10 arcminutes. It is also capable of continuous and intermittent moves,[9] to reposition the aperture opening during observations, and capable of rotating—albeit at a reduced rate—when partially covered by snow or ice.

### 4.2.4. Enclosure Control System (ECS)

The Enclosure Control System (ECS) includes the necessary motion control hardware and software, including drive electronics and insulated thermal enclosures. The ECS controls base rotation, cap rotation, ventilation modules, multi-purpose cranes, top-end service platform deployment, and enclosure-mounted lighting.

The ECS provides command and communication lines internal to the Enclosure and receives motion commands from the Telescope Control Sequencer (TCSe) and Facility Control Sequencer (FCSe), and reports its status and health.

Unlike the telescope, which tracks via combined azimuth and elevation motions, the Enclosure tracks using combined base and cap rotations. The TCSe controls the base and cap rotations and the FCSe controls operation of the ventilation doors, while ESS enforces personnel and equipment safety. Local manual control of maintenance operations is provided by the ECS and safety is enforced by ESS.

### 4.2.5. Enclosure Safety System (ESS)

The Enclosure Safety System (ESS) is a hardware-based safety system with lock-outs and e-stops, which enforces personnel and equipment safety. This product contains all the associated equipment, hardware, and communication lines required to support the overall architecture of the Observatory Safety System (OSS).

[9] It may be desirable to execute small, intermittent repositioning moves, instead of continuous motion, to avoid producing vibration during observations. The aperture opening is oversized to support intermittent moves.

The ESS is the safety “supervisor” of the ECS. The ESS operates autonomously within the ENCL and in concert with the OSS, to ensure local and global safety. Safety equipment includes the safety-rated PLC, input/output devices, local and global e-stops, dedicated slip rings, sensors, and wiring required by the ECS and OSS.

## 4.3. Telescope (TEL)

### 4.3.1. Telescope Overview

MSE's Telescope system (TEL) (Figure 42) is an altitude–azimuth configuration, similar to that of the current 8- to 10-m class telescopes, supported by the Inner Pier. The overall telescope geometry is dictated by the optical configuration. An azimuth structure provides rotation about the azimuth axis and supports the rest of the telescope system. An elevation structure[10] provides rotation about the altitude axis, and supports both the primary mirror and components located at the top end. A Prime Focus Hexapod System (PFHS) carries the top-end components and provides flexure compensation during operations. The hexapod is attached to the telescope's top end, at the apex of six spider legs.

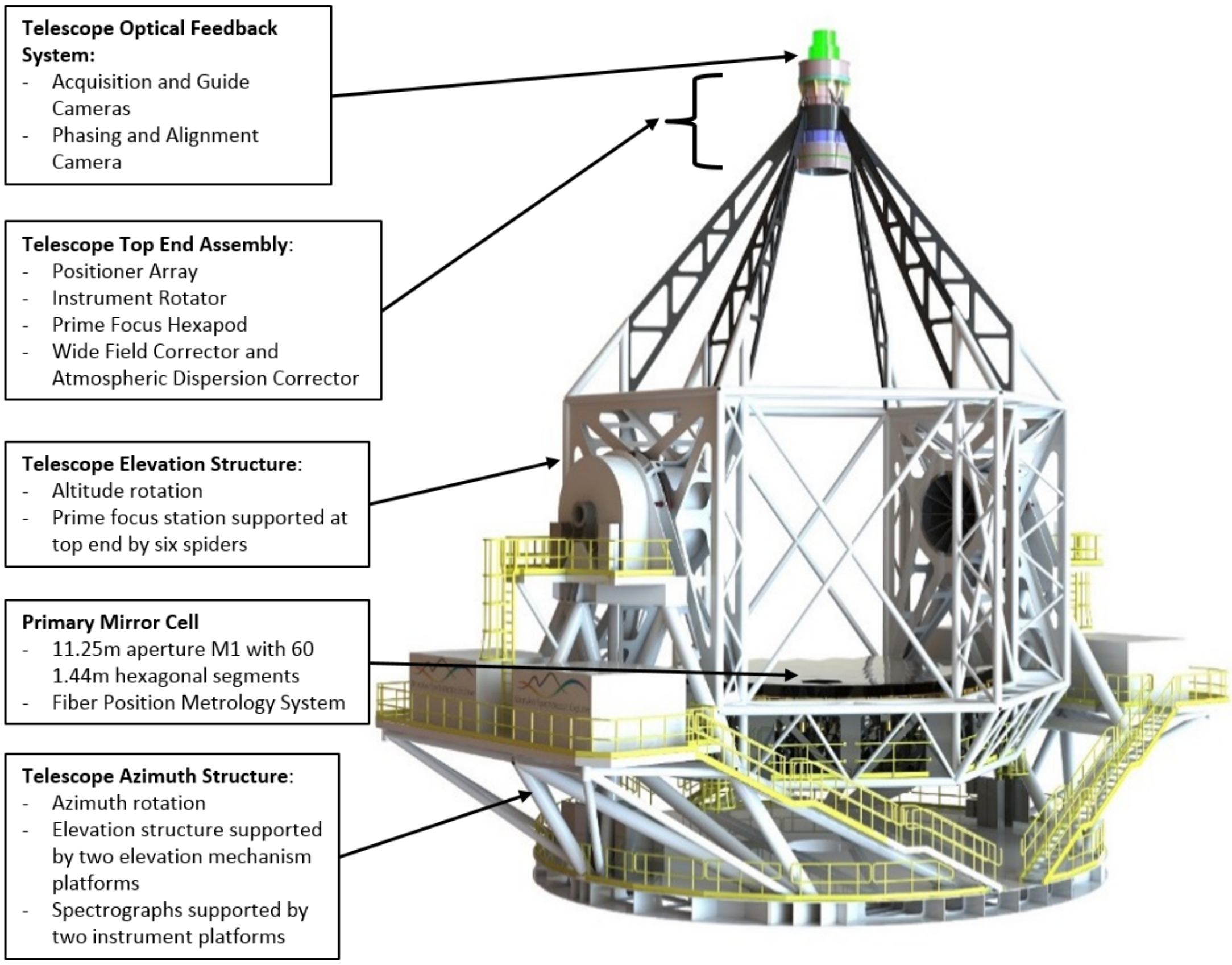

*Figure 42: Telescope—structural components and payload.*

The 11.25 m segmented, primary mirror works optically, in conjunction with a Wide Field Corrector with integrated Atmospheric Dispersion Corrector (WFC/ADC, see Section 4.3.5) to create a prime focus at the top end, where the Fiber Positioner System (PosS, see Section 4.4.3) collects light to send down a Fiber Transmission System (FiTS, see Section 4.4.4) to the spectrographs (LMR and HR, see Section 4.4.5 and 4.4.6). The top-end assembly components are shown in Figure 43.

---

[10] Also known as the altitude structure

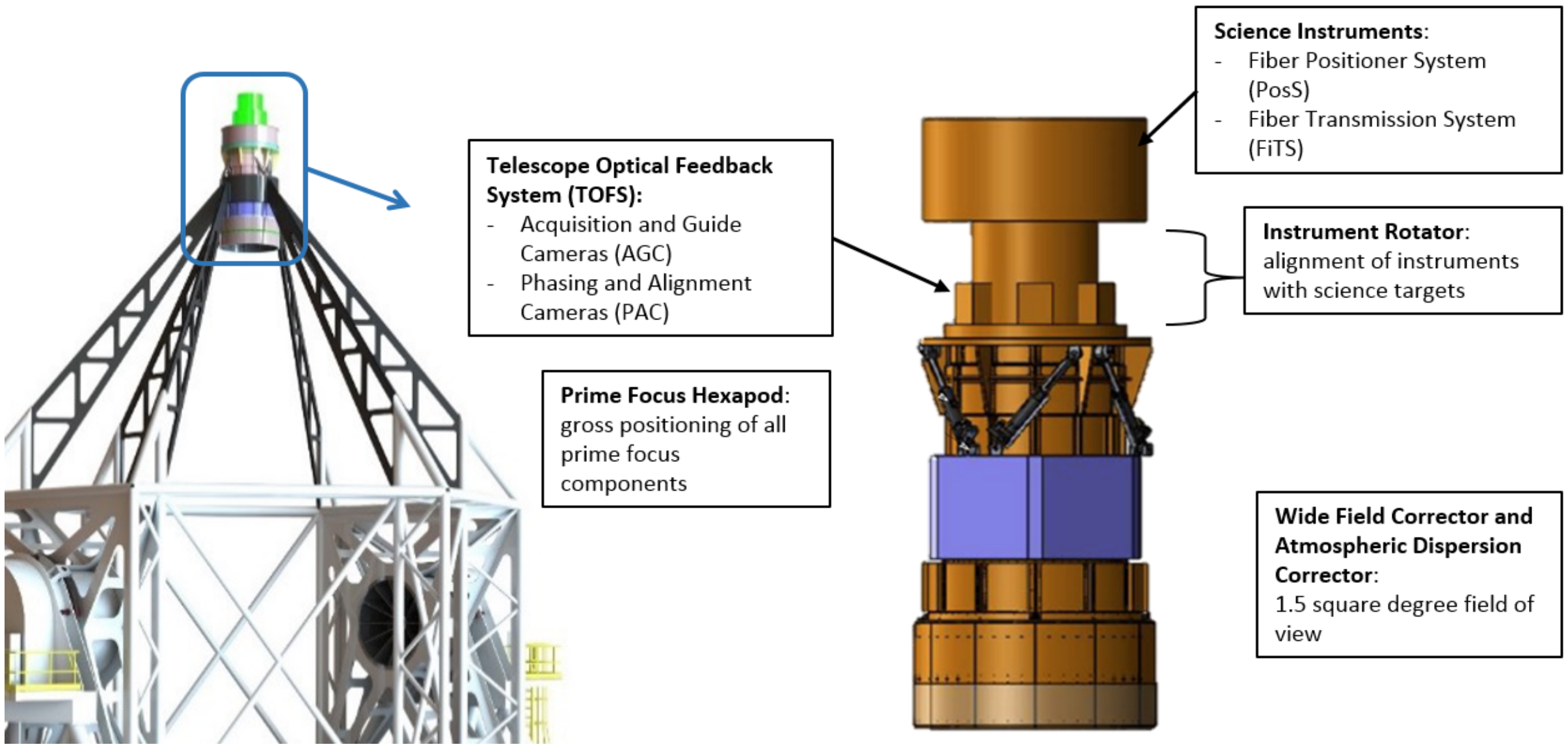

*Figure 43: Top-end assembly components.*

The primary functions of the telescope structure are to maintain alignment of the optical system, and point, track, and guide, as commanded. These functions are aided by focus and pointing models and lookup tables (LUTs), to correct for telescope flexure due to gravity and temperature changes, and to allow for fabrication and installation tolerances. The models and LUTs will be developed during the commissioning phase and refined over the lifetime of the observatory. The telescope also includes a Telescope Optical Feedback System (TOFS), a Phasing and Alignment Camera (PAC) system, and an Acquisition and Guide Camera (AGC) system. They are located at the top-end, prime focus station and measure image quality and provide pointing and closed-loop guiding feedbacks.

The telescope layout is constrained by the Enclosure. A telescope stay-in volume has been imposed, to maintain a 300 mm spherical clearance zone (Figure 44) around the interior of the Enclosure, to prevent collisions and eliminate pinch hazard.

The optical design[iv] was developed by Will Saunders at the Australian Astronomical Observatory (AAO), based on the throughput and image quality requirements, and taking advantage of the natural seeing at the Maunakea site to provide the best system sensitivity (see discussion in Section 3). The primary mirror directs light to the WFC, which has an integrated ADC. This consists of five lenses, combined into one optical ensemble (Figure 45). The telescope optical design delivers a convex focal surface with a radius of curvature of 11.33 m and an optical path length of 19.1 m, measured from the vertex of the primary mirror to the focal surface.

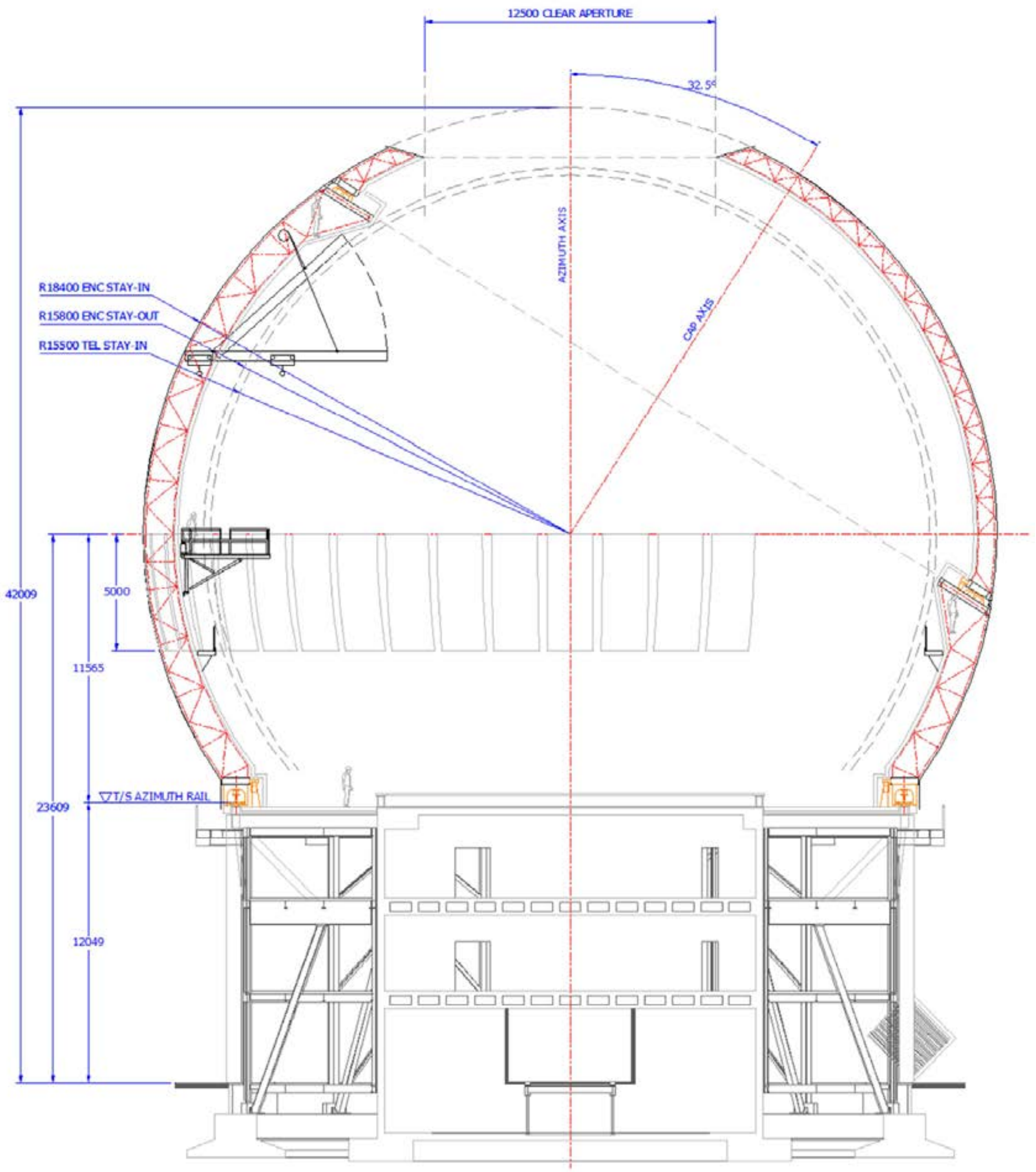

*Figure 44: Telescope stay-in, $R_{TEL}$=15.5 m, and Enclosure stay-out, $R_{ENCL}$ = 15.8 m, radii.*

The overall length of the telescope, including top-end components, is limited to less than 22 m, the maximum length that will fit inside the enclosure with sufficient wind protection. Shorter focal lengths are not desirable, as they require faster primary mirrors, which are more challenging to fabricate and align. Similarly, the image quality requirements on the WFC/ADC are considered attainable in the context of tolerancing, taking into account fabrication errors (figure errors, glass homogeneity) and alignment errors (static from the assembly process and dynamic due to flexure during operations), both for the optical ensemble and for individual lens elements.

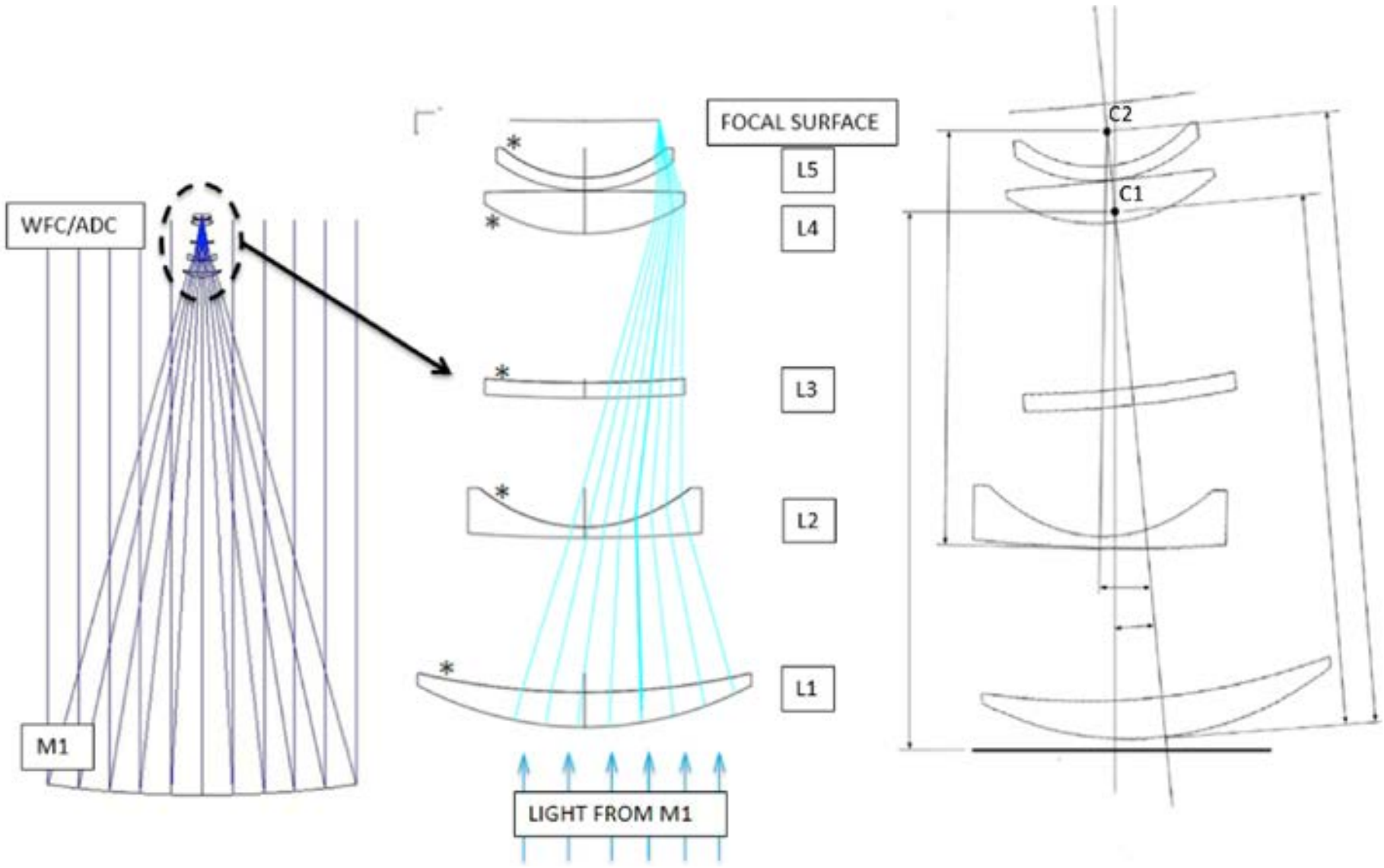

*Figure 45: M1 and WFC/ADC (left); WFC/ADC (center)—Lens 1 to Lens 5, with Lens 2 providing atmospheric dispersion correction; ADC action (right)—the entire WFC rotates about the point C1, as the global tilt via PFHS, while L2 rotates about point C2.*

Atmospheric dispersion correction is provided through the combination of three coordinated actions, as illustrated in Figure 45:

- Second lens (L2) of the WFC/ADC is given a lateral shift and a tilt;
- Entire WFC/ADC ensemble (including L2) is tilted on the Y–Z plane as a rigid body, using PFHS;
- Telescope repoints.

As a baseline, the actions of L2 and the hexapod are set at their "average" optical location at the beginning of an exposure and are not adjusted during the exposure. Plate scale changes have been incorporated into the atmospheric dispersion correction. This halves the differential atmospheric refraction across the field of view over the observing zenith distance range, hence allowing us to maintain the system injection efficiency, without having to reposition the fiber positioners during long exposures with large zenith variations.

Rotation of the field of view is intrinsic to the alt–az mount. To avoid introducing additional optical components when we derotate the field of view (e.g. dove prism, K-mirror), the telescope rotates the payload at the focal surface physically, rather than rotating the focal surface optically. This maximizes system throughput, at the expense of moving more mass at the prime focus.

The delivered field of view is 1.52° diametrically at the prime focus, with a focal ratio of f/1.926. At an average plate scale of 106.7 microns/arcsecond, this translates to a circular field of view of 584 mm in diameter. An inscribing hexagon with a 1.5° square area is defined as the science field of view, in which light is collected by the fibers. The remaining six sectors at the edge of the field of view are available for AGC, as shown in Figure 46. If the field of view for each of the three cameras is 15 mm x 15 mm (142″ x 142″) then guide stars as faint as $i$ = 19.5 mag are available in all three cameras for 95% of the sky near the North Galactic Pole.

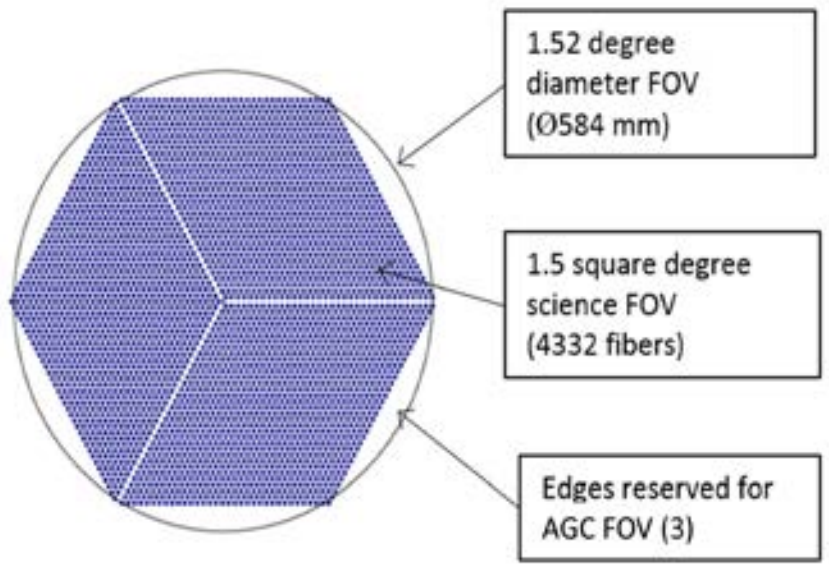

*Figure 46: Field of view at the prime focus.*

Gravity and thermal expansion lead to structural deformation of the telescope, which changes the alignment of the primary mirror (M1) and WFC. To compensate, the hexapod mechanism adjusts through five degrees of freedom at the beginning of every observation, thus correcting for focus, decenter, and tilt errors of the focal surface payloads. The mechanism includes the overall barrel assembly containing the WFC/ADC optical ensemble. The hexapod moves the payloads to the optimal optical focal surface, according the LUTs, focus model, and ADC settings. This maintains the image quality of the optical system, by ensuring that the WFC is aligned with respect to M1.

An Instrument Rotator (InRo) provides rotation of the field of view for guide cameras and science instrument components located at the top end. A phasing and alignment camera for the primary mirror is also mounted at the top end.

The overall telescope configuration supports observations from a 0°–360° azimuth angle and a 0°–60° zenith angle. However, the azimuth structure is able to rotate continuously from -270°–+270°, without unwrapping the azimuth cable wrap. This allows for greater observing efficiency. Maintenance modes are supported at 0° and 90° zenith (as shown in Figure 44), allowing for intermediate positions if required. The telescope will be stowed at a zenith angle of 90°, pointing horizontally. Cable wraps are included to accommodate full altitude and azimuth ranges of motion.

### 4.3.2. Telescope Functions

TEL is organized into products, based on their specific functions, as shown in Figure 47. TEL is a complex set of subsystems. Its primary purpose is collecting and transmitting light to the instrument at the prime focus. This requires the participation of all of the telescope subsystems, working together to ensure science observations are conducted efficiently.

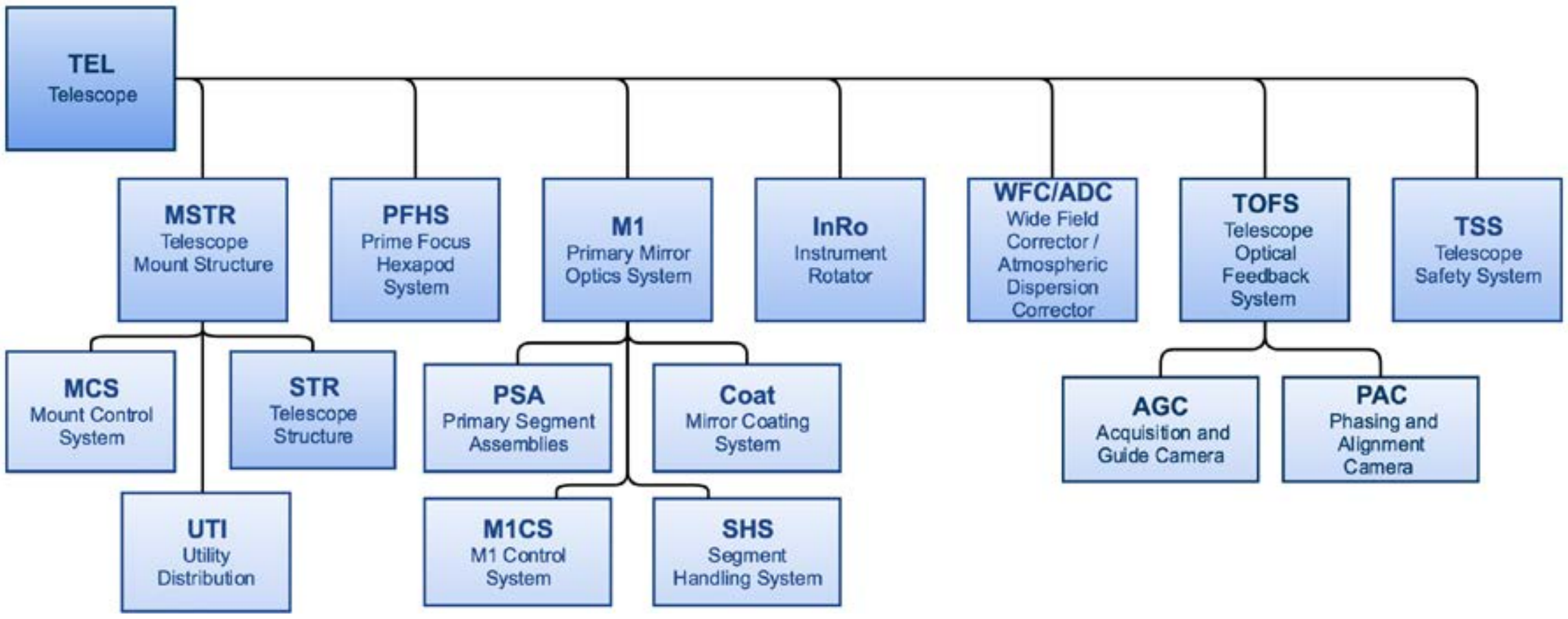

*Figure 47: Telescope Product Breakdown Structure*

The different elements of the PBS structure and their functions are:

- Telescope Mount Structure (MSTR) is the backbone of the telescope, enabling its subsystems, including telescope-mounted subsystems, to function.
  - Telescope Structure (STR) provides structural support for other telescope subsystems.
  - Mount Control System (MCS) points the telescope in altitude and azimuth directions, as commanded.
  - Utility Distribution (UTI) distributes utilities received from the OBF to the subsystems mounted on the telescope, using cable trays and conduits to transmit power, utilities, control, and communications, and cable wraps to accommodate motion of the telescope about its axes.
- Primary Mirror Optics System (M1) collects light and reflects it to the top end.
  - Primary Segment Assembles (PSA) maintains mirror segment position and segment surface figure, based on commands for the M1 Control System (M1CS).
  - Mirror Coating System (Coat) provides the ability to coat mirror segments as needed
  - Segment Handling System (SHS) enables safe exchange of individual segments from the telescope structure and through the observatory, to the coating chamber.
- Wide Field Corrector/Atmospheric Dispersion Corrector (WFC/ADC) provides optical aberration correction and atmospheric dispersion correction for light coming from the primary mirror, and transmits the corrected light to a focal surface at the telescope's prime focus over a 1.5 sq. degree field of view.

- Prime Focus Hexapod System (PFHS) provides rigid body motion of its payload, including the WFC/ADC, InRo, AGC and PAC systems, PosS and FiTS at the top end of the telescope. This action maintains optical alignment of the assembly, especially the WFC/ADC, with respect to M1. InRo provides support and rotational positioning of the PosS/FiTS assembly and an ACG system, to track the observing fields on the sky during observations. The InRo service cable wrap provides mounting, and protects cables and other conduits over its range of motion.
- Telescope Optical Feedback System (TOFS) contains two sets of cameras, which provide control information to relevant TEL subsystems such as MCS, M1CS, InRo, and PFHS, depending on the final MSE observatory control architecture. TOFS also provides the telescope system with feedback on the rigid body aligning errors of the WFC/ADC with respect to M1, and is corrected by the PFHS.
    - Acquisition and Guide Cameras (AGC) system uses guide star positions in the field of view to enable MCS and InRo to correct for guiding errors in real time during observations.
    - Phasing and Alignment Camera (PAC) system measures the position errors of mirror segments in piston, tip and tilt, and the surface figure errors of the mirror segments. These errors are communicated to M1CS, in order to control the actuators and warping harnesses and maintain delivered image quality.
- Telescope Safety System (TSS) supervises the TEL subsystems, to ensure personnel and equipment safety both locally and in concert with the global Observatory Safety System.

The following sections describe the subsystems of TEL in detail.

### 4.3.3. Telescope Mount Structure (MSTR)

The MSTR, designed by IDOM[v], in Spain, consists of a large steel structural assembly and a control system to provide precise motion control and distribution of utilities (electrical power, cooling, etc.) The telescope comprises three separate structures: the central bearing support structure, the azimuth structure, and the elevation structure (Figure 48).

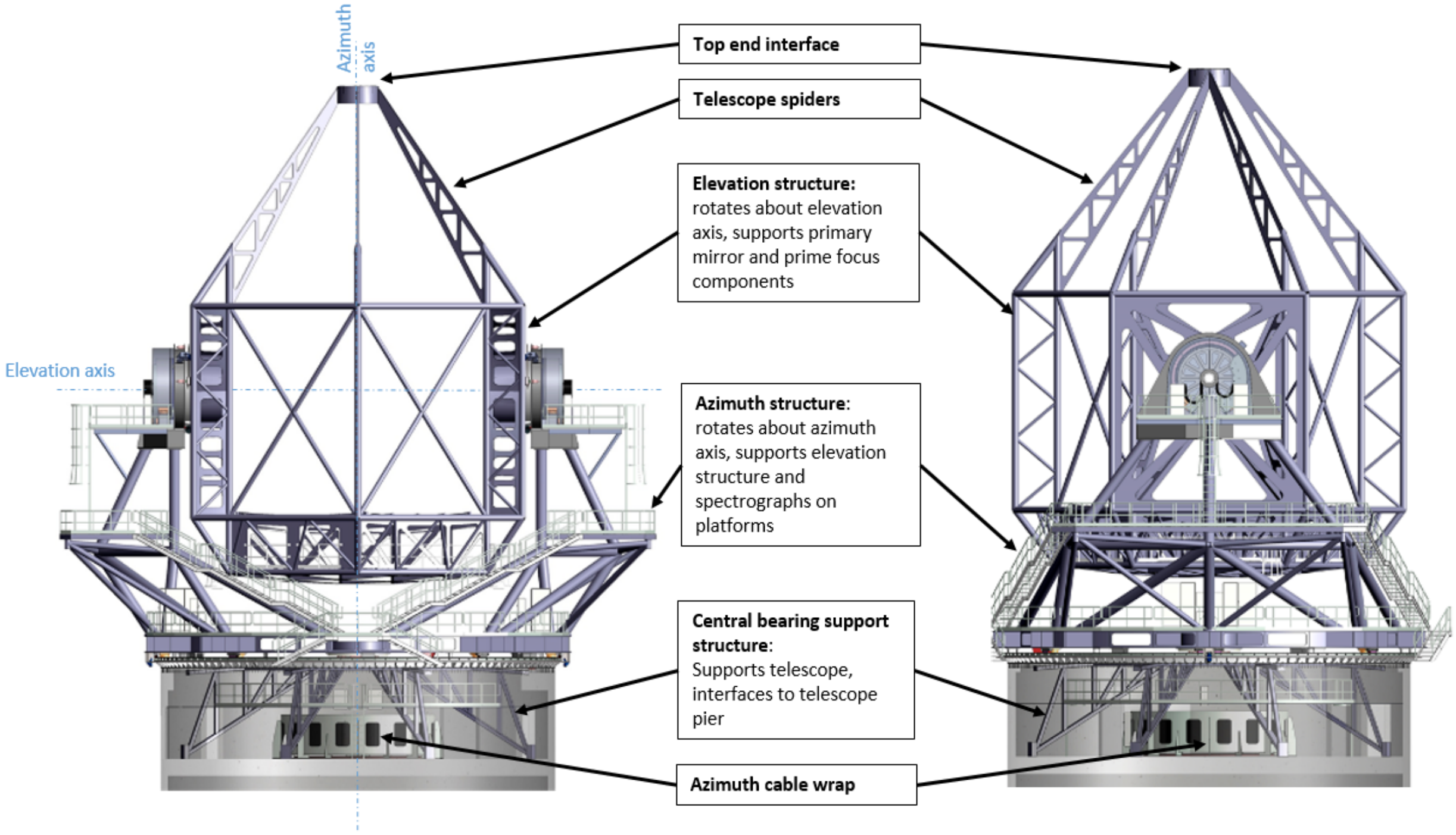

*Figure 48: Telescope Mount Structure (MSTR) layout*

### *(1) Telescope Structure (STR)*

The MSTR design features a "yoke" configuration (Figure 49), in which the azimuth structure resembles a U-shape yoke, whose trunnions support the elevation structure and define the altitude axis. The attributes of two structural concepts—yoke and rocking chair (Figure 50)—were compared. The rocking chair concept involves large "rockers," represented by the large elevation journals, which enable the elevation motion. The trade study compared both designs' structural and aerothermal performances, and anticipated construction costs. For the same structural mass, the yoke configuration performed better both statically—with lower relative deflection along the optical path—and dynamically—with higher modal frequencies. It is also more open to airflow, which facilitates flushing and reduces thermally-induced seeing, and can be constructed at a lower cost, due its modular design.

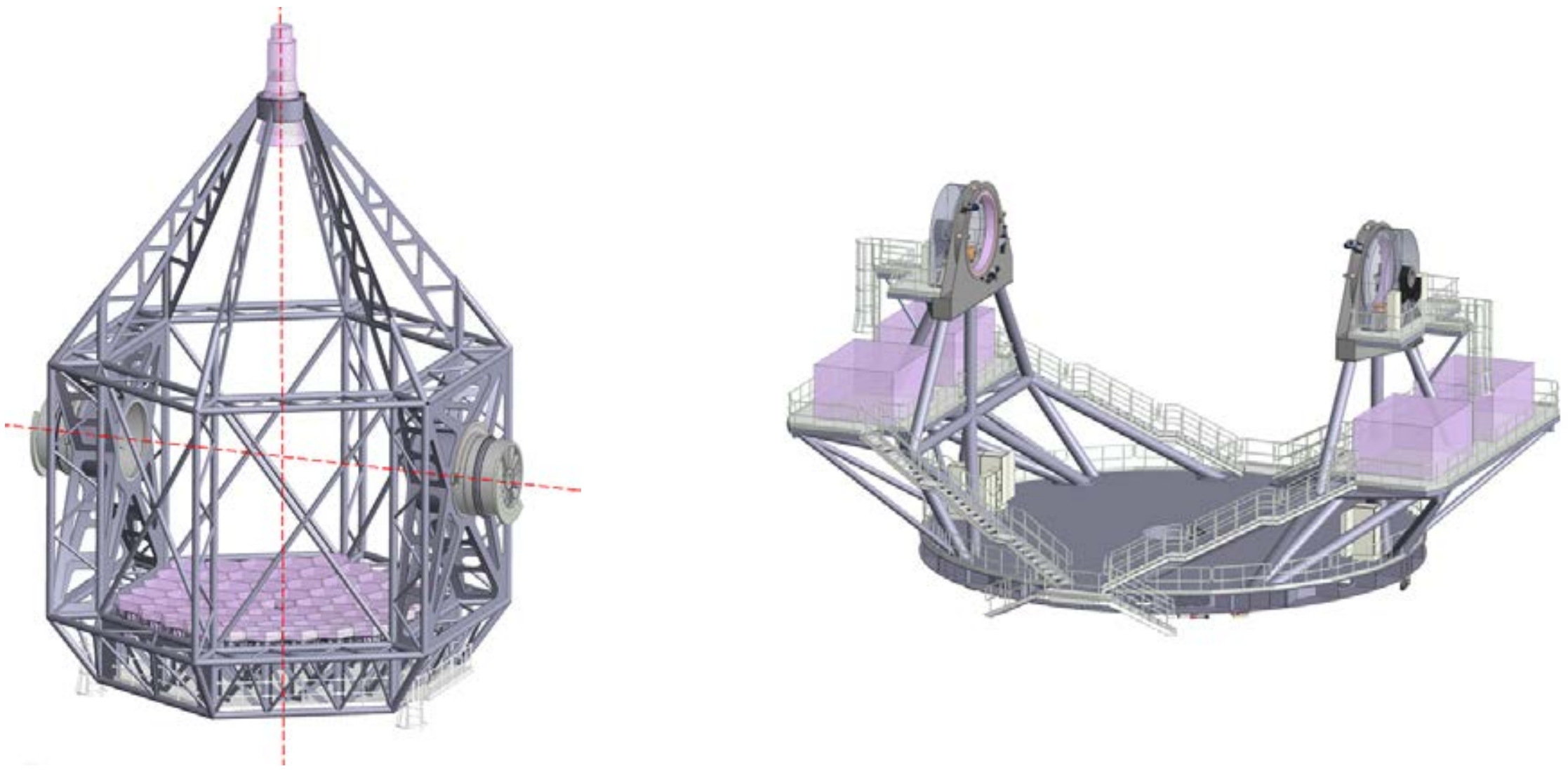

*Figure 49: Telescope mount yoke configuration—elevation structure (left) and azimuth structure (right).*

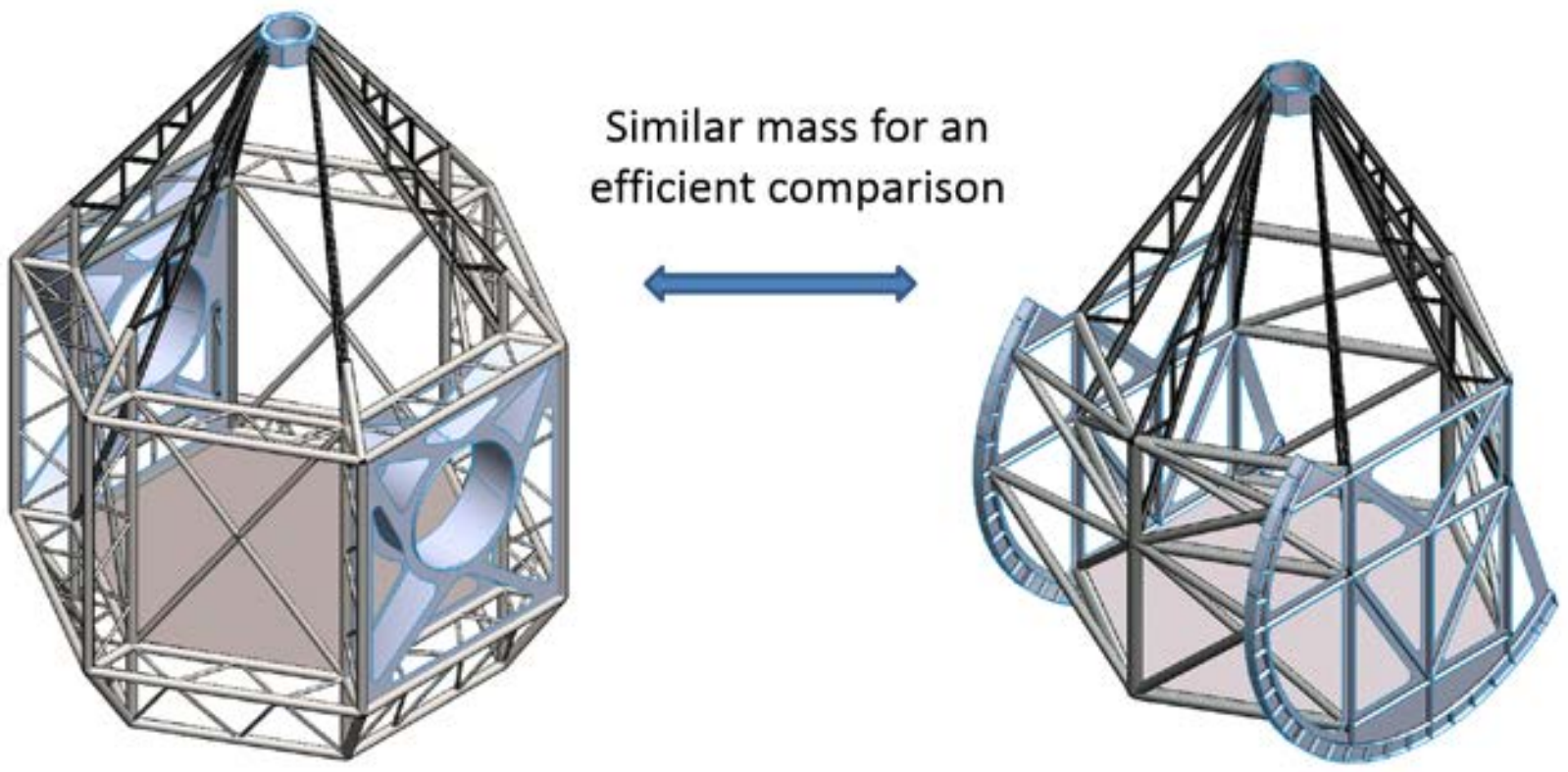

*Figure 50: Altitude or elevation structure—yoke concept (left) and rocking chair concept (right).*

The elevation structure supports a payload that includes the primary mirror, via a large welded plate mirror cell structure, and the top-end components, via a set of six “spiders.” The azimuth structure supports both the elevation structure and spectrographs on two dedicated platforms. Both structures provide cable trays and conduits, which support various cables, utilities, and fiber optic bundles, which are routed through the telescope. Walkways and stairs provide observatory personnel with safe and ready access to all telescope-mounted systems, for the purposes of maintenance and repair. The elevation structure rotates about the altitude axis, on a pair of trunnions, about which the structure and its payload are counterbalanced.

### *(2) Elevation Structure*

The elevation structure is made up of major subcomponents, as shown in Figure 51. The elevation ring, including the elevation–to–trunnion interface weldments, provides the main support for its optical payload. The elevation ring takes the form of a three-dimensional space frame, to facilitate airflow, and consists of slender, tubular profiles. The space frame is designed to transmit loads radially from the payload to the trunnions, through the weldments.

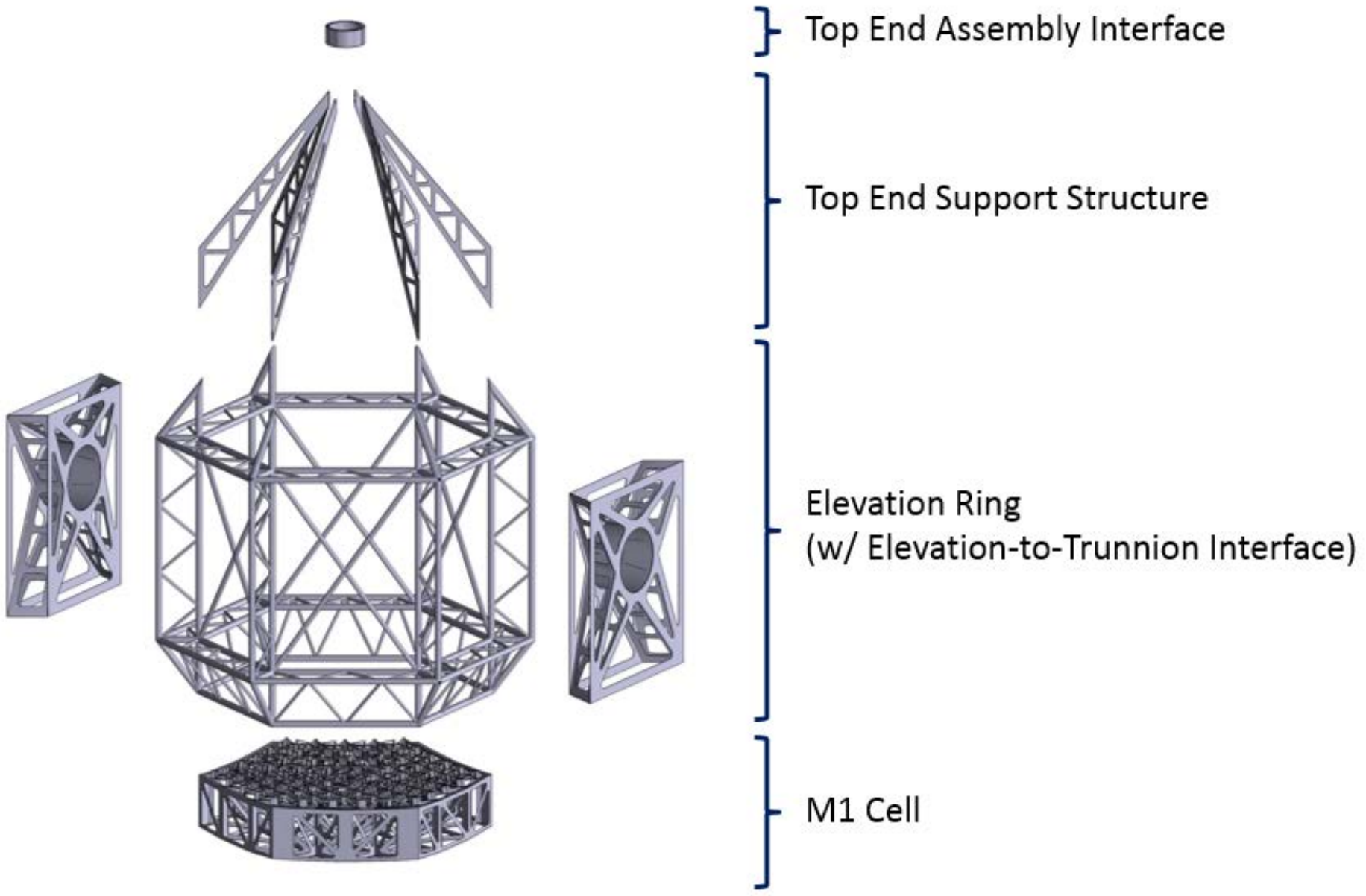

*Figure 51: Telescope elevation structure structural components.*

The top end support structure consists of six legs or "spiders," each made up of a welded H-section truss structure (Figure 52, middle panel), onto which the top-end assembly interface structure is welded, to transfer the top-end loads to the elevation ring. The spiders are not axisymmetric about the azimuth axis: they are arranged in an orientation termed the 1-2-1-2 configuration, as shown in the right-hand panel of Figure 52. This adds torsional stiffness to the system, with minimal compromise to the axisymmetry. In this configuration, the top end structure provides a direct, stiff support for the components at the top end, while minimizing obscuration of the incoming light to less than 3%[11] of the collecting area of the primary mirror.

[11] Including its payload

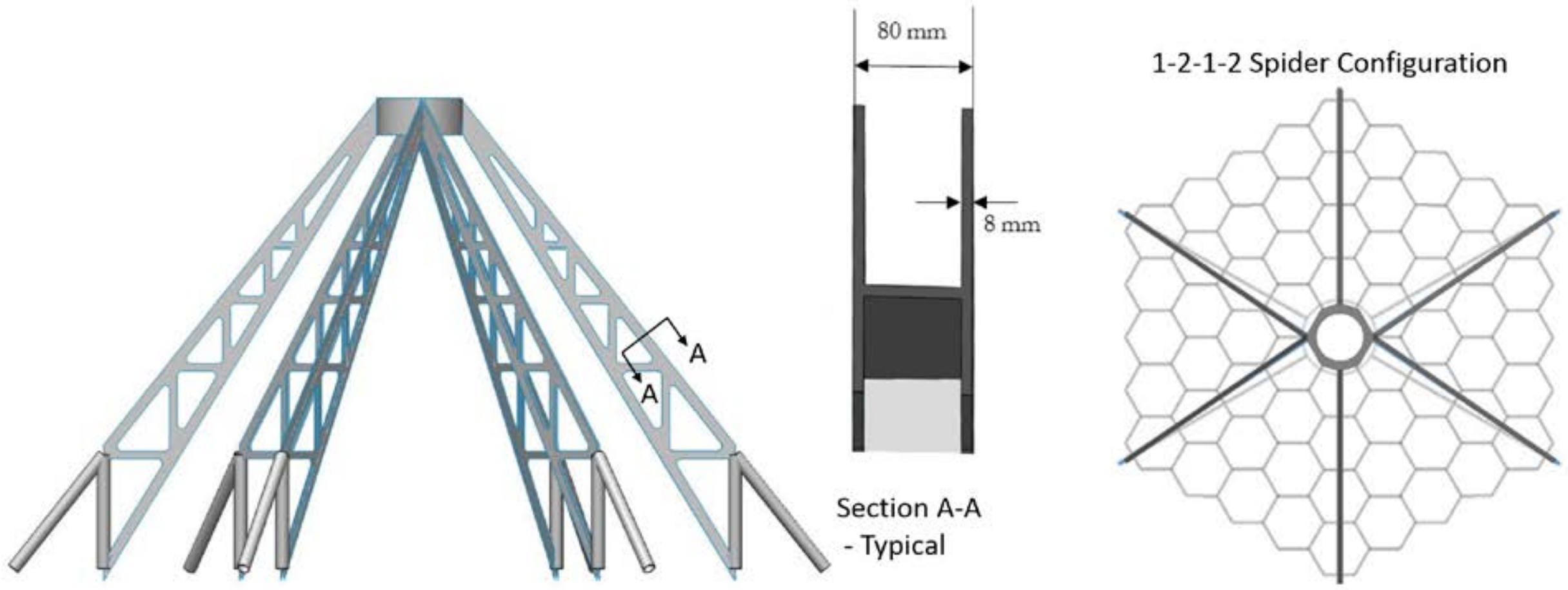

*Figure 52: Telescope spiders and top end. The right-hand figure shows the four welded joints (black lines) between the top-end interface ring and the spiders: two of the welded joints incorporate two spiders each (at 3 and 9 o'clock), while the other two joints have only one spider each (at 6 and 12 o'clock); the middle figure shows a typical section.*

The top-end assembly interface structure supports the top-end components via direct interface with the hexapod (PFHS), which supports the remaining payload at the top end. The top-end interface structure comprises a cylindrical weldment, consisting of two layers of circular plates, connected to each other by a series of radial stiffeners.

Once assembled, the elevation structure will be counterweighted, at the top of the elevation ring, to maintain balance. According to current estimates, this will necessitate counterweights of 26.5 tonnes.

Both azimuth and elevation structures include uplift restraints (Figure 53), to eliminate the danger of lifting and overturning during seismic events. The basic premise is that, during normal operations, there will be several millimeters clearance between the restraint and the structure and no interference with operations. If the hydrostatic bearings lose contact during a seismic event, the uplift restraints will constrain the structures vertically.

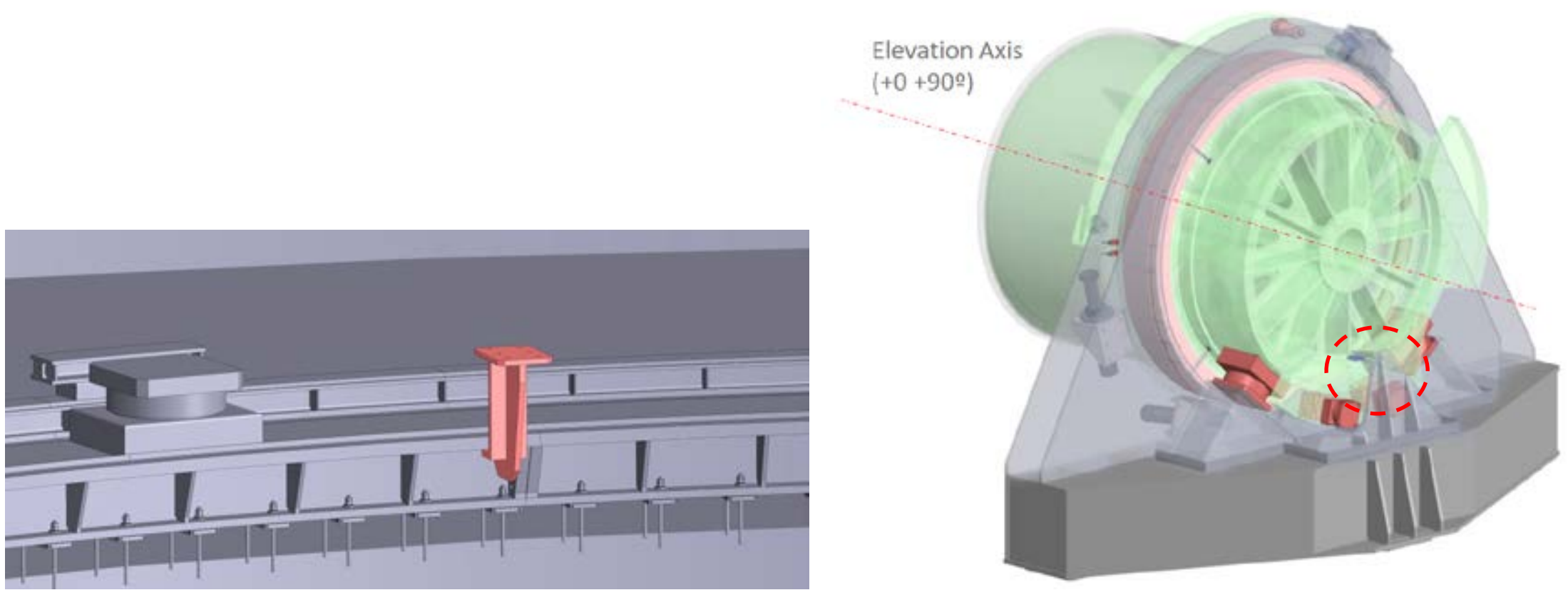

*Figure 53: Telescope seismic restraints. Azimuth restraint acting on azimuth track (left), elevation restraint within the dashed circle (right).*

The M1 cell consists of a welded "isogrid" structure, with a layer of M1 mirror segment interface plates attached to it (Figure 54). The isogrid structure forms a stiff primary mirror cell and contributes to the stiffness of the elevation structure. The structural components of the M1 mirror cell isogrid consist of a bottom plate, a rib structure, and a face sheet: welded plate assemblies, which are simple to fabricate. The M1 cell provides access to the mirror segments, M1CS, and fiber position metrology camera, for maintenance purposes. The bottom plate includes a planar segmented grating made of galvanized steel and provides a "ceiling height" of 2300 mm, to allow personnel access to the back of the mirrors and other components, for maintenance purposes. The "floor" of the bottom plate is smooth, continuous, and free of any obstacles that could present a trip hazard.

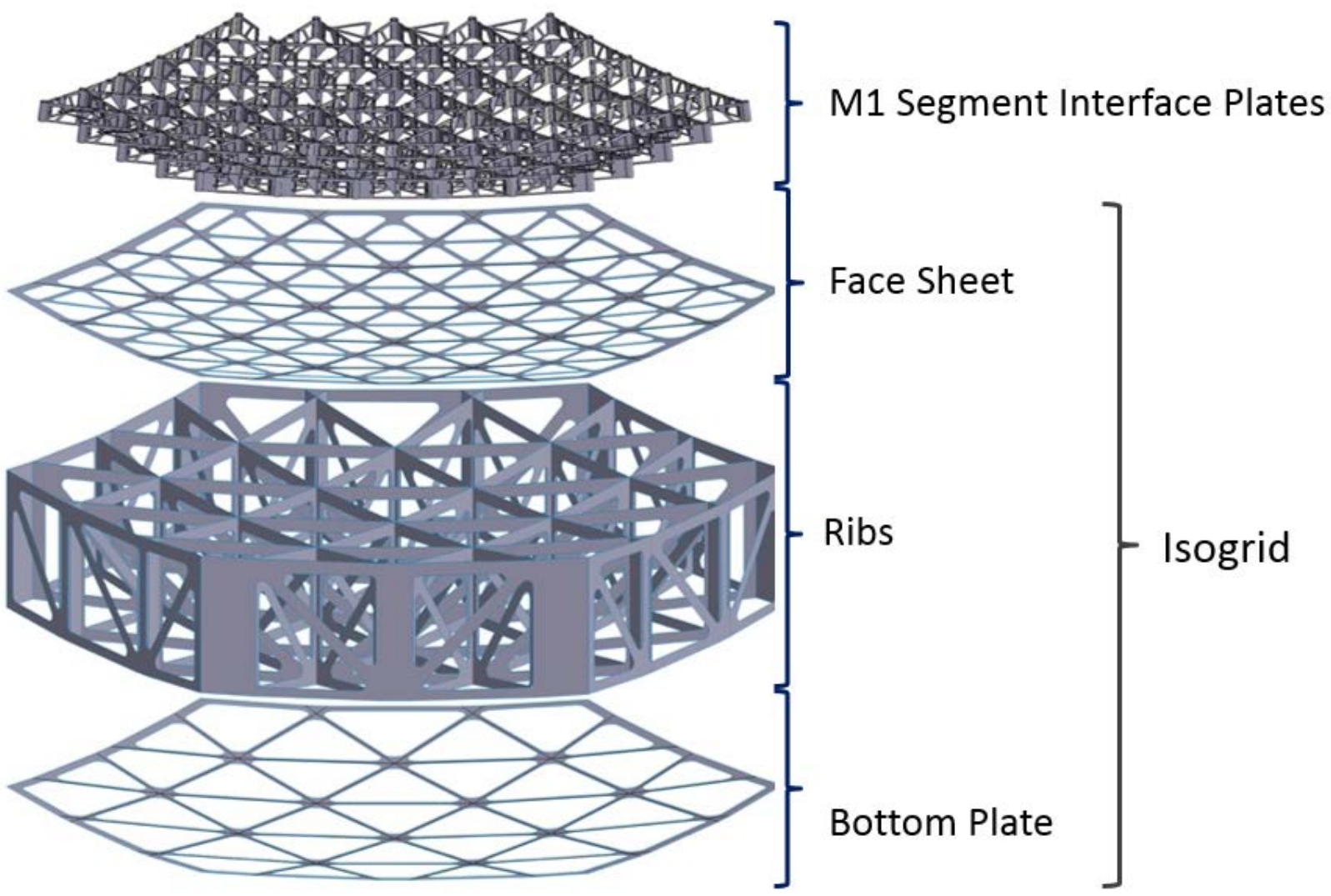

*Figure 54: Telescope structure: M1 cell.*

The segment interface plates provide stiff support for the M1 mirror segment assembles in the vertical direction and are compliance in the lateral direction.[12] The plates have adjustment features, which allow for precise positioning of the mirror segment assemblies at on-site integration, such that errors resulting from the fabrication process can be compensated for (Figure 55). Each mirror segment is supported by three stiff support points, attached to the back of the mirror, with its segment interface plate.

[12] To account for the difference in the coefficients of thermal expansion between the steel mirror cell and aluminum segment support assemblies.

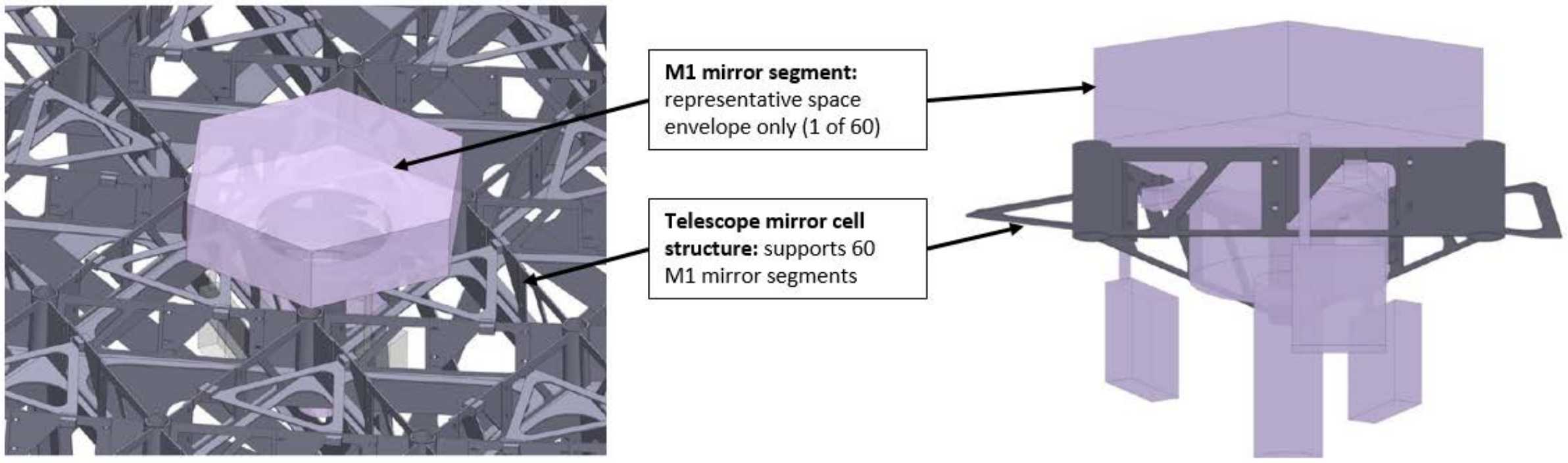

*Figure 55: Attached position of the mirror segment assembly (shown as purple space envelope) relative to the mirror cell can be adjusted.*

Due to its size, the mirror cell consists of several pieces that are bolted together after shipping. Each of the pieces consists of a partial welded isogrid with mirror interface plates welded on top.

### *(3) Azimuth Structure*

The azimuth structure supports the elevation structure and spectrographs on two side platforms. The platforms are designed to accommodate up to six spectrographs, with sufficient structural stability to allow for the anticipated instrument sizes and weights. Vertical loads from the elevation structure and spectrographs are transferred to the interface points on the azimuth floor, through the azimuth pillars, and lateral loads are transferred to the central pintle bearing (Figure 56). The azimuth floor is a welded steel structure with top and bottom plates, 850 mm apart, connected by a series of perpendicular ribs, which act as stiffeners. Three interface points are located on each side of the azimuth floor, directly on top of the hydrostatic bearings, in order to transfer the vertical loads to the Inner Pier

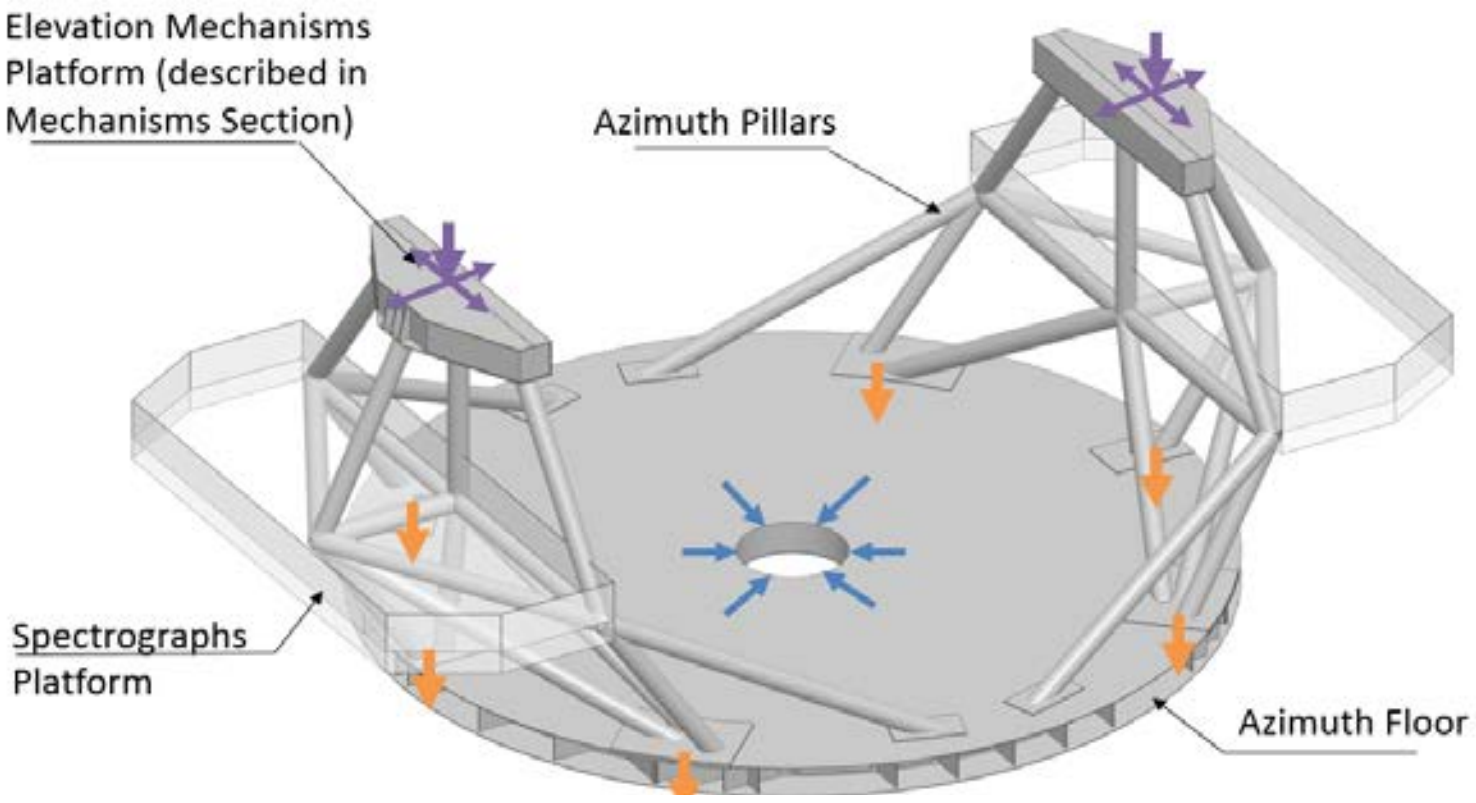

*Figure 56: Telescope azimuth structure vertical and lateral load paths*

The azimuth pillars are three-dimensional space frames, consisting of tubular profiles. The spectrograph platforms are welded structural planer profiles, which provide a stiff interface plane with the azimuth pillars, to support spectrographs. Stairs allow personnel access, for maintenance purposes.

The telescope structure design is constrained by the system's overall earthquake survivability. A two-dimensional seismic isolation system has been incorporated, and is located between the central bearing and the azimuth floor (Figure 57). The system will reduce horizontal seismic loads, transmitted from the telescope pier, and reduce the corresponding seismic acceleration exerted on the telescope and its payload.

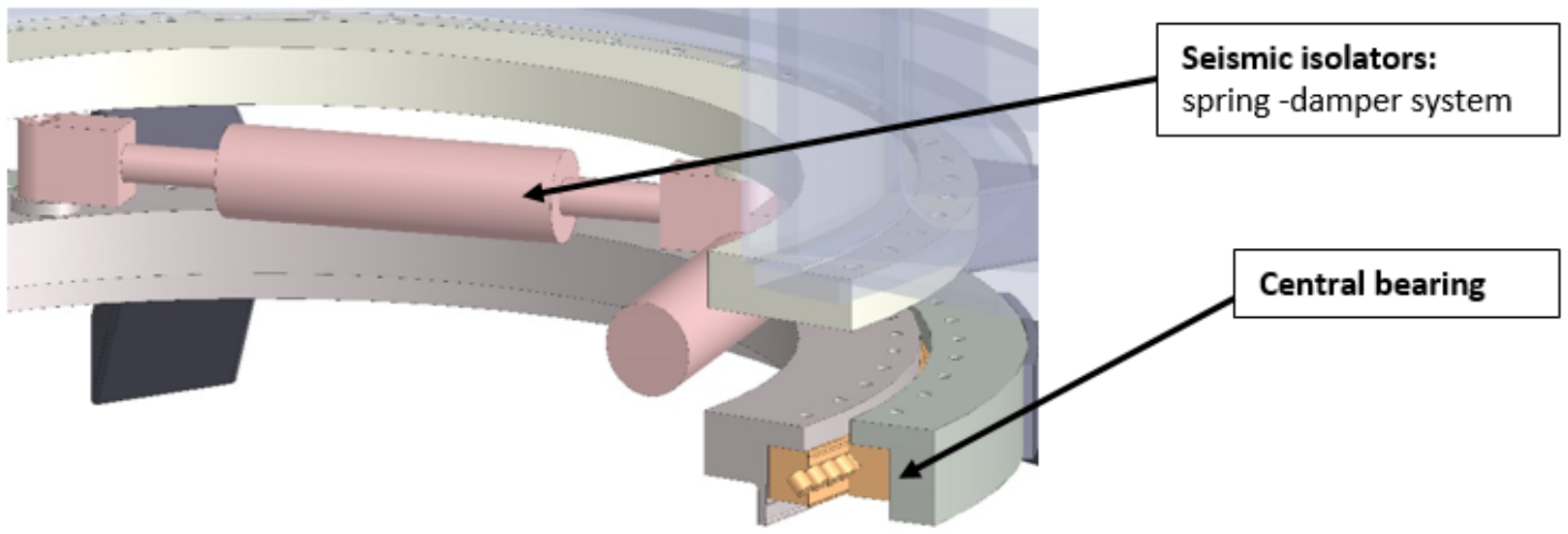

*Figure 57: Telescope seismic isolation system between azimuth floor (top) and central bearing.*

*(4) Central Bearing Support Structure*

The horizontal loads exerted on the central bearing are transferred to the Inner Pier through a central bearing support structure. As outlined in Section 4.1, the top concrete floor of the Inner Pier will be removed and reinforced, to accommodate the azimuth mechanisms and provide optimal maintenance access. The central bearing support structure is an axisymmetric space frame, consisting of eight spokes, which transmit radial loads straight from the central bearing to the telescope pier. Structurally, this replaces the existing 142-tonne concrete top floor with a 19-tonne steel frame, which serves the same function.

*(5) Mount Control System (MCS)*

Trade studies have resulted in the selection of hydrostatic bearings and direct drive motors as mechanisms for azimuth and elevation rotations. The tradeoffs involved in using hydrostatic versus antifriction bearings, and direct versus pinion drives were compared: both in terms of jitter performance, and in terms of installation and operation considerations, such as cost, maintainability, thermal control, and cleanliness. Integrated hydrostatic bearing and direct drive motor systems were selected for both the altitude and azimuth axes, since they offer low friction, zero backlash, and overall system stiffness,[13] thus providing a control system with sufficient bandwidth to meet the current jitter requirement of 0.01 arcseconds RMS.

The azimuth structure rotates around a vertical axis on the azimuth track. Loads are transferred from the structure to the Inner Pier foundation vertically, through six axial hydrostatic bearings.

[13] Provided by the proposed high-frequency, over-constrained, hydrostatic bearing system.

The central bearing uses a crossed roller bearing to transfer the lateral loads from the azimuth structure, through the central bearing support structure, to the Inner Pier.

The direct drive solution for the azimuth drive system is a semi-customized design, composed of eight off-the-shelf, axial-flux, linear motors, together with customized curved magnet sections, placed circumferentially along the azimuth track. Figure 58 shows the azimuth axis motion mechanisms, including the disc brake system, limit switches, and end stops with shock absorbers. A commercially available full-circle tape encoder system is integrated into the outer surface of the central bearing, with a small runout of 20 microns, to maximize the accuracy of the encoder system.

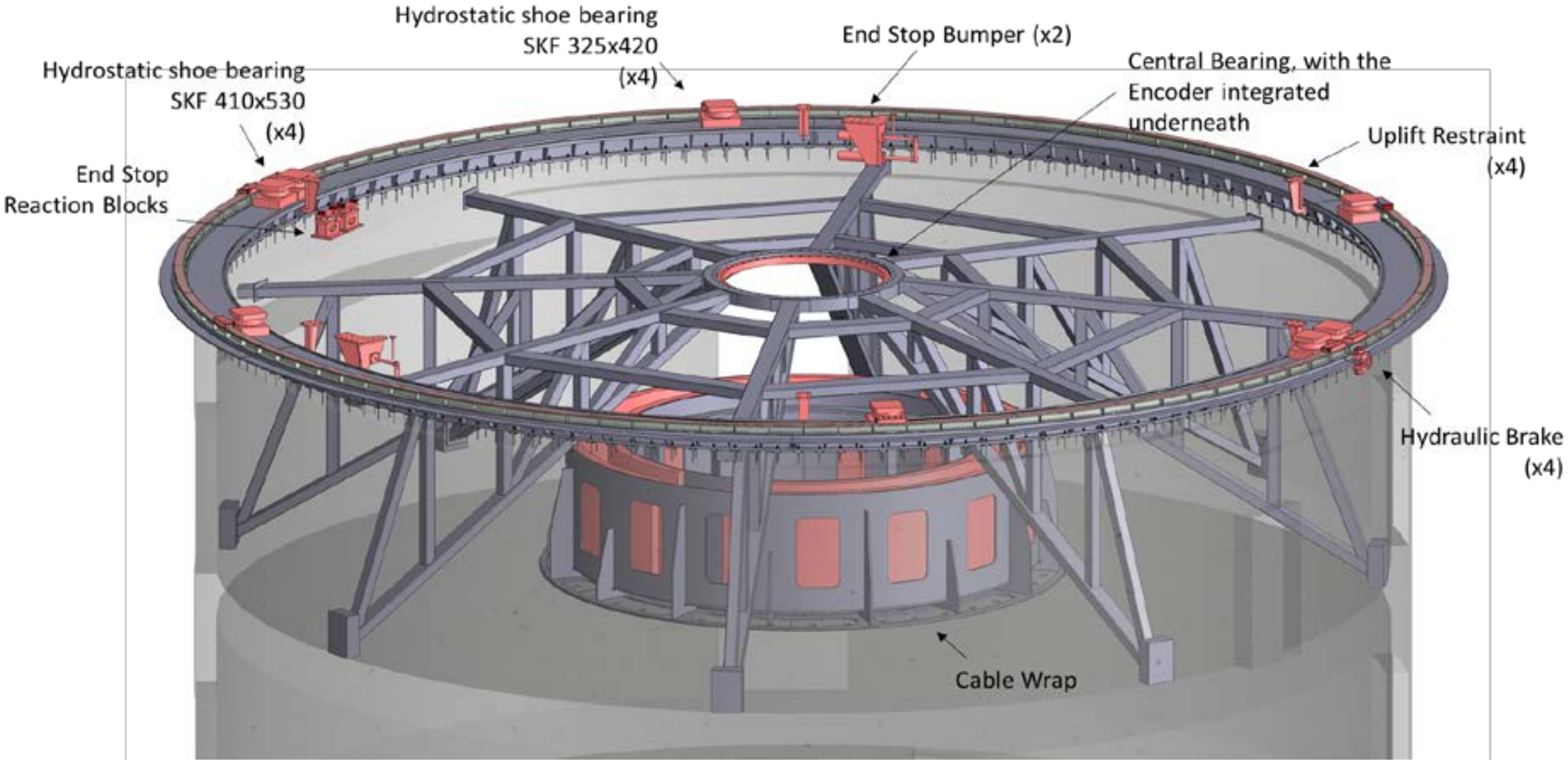

*Figure 58: Telescope azimuth axis mechanisms*

The elevation rotation uses radial, hydrostatic shoe bearings. In the yoke configuration, the trunnions allow for compact packaging of the elevation axis mechanisms, such as the brake system, limit switches and the end stops with in-built shock absorbers (Figure 59). The trunnions provide both structural and mechanical interfaces with the altitude mechanism. Mechanically, the trunnions' geometry defines the accuracy of the elevation axis, through the full zenith range of the motion. They are subject to deformation, resulting from the load transfer from the altitude structure to the azimuth structure, via the radial hydrostatic bearings.

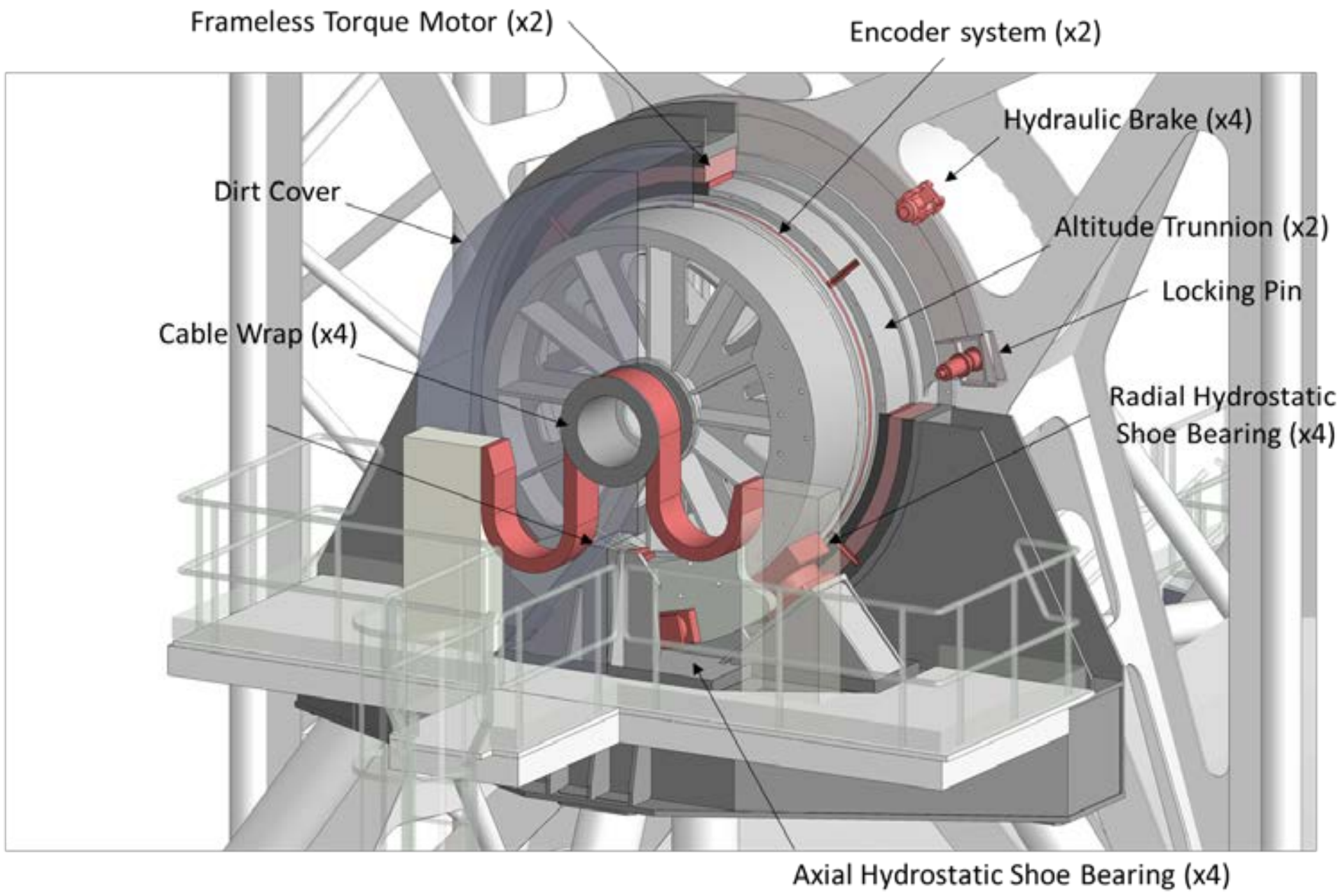

*Figure 59: Telescope elevation mechanism*

The drive system comprises two circular, synchronous, frameless, direct-drive, torque motors, each equipped with a stator assembly on the elevation mechanism platform of the azimuth structure side, and a rotor assembled on the trunnion side.

Like the azimuth mechanism, the elevation mechanism uses a tape encoder with four scanning heads. The drive control system synchronizes both motors in torque. The drive system is reconfigurable, to allow continuous operation if one unit fails.

The MCS informs TSS of the status and health of the drive system and protects against runaway over-speed and over-travel conditions.

*(6) Utilities Distribution (UTI)*

Utilities required to operate the telescope (power, compressed air, chilled water, etc.) are supplied by the OBF at a main connection panel, for distribution to the telescope and instrument subsystems throughout the telescope structure.

Three cable wraps are included: one at the base of the central bearing support structure and one on either side of the elevation axis (Figure 58 and Figure 59). The proposed cable wrap designs have been adapted from previous IDOM projects and are expected to support the full range of MSE motion.

### 4.3.4. Primary Mirror Optics System (M1)

*(1) Primary Mirror Overview*

The M1 Optics System is a complex collection of optics, opto-mechanics, and controls, coating, and handling systems, which work in combination to collect light from astronomical sources and reflect it onto a focal point. M1 provides a reliable and maintainable high performance primary mirror, delivering throughput and image quality performance that meet the sensitivity demanded by MSE science requirements.

The primary mirror has a serrated 11.25 m[14] entrance pupil, an 18.81 m focal length, a -1.113 conic constant, and a radius with a curvature of 37.70 m.

Design considerations for the M1 segments are mostly driven by the sensitivity requirements, as defined in the throughput and image quality budgets. To attain the best possible image quality, the Primary Mirror Segment Assemblies (PSAs) and the M1 Control System (M1CS) subsystems work in concert, using actuators and edge sensors to maintain the overall shape of M1 when subjected to structural deformations caused by temperature and gravity (due to changing zenith angles) or disturbances from wind and (observatory generated) vibrations. To maintain optimal throughput, the Mirror Coating System (Coat) and Segment Handling System (SHS) subsystems work together to enable periodic mirror segment recoating, routine maintenance, and segment exchanges, in order to maintain optical performance over the life of the observatory. SHS facilitates the safe transportation and handling of segments: from their position in the telescope mirror cell, through the observatory to the coating chamber and storage facility, and back to the telescope.

In accordance with the symmetry of the hexagonal array, the primary MSE mirror segments are distributed in six identical sectors of ten segments each, in accordance with their (hexagonal) orientation with respect to the center of M1. As shown in Figure 60, there are thus ten unique types of segments, and each segment is interchangeable with any other of the same hexagonal shape and optical prescription, according to its placement in the mirror cell. For example, segments A1, B1, C1, D1, E1, and F1 in Figure 60 are interchangeable. For operational reasons, one spare segment of each type—i.e., a total of ten additional mirror segments—is included in first light operations.

[14] 11.25 m is defined by the diameter of the circumscribing circle.

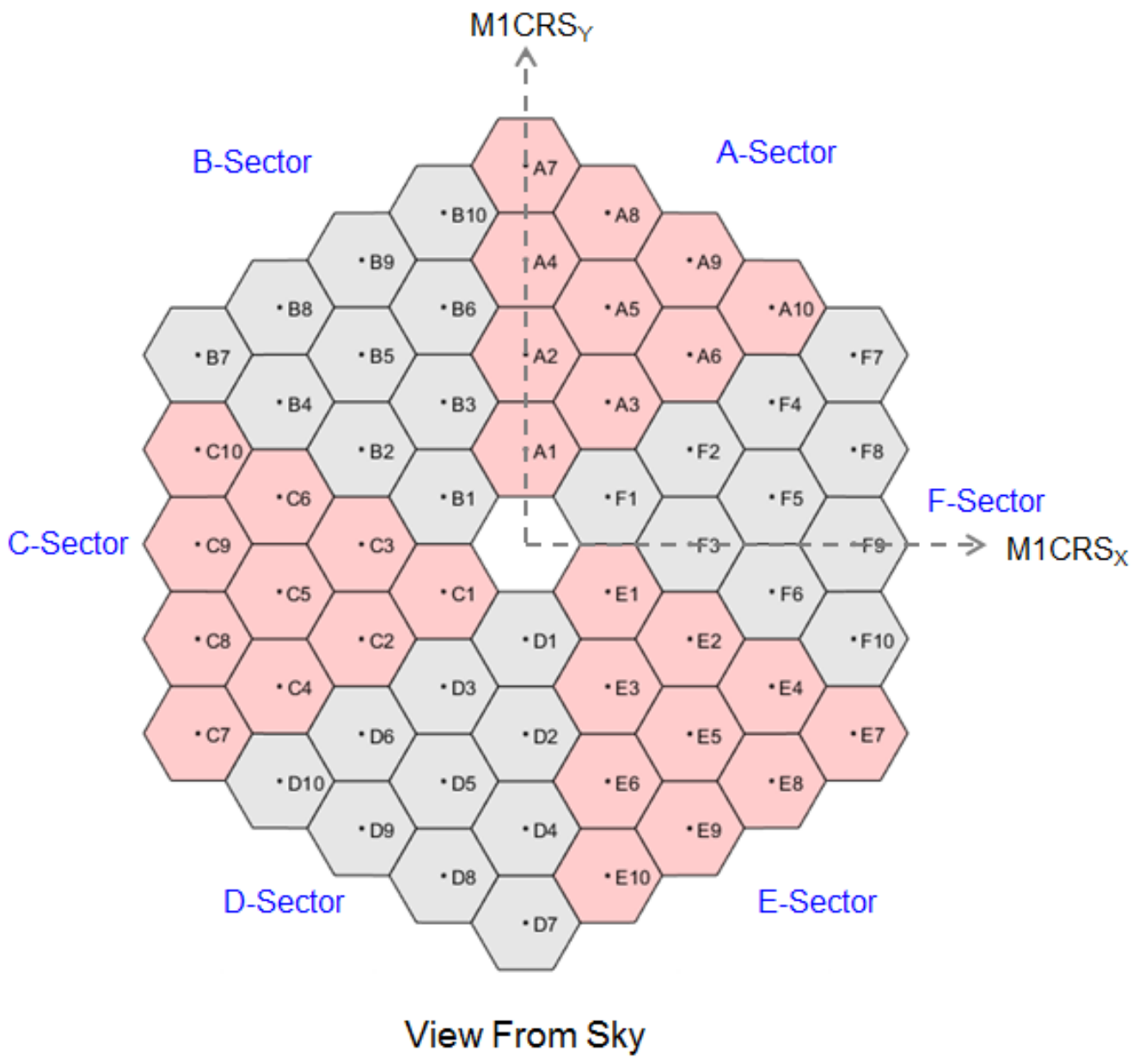

*Figure 60: Diagram of the primary MSE mirror segments with the 60 Segments numbered and arranged in six identical sectors (A–F).*

### *(2) Primary Mirror Segment Assemblies (PSAs) and M1 Control System (M1CS)*

MSE makes use of mature segmented mirror technology (Figure 61) to achieve the effective collecting area of a 10-m diameter circular aperture. Active segmented mirrors are a proven technology, which has been in use for many years at astronomical facilities, including the twin Keck and the Gran Telescopio Canarias (GTC) telescopes. This technology has been further refined and prototype tested by the Thirty Meter Telescope (TMT) and ESO Extremely Large Telescope (ELT) projects.

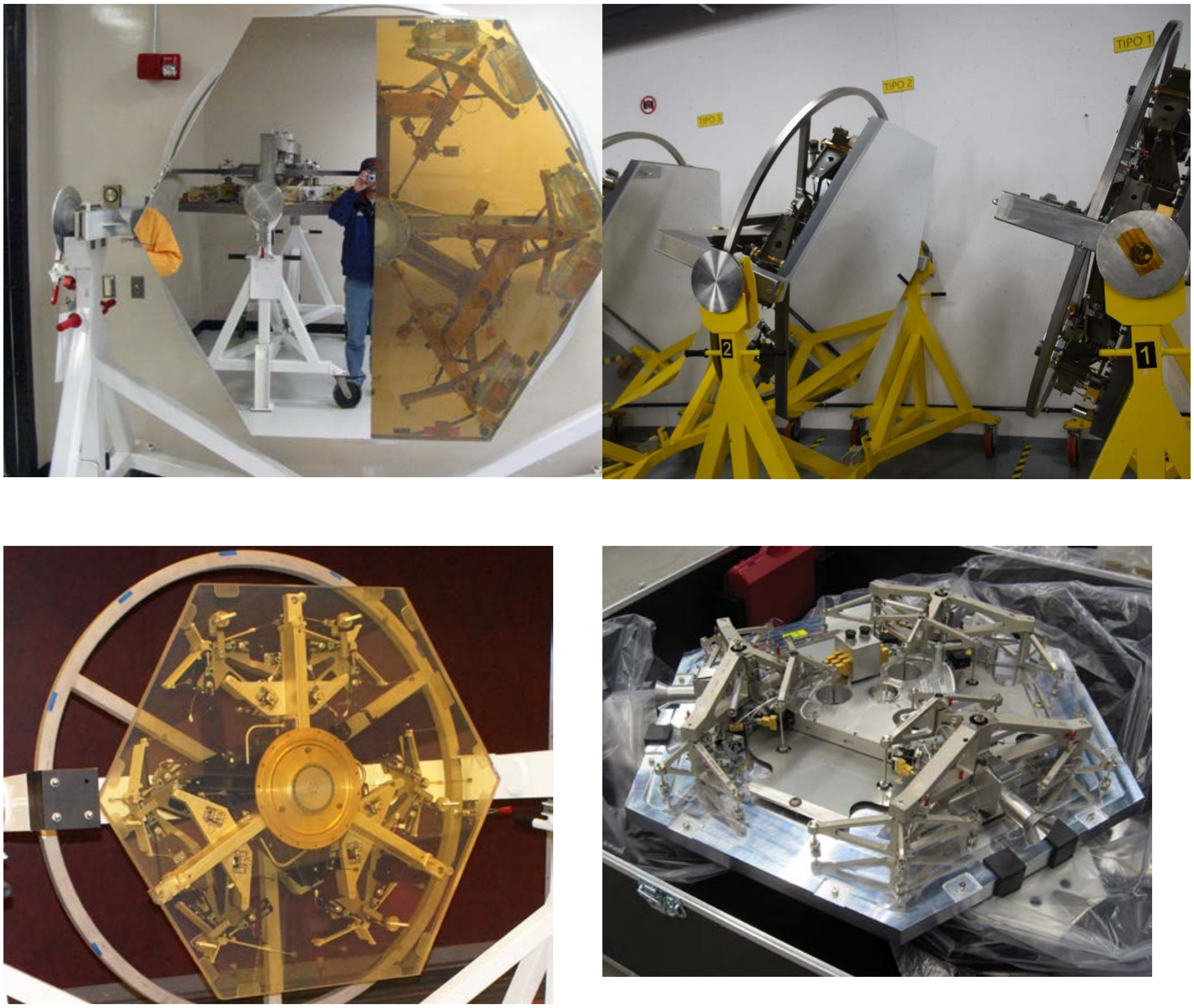

*Figure 61: Examples of M1 designs clockwise from top left: Keck segment display (half coated) on handling cart, GTC segments on handling carts, ELT segment support structure at the vendor assembly facility, TMT segment and its support assembly on display at SPIE.*

The Polished Mirror Assembly (PMA) includes hexagonal mirror segment of the same size (1.44 m from corner to corner) as TMT and ELT. Mirror blanks are available from two suppliers: Schott (Zerodur) and OHARA (ClearCeram). MSE segments can potentially be polished by any of the TMT or ELT suppliers/partners, using the same processes and testing procedures. Since the maximum aspheric departure of the MSE segments is only 60% that of those of TMT, the segments are proportionally easier to polish and test optically.

The realization of the envisaged Primary Segment Assembly (PSA) and its subsystems—organized by parts, components, subassemblies, and assemblies—is illustrated by the colored blocks in Figure 62. Once installed in the telescope mirror cell, each PSA assembly can be separated into a subcell, which is fixed to the mirror cell, and a Mounted Segment Assembly (MSA), which is actuated by M1CS, in piston, tip and tilt. The subcells are installed and aligned on the telescope mirror cell during integration and are not disturbed again. The entire MSA is removed during segment exchanges and is compatible with a coating process in a vacuum chamber. In Figure 62, the dashed lines indicate which components within the PSA are attached to the MSA and removed for recoating, and which are attached to the subcell and fixed to the telescope mirror cell.

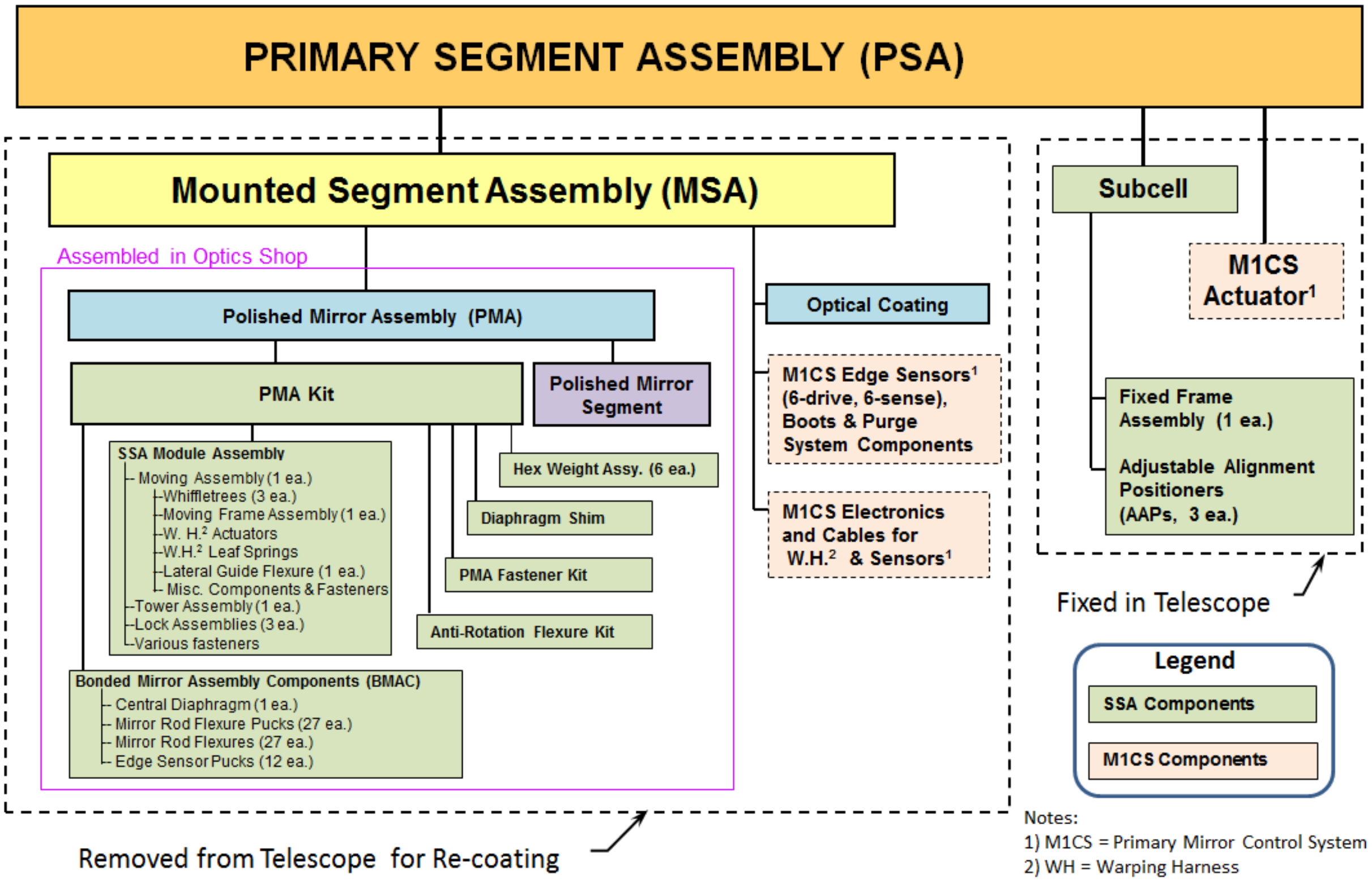

*Figure 62: PSA block diagram of the assemblies. Green blocks represent the SSA, pink blocks the M1CS, and blue blocks the PMA.*

Each MSA includes a Segment Support Assembly (SSA), which supports both the polished mirror segment and the M1CS edge sensors and electronics, and interfaces with three M1CS position actuators (for piston, tip, and tilt adjustments). The subcell fixed frame is permanently attached to the telescope structure and supports the MSA at a repeatable interface. A total of ten additional MSAs are delivered at telescope first light, making a total of 70.

Each SSA includes a moving whiffletree assembly, bonded mirror assembly components, and the tower structure that interfaces with the fixed frame in the subcell. Each PMA has a polished mirror segment mounted on the SSA, with an integral warping harness system, which allows for the removal of low spatial frequency surface errors, by using force actuators to gently bend the segment.

To maintain M1 phasing during observations, M1CS uses edge sensors to detect segment-to-segment alignment at the inter-segment edges, and the three position actuators to correct piston, tip, and tilt errors in real time. This compensates for wind disturbances, and changes in the gravity vector and temperature of the telescope, in order to maintain the overall shape of the primary mirror and allow the 60 individual segments to function as a single monolithic mirror. Once commissioned, the M1CS will operate in closed-loop mode during observations, and will require little intervention.

The Phasing and Alignment Camera (see section 4.3.8) measures segment alignment, surface shape errors, and segment-to-segment phase errors. The PAC uses starlight to determine global and local M1 errors. On the basis of its measurements, the M1CS commands the position and warping harness actuators to align the segments, reduce segment shape errors, and, finally, phase M1. This process is performed both monthly and after each segment exchange.

The design and development of the M1 system are already sufficiently mature. However, the Project Office has not yet chosen whether to use TMT or ELT as the M1 technology. The choice will depend on intellectual properties, licensing and programmatic considerations, and contributions from the MSE participants.

*(3) Mirror Coating System (Coat)*

The baseline coating recipe for the primary MSE mirror segments uses ZeCoat[vi] protected silver with enhanced UV reflectivity. The backup coating is composed of Gemini Observatory[vii] protected silver coating.

Our operations concept requires replacing every mirror segment on the telescope with a freshly coated segment every two years, in order to maintain an M1 throughput budget of more than 94%, for all wavelengths between 370 nm and 1800 nm, with an expected degradation of 1% per year of reflectivity per segment. Every segment will therefore be exchanged and recoated, on a two-year cycle.

The M1 Coating System includes access to the optical coating recipe and the actual coating process, in addition to the equipment required for its deposition. The coating system also includes the cleanroom and coating chamber with its auxiliary equipment; facilities used to remove the reflective coating; the equipment required to wash, clean, and prepare mounted mirror segments for the coating chamber; storage facilities for consumables and waste water, and for both in-process and freshly coated segments; the coating of laboratory instruments, including those used in testing and quality control; handling fixtures to move and support the mounted mirror segments during washing and in the coating chamber, etc.

*(4) Segment Handling System (SHS)*

The SHS is a set of equipment used by the observatory staff to install, remove, and transport segments to and from the telescope safely and efficiently.

The Project Office has adopted the segment exchange process developed by the Keck telescopes as the MSE baseline (Figure 63). The process utilizes a segment-lifting jack to raise the segment slightly from its installed position in the mirror cell, so that the segment-handling crane can access it from above. Then the segment is grasped via the segment-lifting fixture, which is lowered by the segment-handling crane. Once lifted, the segment is moved, lowered, and placed onto a segment-handling cart at the observatory floor level. To install a segment back into M1, follow the same process in reverse.

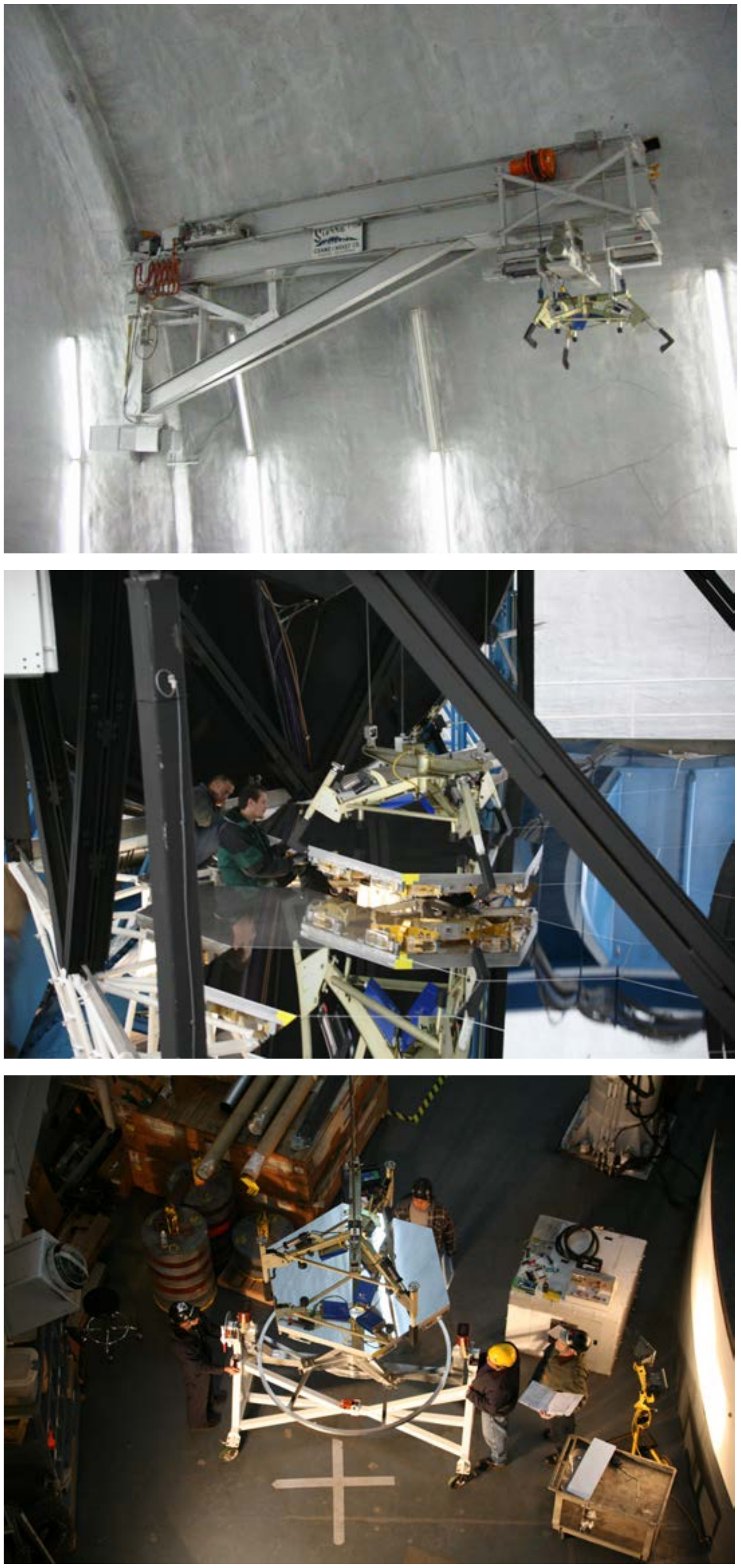

*Figure 63: Keck segment-handling system—Enclosure-mounted, segment-handling crane with segment lifting-fixture (top), segment secured on lifting fixture on top of M1 (middle), and segment lowered to handling cart on the observatory floor (bottom).*

Since the current Keck process enables three segment exchanges per day, the baseline plan for MSE involves ten exchanges per year, which will allow us to remain consistent with the proposed two-year segment replacement cycle. Segment exchanges are a daytime operation and do not impact nighttime operations, except for the necessary post-exchange M1 alignment, warping, and phasing processes, which take place at twilight. Each alignment is currently estimated to take 120 minutes, based on the PAC feasibility design study.

The system also includes the auxiliary equipment required to initially assemble the mirror and to maintain it throughout the life of the observatory. This includes the equipment used for regular daytime $CO_2$ snow cleaning.

#### 4.3.5. Wide Field Corrector and Atmospheric Dispersion Corrector (WFC/ADC)

The WFC/ADC optical design (see Section 4.3.1) consists of five lenses, ranging in diameter from L1 = 1340 mm to L5 = 800 mm. Three elements (L1, L2, and L4) are made of fused silica and the other two elements (L3 and L5) are made of Ohara PBM2Y. Optically, L1, L2, and L4 are strongly powered; L3 is thin and weak; and L5 is thin, with strong curvature on both sides (Figure 64).

The ADC action is provided by a shift and tilt of L2, combined with adjustments to the hexapod orientation and telescope pointing. The action also changes the plate scale, halving the target motions caused by differential atmospheric refraction across the focal focus.

The optical design is constrained by throughput and image quality requirements. “Simpler,” alternative designs have been proposed and explored by the design team, but none of them were able to achieve the required image quality. The design uses the biggest blank sizes commercially available to boost throughput, but L1 and L3 are “undersized,” due to availability issues. Although the smaller blanks cause vignetting, the design uses this to block out unwanted rays that deviate from normal incidence to the focal surface beyond the acceptance cone angle of the fibers.[15] Vendors have confirmed the availability of all blanks of the requisite sizes and with the anticipated optical qualities. A Sol-Gel plus $MgF_2$ anti-reflective coating has been adopted as the baseline, due to its high throughput performance over the full wavelength range. The spin coating technique will be used to apply Sol-Gel over the $MgF_2$ coated optics, and then it will be hardened to facilitate handling. Some development work is needed, as the largest current spin-coated lens is only 1200 mm in diameter. However, this is not expected to be technically challenging.

The five-lens design of the WFC/ADC is housed in an opto-mechanical barrel assembly, designed by the Division Technique de l’INSU (DT-INSU) in France. The main components include an overall optical barrel; independently removable and adjustable cells for each of the lens elements; a mechanism for the L2 ADC actuation; a structural interface, for mounting onto PFHS; and a baffle structure.

[15] The vignetting at L1 efficiently cuts out these rays. This makes the design effectively pupil-centric, in the sense that the fiber acceptance angle on axis at zenith suffices at all other field positions and zenith distances.

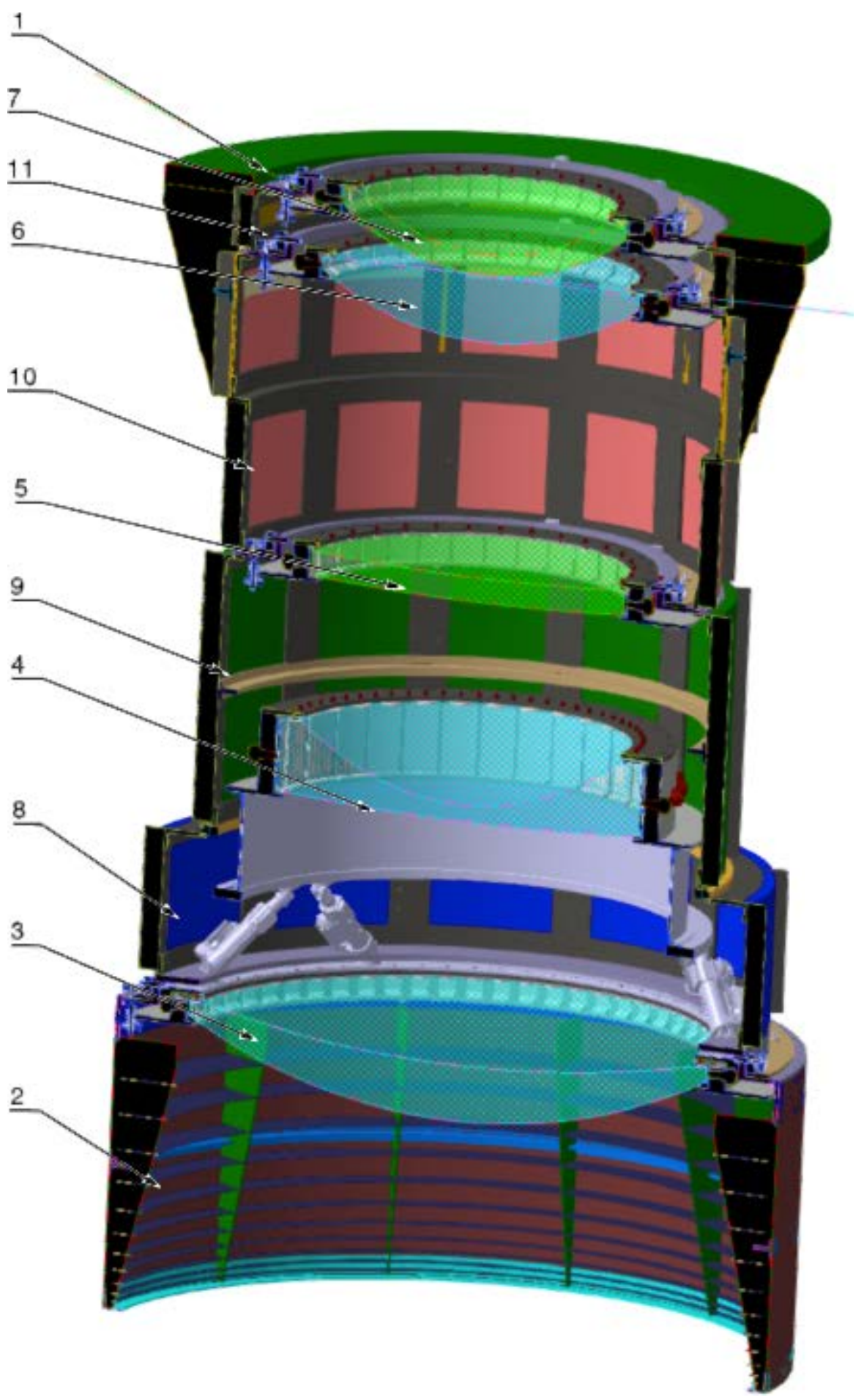

*Figure 64: WFC/ADC Opto-mechanical design—(1) Optical barrel support, (2) Light baffle, (3) L1 cell set assembly, (4) L2 cell set assembly, (5) L3 cell set assembly, (6) L4 cell set assembly, (7) L5 cell set assembly, (8) Optical barrel spacer L1–L2, (9) Optical barrel spacer L2–L3, (10) Optical barrel spacer L3–L4, (11) Optical barrel spacer L4–L5.*

The optical barrel is intrinsically a support structure (Figure 64), which provides the required stiffness to maintain optical alignment through all zenith and gravity orientations. The optical barrel encloses, supports, and protects the components from dust and during handling. Where needed, it also provides internal and external light baffling. The outer size is 1750 mm diameter, which is the largest allowable central obscuration size specified in the system throughput budget.

In addition, the lens cells and optical barrel are strong enough to support the entire cantilevered load of the WFC/ADC while the telescope is in its “stow” and maintenance position, at 90° zenith. As stated above, the barrel maintains the alignment of the lenses with respect to each other, within an acceptable amount of deflection, as defined by an image-quality-based tolerance analysis, conducted during the optical design process. Over the 0°–60° zenith observing range, overall tolerances on individual elements must meet the overall image quality requirement range of between ±0.1 mm and ±0.5 mm for decenter, between ±100 microradians and ±7000 microradians for tip and tilt, and about ±0.5 mm for defocus. A baffle structure at the entrance to the WFC/ADC blocks stray light from reaching the focal surface. The entire assembly remains within the central obscuration diameter limit.

The WFC/ADC opto-mechanical design facilitates the optical alignment process by mounting lenses in independently adjustable cells. Each lens can be individually aligned and handled, without damaging its optics or AR coating, which can be scratched if mishandled.

The concept involves mounting optical elements individually on compliant pads in a cell assembly, by capturing the lens between two rings (Figure 65). The cell assembly is then positioned, using small actuators within a larger diameter “setting” assembly (Figure 66), which can adjust the focus, tip, tilt, and decenter of the lens (in its cell assembly) with respect to the setting assembly. The lenses can be adjusted independently, to achieve overall alignment with each other within the optical barrel, using five setting assemblies. The assemblies include provision for radial thermal compensation due to coefficients of thermal expansion mismatch. Future work on this concept is required: to simplify the actuators and allow for the possibility of removing them entirely and relying on machining and fabricating tolerances to align the barrel components within the allowable optical alignment tolerances—using, for example, a machined spacer to control the lens separations between the lens cells in the barrel. Future development will also investigate whether the lens can be mounted in its cell using room-temperature-vulcanizing (RTV) silicone or a similar compliant adhesive, rather than mechanical clamping.

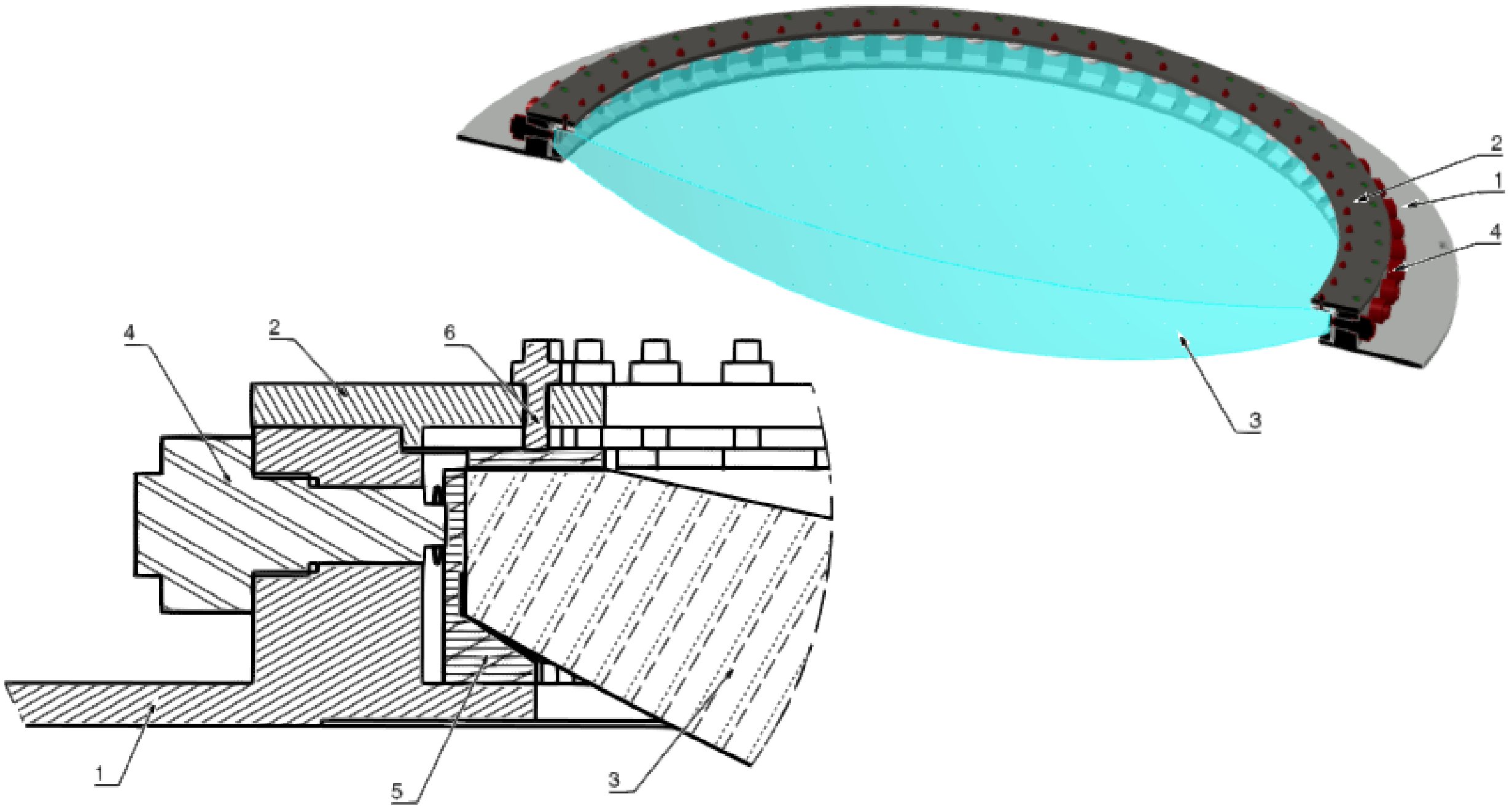

*Figure 65: WFC lens cell assembly, typical: (1) L1 cell base ring, (2) L1 cell clamp ring, (3) Lens 1, (4) L1 cell-athermalized radial restraint assembly, (5) L1 cell radial pad, (6) L1 cell axial pad.*

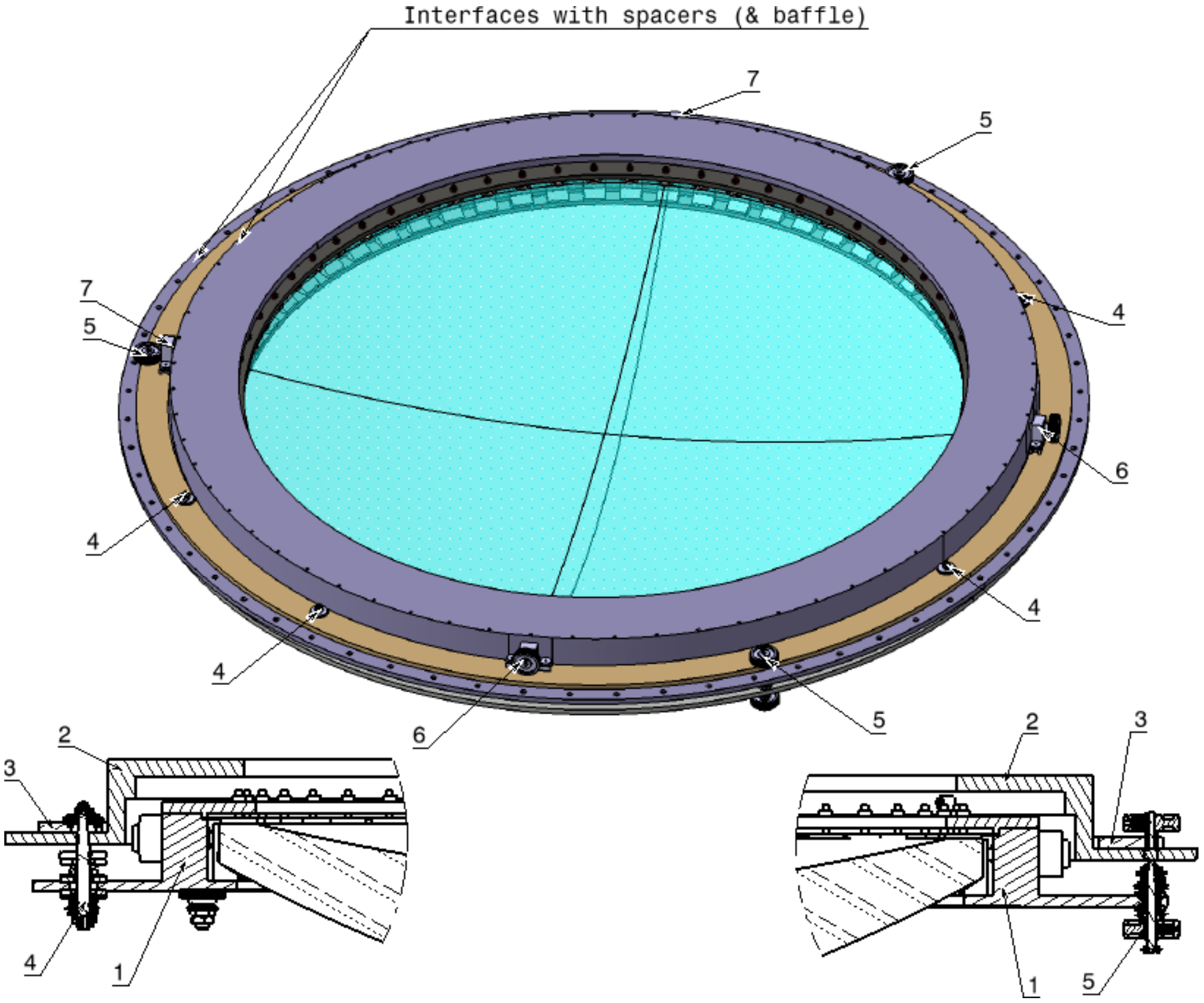

*Figure 66: WFC lens cell design concept, typical—Left-hand panel illustrates overall setting assembly: (1) L1 cell assembly, (2) L1 spacer ring, (3) L1 set setting ring, (4) Lens setting clamp, (5) Lens setting adjustment vertical, (6) Lens setting adjustment horizontal, (7) Lens setting guide horizontal.*

Alternate ADC mechanisms have been considered, including a small hexapod unit, which fits in the available space, and a linkage assembly, which is less complex, but which does not fit within the optical barrel and may cause unwanted obscuration. Both options are shown in Figure 67. A selection will be made during the next design phase.

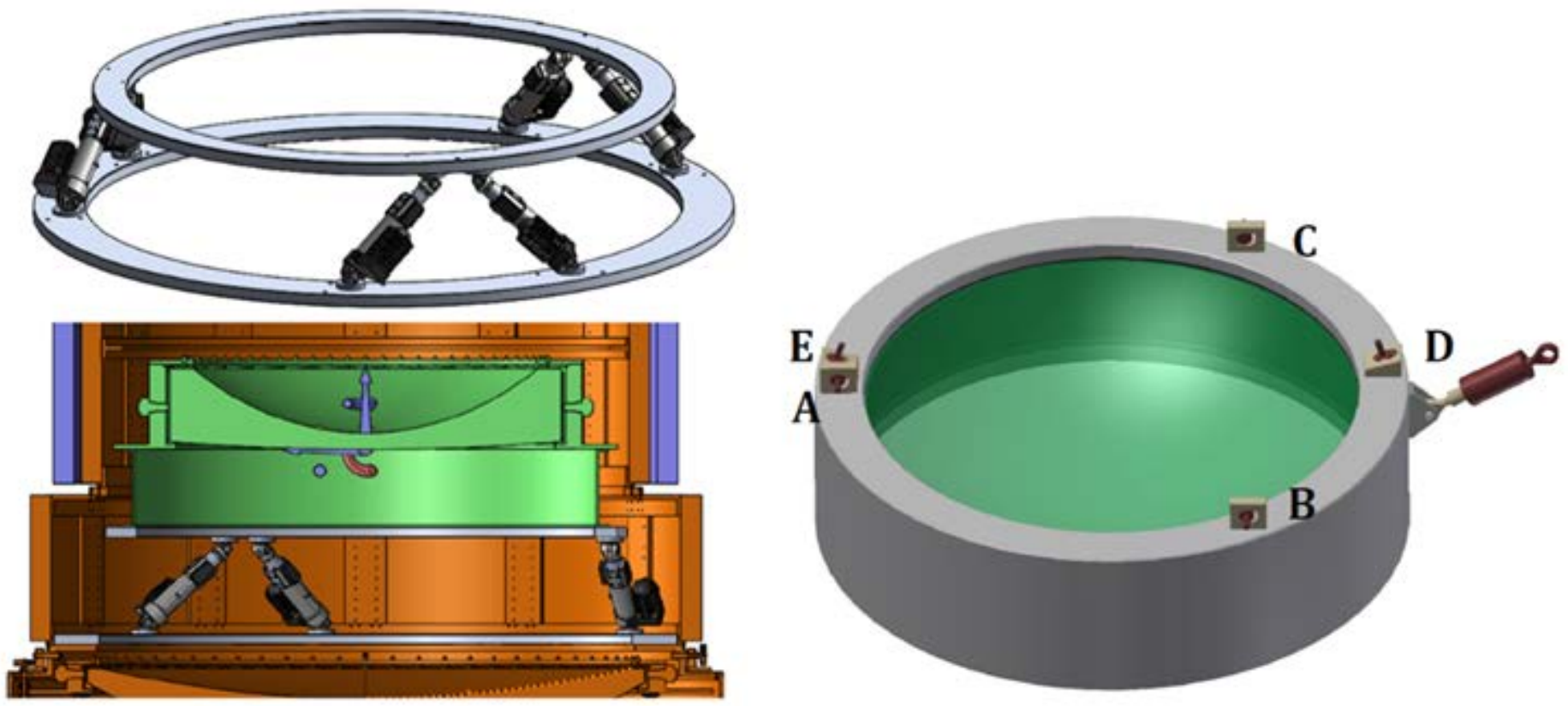

*Figure 67: L2 mechanism options—hexapod (left) and linkage assembly (right). The linkage assembly uses a single actuator to move L2 on a prescribed trajectory, guided by three horizontal cam followers (A, B, and C), and two vertical cam followers (D and E), which constrain the rotation of L2.*

### 4.3.6. Prime Focus Hexapod System (PFHS)

The Prime Focus Hexapod System (PFHS), designed by the Division Technique de l'INSU (DT-INSU) in France, is a mechanism mounted onto the elevation structure's top-end interface (Figure 68), which provides precise positioning of the top-end components with respect to the effective focal surface in five degrees of freedom: focus, tip, tilt, and decenter (i.e. lateral X–Y).

PFHS consists of six actuators, two "rings," and a motion controller. The base ring is fixed to the telescope top-end assembly interface structure and the top ring is the moving ring that supports its payload, including the Instrument Rotator (InRo), the WFC/ADC, the Telescope Optical Feedback System (TOFS, i.e., AGC and PAC), and the fiber Positioner System (PosS) and Fiber Transmission system (FiTS).

During the set-up of each observing field, the observatory control system determines the position of the optimal focal surface,[16] based on the dimensional changes of the telescope structure due to gravity and temperature effects and the offset required for atmospheric dispersion correction, and moves PFHS (with its payload as a rigid body) accordingly, to ensure the best possible injection efficiency for the system.

When it receives the appropriate commands from the control system, PFHS attains and locks its position to within 0.15 mm laterally, 0.05 mm in focus, and 100 microradians in tip and tilt over the course of any given observation. The operational concept is that PFHS should move to the instructed position and stay there until the next observation. The residual error in the PFHS position over the range of typical observations is expected to contribute very little to either image quality or injection efficiency degradations, compared with other effects within the partitioning of our system budgets. However, for observations of long duration and/or spanning a large zenith range, it may be beneficial for PFHS to make small corrective moves in mid-observation, in order to enhance injection efficiency while observing. Future work will assess the impacts on the PFHS design and operations in terms of additional cost, risks, and complexity.

---

[16] Future refinement of the optimal focal surface position may include offsets to compensate for average defocus due to spine tilts and wavelength-dependent chromatic effects.

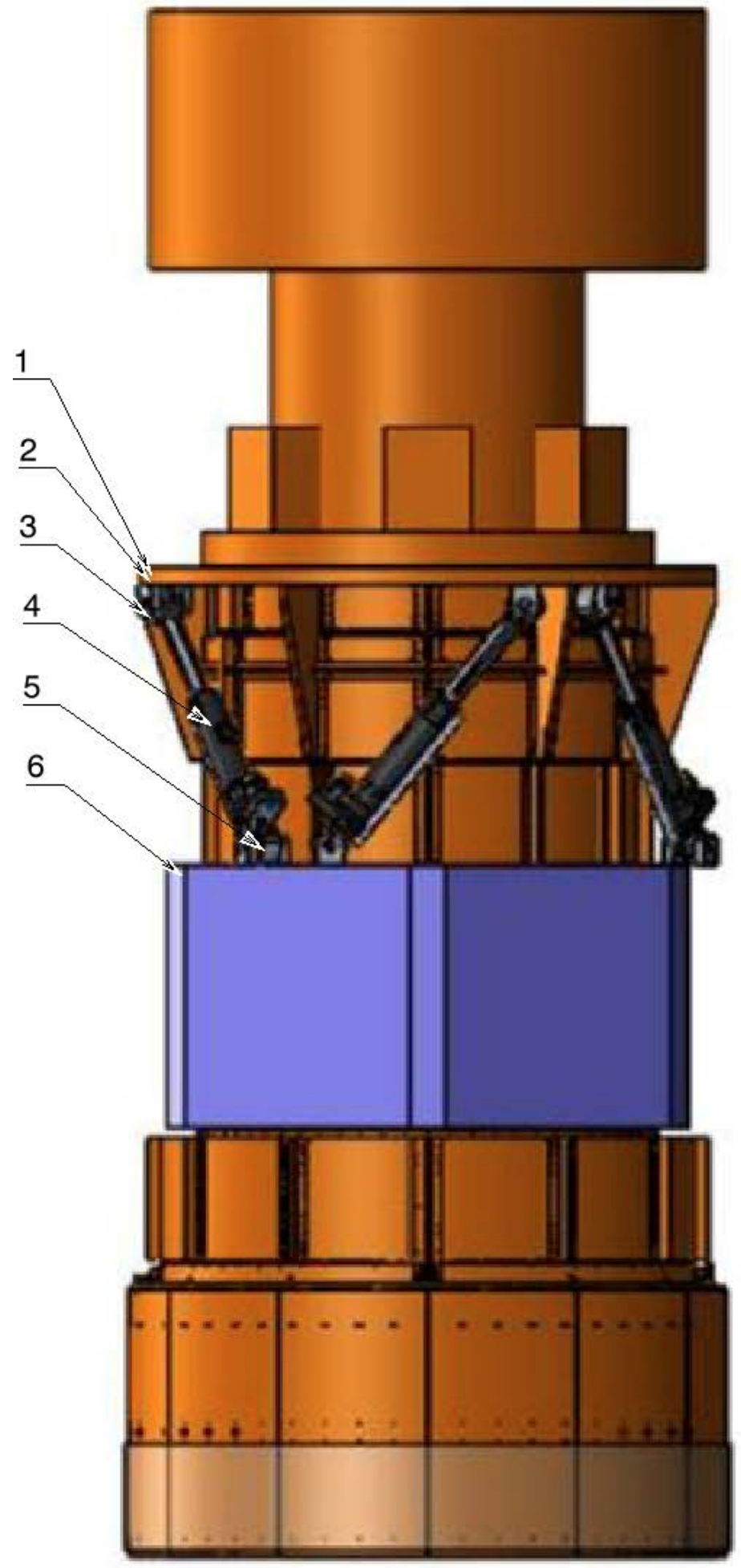

*Figure 68: PFHS assembly with illustrative payload in orange—(1) Payload attachment spacer, (2) Hexapod top ring, (3) Top ring joint, one per actuator, (4) Actuator, six total, (5) Bottom ring joint, one per actuator, (6) Hexapod base ring.*

During CoDP, commercial vendors were approached by DT-INSU. Symetrie,[viii] in France, has proposed a modification of their commercial JORAN hexapod (Figure 69), deployed on the Large Millimeter Telescope Alfonso Serrano (LMT),[ix] which is similar in size and capacities. In the JORAN version of the hexapod, the actuators are brushless, motor-driven, precision ball-bearing screw jacks with preloaded nuts. It includes two sensors: an absolute sensor with a linear scale and an additional encoder in the motor. This arrangement does not require an external braking mechanism.

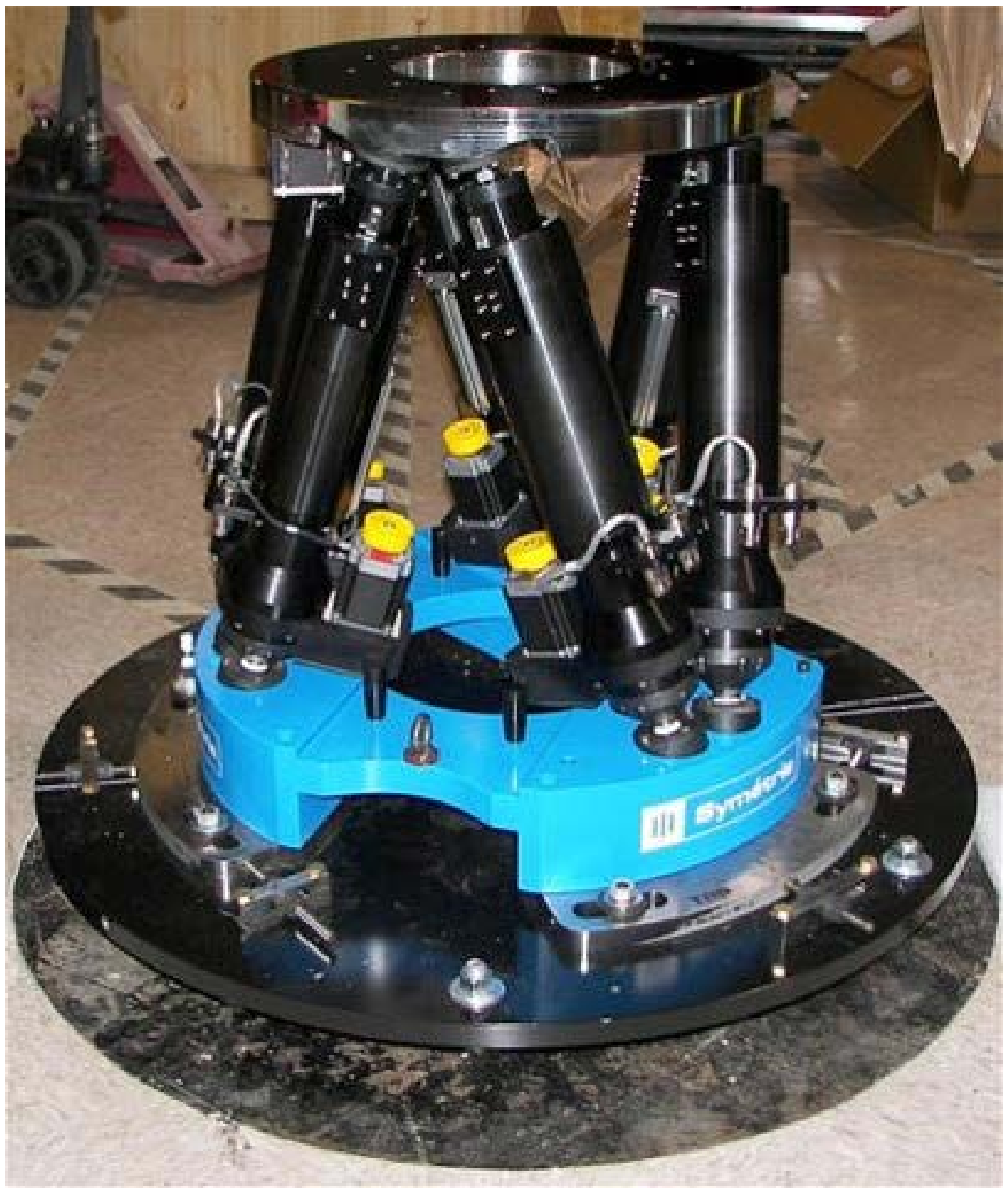

*Figure 69: Example of the Symetrie JORAN hexapod.*

Future work will include confirming whether the cross coupling (a parasitic movement that appears when the commanded trajectory length approaches the precision of the actuation system) dominates the system error and, in particular, ensuring that ±5 microns of focus accuracy can be reliably attained during operations. Potentially, this could be the limiting factor that prevents the aforementioned mid-observation adjustments of the system in the future.

Future work will also confirm whether the JORAN hexapod can maintain its positional stability during an observation when unpowered, without continuous closed-loop control, or whether the control system and its sensors must remain powered. In the latter situation, there is cause for concern, since image quality may degrade, due to power-on heat dissipation. However, this concern will probably only apply to positional stability during an hour-long observation. PFHS, without closed-loop control, may be sufficiently stable for shorter observations. We will be revisiting this question by designing a modified duty cycle for PFHS, based on realistic observation lengths.

Under all conditions, PFHS with payload must withstand the anticipated seismic loads, without sustaining structural failure that damages its payload. Its ability to do so will be confirmed in future work at the overall system level, including an examination of the telescope structure's seismic response characteristics.

However, PFHS is a well-established technology and has proven reliable and effective for similar telescope applications.

### 4.3.7. Instrument Rotator (InRo)

Since MSE is an altitude–azimuth telescope, its field of view rotates as the telescope tracks the sky. To compensate, the integrated PosS and FiTS, and AGC are de-rotated about the telescope's optical axis. The Instrument Rotator (InRo), designed by the Division Technique de l'INSU (DT-INSU) in France, provides this rotation, using a precision rotary bearing system (Figure 70). AGC monitors the guide star positions and provides positional feedback to the telescope's guiding control, based on the angular position and rotating speed of InRo. This system ensures the ultimate goal of keeping the array of fiber inputs aligned with their sky targets during observations.

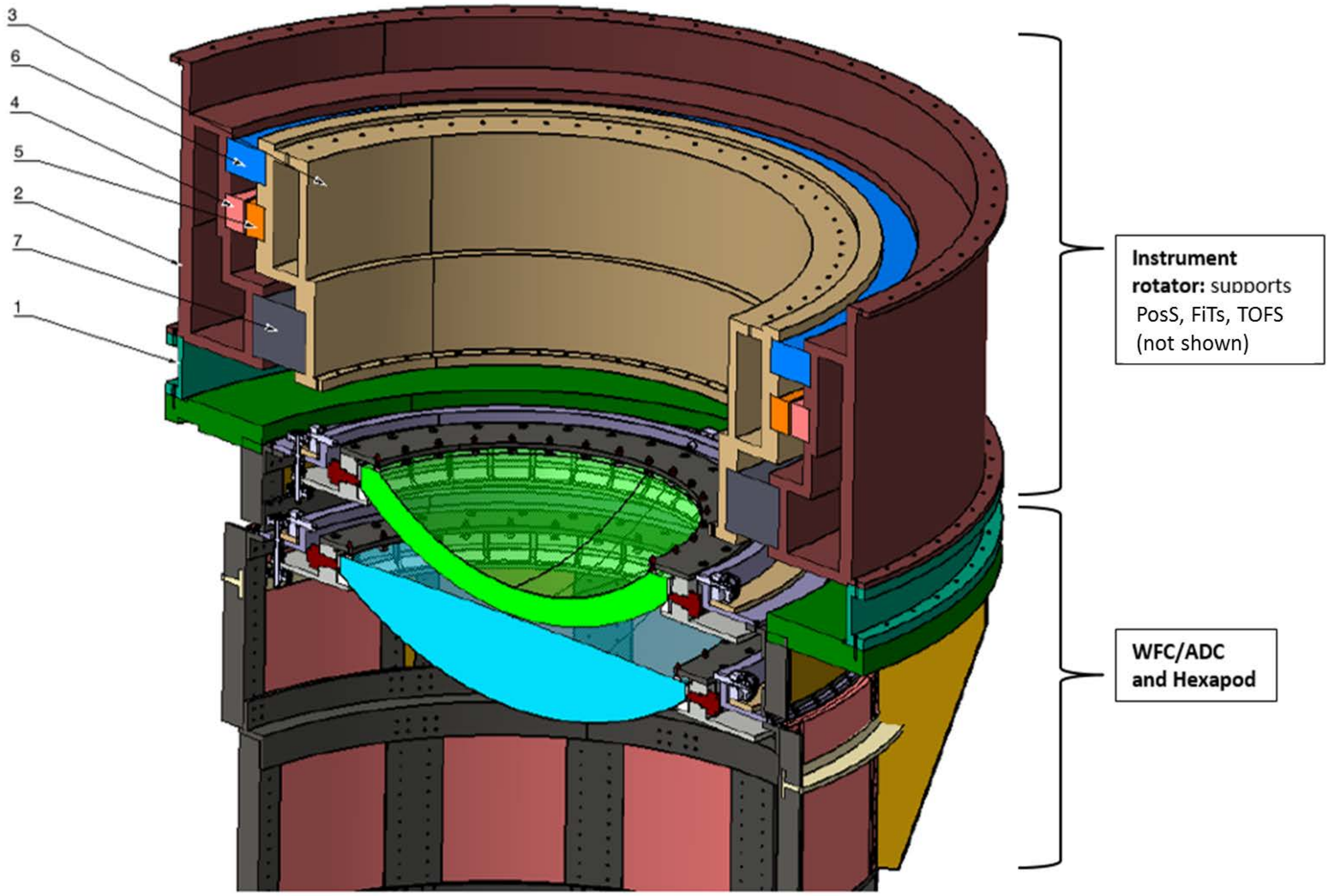

*Figure 70: Instrument Rotator concept, with torque motor mechanism – (1) Payload attachment spacer, (2) Stator support, (3) Rotor support, (4) Stator, (5) Rotor, (6) Brake and encoder system, (7) Precision rotation bearing*

The InRo subsystem includes a large diameter rotary bearing, drive- and motion-control system, as well as a structural interface with PFHS, which supports it and its payload. The InRo payload consists of PAC and ACG, as well as PosS and FiTS. InRo also has a service cable wrap (SCW), which routs the necessary electrical (power and data) and utility services (coolant and dry air) from the telescope to its payload.

The required rotation range of InRo is ±180°, which includes extra range to facilitate maintenance. During the Conceptual Design Phase, this was reduced from ±270°, to allow a more feasible design for the SCW, given the space available. In practice, the fiber system may only accept a more limited range of motion, likely ±90°, due to the risk of damage to fiber optics when using the larger range. In the next phase, the impact of a more limited range on maintenance, operations, and observing efficiency (due to more frequent unwrapping of the SCW) will be examined, and the safe rotation range of FiTS will be determined.

A commercial vendor may supply the InRo rotation mechanism. DT-INSU has approached several commercial vendors in Europe, with a focus on the concept design of the critical drive components of InRo, including its motor and bearing systems.

A torque motor (Figure 71) or gear slew motor would be suitable for this size and payload capacity. Both options will be explored going forward, and a decision will be reached early in the next phase of the project:

- The torque motor mechanism has an integrated actuator, bearing, and encoder and offers precision control and high torque, with fewer mechanical parts. However, it may exceed the allowable heat dissipation for components operating near the focal surface (i.e. more than 50W after active cooling) because it requires continuous operation. Designing a custom motor that minimizes dissipation may mitigate this effect, but it will increase the mass and size in this already crowded space.
- The gear slew motor mechanism includes double roller gear bearings, motors, and pinions, and an encoder. Conceptually, we envision the inclusion of two counter-rotating motors to provide motion, braking, and anti-backlash control. This would likely provide a less massive solution than the custom torque motor solution, but may not provide sufficient accuracy.

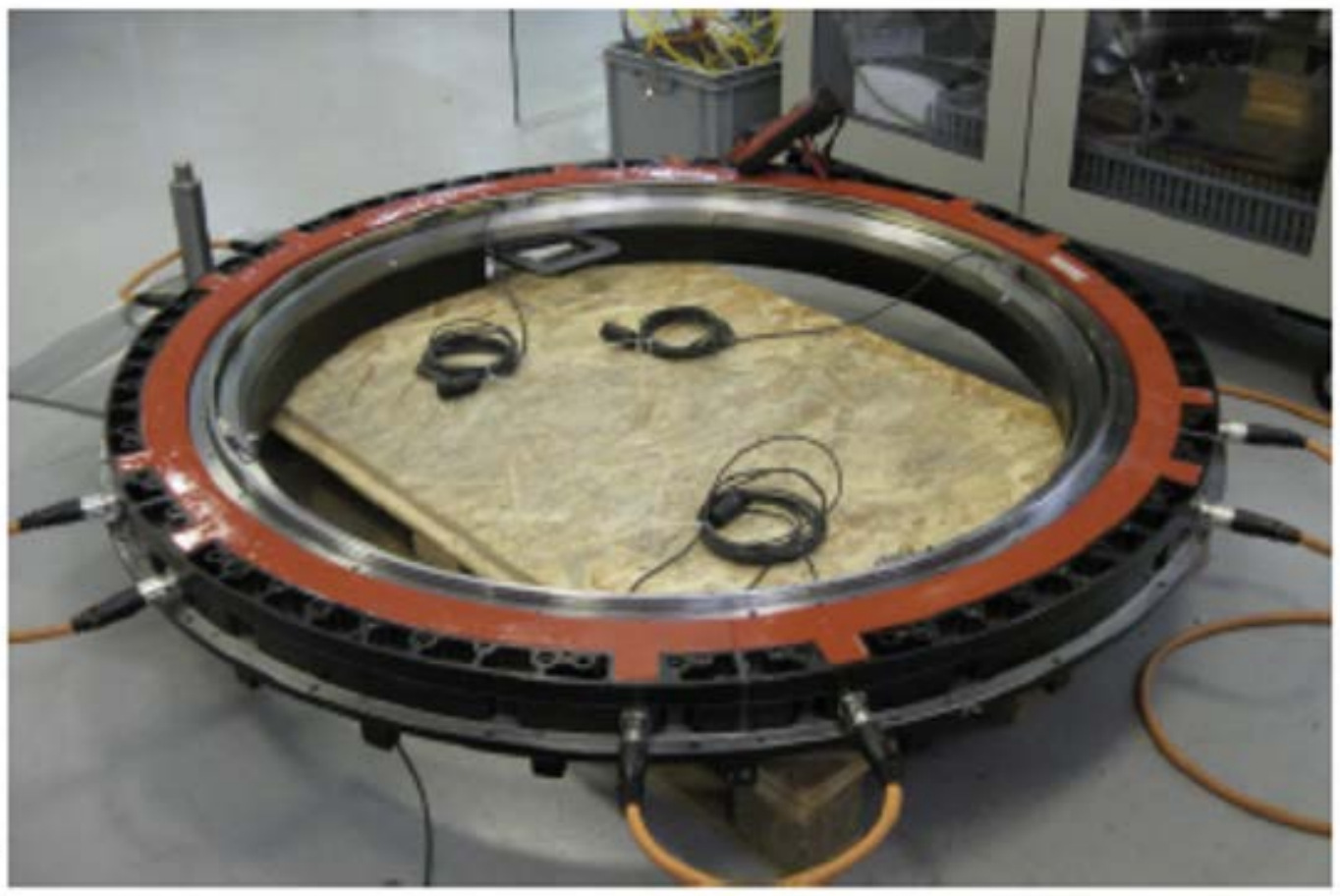

*Figure 71: InRo – Example of a dedicated, integrated, low-voltage torque motor, made by PHASE MOTION CONTROL, for rotator application.*

The main constraint on the bearing is the need to ensure bearing runout remains as low as possible and the rotational axis of InRo is coaxial with the optical axis of the telescope over the full rotation range. The only appropriate technology we have identified is a precision roller bearing. Two types of bearing configurations have been deemed suitable: axial-radial cylindrical roller bearings or crossed roller bearings.

With these technologies, InRo will meet its 3.5 arcsec angular (rotational) accuracy requirement, which corresponds to 0.05 arcseconds on-sky, at the edge of the field of view. The angular rate of rotation varies as a function of the azimuth and elevation positions of the telescope, during an observation. Near zenith, there is a 1° diameter “keyhole,” in which the telescope does not observe and the rotator is not required to provide the rotation speed needed to track the field.

Based on the technical specifications, it is assumed that the bearing will have a residual tilt error of 50 microradians, resulting in a maximum defocus of 30 microns at the edges of the field, during an observation. These errors have been accounted for in the pointing, tracking, and guiding error budgets, and are deemed acceptable, if we take their effects on the overall system injection efficiency into account.

Like PFHS, InRo supports its payload (PosS, FiTS, AGC, and PAC), during both normal observing and daytime maintenance. The payload is cantilevered for a single bearing system. Under all conditions, InRo and its payload must withstand the anticipated seismic loads of earthquakes, without sustaining a structural failure that damages the payload. This resilience will be confirmed in future work, at an overall system level, which will include the telescope structure’s seismic response characteristics.

Rotator systems with similar size and mass requirements are under development as part of several wide-field, multi-object spectrograph projects, being undertaken in collaboration with commercial vendors.

#### 4.3.8. Telescope Optical Feedback System (TOFS)

*(1) Acquisition and Guiding Cameras (AGC)*

The Acquisition and Guide Cameras system is a set of three off-axis cameras, with individual field of view large enough to facilitate guide star acquisition and guiding during observations (Figure 72). The cameras provide control signals for real-time feedback to telescope guiding and measurements of top-end misalignment with respect to M1. The top-end measurements are used to develop focus and pointing models, and lookup tables. This allows for compensation for fabrication and installation tolerances, and adjustment for gravity and temperature effects.

The cameras are mounted on the Instrument Rotator, within the field of view, along with the Fiber Positioner System, and they link the fibers’ positions from the hexagonal science focal surface to the sky positions. The precise locations of the cameras with respect to the science focal surface are essential. Each camera therefore contains self-illuminating fiducials—which are viewed and measured by the fiber positioner metrology system—together with self-illuminating reference fiducials within the science field of view and the back-illuminated fiber input location of the positioners. The fiber input locations are back illuminated by the spectrographs.

In the MSE baseline, guiding and tracking the rotating field of view will be accomplished using three 1,024 x 1,024 pixels (15 microns) CCD guide cameras, located at 0.709 degrees radially from the optical axis, at 120 degree intervals around the MSE field of view. The field of view of each camera is 142 arcseconds x 142 arcseconds (15 mm x 15mm). Analysis[17] has shown that the combined field of view is sufficient to acquire three guide stars, one at each camera, at magnitude *i*=19, for 95% of the simulated pointings near the North Galactic Pole, where the density of stars is expected to be lowest. Light from the *i*-band stars is used in the AGC, in order to minimize the effects of chromatic dispersion, while remaining within a relatively high quantum efficiency region of the camera's CCD detectors.

The overall guiding accuracy is 0.055 arcseconds. Error contributions from the telescope mount control system, instrument rotator, and AGC have been included in the guiding budget allocations. The centroiding error analysis, and the expected star profile on the detectors, based on the delivered image quality and noise contributions (Poisson noise, sky flux, and detector noise), suggest that the *i*=19 magnitude limit will also meet the AGC guiding error allocation for a one-second exposure time.

Moreover, we have performed alignment analysis, using the guide cameras to ascertain the WFC rigid body misalignment in decenter, tip and tilt, with respect to M1. Using the Zemax to simulate the inside and outside focus images[18] and post-alignment errors typical of opto-mechanical process as test cases, the analysis was able to discern the misalignment and determine the PFHS correction required.

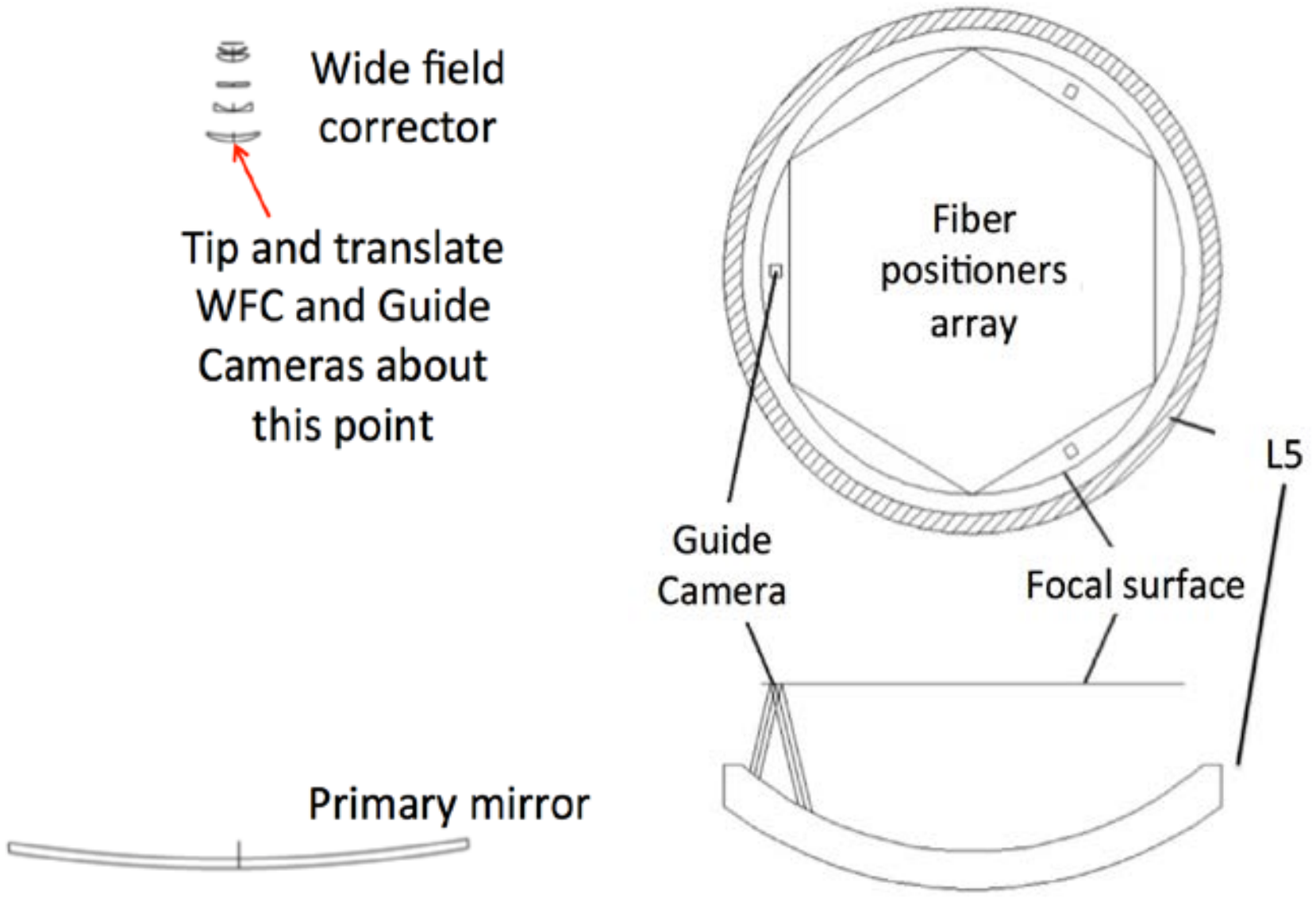

*Figure 72: Left side view of the origin of misalignment in the Zemax simulation and (right) top view of the nominal locations and sizes of the Acquisition and Guide Cameras with 1k x 1k CCD guide camera in the telescope-delivered circular field of view.*

The conceptual design of the AGC system's opto-mechanical packaging is planned for the next phase. This will include the incorporation of larger 2k x 2k detectors, which will ensure brighter guide stars are available ($i$=17) in 95% of the sky near the North Galactic Pole. This will enhance the guiding accuracy, by increasing the guide loop control bandwidth, with shorter exposure time.

---

[17] Using 1000 random telescope pointings near the North Galactic Pole (HA 194–95 degrees, Dec +27 to +28.0 degrees) from the SDSS star catalog. Star counts were made in each of the three boxes, defined by the size and positions of the guide camera fields, with 142 arcseconds on a side.

[18] Analogous to curvature wavefront sensing, but not solving for wavefront errors or Zernike polynomials—instead, directly solving for misalignments, by decomposing the images into circular functions.

In addition, the optical alignment strategies of the WFC will be formalized in parallel with the development of the Phasing and Alignment Camera system.

*(2) Phasing and Alignment Camera (PAC)*

A feasibility study of the Phasing and Alignment Camera system, conducted by the Jet Propulsion Laboratory and Synopsys, Inc., has determined that a system similar to those at Keck and TMT is applicable to MSE. The PAC design concept is based on both the Phasing Camera Systems (PCS) operating at Keck I & II, and the proposed design for the TMT Alignment and Phasing system (APS). The two PCS instruments have been in operation for over forty years in total, and have successfully aligned the Keck telescopes hundreds of times. PCS is used to measure and correct for segment piston tip and tilts, and for secondary mirror piston tip and tilt, as well as to measure the surface shape of the segments, for correction via the manual warping harnesses.

For MSE, PAC maximizes the M1 delivered image quality by performing the following adjustments:

- M1 segments in piston tip and tilt,
- M1 segment surface figure, via warping harness,
- PFHS rigid body motion with three degrees of freedom: piston, decenter, tip and tilt.

The PAC system is a Shack-Hartmann (SH) wavefront sensor system, with lens slit arrays to generate three progressive sub-aperture geometries to “phase and align” the M1 system:

- SH-0: capturing segments in tip and tilt after a segment exchange, using one spot per segment at the center of each segment to stack the segments
- SH-3: measurement of segment tip and tilt and PFHS piston tip and tilt alignment using 37 sub-apertures per segment, in a sparsely sampled hexagonal pattern
- SH-5: measurement of segment shapes as well as segment tip and tilt and PFHS piston tip and tilt using 91 sub-apertures per segment, in a densely sampled hexagonal pattern

Like AGC, the PAC camera is located at the top end of the telescope on InRo. The PAC-deployable pickoff mirror intercepts the light on axis immediately after the WFC and generates an optically high quality re-imaged pupil of M1, as required for phasing at the SH lenslet array (Figure 73). The phasing procedure uses 120 mm sub-apertures, projected on M1, and the spot size specification is better than 10 mm, which corresponds to 22 microns at the 25 mm diameter re-imaged pupil. Using a nominal design demagnification ratio of 450, a pupil size of 25 mm was selected, in order to minimize the instrument size, since the available top-end space for PAC is limited. The image quality requirement applies across the operating spectral band of 600 nm to 900 nm and over the 10″ field of view.

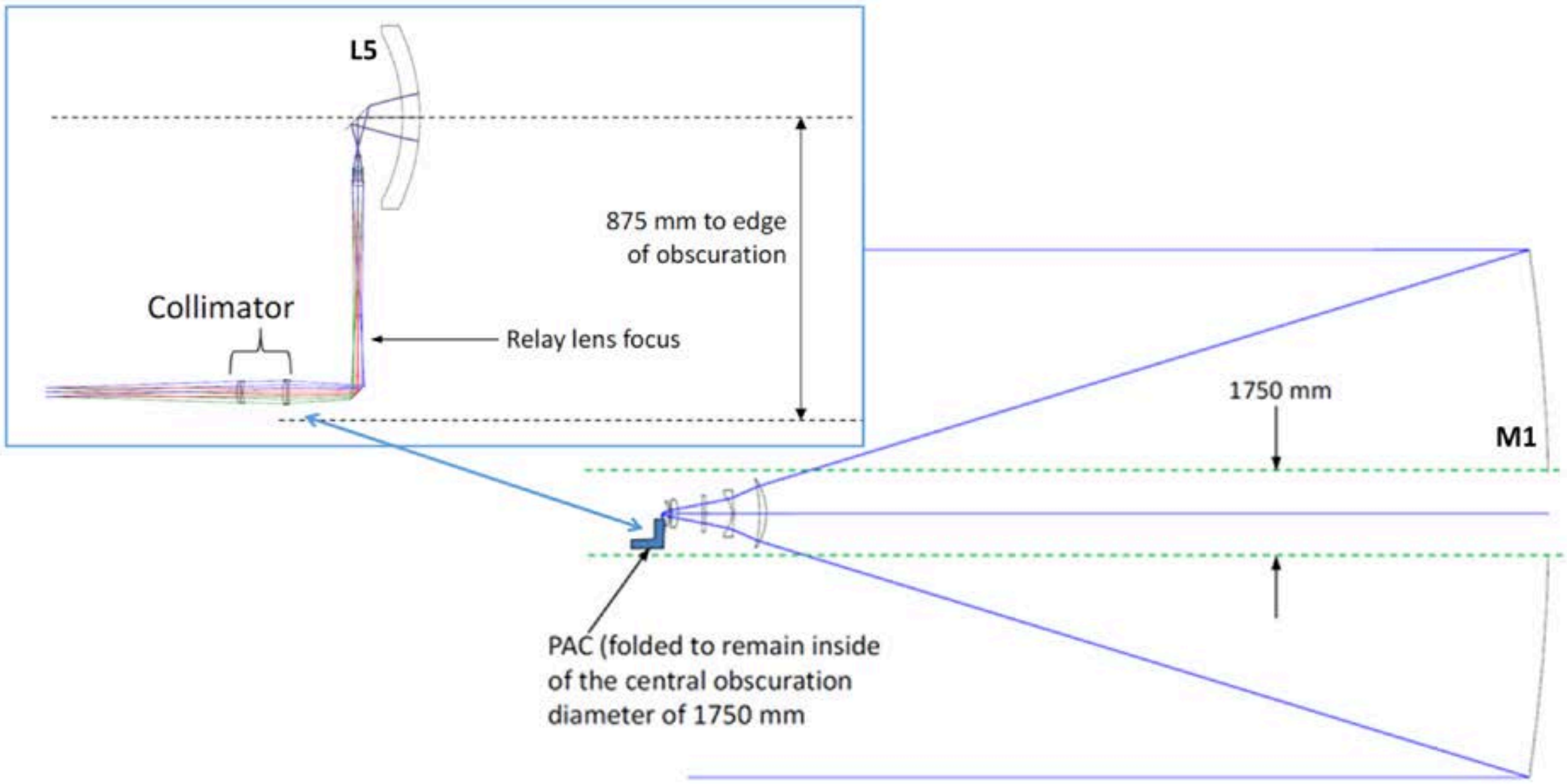

*Figure 73: Phasing and Alignment Camera optical layout*

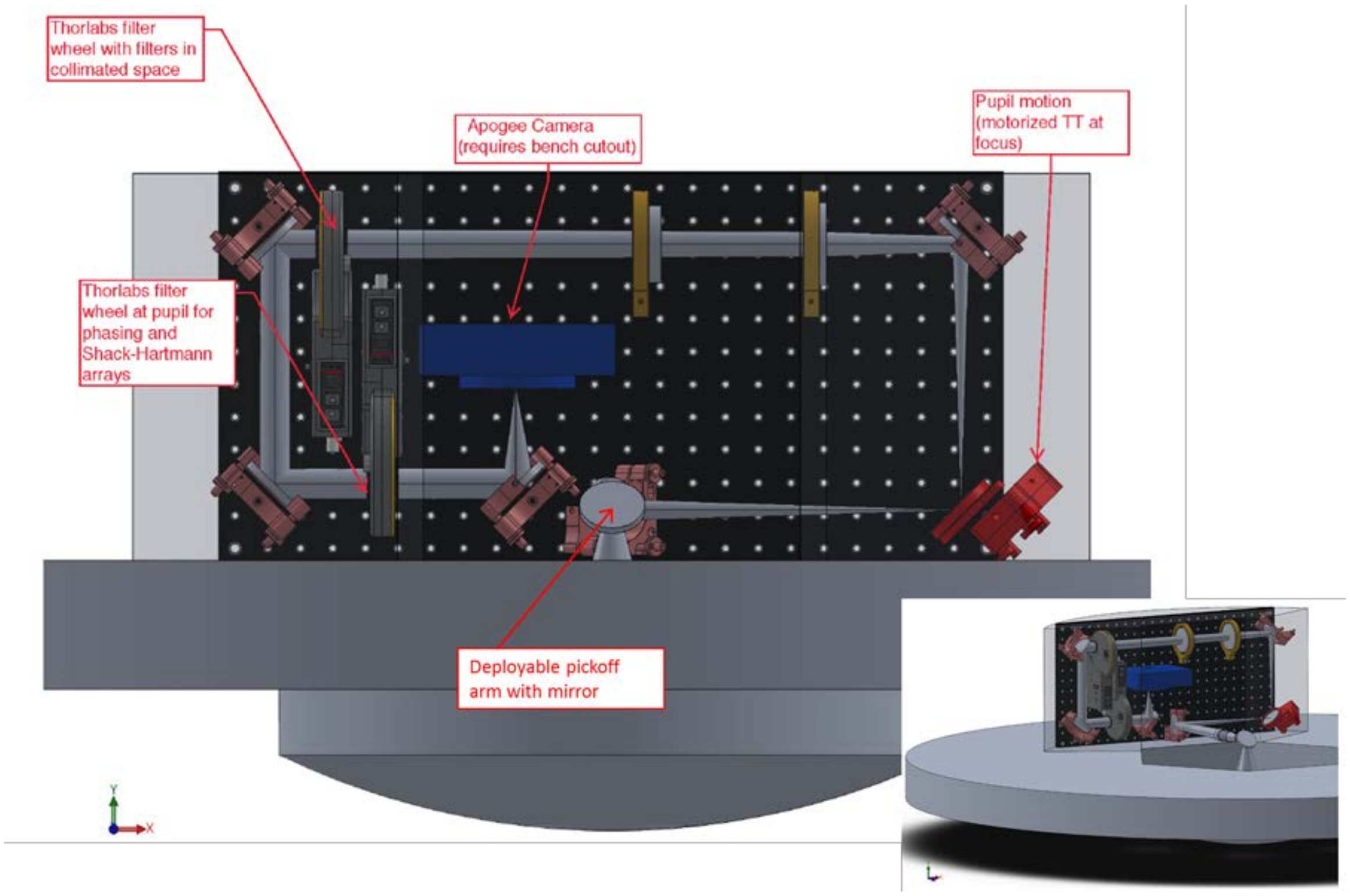

*Figure 74: PAC opto-mechanical layout—the top of gray disk (solid color) represents the interface plane for the optical bench and the semi-transparent gray box represents the allowable space envelope.*

The PAC design (Figure 74) includes an optical bench, to support commercially available opto-mechanical components, such as optical mounts, filter wheel mechanisms, and CCD cameras. A pickoff arm (Figure 74, left) is deployed on axis, to pick up the light, which is then folded into the optical bench. A steering mirror has been positioned at the re-imaged optical focus, to provide pupil translation control. Afterwards, the beam is re-folded and passes through the two collimating elements, to form a collimated beam, nominally 25 mm in diameter. A Thorlabs filter wheel holds the six filters needed for broadband and narrowband phasing. The beam is folded once again and passes to the next Thorlabs filter wheel, which holds the required phasing and Shack-Hartmann arrays. The light is then relayed onto an Apogee CCD camera, via the relay lens. In the current concept, the Apogee camera (blue box) is mounted through a hole in the optical bench, in order to obtain the correct camera height with respect to the incoming light. Not shown in the illustration: the relay optics between the pickoff mirror and the steering mirror, and the deployable field stop at the re-imaged focus, which limits the field to 10″ in SK mode, down from 30″ in imaging mode. The larger 30″ field of view is specified for the imaging mode, to facilitate acquisition, when exquisite image quality over the full field is not required.

The PAC system includes its own Instrument Control System (ICS) for all components on the optical bench. The PAC ICS includes the computers, electronics, software, and cables needed to interface and control the opto-mechanical components. Components controlled by the ICS include the motors, CCD camera, shutters, photo-diodes, light sources, and temperature controllers.

In addition, the PAC Procedure Executive and Analysis Software (PEAS) provides the central interface for all alignment and phasing activities. PEAS will interact with ICS and the appropriate MSE telescope software interface (probably the Telescope Control Sequencer) to analyze and correct misalignments, using a set of predefined procedures. PEAS is the software framework within which the analysis computations and alignment procedures are executed, and it is similar to the APS PEAS recently deployed at Keck, which provides the operational software for both the Keck I and II PCS and is routinely used by the telescope operators.

The Project Office has expressed a high degree of confidence in the proposed PAC design. In the next phase, we will work to finalize the design options for the pupil translation mechanism, tilt plate in the collimated beam vs. steering mirror near the re-imaging focus, and the opto-mechanical packaging of PAC, according to the interface control requirements. The PAC performance requirements will also be reconciled with and incorporated into the system performance budgets.

#### 4.3.9. Telescope Safety System (TSS)

The Telescope Safety System (TSS) is a hardware-based safety system, with lock-outs and e-stops that enforce personnel and equipment safety. The TSS is the safety component of the telescope and the safety “supervisor” of the Telescope Mount Structure (MSTR) and Mount Control System (MCS). The TSS operates both autonomously and in concert with the Observatory Safety System (OSS), to ensure local and global safety. Safety equipment includes a safety-rated, programmable logic controller, remote input/output devices, local and global e-stops, sensors, and the wiring required by the MCS and OSS.

TSS includes the safety related components, software, and hardware—such as a motorized elevation structure counterbalance system and deployable locking pin system, to hold an unbalanced elevation structure at predetermined zenith angles for maintenance and servicing—and the input/output devices required to interface with both systems. TSS also monitors the health and status of the MSTR components—such as the MCS, motorized cable wraps, and hydrostatic bearing system—to ensure operational and equipment safety. TSS also includes data and communication lines and liquid-cooled, insulated electronic cabinets.

## 4.4. Science Instrument Package (SIP)

### 4.4.1. Science Instrument Package Overview

MSE's Science Instrument Package (SIP) takes advantage of Maunakea's excellent site seeing, large aperture, and large field of view, by implementing a highly multiplexed, fiber-fed, spectroscopic system, capable of collecting tens of thousands of spectra each night. SIP is a collection of subsystems, a.k.a. products, which work together to collect the light delivered to the prime focus by the telescope, transmit it to the spectrograph suites, and convert it into detector counts for millions of science targets per year. The science calibration hardware needed to produce science-ready data products by enabling the necessary calibration procedures, is also included in the package. The SIP subsystems were designed by various institutions in MSE participants' countries and are shown in *Figure 75*.

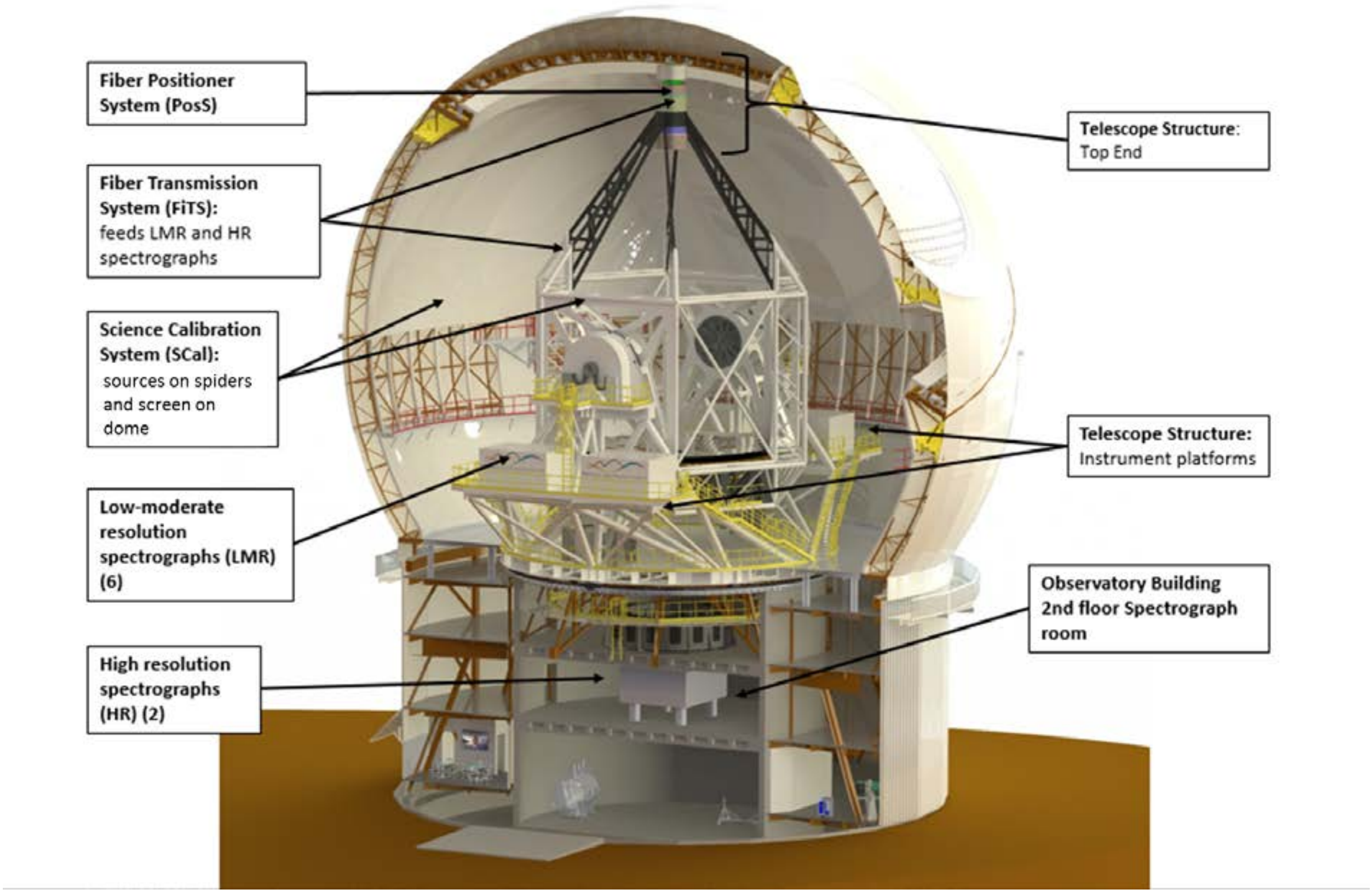

*Figure 75: Components of Science Instrument Package.*

The subsystems included in SIP are a Fiber Transmission System, which comprises an array of 4,332 fiber optics at the focal surface; a Fiber Positioner System, which contains an array of 4,332 piezo actuators and their control and metrology systems; two banks of High Resolution (HR) and Low/Moderate Resolution (LMR) spectrographs, several tens of meters away; and the Science Calibration System (SCal), comprised of the calibration hardware that ensures quality observations. The second generation Integral Field Unit (IFU) capability (see Section 3.4.6) will be part of the SIP system, once implemented.

The spectral resolution, degree of multiplexing (i.e., total number of spectra), and wavelength range needed in each of the LR, MR, and HR configurations are prescribed by high-level science requirements. Observations using the HR spectrographs involve 1,083 fibers of 0.8 arcseconds (85 microns) in diameter, which feed two spectrographs, located in the Inner Pier spectrograph room. Observations using the LMR spectrographs involve 3,249 fibers of 1.0 arcsecond (107 microns) in diameter, which feed a suite of six spectrographs, located on two spectrograph platforms, near the elevation axis of the telescope. The light enters the fibers at f/1.926, as delivered by the telescope, and exits at a faster focal ratio, due to the effects of focal ratio degradation. FRD from the fibers is assumed to be 5%. Both arrays of fibers (HR and LMR) are independently capable of full field coverage at the focal surface, enabled by the large patrol area of the Sphinx positioners, which use tilting spine technology. During the integration of the fibers with the positioners, the HR and LMR fibers are distributed in the field of view (*Figure 76*), with their tips arranged on the convex focal surface.

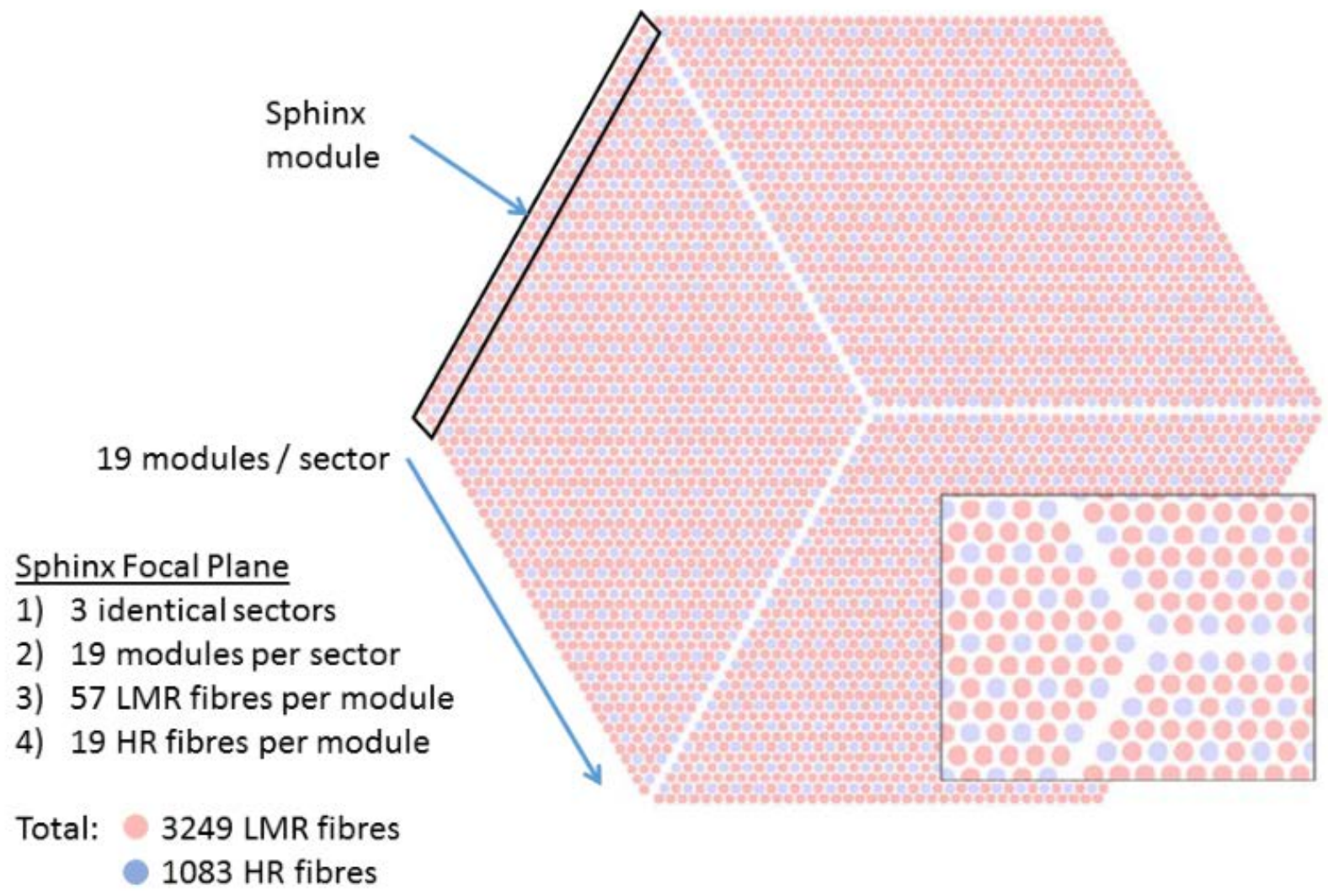

*Figure 76: Focal plane arrangement—full field coverage with distribution of LMR fibers (pink) and HR fibers (blue).*

At the beginning of an observation, PFHS moves through five degrees of freedom, to position the entire assembly of InRo, AGC, FiTS, and PosS in the optimal position with respect to the focal surface. During observations, InRo ensures that the entire fiber array is precisely rotated with respect to the telescope structure, in order to track the field and compensate for the rotation of the Earth.

### 4.4.2. Science Instrument Package Functions

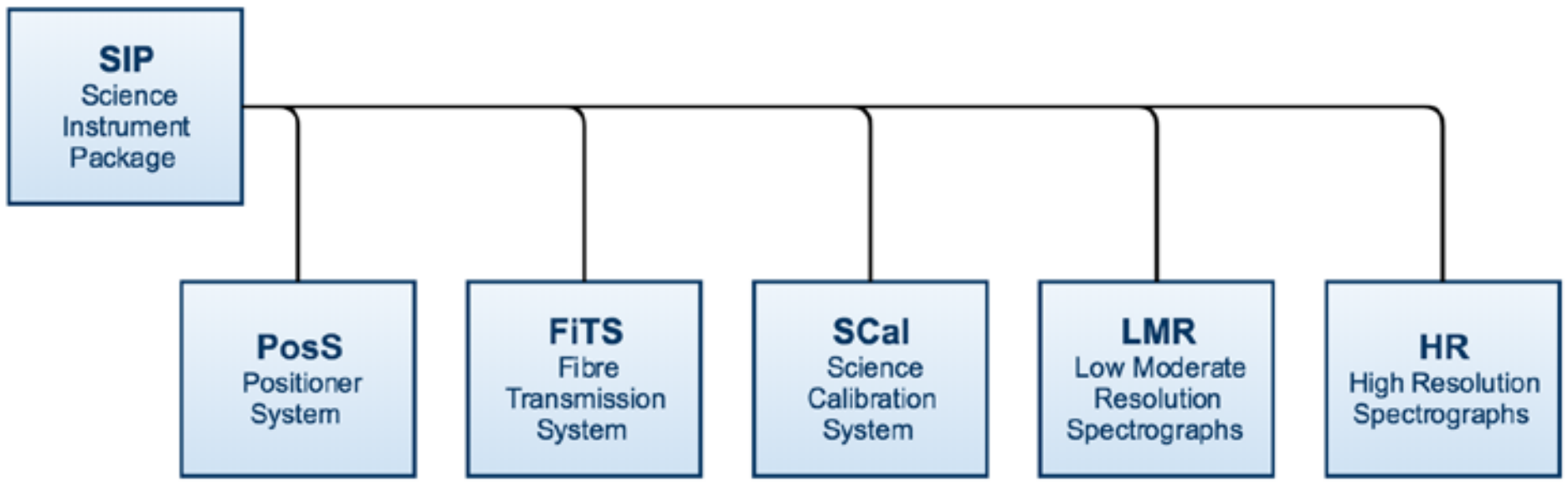

*Figure 77: SIP Product Breakdown Structure*

SIP is organized into the following PBS elements, based on their functions (*Figure 77*).

- The Positioner System (PosS) provides for the positioning of individual fiber input ends in the field of view, which includes the associated motion control and metrology cameras.
- The Fiber Transmission System (FiTS) collects light and transmits it to spectrographs, as well as protecting, routing, and handling the fibers along the telescope structure.
- The Science Calibration System (SCal) generates the “reference” photons needed to ensure that the science data is of the best possible quality.
- The Low and Moderate Resolution Spectrographs (LMR) receive light from the low/moderate resolution fibers and convert it into raw science data.
- The High Resolution Spectrographs (HR) receive light from the high resolution fibers and convert it into raw science data.

### 4.4.3. Positioner System (PosS)

The Sphinx positioner system is an Australian Astronomical Observatory (AAO) design. It is an evolution of a piezo-actuated tilting spine technology, first implemented in FMOS-Echidna (Subaru). Over the course of its on-going development and prototypes, titling spine positioner technology has been proposed for WFMOS (Gemini), WFMOS-A (AAT), Mohawk (Blanco), and 4MOST (VISTA). It represents a mature, low risk, high performance solution to MSE positioner requirements.

The Sphinx design includes two main components: the array of positioners, which corresponds to an assembly of actuators, together with their support structure and electronics, located at the top end of the telescope structure; and a fiber positioner metrology system, located in the empty central segment position of the primary mirror (*Figure 78*).

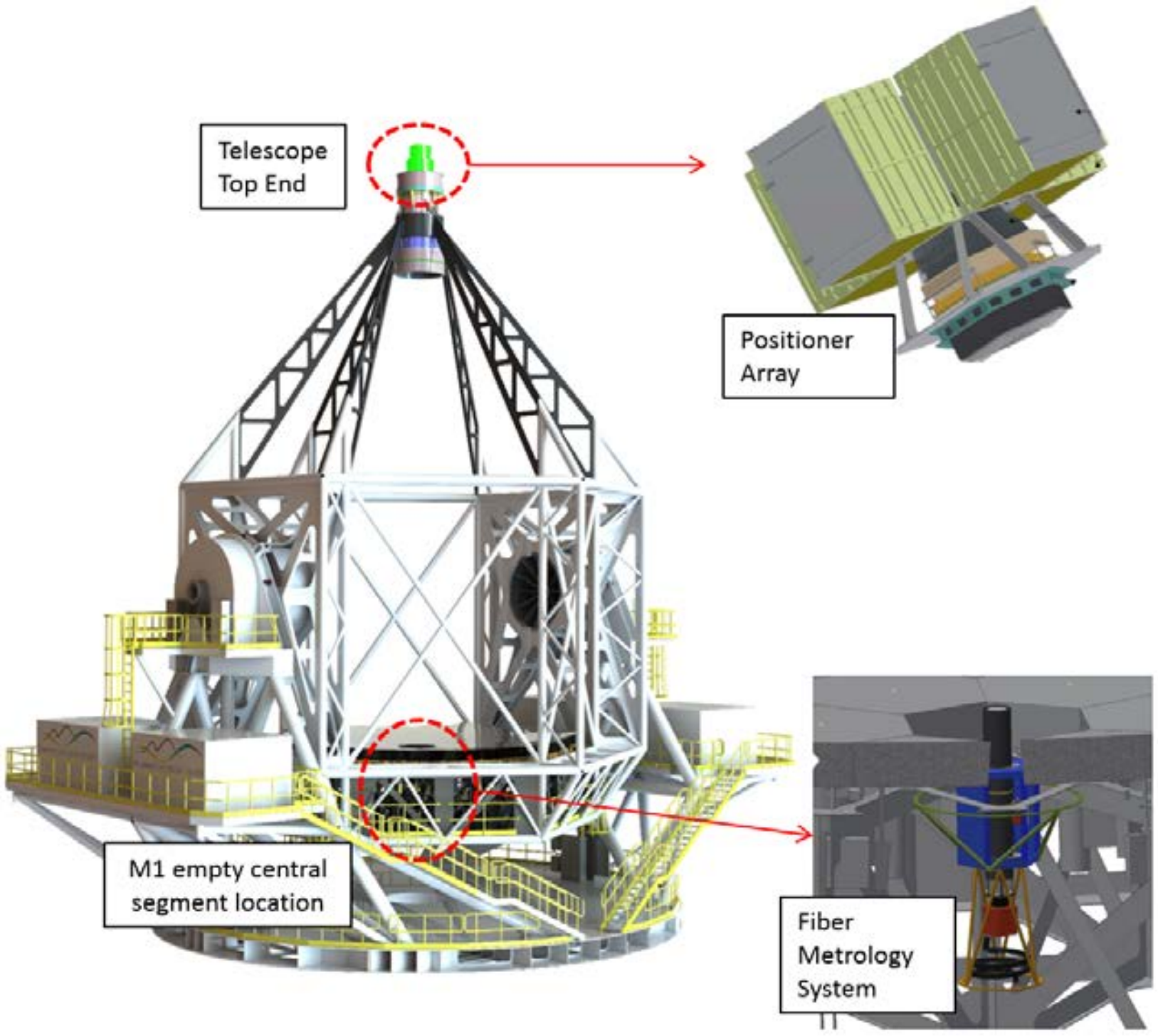

*Figure 78: PosS components in the MSE telescope structure—Positioner array and control electronic cabinets are located at the top end (top right) and the metrology system is located in the empty central M1 segment location (bottom right).*

### *(1) Positioner Array*

The positioner array is located at the focal surface of MSE and composed of an arrangement of individual "spines." These spines carry the FiTS fibers. Piezo actuators move them into position to collect light from science targets. Each spine assembly (*Figure 79*, top left- and right-hand panels) includes several components. A telescopic arrangement of two carbon fiber tubes forms each spine; each spine houses a FiTS fiber, together with its ferrule, and is supported by a pivot ball, held in place by a magnetic "cup," which acts as the fulcrum for the tilt motion. Driven by its control system, the piezo actuator contracts and expands. This produces a stick–slip motion at the cup and ball interface, which translates into an angular displacement of the spine in tilt. The tilt of the carbon fiber tube produces lateral motion at the fiber end, moving it toward the target position on the focal surface.

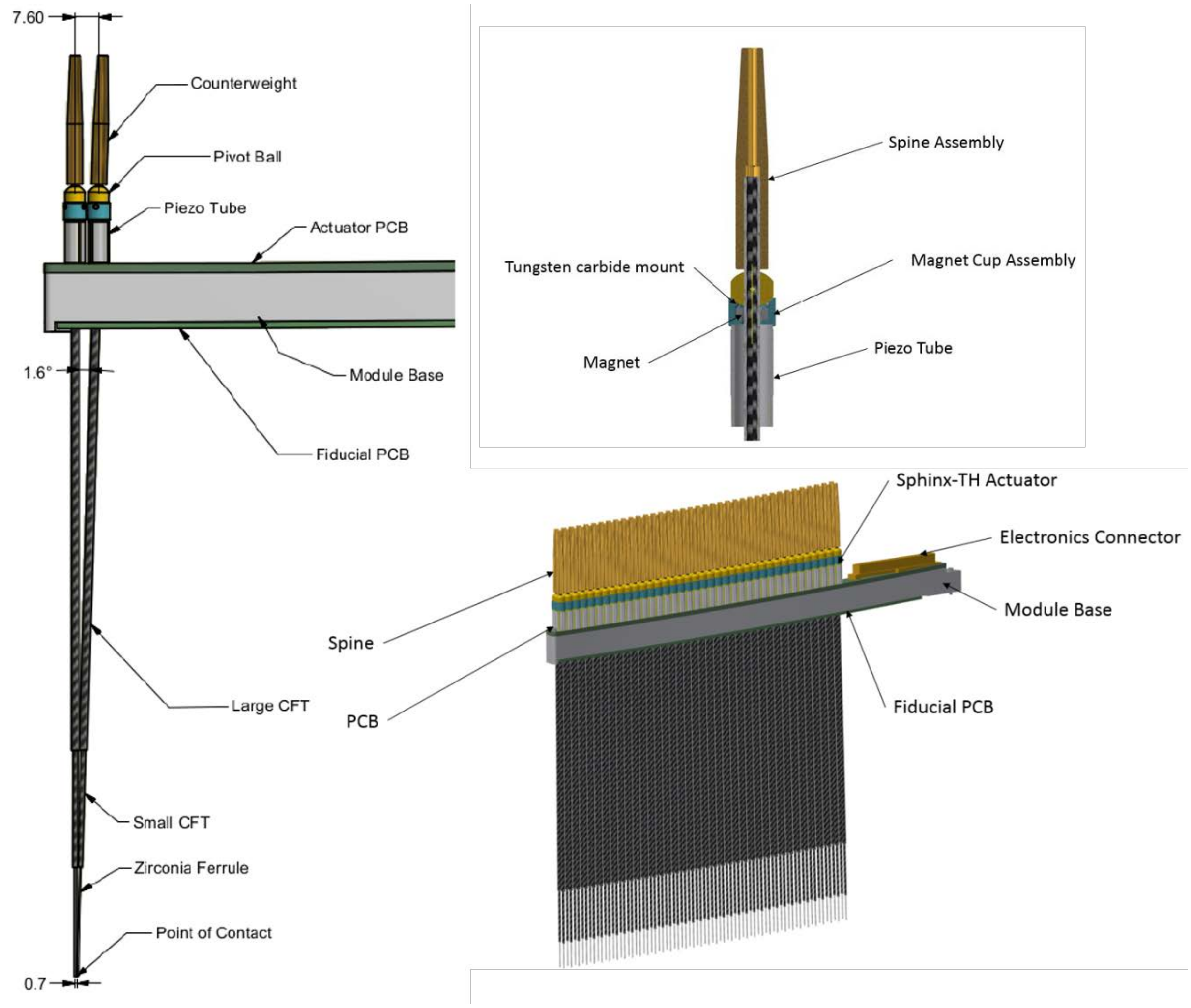

*Figure 79: Sphinx fiber positioner module—Each module carries 76 spines, with 57 LMR and 19 HR fibers (bottom right), two adjacent spine assemblies (left) and a close-up of the piezo actuator (top right).*

To aid assembly and maintenance, spine assemblies are mounted on identical modules (*Figure 79*, bottom right panel). Each module carries 76 spines, with 57 LMR and 19 HR fibers. These are integrated into a hexagonal reference support structure, to form the main assembly (*Figure 80*). The entire assembly consists of 57 modules, arranged in a Triskele pattern, which fills the hexagonal field of view, making a total of 4,332 spines (3,249 for the LMR fibers and 1,083 for the HR fibers). With Sphinx, the pitch of the actuators (the distance *p* between any two positioners) is 7.77 mm and their patrol radius is 1.24 p. This is enough to allow both HR and LMR fibers to simultaneously cover the full field (*Figure 81*), with the exception of a very small area at the center of the field (less than 0.01% of the total area), where the modules are attached to three support struts on the reference structure. The reference structure provides an accurate and repeatable interface with InRo. Once mounted onto InRo, the fiber inputs conform to the spherical focal surface.

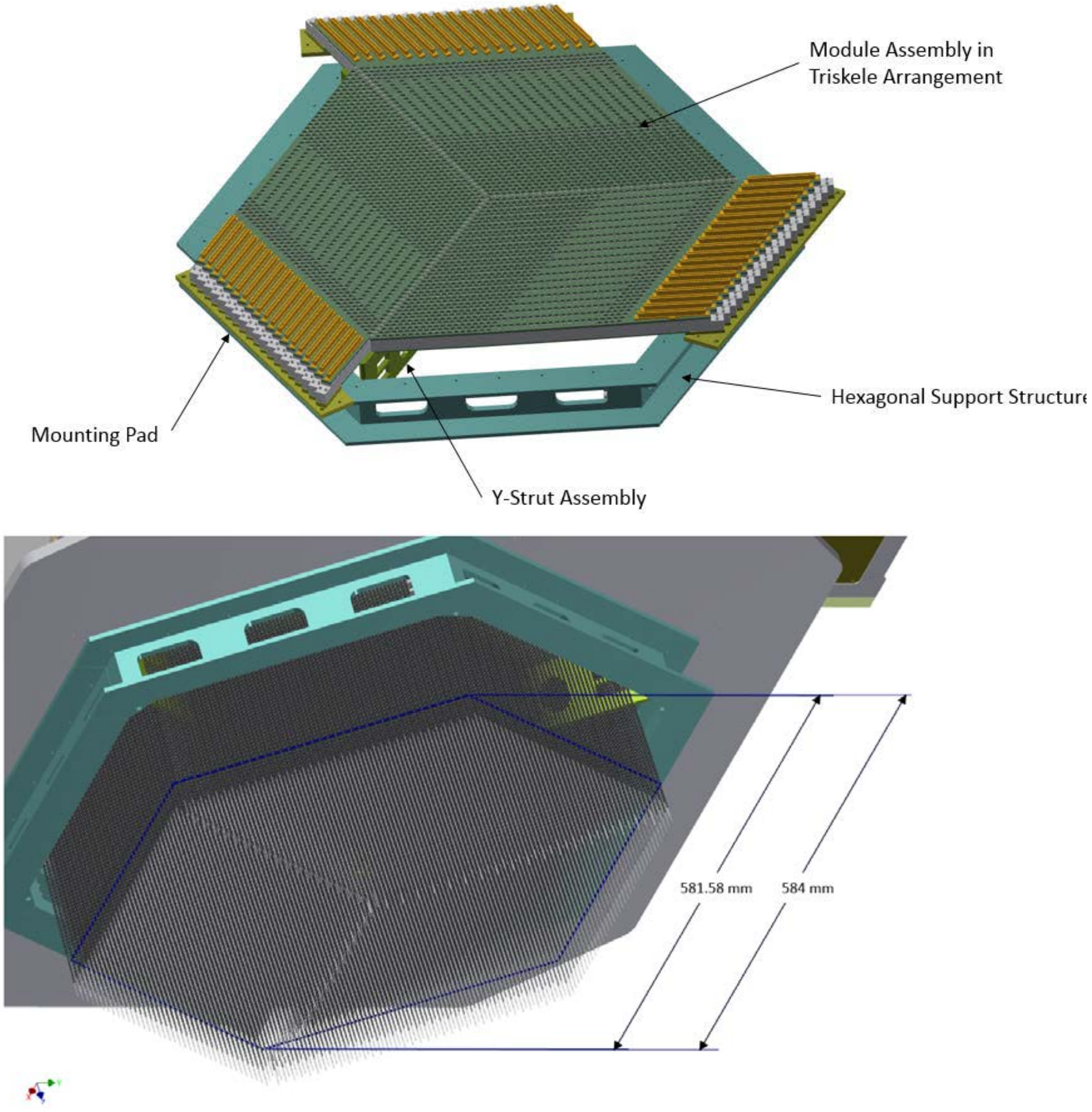

*Figure 80: 57 Sphinx positioner modules mounted onto the reference support structure in the Triskele pattern.*

The entire positioner assembly, including control electronics and insulated enclosures, is mounted onto InRo, which rotates the input fiber array with the observing field, as the telescope tracks the sky. InRo is in turn mounted onto PFHS, which places the input fiber array at the optimal focal surface position at the beginning of every observation, thus maximizing injection efficiency.

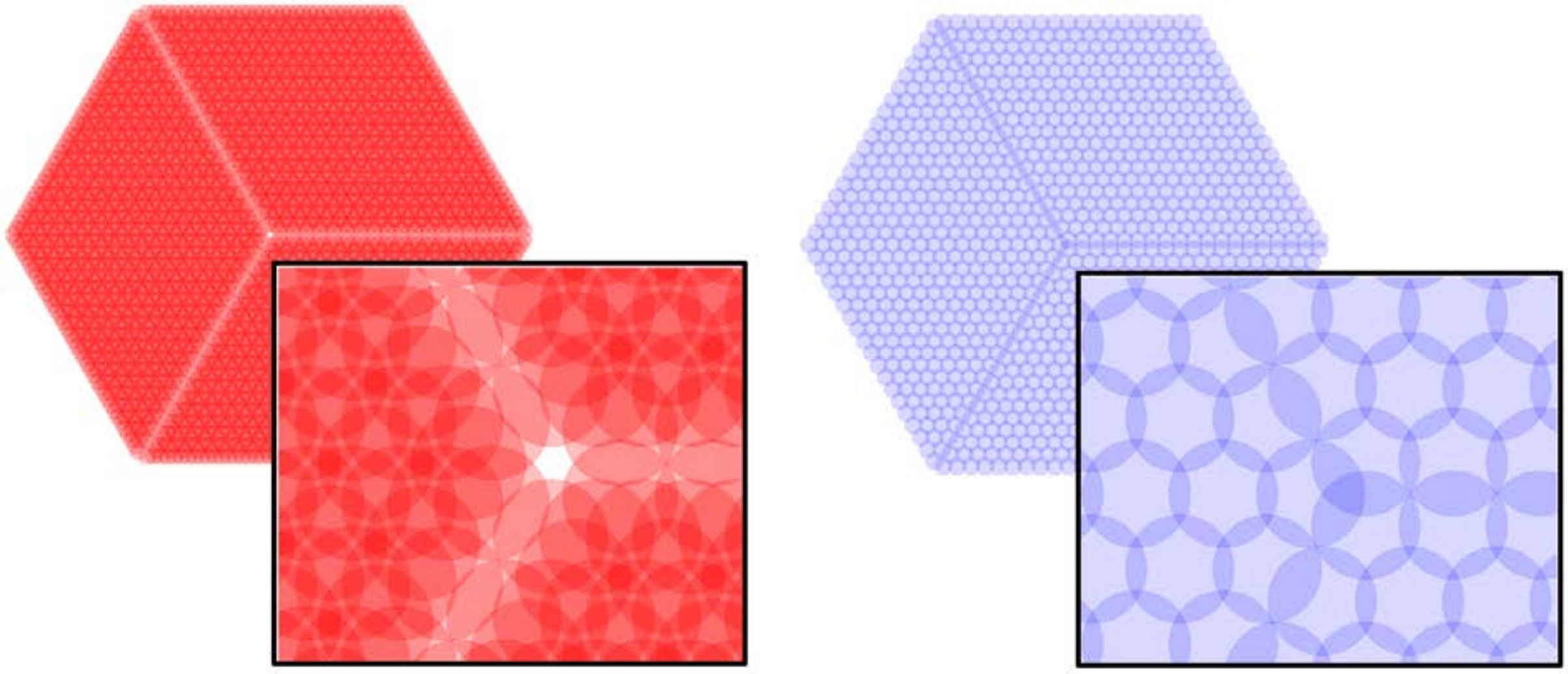

*Figure 81: Positioner patrol areas and field coverage—LMR (left) and HR (right).*

### *(2) Fiber Positioner Metrology System*

In order to accurately position the fibers at the focal surface, each fiber is backlit at the spectrograph slit input. This allows the metrology system to measure the position of each fiber input in the focal surface, as seen through the WFC/ADC, using the metrology camera located at the center of M1. The fibers are directed to their individual targeted positions and the metrology system provides the positioner control system with position feedback correction, by comparing each fiber's current position with its targeted position. This procedure is repeated until all the fibers have reached their targeted positions, within allowable tolerances. It nominally takes an average of four iterations, lasting less than 70 seconds, to position all 4,332 fibers to within an accuracy of 6 microns RMS.

The metrology system includes a large-format, commercial CCD camera, with optics which can image the full array of fibers and the surrounding focal surface fiducials, including those of the acquisition and guide cameras. The metrology system mounted on the mirror cell structure rotates synchronously with the InRo rotation. This allows for metrology exposures as long as 1 second, while InRo slews to the new angular position, ready for the next science observing. Since angular reconfiguration of InRo can take up to one minute, allowing PosS to reconfigure in parallel minimizes configuration time and improves observing efficiency. Once reconfigured, the metrology system takes a final high accuracy measurement, making a record of the as-observed fiber positions by averaging several exposures over a period of up to 10 seconds, to eliminate seeing-induced errors.

### *(3) Tilt-induced Effects*

One of the downsides of the spine design is the tilt-induced effect on defocus and Point Spread Function (PSF), especially near the limit of their patrol radius, i.e., at maximum tilt. An analysis of the injection efficiency of the entire MSE system, which also took effects such as image quality into account, deemed the worst case tilt-induced defocus of 80 microns to be acceptable. In addition, the mean defocus of all positioners within an observing field can be estimated and partially compensated for, by slightly offsetting PFHS by the same amount, to further reduce the range of the defocus. Simulations of the fiber-to-target allocation have shown that, in a typical tilt distribution, very few spines are at the extreme edge of their patrol regions.

The tilt-induced PSF variations affect the wavelength solution and are accounted for by the proposed science calibration procedure (see Section 3.5.4 and 4.4.7).

*(4) Control Software*

The Sphinx control system software interfaces with the Instrument Control Sequencer (ICSe, see Section 4.5.3) and moves fibers to precise positions on the focal surface. The software controls the amplitude of these moves, by applying precise control waveforms[19] to all spines simultaneously. The number of steps and the direction of each move are both spine-specific. Key requirements for the software include conforming to the ICSe, providing calibration functionality, permitting spines to be assigned to clustered targets, and quickly positioning all spines at their assigned targeted positions, within a specific tolerance.

The control software can ensure that the spines do not collide, cross, or otherwise interfere with each other, by correctly routing them between current and targeted positions. The software is heavily reliant on the measurements of the metrology system and the self-illuminating fiducials, located at the focal surface, for fiber identification. It can distinguish nearby fibers by tracking their starting positions and the trajectories of their moves. A disaster recovery procedure has also been incorporated. In the event of a serious fault (e.g., a power failure or earthquake), it provides a fully automated way of reliably moving all spines to their safe positions, without the need for human intervention. Once they have reached the safe positions, the PosS system will be reinitialized.

*(5) Sphinx Design Alternatives*

The proposed actuator for MSE is the Sphinx-TH, which uses piezo tube actuators driven at high voltages of up to 400 Vdc, a design based on that used by FMOS/Echidna. The high voltage drive results in a large electronics package—an undesirable effect, given the highly constrained top-end volume. Two alternative versions of the actuators, each with different piezo arrangements— the Sphinx-S and Sphinx-TL designs—are currently under consideration. The Sphinx-S uses an arrangement of three piezo stacks and the Sphinx-TL uses the same piezo tube geometry as the Sphinx-TH. Both the Sphinx-S and the Sphinx-TL piezo actuators are driven at a lower voltage, +/-15 Vdc. Because of their lower drive voltage, smaller electronics and potentially better accuracy are attainable. These actuator technologies are currently under development at AAO and their technical readiness for MSE implementation will be evaluated during the next design phase.

[19] The waveform is a timed series of voltages, applied to the piezo spine actuator in a “saw-tooth” profile.

### 4.4.4. Fiber Transmission System (FiTS)

The Fiber Transmission System (FiTS) is a fiber optic relay, which supplies 4,332 fibers, which collect the light at the focal surface, in concert with the fiber positioners system (PosS, see Section 4.4.3), and deliver it to the LMR and HR spectrographs (see Section 4.4.5 and 4.4.6 ), located on the instrument platforms and in the second floor spectrograph room, respectively (*Figure 82*). The FiTS design team is led by Herzberg Astronomy and Astrophysics (HAA) of the National Research Council (NRC) in Canada. The HR and LMR fiber diameters are 0.80 arcseconds and 1.0 arcsecond respectively, in order to take advantage of the exquisite image quality delivered by the MSE system.

FiTS provides routing and handling for the fibers, through the Observatory and over the full range of the telescope motion in two axes, and the rotation of InRo. Importantly, FiTS is designed to minimize fiber stress during observations. This serves to reduce differential focal ratio degradation and far-field effects as the telescope moves across the sky, as well as to maintain uniform and stable throughput among the thousands of fibers.

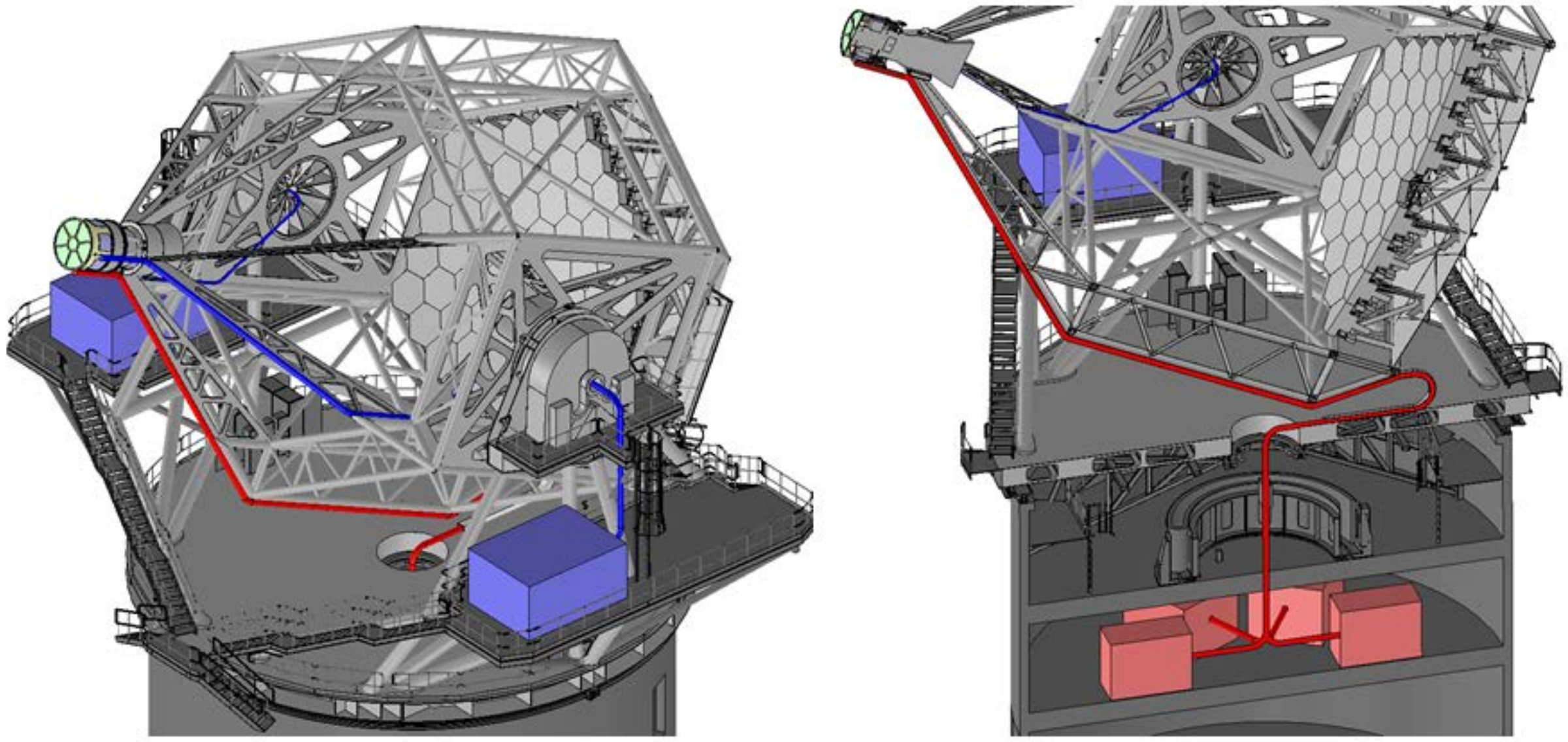

*Figure 82: FiTS cable routing—LMR fibers (blue lines) and HR fibers (red lines). The colored boxes illustrate the conceptual locations of the spectrographs.*

The design for FiTS builds on a collaboration between HAA NRC Canada and FiberTech Optica (FTO), which produced the extremely successful Gemini Remote Access Echelle Spectrograph[x] (GRACES) fiber system. GRACES is a fiber optic link between the Gemini North Observatory and the Echelle Spectro Polarimetric Device for the Observation of Stars (ESPaDOnS) spectrograph, located in the CFHT Coudé room (*Figure 83*). Despite the fibers' 270 m length, the system's FRD is less than 14% and transmittance is 85% at 800 nm, making GRACES more efficient than the High Resolution Echelle Spectrometer (HIRES) at Keck at wavelengths longer than 600 nm, despite a significant collecting area disadvantage (Gemini's collecting area is only approximately 64% that of Keck). The GRACES fiber link uses a high numerical aperture fiber (NA = 0.26), whose performance has proven extremely stable during observations.

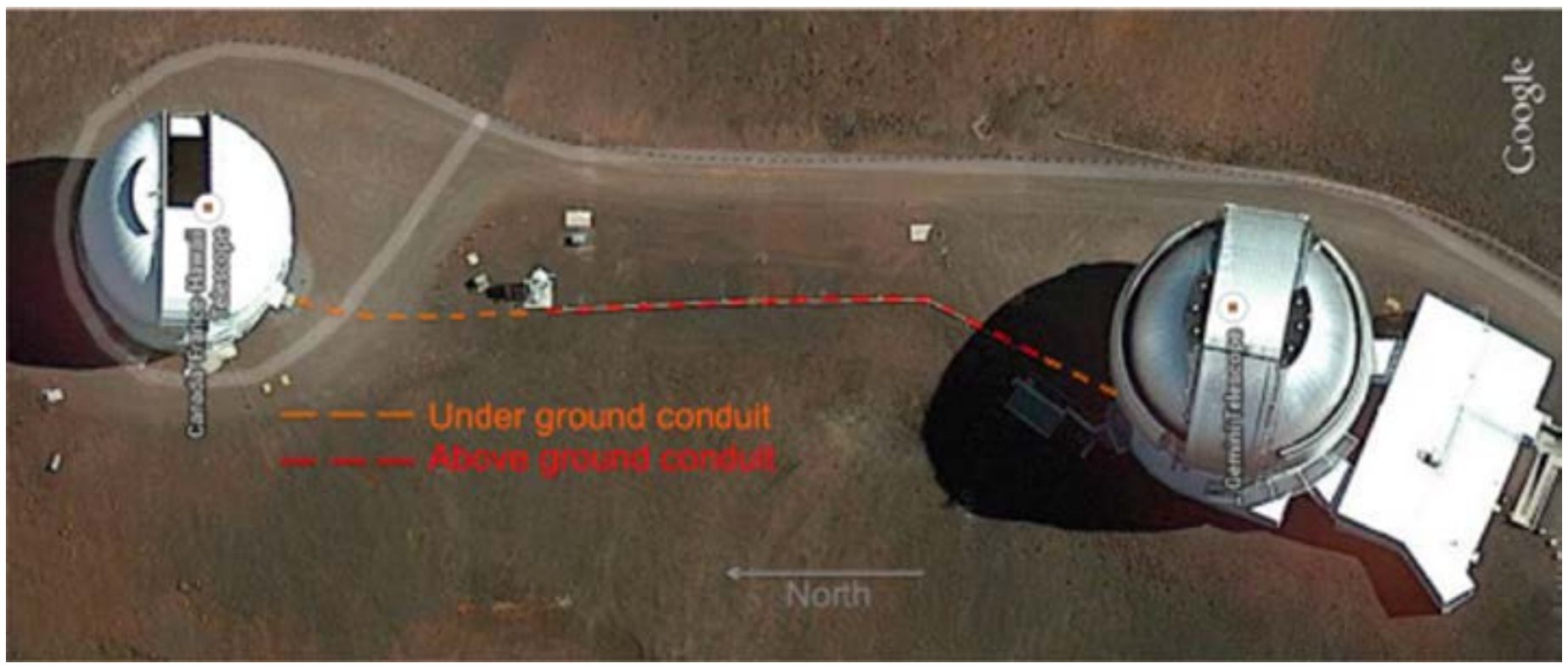

*Figure 83: GRACES fiber optic link*

FiTS is designed to maintain the highest possible throughput for the overall MSE system. The fibers are provided as a continuous link, without any connectors. High numerical aperture fibers (NA = 0.26–0.28) have been selected, capable of accepting the f/1.926 beam delivered by the telescope optics, in order to obviate the need for additional optics at the fiber input and avoid incurring the associated throughput losses. The fibers have an anti-reflection (AR) coating[20] on the output ends, as the baseline.

Throughput is affected by the fibers' length and transmission characteristics,[21] particularly at the blue end of the wavelength spectrum. The current configuration (shown in *Figure 82*) houses the HR spectrographs in the inner pier (50 m in length) and the LMR spectrographs on instrument platforms (35 m in length). However, this will form the subject of an upcoming trade study (see Section 3.5.2): high spectral resolution observations with MSE will benefit from a shorter fiber length, since blue throughput is of particularly important for the high-resolution spectroscopic science.

At the top end of the telescope, FiTS and PosS are closely integrated subsystems (*Figure 84*). Fibers are individually inserted into each spine of PosS. This process is currently being developed by the two design teams. FiTS provides a rotation guide system to protect and organize the fibers from the fiber combiner to the loop boxes, while InRo rotates both FiTS and PosS about the optical axis during observations.

---

[20] The methodology for applying AR coating at the input end will be explored in the next phase, but it is known to be challenging with a continuous fiber link system.

[21] The baseline fibers under consideration are Polymicro FBP and FBPI.

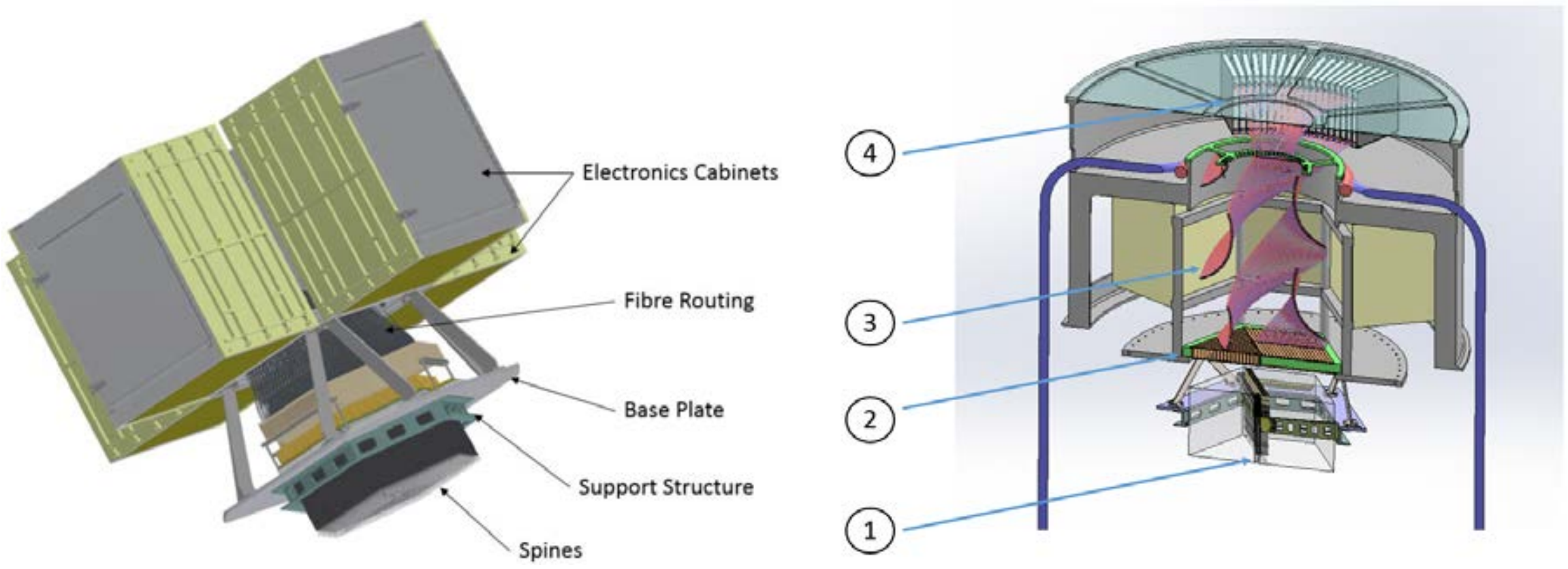

*Figure 84: PosS (left) and FiTS (right) at the top end: (1) Positioning system (simplified), (2) fiber combiner, (3) helical tubes, (4) loop boxes, which share space with the electronic PosS cabinets.*

The fiber cables are modular, which facilitates fabrication, testing, and maintenance. In accordance with this philosophy, each Sphinx positioner module (see Section 4.4.3) has a corresponding fiber cable. This results in 57 identical cables: one branch of which feeds one or more LMR spectrographs and the other one or more HR spectrographs. Figure 85 presents a systematic illustration of a fiber cable and its components.

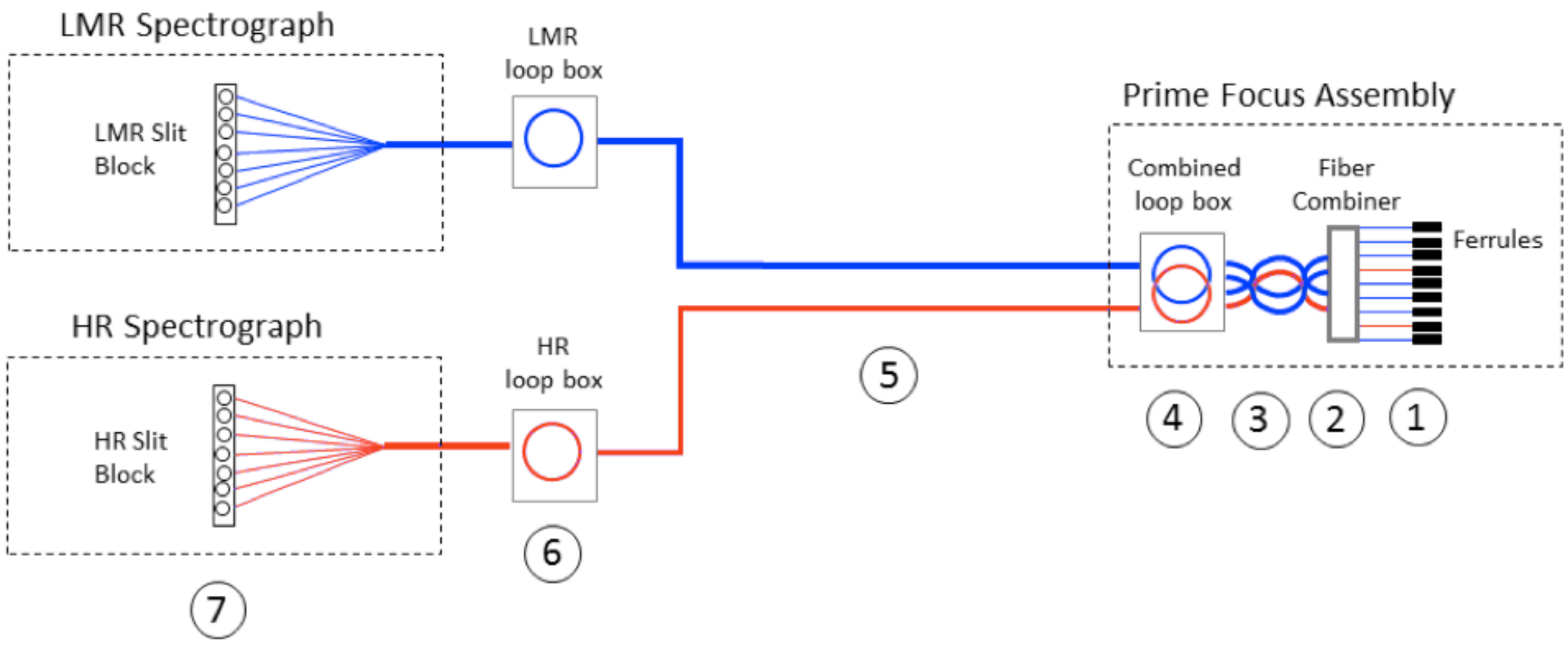

*Figure 85: Cable schematic: the photons enter the fibers in the Prime Focus Assembly from the right. See below for descriptions of components 1–7. To feed multiple spectrographs, additional fiber slit blocks can be added after each loop box (component 6).*

We can trace the fiber cable by following light through the system (*Figure 85*).

1) Individual fibers are grouped into four protection tubes at the fiber combiner.
2) Three tubes contain 19 LMR fibers each, while a fourth tube contains 19 HR fibers. The four tubes exit the back of each fiber combiner.
3) The protection tubes are wound through a helical section, to provide rotation compliance with the end containing the positioners, ferrules, and fiber combiners, as rotated by InRo. Fiber tubes enter the first set of loop boxes at the fixed end of the helical section.

4) Loop boxes provide access to bare fibers for splicing, in case repairs are needed.
5) Fiber cables are routed across the telescope structure (there are 57 fibers plus spares in each LMR cable tube; 19 fibers plus spares in each HR cable tube).
6) A second set of loop boxes is located near the spectrographs, providing access to bare fibers should repair be necessary.
7) Individual fibers are arranged into an array, using slit blocks. The slit blocks act as the entrance slits to the spectrographs (*Figure 86*).

At the inputs of the spectrographs, the fibers terminate in slit blocks, which provide the interface between FiTS and the spectrographs. The mechanical interface to the spectrographs accommodates the fiber output slit geometry determined by the LMR and HR spectrograph designs. For example, the shape of the slit compensates for optical distortion, such that spectra are projected onto the spectrograph detectors in straight columns (to maximize detector utilization). We expect the MSE slit blocks to leverage the three-dimensional AAT/HERMES[xi] curved slit block design and fabrication experience, and incorporate design concepts such as V-groove attachment and captive strain-relief, as proposed by HAA (*Figure 86*).

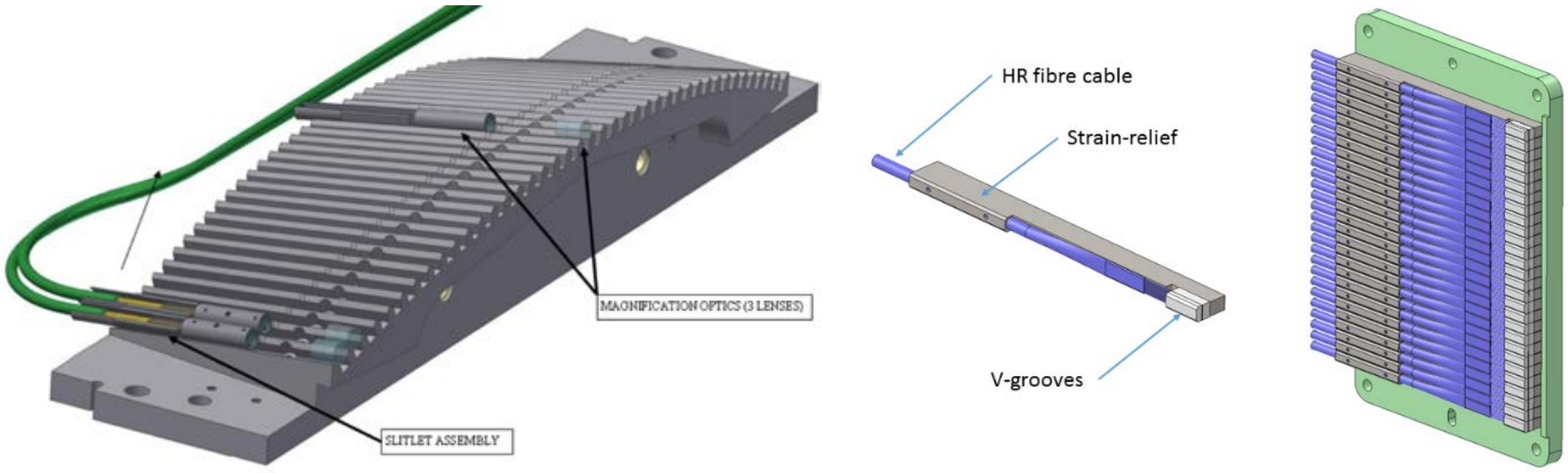

*Figure 86: Left: Curved slit block from AAT/HERMES (Note: Magnification optics will not be used in MSE). Right: V-grooves and strain-relief concepts for a straight slit input unit, as proposed by HAA.*

Once the cables have been fabricated, they will undergo detailed, end-to-end optical performance testing, to measure the throughput and FRD characteristics and stability of each fiber under simulated telescope conditions. To accomplish this, the University of Victoria (UVic) in Canada is building a fully automated, opto-mechanical test facility.[xii] The facility will include sources and detectors that can measure the optical performance of the fiber cables under operational conditions. This will result in dynamic performance measurements, including measurements of each cable’s repeatability and stability. All tests will be performed under closed-loop computer control, with minimal human intervention (only what is necessary to install and remove the cable). By fully automating the metrology process, the testing can support a one-cable-per-week production schedule for the 57 cables required.

The facility is currently under development, but preliminary fiber performance results are already available. *Figure 87* shows the output image of the ring test, used to measure FRD. The image on the left shows the full 2D illumination pattern, while the intensity profile on the right displays a cross-section of the central row of pixels, used to measure FRD. The post-processed results shown in *Figure 88* demonstrate that an FRD of less than 5% is possible when using high-numerical aperture fibers—a result consistent with other published measurements.[xiii] The facility is proceeding with the set-up of a full-aperture FRD test, to be followed by the use of the planned computer controlled fiber motion simulator.

Once fully operational, the fiber test facility will provide on-going performance tests to support design development of the fiber cable fabrication process. One of the challenges facing the FiTS system is the need to verify the stability and repeatability characteristics necessary to meet calibration requirements. The testing process at UVic will be essential to this, since it will allow end-to-end testing of the fiber characteristics under observing conditions.

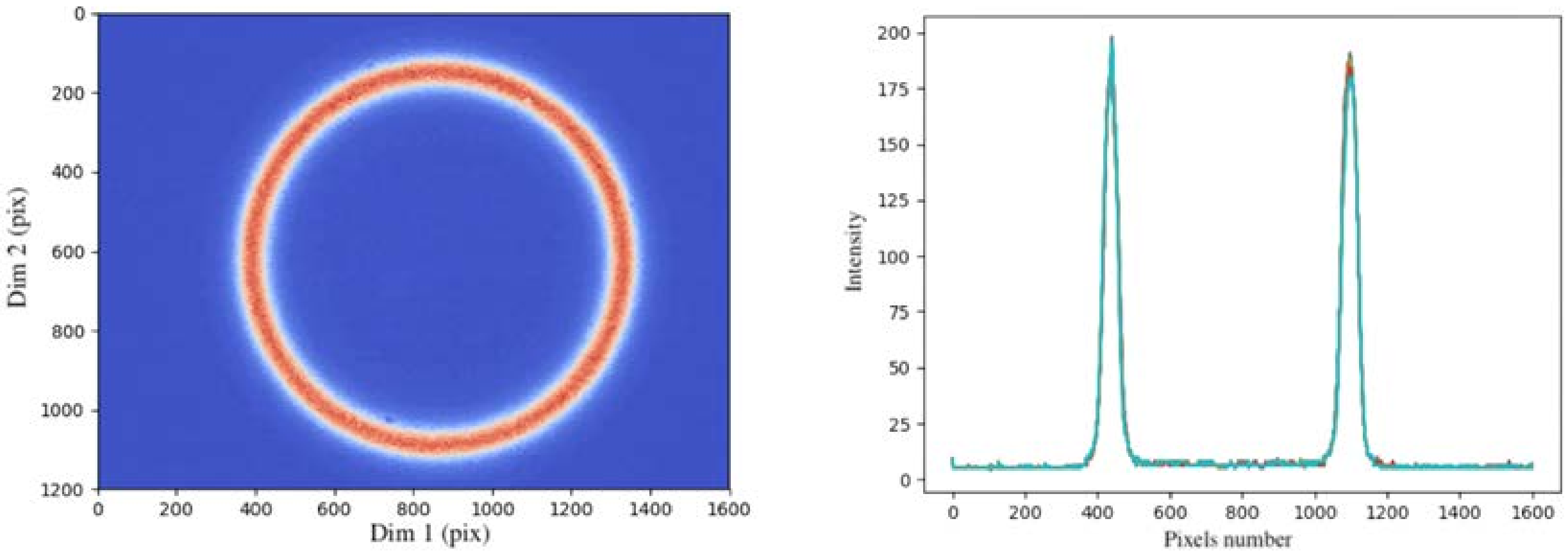

*Figure 87: Ring test for FRD measurement and the corresponding processed image.*

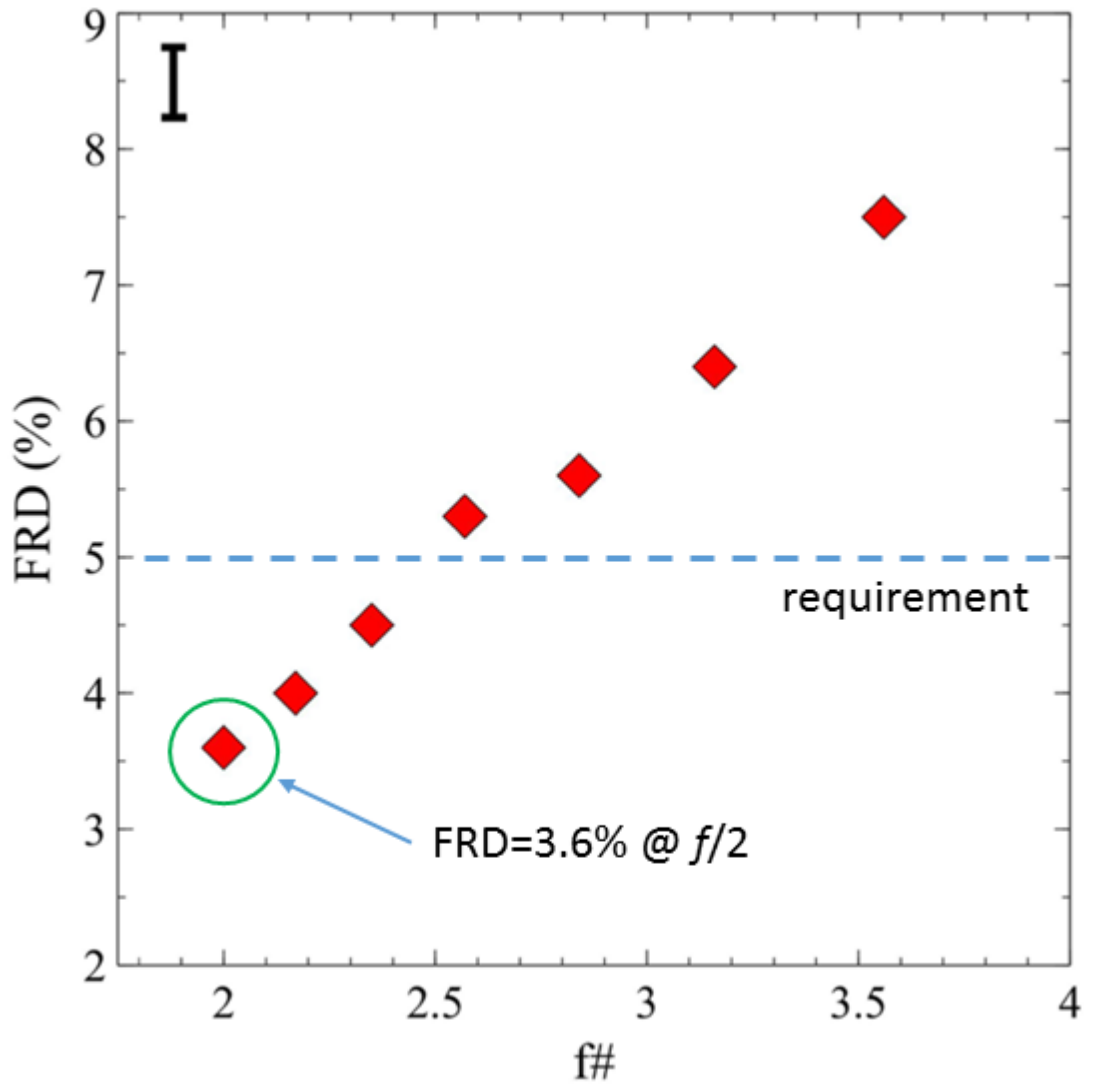

*Figure 88: FRD results according to input beam f# with candidate Polymicro fiber, using the ring test.*

Another challenge for FiTS is the need to industrialize the fabrication process, so we can build cables for thousands of fibers cost effectively, and to find a reliable process for AR coating the input ends of the fibers. The fiber cable design and production experts at FTO are expected to address these concerns during the next design phase.

Future work also includes characterizing the performance of spliced fibers, should they get damaged and require in-situ repairs using fusion bonding. The FiTS team has determined that it is possible to achieve very good consistent performance with fusion bonds under controlled conditions. The in-situ repair equipment and process that ensures consistent performance will be developed during the next design phase.

### 4.4.5. Low and Moderate Resolution Spectrographs (LMR)

MSE includes a bank of Low/Moderate Resolution spectrographs (LMR), designed by Centre de Recherche Astrophysique de Lyon (CRAL) in France.[xiv] Six identical LMR units, located on two instrument platforms on the telescope structure (*Figure 89*), accept light from the science fibers which are arranged in a slit assembly, and generate spectra, which are recorded by the detectors. The raw data are exported to an appropriate repository in the Program Execution Software Architecture system for further processing, distribution, and storage (see Section 4.6).

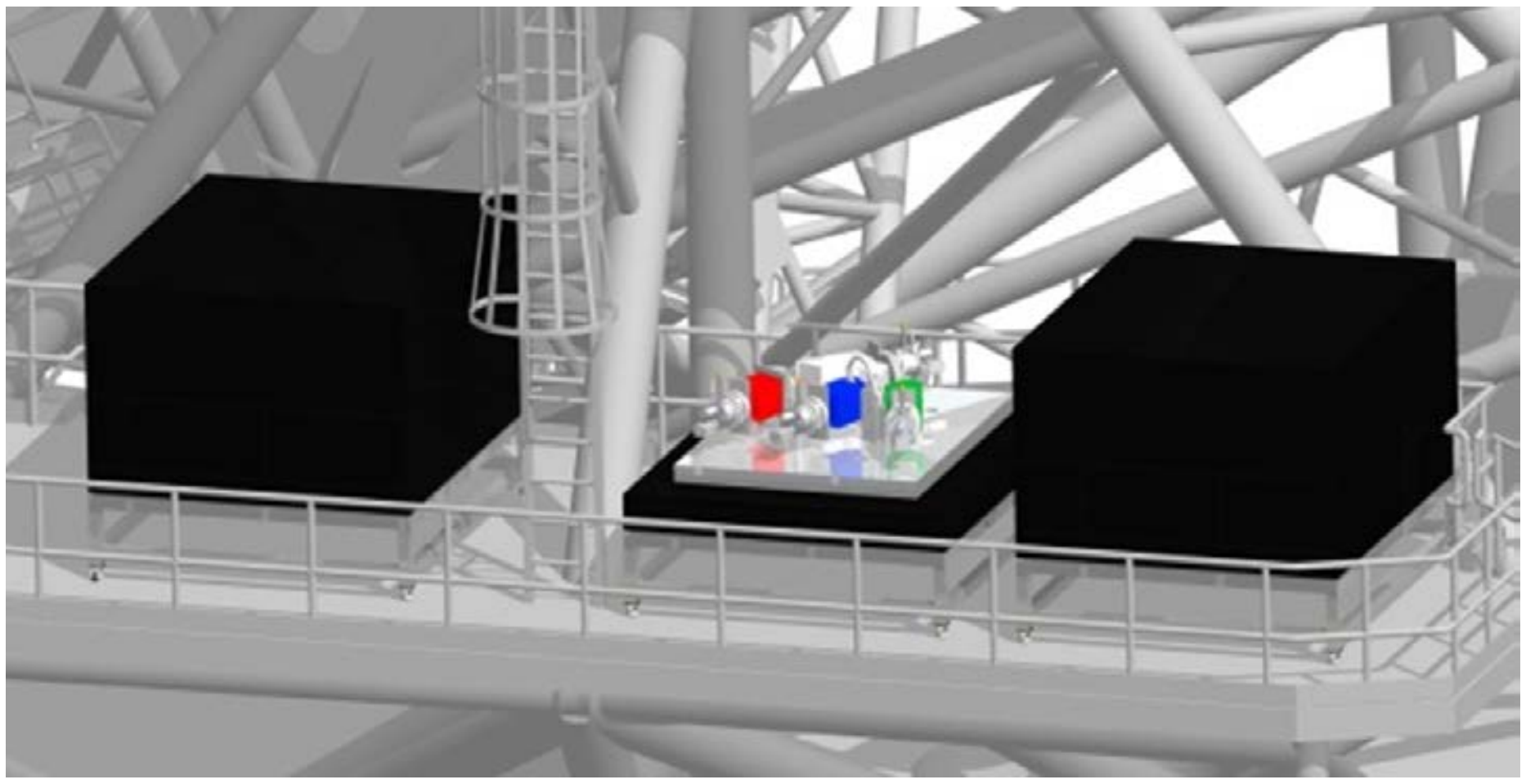

*Figure 89: LMR spectrographs on instrument platforms, three units per platform, the middle unit with cover removed.*

*(1) Overview and Performance*

The LMR spectrograph suite must accommodate 3249 spectra. At low spectral resolution, the required wavelength coverages are 360–1300 nm and 1500–1800 nm (optical + J, H). A three-arm design is preferable for this optical range (360–950 nm), given the design space and the detector real estate available, especially considering the number of resolution elements required. This also means that the wavelength range covered by each VPH grating must be sufficiently narrow to ensure good efficiency is obtained over the range of each grating. At NIR wavelengths, the entire YJ wavelength range (950–1300nm) can be covered by a single detector, given the fiber size (1 arcsecond = 106.7 microns) and a sufficiently fast (f/1.2) NIR camera (the largest NIR detectors available are Hawaii 4RG15, 61 mm x 61 mm). This helps minimize costs, since the NIR detector represents a major fraction of the cost of the spectrograph system. These considerations have led the CRAL design team to investigate a conceptual design of a four-arm system, in which the NIR channel is switchable between YJ and H bands.

At moderate resolution, the wavelength coverage required is in the range of 360–950 nm i.e., it only makes use of the optical arms. All the arms are designed to be switchable between two configurations: the optical arms switch between low and moderate spectral resolutions, and the NIR mode switches wavelength range between YJ and H. Given the detector sizes and f-numbers, one LMR unit can accommodate ~550 fibers. The spacing between spectra is 170 microns (or around 11 pixels, center to center). Six LMR units are therefore needed in total. The overall configuration of the spectrograph is shown in *Figure 91*. The wavelength coverage and nominal spectral resolution for each arm is shown in *Table 7*. *Figure 90* shows the spectral resolution as a function of wavelength for every configuration of every arm, and compares it with the requirements.

*Table 7: Wavelength range and spectral resolution of the LMR arms in each configuration.*

| | LR | | | | MR | | | |
|---|---|---|---|---|---|---|---|---|
| | **Blue** | **Green** | **Red** | **NIR** | **Blue** | **Green** | **Red** | **NIR** |
| $\lambda_{min}$ | **360nm** | **540nm** | **715nm** | **960nm** | **391nm** | **576nm** | **737nm** | **1457nm** |
| $\lambda_{max}$ | **560nm** | **740nm** | **985nm** | **1320nm** | **510nm** | **700nm** | **900nm** | **1780nm** |
| **Resolution (Å)** | **1.78Å** | **1.75Å** | **2.36Å** | **3.15Å** | **1.02Å** | **1.02Å** | **1.35Å** | **2.68Å** |
| $R_{min}$ | **2000** | **3100** | **3000** | **3000** | **3800** | **5600** | **5500** | **5400** |
| $R_{max}$ | **3100** | **4200** | **4200** | **4200** | **5000** | **6800** | **6700** | **6600** |

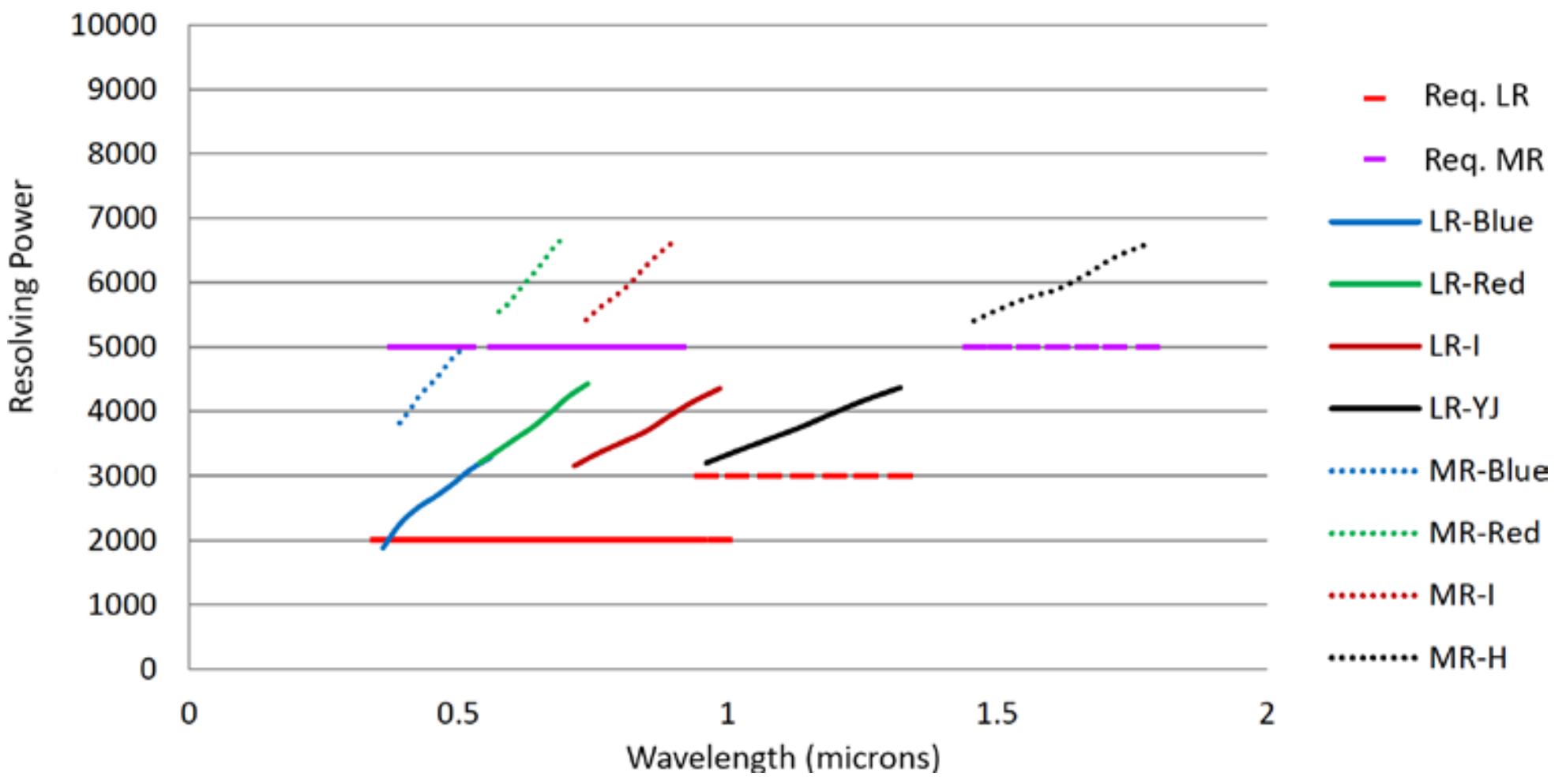

*Figure 90: Spectral resolution as a function of wavelength for all configurations. Bold horizontal lines indicate the science requirements at both low (red) and moderate (purple) spectral resolutions (solid lines for optical channels, dashed lines for NIR channels).*

The current design of LMR evolved from a design for AAT/Hector, which has similar resolution requirements over the optical wavelength range. The pupil is not on the grating, but just inside the camera, which improves image quality and reduces the required lens sizes. The demagnification is 0.58, and the beam size is 175 mm. The layout is shown in *Figure 91*. The slit is 104.5 mm long, and curved laterally (with 1.3 mm sag), to straighten its image on the detector. A rectangular field lens is gelled to the fibers—as in the AAT/AAOmega model—with an anti-reflection coating on the exit side. This increases throughput and allows the spectrograph pupil to be placed where needed. The design also incorporates an off-axis Schmidt collimator, with dichroic beamsplitters located between the collimator mirror and the corrector lenses in each arm.

For low-spectral resolution, each arm has an off-axis Schmidt corrector lens, followed by a VPH grating. For moderate spectral resolution and H-band, these are exchanged for a large sapphire grism. The grism has a higher dispersion VPH immersed within it and its collimator is bonded to the first surface. All the cameras work at f/1.2. Each arm has a single 61 mm x 61 mm detector and an e2v 231-series CCD for the optical arms—back-illuminated in the blue arm; deep depletion in the green arm; possibly high rho/bulk silicon for the red arm—and a Teledyne Hawaii 4RG15 for the NIR. The pixel sizes are all 15 µm, and the images are well sampled at ~3.5 pix/FWHM.

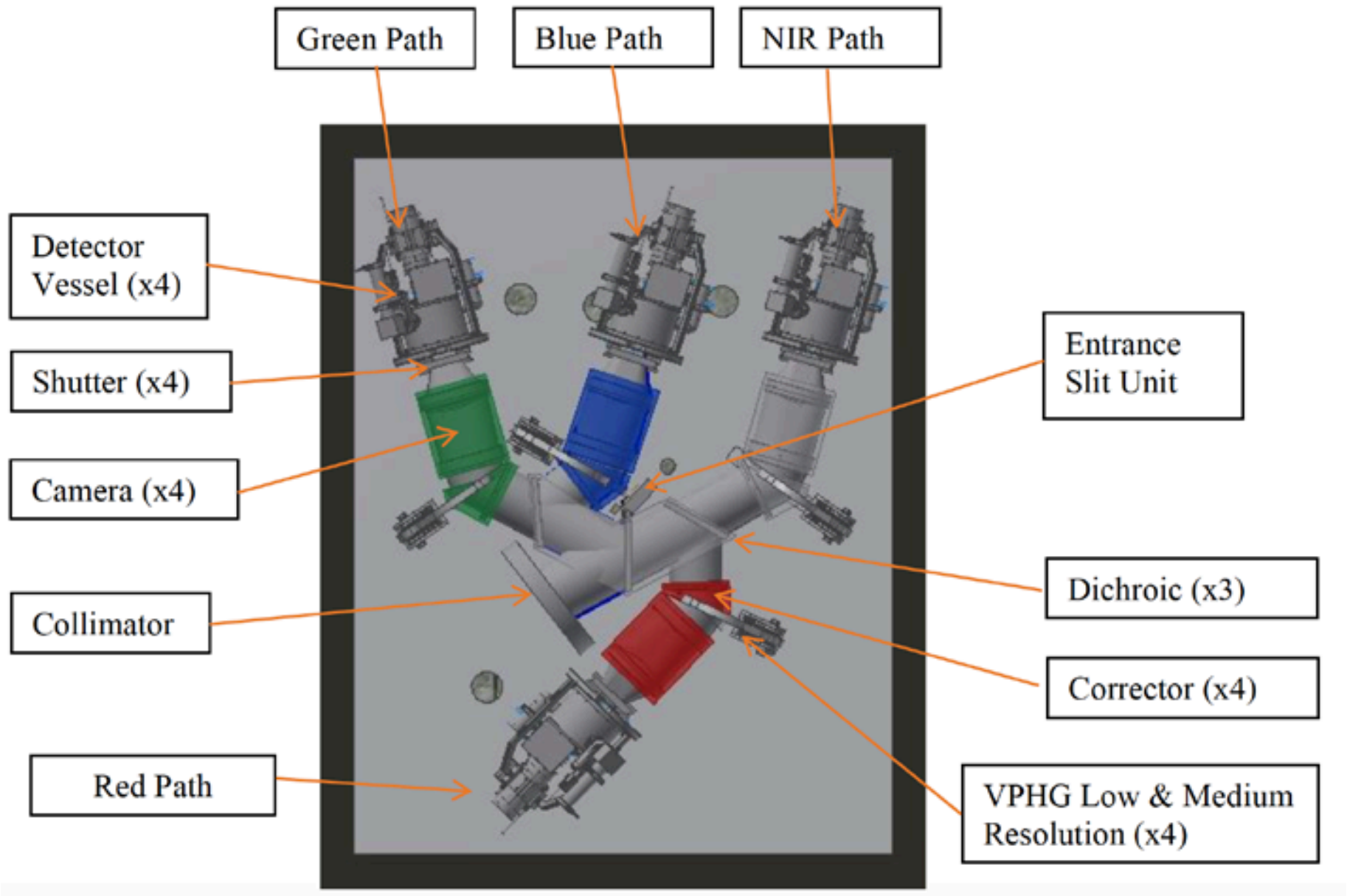

*Figure 91: Layout of the LMR with major elements highlighted.*

Representative spot diagrams of the image quality are shown in *Figure 92*. The goal was to maintain an RMS radius of less than a quarter of the projected fiber radius (~60 microns) for all configurations, wavelengths, and slit positions. The optical design's worst RMS value is less than 12.3 microns. *Figure 93* shows the theoretical performance of the low-resolution MSE gratings. In each case, peak efficiency is ~93%, minimum efficiency ~72%, and average efficiency 86%. Real performance is estimated to be no more than ~1% worse than this. The moderate resolution gratings are somewhat less efficient (*Figure 94*) because the larger grating angle increases the polarization splitting. In the low spectral resolution configuration, overall throughput is 42% at 360nm, and exceeds 50% for wavelengths longer than 370nm, with an average of >60%. In the moderate resolution configuration, overall throughput exceeds 45% everywhere, with an average of >55%.

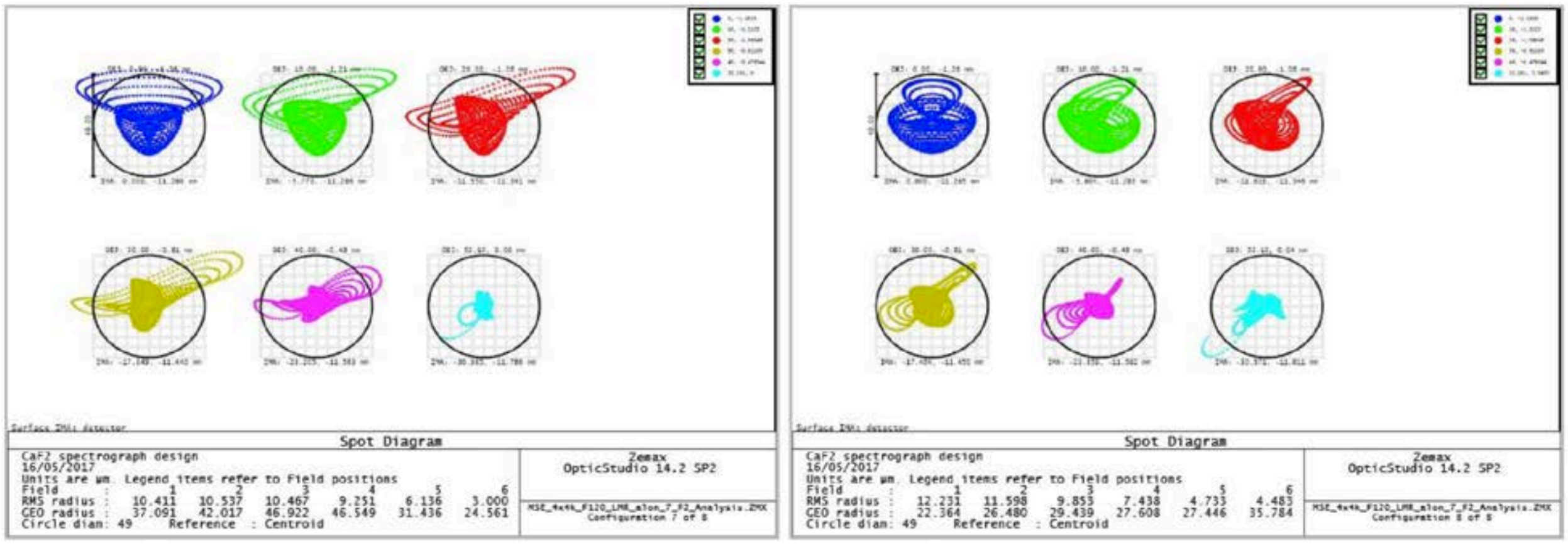

*Figure 92: Representative spot diagrams versus field position and wavelength for LMR spectrograph optical design.*

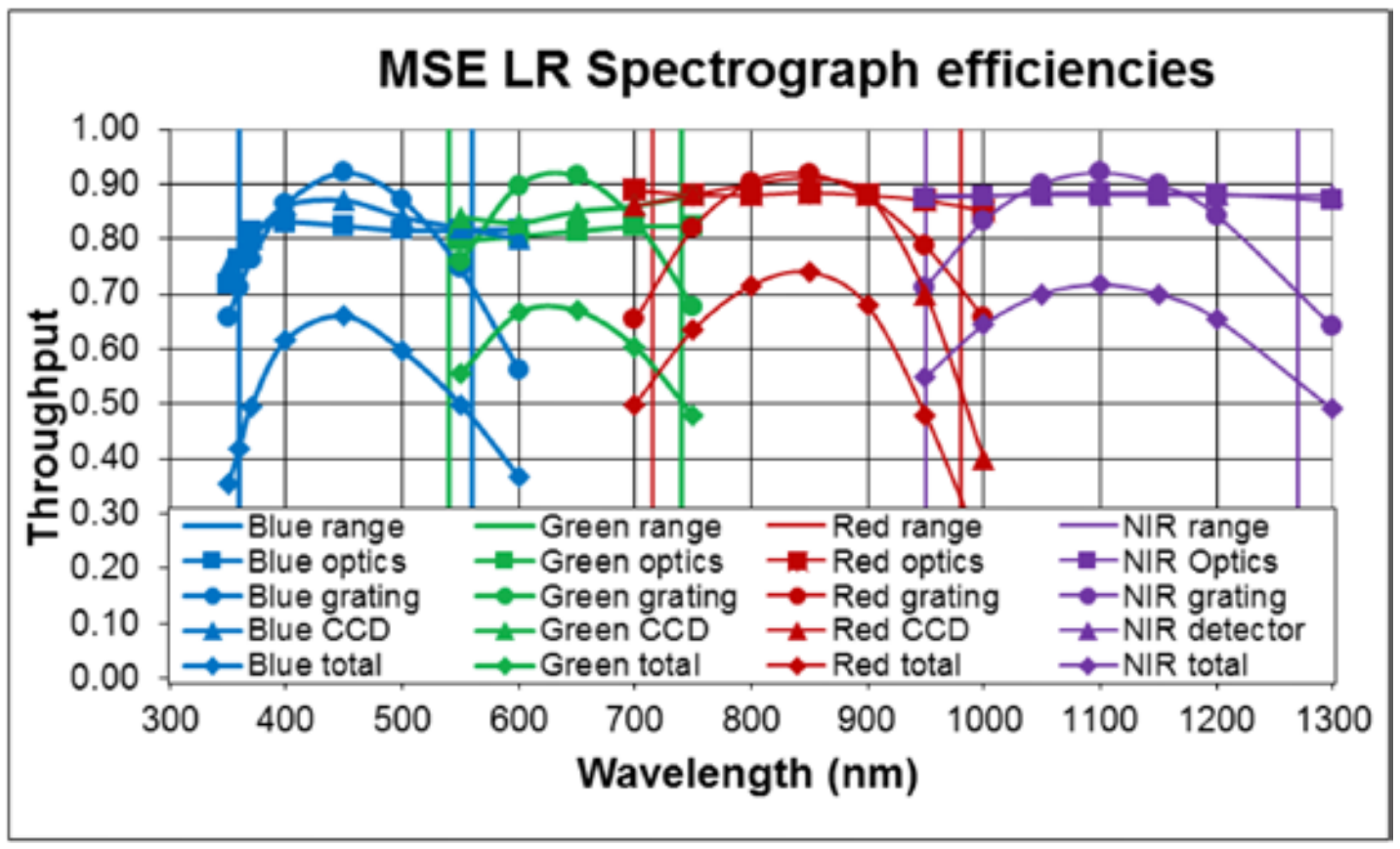

*Figure 93: Throughput as a function of wavelength for low spectral resolution.*

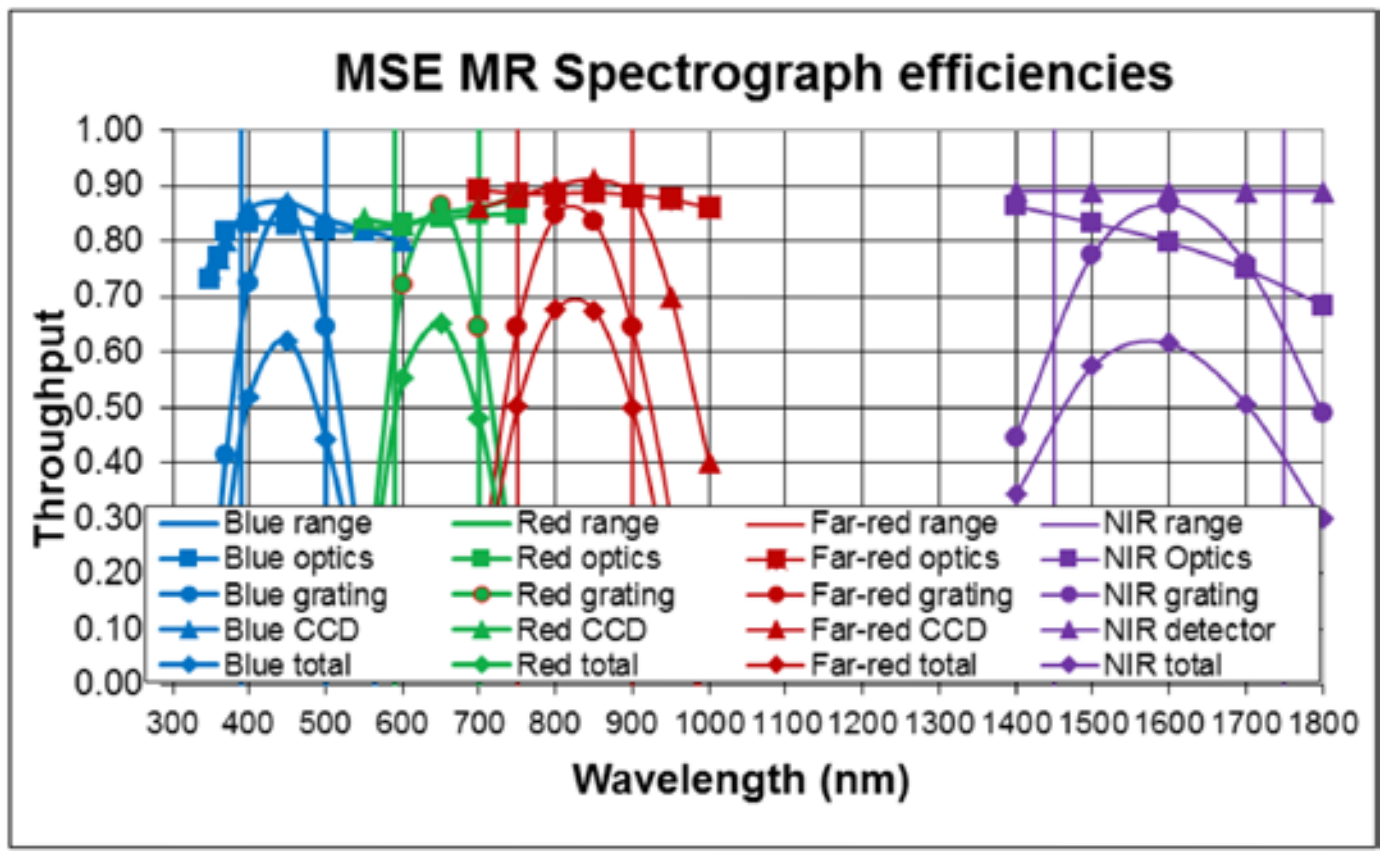

*Figure 94: Throughput as a function of wavelength for moderate spectral resolution.*

### (2) Components of the Design

*Figure 91* shows a schematic top view of an LMR unit, highlighting major design components. Some of the major optical components are discussed below.

**Collimator:** The light exits the fibers in a strongly apodized beam because of the non-circular M1, Focal Ratio Degradation within the fibers, and geometric FRD (caused by fiber tilt). The demagnification between cameras and collimator is set by the fiber size and resolution requirements, and there is a trade-off between camera and collimator speeds. A slightly slower collimator (f/2.083) was selected and the resulting light loss (*Figure 95*) was compensated for by simplifying the camera design, making it optically more efficient.

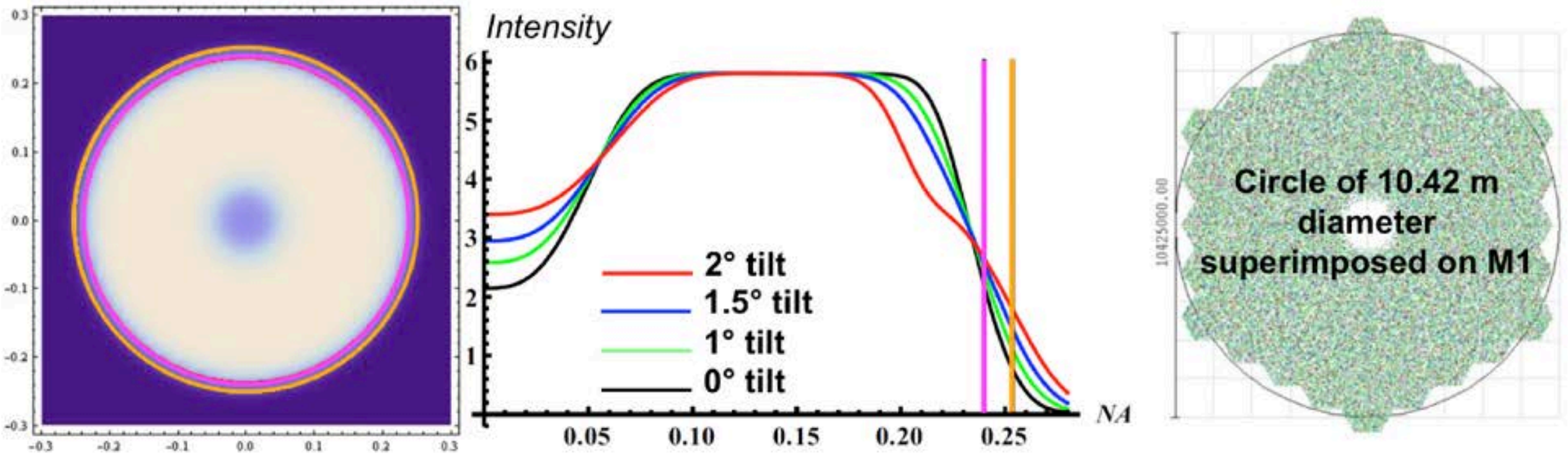

*Figure 95: (Left) Simulated far-field image at the fiber output for an on-axis untilted fiber, showing beam cut-off, with collimator acceptance speeds of f/2.083 (magenta circle) and f/1.97 (orange circle). (Center) The corresponding radial intensity profile through the same beam (black) and for tilted fibers. The radial positions of the beam cut-offs at the same collimator speeds of f/2.083 (magenta) and f/1.97 (orange) are also shown. The corresponding collimator overfilling light losses for an untilted fiber are 4.5% and 2.2%, respectively. (Right) The selected collimator speed of f/2.083 corresponds to a circular primary mirror diameter of 10.42 m.*

The selected collimator speed of f/2.083 corresponds to a diameter of 10.42 m at M1 (*Figure 95*, right). The collimator overfilling loss, for a fiber near the center of the field—assuming non-tilted fibers and FRD—is 4.5%.[xv] An f/1.97 collimator that matches the telescope input beam f/# would still have losses of 2.2%. These numbers correspond to central fibers: fibers away from the center of the field will have smaller losses.

The off-axis design requires only a single corrector in each arm, each with a slightly different prescription. The collimator corrector lens is a freeform asphere, since it is an off-axis design, but the optical axis is included within the aperture, which makes testing and alignment straightforward. The off-axis design also means that the fiber face is not orthogonal to the optical axis and there is therefore some defocus across the face of the fiber. However, the effect is small: adding a mere 1.25 µm to the RMS spot radius, in the spectral direction only. The off-axis collimator allows enough room for the slit, dichroics, dispersers, and cameras, although some clearances are small and mechanical packaging presents a risk. The design will be retired during the next phase.

**Disperser:** The gratings are straightforward VPH gratings of reasonable (200 mm) size, angles, line densities, etc. The grating fringes are slanted at 4.5°, in order to completely eject the Littrow ghost from the detector. The resulting anamorphism provides useful resolution gain and compensates for the increase in beam size in the spectral direction caused by the pupil's location below the grating.

The low-resolution VPH gratings are mounted on fused silica substrates, an arrangement which provides excellent thermal stability (CTE = 0.5 ppm/°C). The prisms used in the moderate-resolution (and H-band) grisms are made of sapphire, and therefore large and heavy (8 kg). The birefringence has no effect on the collimated beam. The prisms provide a 1.75 factor increase in resolution over the low-resolution configuration. A larger increase in the spectral resolution will require placing reverse prisms onto the low-resolution gratings. This will increase the required clearances on both sides of the gratings, thus increasing lens sizes and decreasing image quality. There is a further asphere on the exit surface of the grism, which provides a small improvement in image quality. The grisms are a high-risk component of the LMR design.

**Camera:** An f/1.2 camera is required for the NIR arm. This is a very challenging speed for a transmissive camera and a catadioptric design was considered. However, a transmissive design was adopted, since it offers multiple advantages over a catadioptric design:

- increased throughput because of the unobscured pupil,
- reduced ghosting and scattering for the same reason,
- much easier detector mounting and access, and
- a more compact design.

The optical cameras can be either the same speed and detector size as the NIR camera, or use larger, 92mm x 92mm detectors, operating at ~f/1.8. It is not desirable to use two fundamentally different camera designs, and slower designs represent a significant increase in cost and volume. Hence we decided to operate all cameras at f/1.2.

Each camera barrel contains a pair of doublets of Ohara S-FSL5 (S-FSL5Y for the blue camera) and PBL35Y. They are bonded with Norland NOA88, which has excellent UV transmittance and closely matches the CTE of the glass. FEA tests at AAO suggest that a relative change in scale of 50 ppm is acceptable. The largest difference for MSE is several times smaller, even for the NIR camera.

One risky element of this camera design is that it depends on strongly aspheric surfaces. There are aspheres on each face of the first doublet, on the first face of the second doublet, and on the rear face of the blue camera. The largest aspheric departure is 2.4 mm. The gull-wing nature of L1 means that it must be tested in transmission. The lens sizes are modest: the largest aperture is equivalent to 216 mm. Image quality can be improved by increasing the lens sizes.

**Field-flattener / Dewar window:** The ideal material for field flattener use in fast, low-dispersion spectrograph cameras should be affordable, easily polished, non-radioactive, high-transmission, high index and low dispersion. The best glass the AAO has tested to date is Nikon 7054, an i-line glass with superb transmission, which is mildly radioactive (alpha particle radiation interacts with e2v 231-series detectors to form ~30 trails per detector per hour).

Several solutions that have been explored for AAT/Hector are also relevant to MSE. The preferred solution is Aluminum OxyNitride (AlON), a strong, clear ceramic, whose main drawbacks are that it produces a slight haziness, which reduces throughput by a few percent at all wavelengths, and that it has relatively high polishing costs, and a microroughness limited to ~5–10 nm. It produces roughly an order of magnitude more scattered light per surface than typical glass lenses. However, the scattered light level is still lower than any realistic value for the VPH gratings. It is also much closer to focus, meaning that a smaller fraction of the scattered light is outside the core of the PSF.

In all cases, the field-flattener is aspheric on the front concave surface, and planar on the rear surface. The clearance between Dewar window and detector is 1 mm. SpecInst[xvi] have confirmed they can accommodate it into their 1100S series Dewars with e2v 231-series detectors.

*(3) Design Development*

The LMR spectrograph suite has multiple configurations and operates over a broad wavelength range. The proposed spectrograph design meets the requirements for spectral resolution everywhere except in the blue arm of the moderate spectral resolution configuration (see *Figure 90*) and is a reasonable solution for this challenging design space. No tolerancing analysis has yet been performed. However, judging by analogous designs—in particular for AAT/Hector—it is expected that the most difficult tolerances will involve the centering of the lenses and surface slope errors. More generally, major risks identified in the current design include:

- The optical design's reliance on highly aspherical elements, which may be difficult to manufacture, test, and align;
- Challenges involving the mechanical packaging of the optical design, in particular the tight space constraints on various optical elements;
- Mechanical challenges associated with the mechanisms necessary to change configurations between low and moderate spectral resolution (at the dispersers and detectors), involving precision movements of large elements;
- Incorporation of H-band capabilities, in particular the requirement of spectrograph cooling, to reduce thermal background.

Design work early in the next phase will be focused on obviating the first three of these risks. In preparation for the next stage, the Project Office is conducting a trade study to determine alternative scenarios for providing H-band capabilities. The study involves a science-based analysis of the target density of H-band sources, to determine whether it would be sufficient if only a subset of fibers fed a dedicated H-band spectrograph—a solution that could significantly ease the design challenges for those conducting the remaining work on the LMR spectrograph.

### 4.4.6. High Resolution Spectrographs (HR)

MSE includes a bank of high spectral resolution spectrographs (HR), designed by the Nanjing Institute of Astronomical Optics & Technology (NAIOT) in China.[xvii] The two identical HR spectrographs are located in the environmentally stable spectrograph room in the observatory building. Each accepts light from 542 science fibers, arranged on a slit assembly, and generates spectra, which are recorded by the detectors. The raw data are exported to an appropriate repository in the Program Execution Software Architecture system, for further processing, distribution, and storage (see Section 4.6).

### *(1) Overview and Performance*

The HR spectrograph suite must accommodate 1084 spectra, within a wavelength range from 360 to 900 nm. A three-arm design is under consideration. This would involve dividing the wavelength range into three wavelength windows, one per arm. The opto-mechanical arrangement of all the elements on the optical bench is shown in *Figure 96*. At wavelengths shorter than 500 nm, the required spectral resolution is R~40,000, in order to be able to study the many (weak) lines at blue wavelengths, without being detrimentally affected by line blending. At wavelengths longer than 500 nm, the required spectral resolution is R~20,000: since the effects of line blending are reduced and many of the lines of interest are relatively strong at these wavelengths, this lower spectral resolution is acceptable. Regardless of the resolution, only relatively small selected sections of the wavelength range (termed "working windows") can be accessed, given the high spectral resolution. The sizes of these working windows are approximately $\lambda_c/30$ at R=40,000 and $\lambda_c/15$ at R=20,000, where $\lambda_c$ is the central wavelength of the working window.

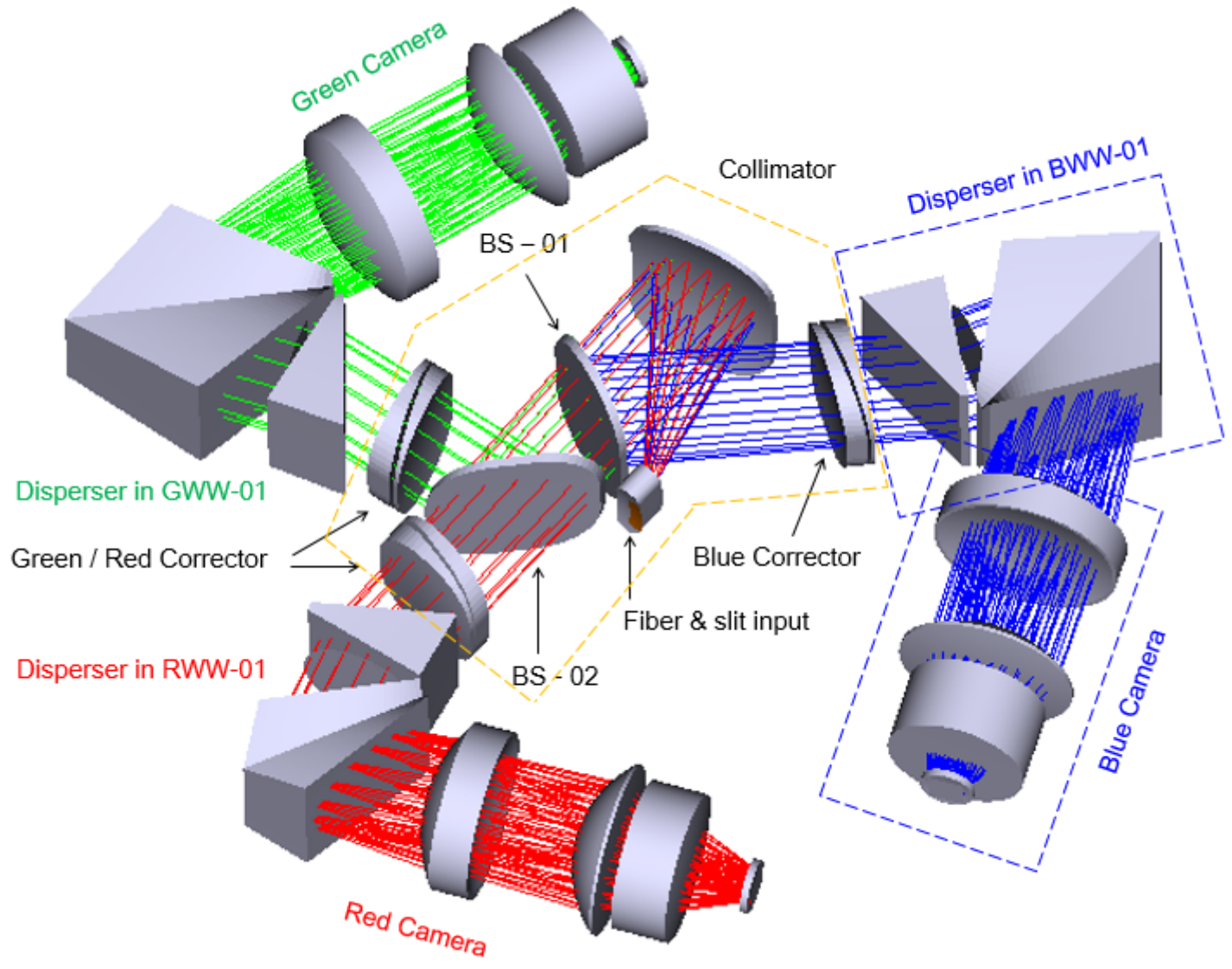

*Figure 96: The optical layout of HR, with major elements highlighted.*

The HR collimator accepts light from FiTS from the slit assembly at f/2.05. The slit assembly holds the fibers, arranged in a spherical slit, with a spacing of 220 microns (center to center). An off-axis Houghton collimator reflects light into the three optical arms (blue, red, and green) through a series of dichroic beamsplitters. The optics are quite large, with a 300 mm pupil at the collimator. The maximum clear aperture in the refractive cameras is 500 mm.

The detector system sits on a linear stage, which provides focus adjustment over ±1 mm and is cooled by closed-cycle helium coolers. Each arm requires a 92 mm x 92 mm detector and the optical design can accommodate pixel sizes of either 10 or 15 microns.

*Table 8: Wavelength range of each arm of HR, the working window within each arm, and the spectral resolution at the central wavelength of the working window.*

| | Blue | Green | Red |
|---|---|---|---|
| **λ min** | 360 nm | 440 nm | 600 nm |
| **λ max** | 460 nm | 620 nm | 900 nm |
| **Working Window** | 401 - 415 nm | 472 - 488.5 nm | 626.5 - 672 nm |
| **R** | 40000 | 40000 | 20000 |

The wavelength range, working window, and spectral resolution of each arm of HR are shown in Table 8. The working windows within each arm can be changed over the lifetime of the spectrograph, by replacing the dispersers. The image quality of the complete optical design is very good: Figure 97 shows representative spot diagrams over the full field of view in the blue working windows. Each box scale is 62 x 62 microns and the RMS values of the spot radius are less than one quarter of those of the geometric image radius. Indeed, the image quality in each of the working windows is essentially identical, which guarantees a similar scientific performance.

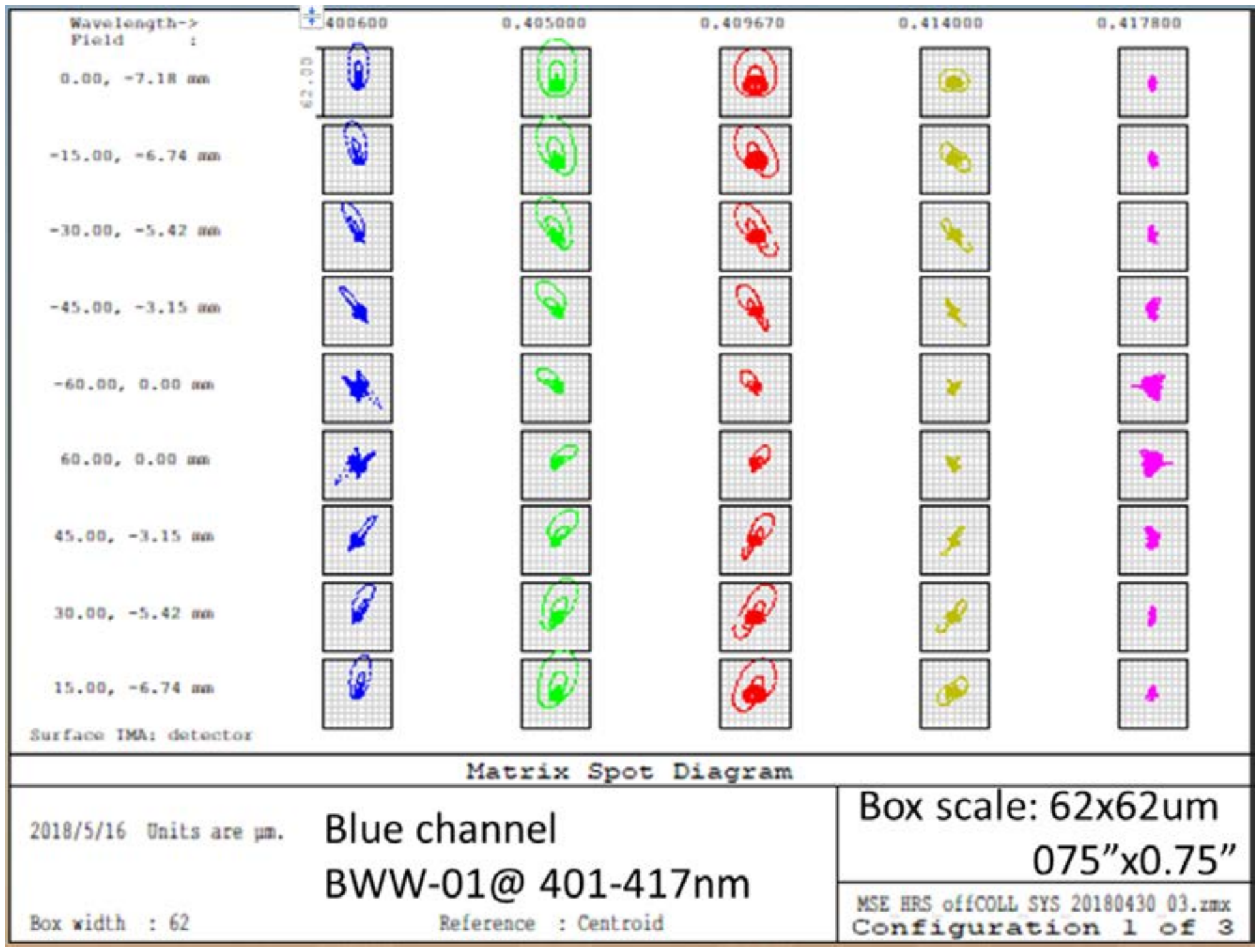

*Figure 97: The image quality of HR, as represented by spot diagrams across the field for the blue working windows.*

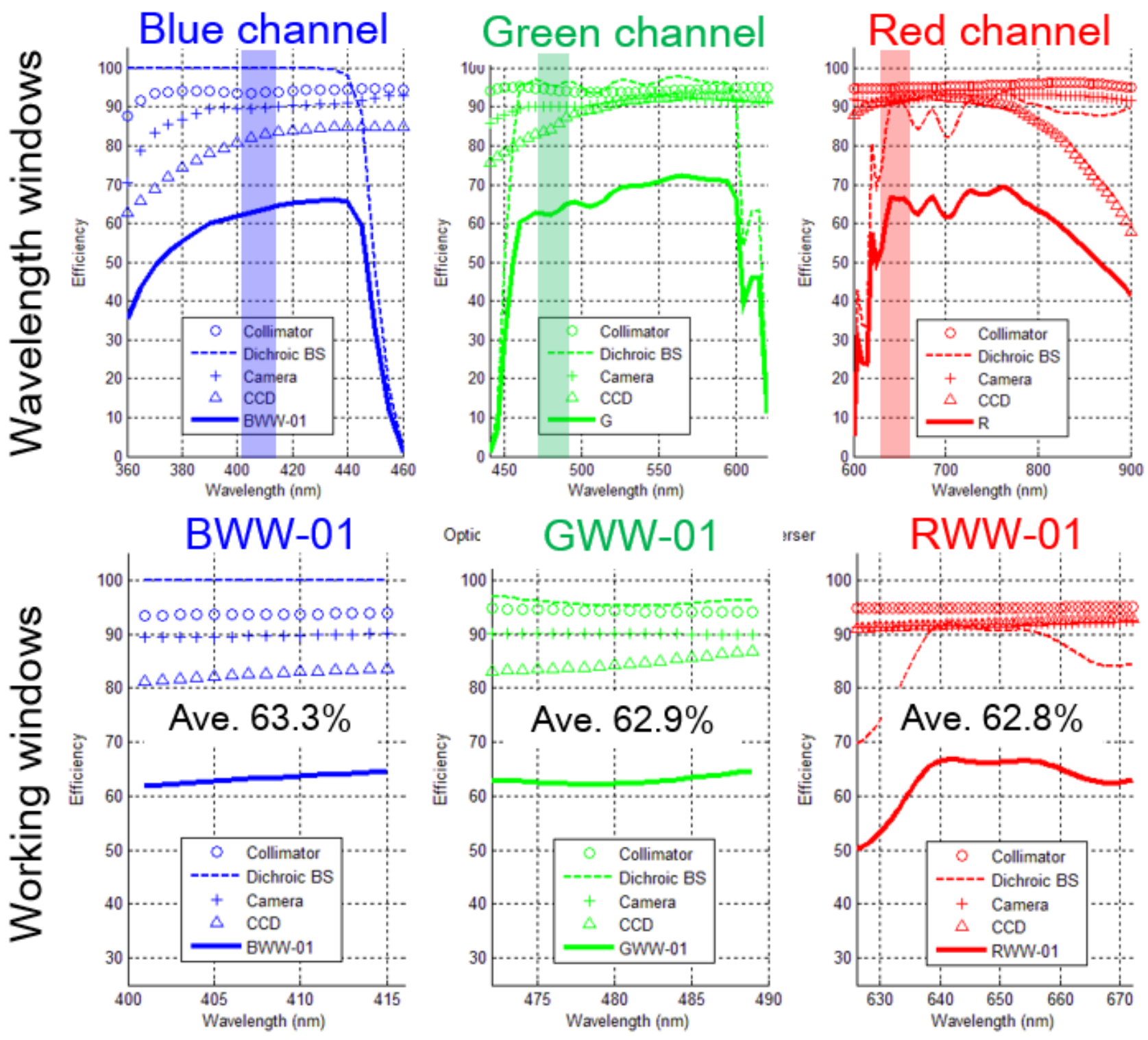

*Figure 98: The throughput of the HR in each of the three channels (top panels) and in the three working windows in each channel (bottom panels).*

*Figure 98* shows the overall throughput of the HR spectrographs for all optical elements except the disperser. The top panel shows the throughput in the three channels, and the bottom panel shows the throughput at the specific locations of the working windows. Within each working window, the average throughput is in excess of 60%.

### *(2) Components of the Design*

The optical layout is shown in *Figure 96*, with major elements labeled. Some of the major optical design elements are discussed below.

**Slit Assembly:** The decision to use an off-axis collimator design has major advantages: the absence of central obscuration and more flexibility in the placement of the fiber slit assembly and other opto-mechanical components. However, it also necessitates a more complicated fiber assembly geometry, to ensure that the spectra are appropriately projected onto the detector. The proposed fiber layout at the slit is shown in *Figure 99*. Viewed from the front (*Figure 99*b), the fiber layout at the slit resembles a “smile,” which compensates for the spectral deviations along the spatial dimension on the detector. The “smile” has a projected radius of 254 mm, and a “slit length” of 120 mm. It accommodates 542 fibers, with central spacing of 220 microns. Each fiber is 80 microns in diameter.

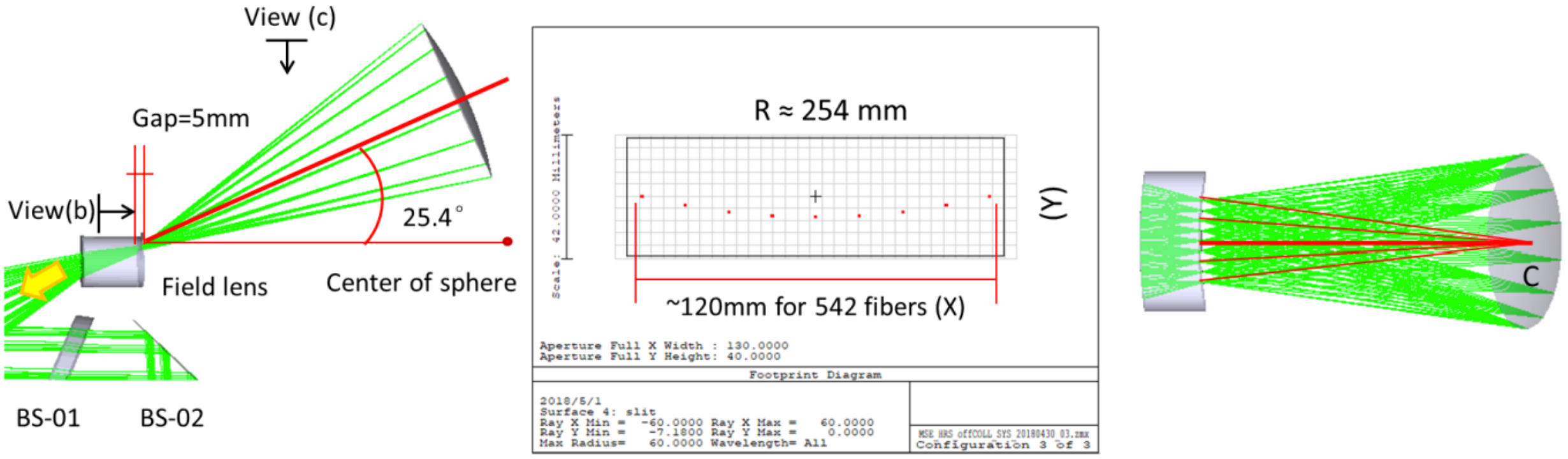

(a) Side view of fiber layout (b) Front view of fiber layout (c) Plan view of fiber layout

*Figure 99: Proposed fiber layout at the slit. The directions of views (b) and (c) are indicated in (a).*

To fully correct for the field curvature of the collimator, hundreds of fibers have to be evenly distributed across a spherical slit surface. The “parent” sphere, which has a curvature radius of ~549 mm, is indicated in *Figure 99*a . Due to the off-axis optics, an incident angle of 25.4º has been added, to match the optical path of the off-axis collimator. A plan view of the fiber layout is shown in *Figure 99*c. A similar fiber slit assembly design has been successfully produced for AAT/HERMES and forms the basis of the HR design that accommodates the faster collimator.

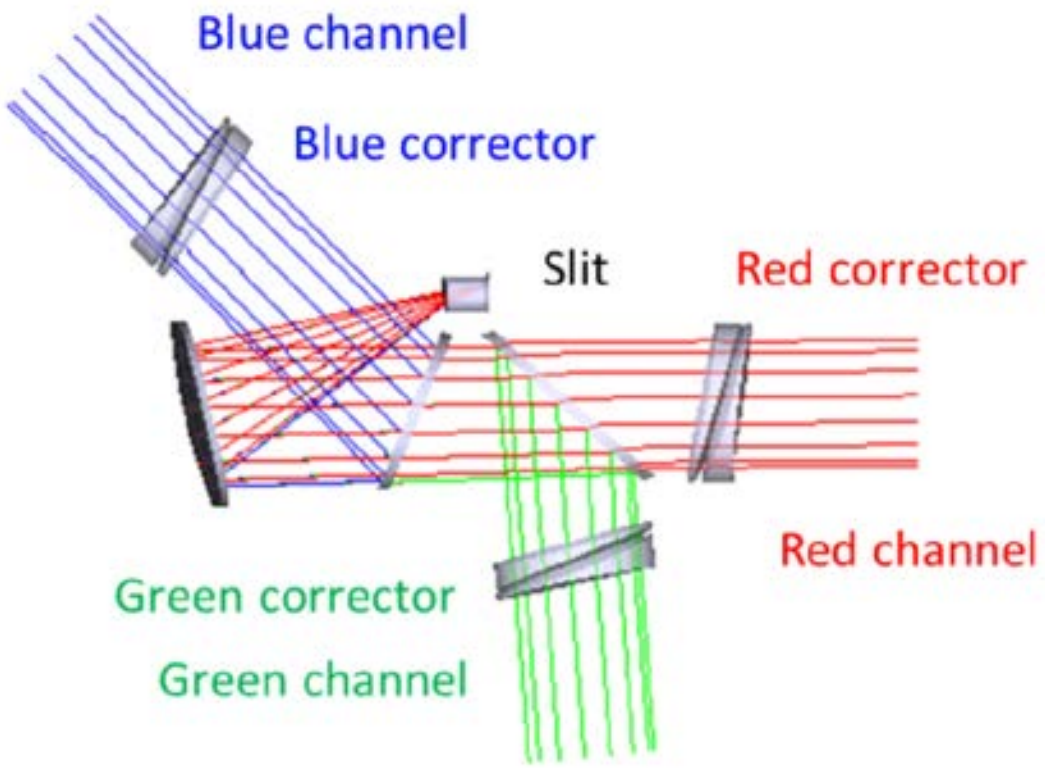

*Figure 100: The f/2.05 off-axis collimator design.*

**Collimator:** The off-axis collimator uses the Houghton system to obtain good optical performance, using a pair of aspherical correctors (*Figure 100*). Two dichroic beamsplitters are located between the collimating mirror and the correctors. An on-axis collimator was also considered. Both options produced similar image qualities, but the off-axis design requires additional aspherical surfaces and higher asphericities than the on-axis design. Discussions with vendors about the fabrication of these highly aspheric lenses are currently underway. However, the spectrograph throughput of the off-axis design is 5.8% higher, and provides considerably more space for mechanical accessibility (at least 500 mm of space along the main optical axis).

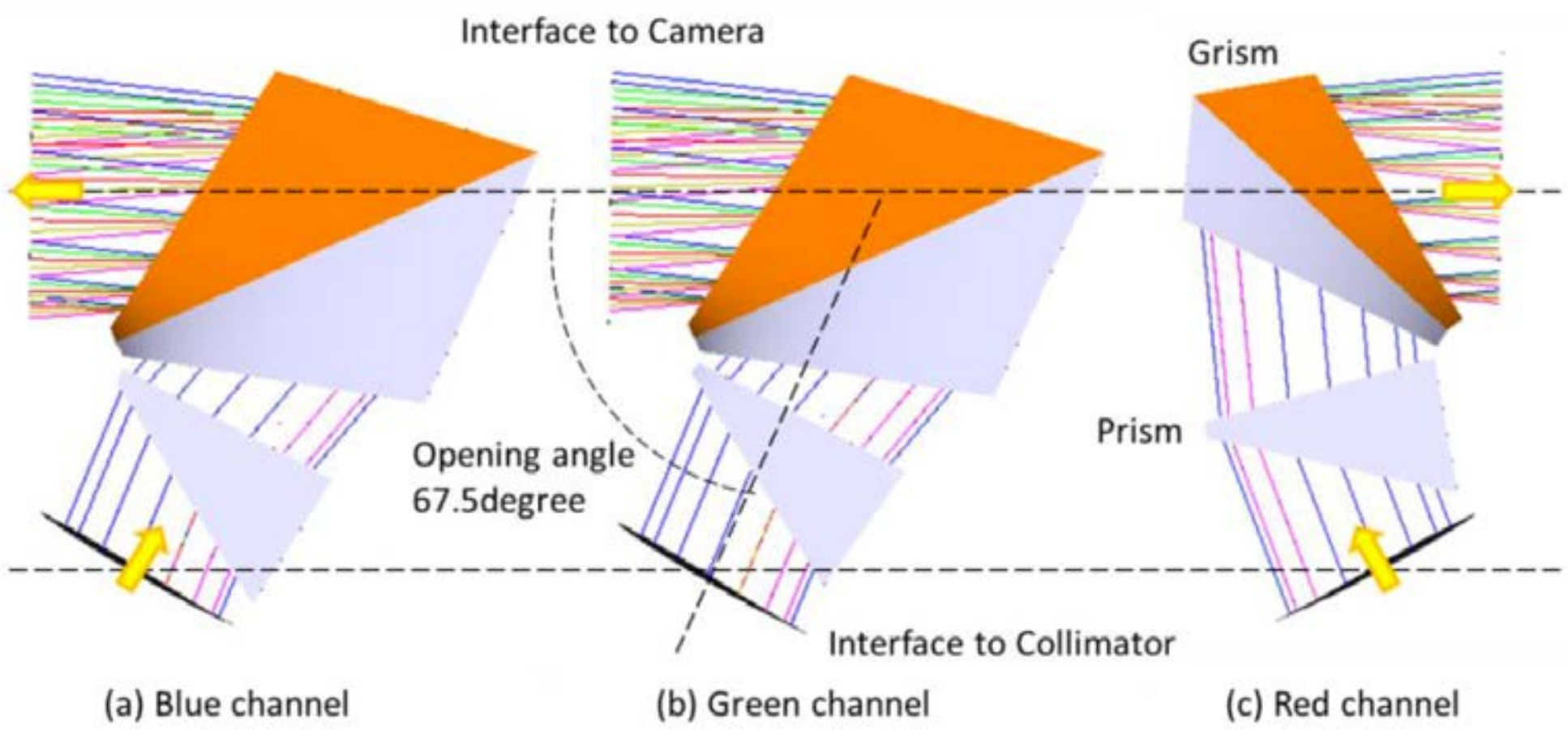

*Figure 101: HR disperser configuration for each of the three channels.*

**Disperser:** Producing a good HR optical design is fundamentally challenging, due to the large aperture of the telescope and the necessary spectral resolution of R=40,000 in the blue and green arms, which requires dispersers with a very high line density. This is challenging in terms of both the efficiency of the dispersers and their ability to be manufactured.

The current disperser configuration for the three channels is shown in Figure 101. It consists of a grism and a prism in each spectral channel. The orientation of the prism in the red R=20,000 channel is different from that of the two R=40,000 channels, to compensate for the difference in opening angles. The grism is a “sandwich-like” combination of two non-right-angle prisms, divided by a diffraction grating. The non-right-angle prism takes advantage of the vertex angle to reduce the incident angle on the air–glass interface and amplify the collimated aperture in the dispersed dimension, using the refraction effect on the entrance surface. This counteracts the optical effect of the off-axis collimator, which compresses the exit pupil in the same direction and results in a circular pupil at the disperser.

The required line densities for the larger dispersers present challenges in the blue and green channels. The required densities are 5,800 lines/mm and 4,920 lines/mm, respectively, on a very large substrate (~700 mm x ~400 mm for a clear aperture of 580 mm x 300 mm). Two different disperser technologies are being considered and both require prototyping for demonstration purposes. The Project Office is considering using the surface relief grating technology proposed by Fraunhofer[xviii] as the baseline. Surface relief gratings are made from a lithographic pattern, using ion beam etching, and provide significant gains in efficiency over conventional VPH gratings. The fallback technology is VPH gratings. KOSI[xix] has previously produced smaller gratings with similar line densities. However, those dispersers have intrinsic characteristics that cause efficiency to drop off steeply outside of a narrow wavelength range. Both technologies require mosaicking with smaller gratings. Grating size is limited by the current production facilities. Grating handling and alignment during the mosaicking process need to be developed and prototyped. The VPH grating poses even greater challenges, since the “printed” dichromated gelatin cannot be handled directly.

**Camera:** Each spectral channel feeds a transmissive camera, composed of a doublet, two singlets, and a powered vacuum window (*Figure 102*). Three aspherical surfaces are placed on the first surfaces of the three lenses, marked by an asterisk in *Figure 102*. The large camera aperture limits the available types of transmission glass. The glass types used in the current design consist of S-FSL5Y, S-BSL7, PBM2Y, BSM51Y, and fused silica. The total physical lengths of the cameras are around 900 mm. The central spacing between the third lens and the front of the vacuum window is larger than 60 mm, to accommodate the mounting interface of the detector cryostat. The central spacing between the back of the vacuum window and the CCD chip is 7 mm. When changing the dispersers to observe different working windows, each camera can be quickly realigned by adjusting the detector's focus and tilt angle by ±1.25 mm and ±0.05 degrees, respectively.

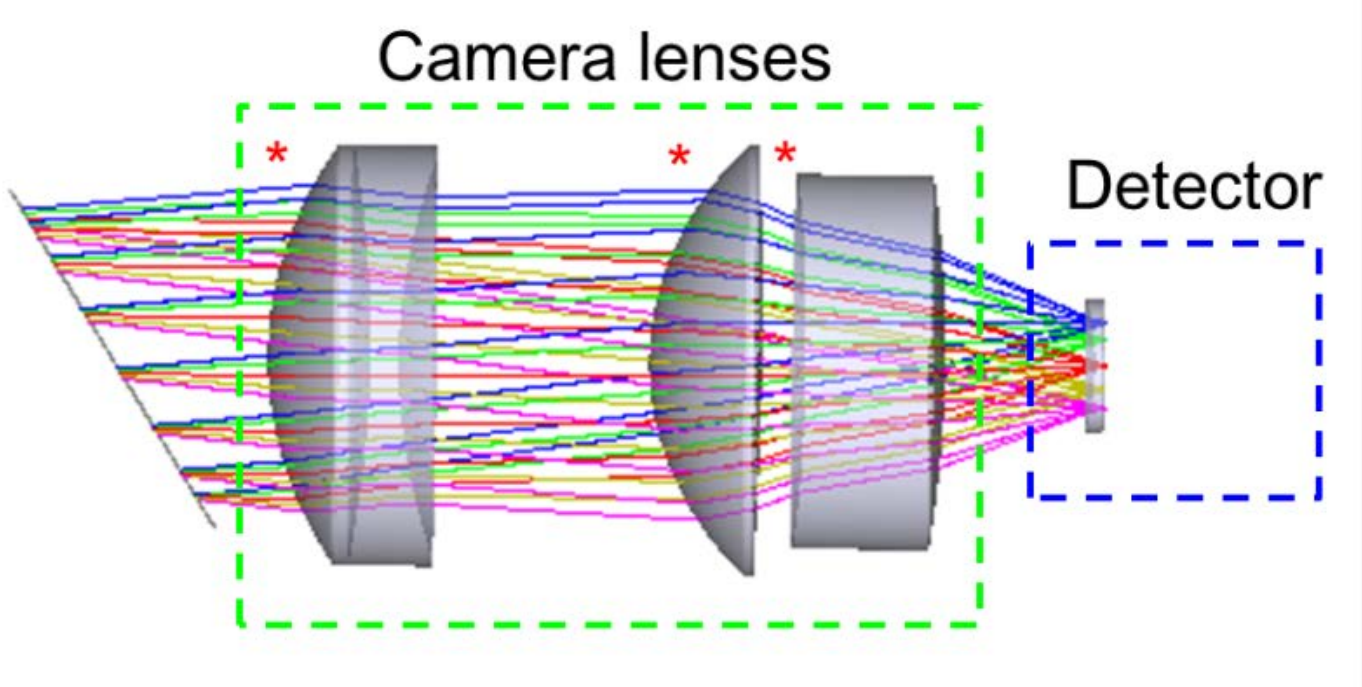

*Figure 102: HR camera lenses and detector. Aspheres are marked by an asterisk.*

### *(3) Design Development*

The HR spectrograph system is one of the most technically challenging components of the MSE observatory. It pushes the limits of optical design and fabrication, especially with regard to disperser technology. Any reduction on the demands placed on the dispersers could improve their manufacturability and efficiency:

- The Project Office will conduct a science-based analysis to determine whether the spectral resolution of the blue and green arms can be reduced by ~10% (i.e. R~35,000), without adversely affecting the science capabilities.
- The Project Office will conduct a trade study prior to the next phase to determine whether a smaller fiber aperture could be used, without significantly affecting sensitivity and observing efficiency (in the context of the analysis described in Section 3.5.1).

In addition, the Project Office will fund prototyping of disperser technologies (surface relief and VPH) in mosaic form, to obviate fabrication and alignment risks.

Despite these technical challenges, the design of the HR spectrograph provides remarkable image quality and throughput. In particular, the throughput of the design described here is ~15% higher than the value used in the overall sensitivity analysis described in Section 3.5.2. This improvement is due to extensive efforts by the design team, since the completion of the conceptual design of the HR spectrograph.

### 4.4.7. Science Calibration System (SCal)

Science Calibration system (SCal) refers to the physical hardware required to obtain science lamp[xx] calibration exposures (continuum and arcs). It is one component of the overall science calibration plan for MSE, discussed in more detail in Section 4.4.7. SCal has been developed to the feasibility study level and will undergo conceptual design development during the next design phase.

MSE has several high level science requirements, relating to the scientific calibration of the data, especially with respect to velocity accuracy, relative spectrophotometry, and sky subtraction. Accurate science calibration of a fiber-fed spectrograph essentially involves a precise understanding of the system throughput, wavelength solution, and point-spread function of each fiber during the science observations. It is also important to understand the fibers' relative behaviors as a function of all physical observing (e.g., wavelength, field position, telescope pointing) and environmental (e.g., temperature, humidity) conditions.

SCal provides illumination sources that are particularly important for throughput and wavelength solution determination. In particular, continuum sources are important for the measurement of the differential throughput of all the fibers as a function of wavelength. Arcs provide the wavelength solution for all fibers, in addition to helping us estimate the point spread function (PSF) from unresolved spectral lines.

The best calibration of throughput, PSF, and wavelength solution for MSE is obtained from configurations as close to the science observations as possible (both physically and temporally). However, in order to preserve the observing efficiency of MSE (see Section 3.5.3), it is better to ensure that nighttime calibration exposures only add minimal overheads to the science observations.

The hardware for SCal consists of continuum lamps and arc lamps. The arc lamps include both nighttime arcs and daytime dome arcs. An appropriate flat screen must therefore be included in SCal. While the baseline calibration plan does not foresee the need for dome flats, this capability is viewed as potentially useful and desirable. Lamp flats and arcs sources must provide enough photons for all resolution modes, fibers, and wavelengths, within a matter of seconds, to allow exposures to be obtained at a high SNR, within a short time. In addition, the sources must be stable and repeatable over time, in order to fully characterize the MSE system, with minimal calibration overheads.

During nighttime observations, the need to minimize overheads precludes deploying a screen or closing the dome. The baseline for the nighttime calibration is to mount the lamps on the underside of the telescope's top-end spiders and illuminate the primary mirror directly. Azimuthal scrambling in the fibers can produce a uniform near-field illumination pattern at the fiber outputs. Lamps may be shuttered when not in use, if continuous sources are used to maintain stability. The baseline for the dome arcs system consists of a dome screen mounted on the inside of the enclosure and light sources mounted at the top of the telescope. The dome arcs system reflects light sources off the dome screen to illuminate the primary mirror. It is highly desirable to use the same light sources for both systems, to ensure flux and wavelength consistency.

The envisioned generic sequence of MSE calibration exposures that involve SCal includes:

- Continuum exposures:
    - Twilight flats are taken at the beginning and end of every night at a fixed, reference observing configuration. The median of many of these exposures provides a high degree of uniform illumination for every fiber, with the same far field illumination pattern as the science exposures;
    - Science lamp flats are taken at the beginning and end of every science exposure, after system configuration. They are provided by SCal, and do not provide the same far field illumination pattern as the science exposures. These exposures are kept short, to minimize nighttime overheads, and are essentially coeval with the science frames.
    - Daytime science lamp flats are taken in the same system configurations, though with longer exposures, along with reference lamp flats in the reference observing configuration.
    - The twilight flats (with the correct far field illumination) are connected to the science lamp flats (with the correct system configuration) via a set of reference lamp flats.
- Arc exposures:
    - Daytime dome arcs are taken in a fixed reference configuration. These arcs have the correct far field illumination pattern that matches the science exposures;
    - Science lamp arcs are taken at the beginning and end of every science exposure after system configuration. They are provided by SCal, and do not provide the same far field illumination pattern as the science exposures. These exposures are kept short, to minimize nighttime overheads, and are essentially coeval with the science frames.
    - Daytime science lamp arcs are taken in the same system configurations, though with longer exposures, along with reference lamp arcs in the reference observing configuration.

The dome arcs (with the correct far field illumination) are connected to the science lamp arcs (with the correct system configuration) via a set of reference lamp arcs.

### 4.5. Observatory Execution System Architecture (OESA) Overview

The Observatory Execution System Architecture (OESA) includes the software and hardware architecture for the control systems that support daytime and nighttime MSE operations. OESA configures all subsystems involved in observations, stores operational data, and monitors safety in the observatory. OESA is distinct from the Program Execution System Architecture, which is responsible for planning, executing, reducing, and distributing science survey programs. Hierarchically, OESA receives commands from PESA and executes observations, using the observing interface (*Figure 103*). OESA contains both software and hardware products. The OESA design team is currently lead by CFHT staff in Waimea.

OESA divides its principal functions into software control sequencers, for the purpose of efficiently controlling all observatory subsystems. All observatory functions are principally controlled by the top-level Observatory Control Sequencer (OCSe). OCSe interacts with the observing interfaces, autonomous rules, and engineering interfaces to coordinate and conduct observations, as shown in *Figure 103*. This architecture is heavily dependent on the current CFHT queue observing and remote observing modes, which have proven very effective, ever since they were implemented in 2001 and 2008, respectively.

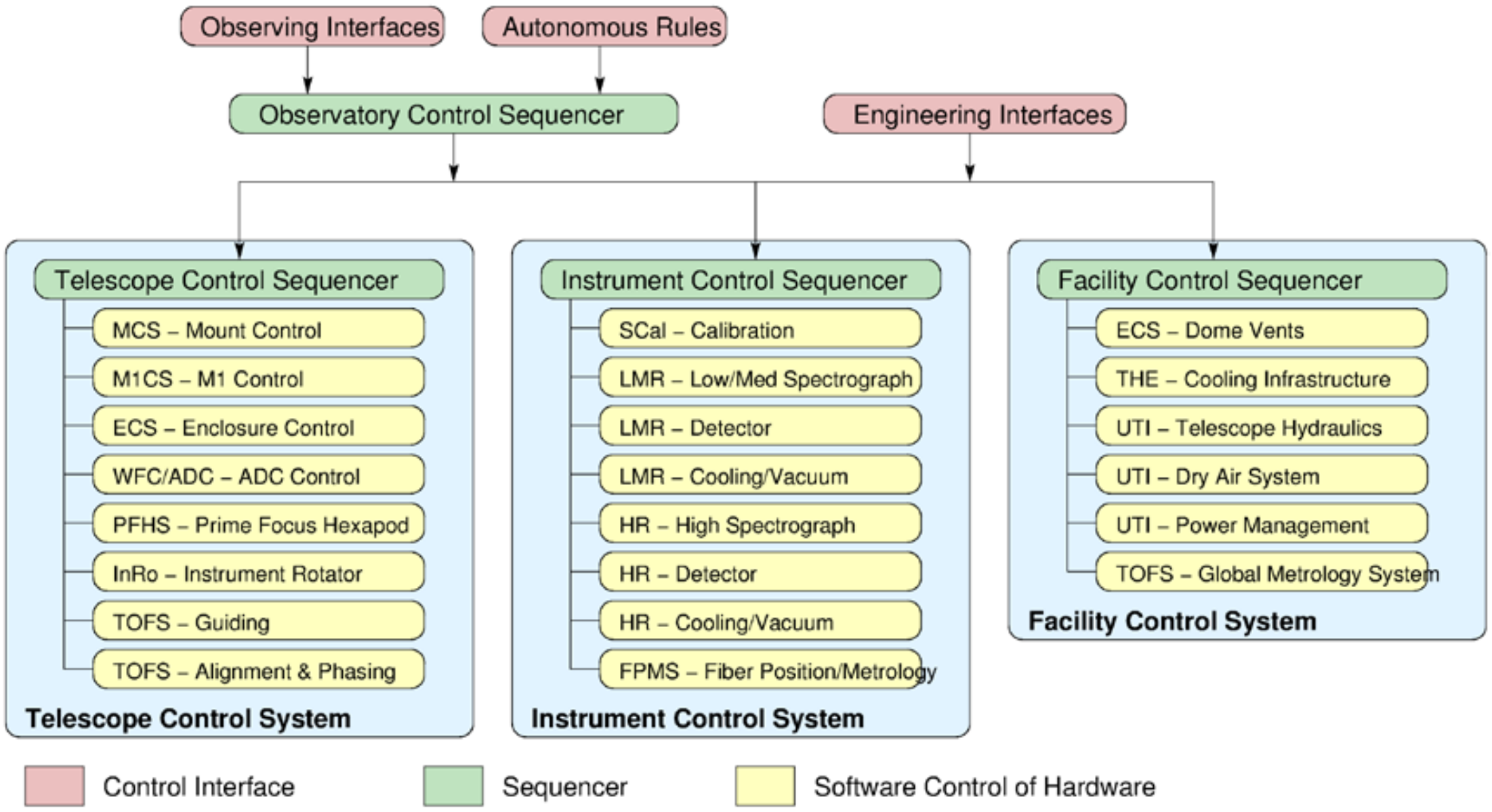

*Figure 103: OESA and its principal branches in a hierarchical tree structure*

The central OCSe controls and coordinates the activities of the subsystems, through three principal branches of the OESA architecture: the Telescope Control Sequencer (TCSe), the Instrument Control Sequencer (ICSe), and the Facility Control Sequencer (FCSe). The OCSe initiates hardware actions and acquires science, calibration, engineering (telemetry and diagnostic) data from these branches. Each of these three principal control sequencers in turn interfaces with the individual control systems of the lower level, such as the telescope Mount Control System (MCS). The data flow principle is top down: for example, a command is passed from the OCSe, downward through the TCSe, and into the MCS (see yellow box, *Figure 103*). The individual subsystem agents are then responsible for executing, rejecting, or pausing the command, in accordance with their operational rules. The agent is the lowest level of the control sequence within OESA.

### 4.5.1. Observatory Execution System Architecture Functions

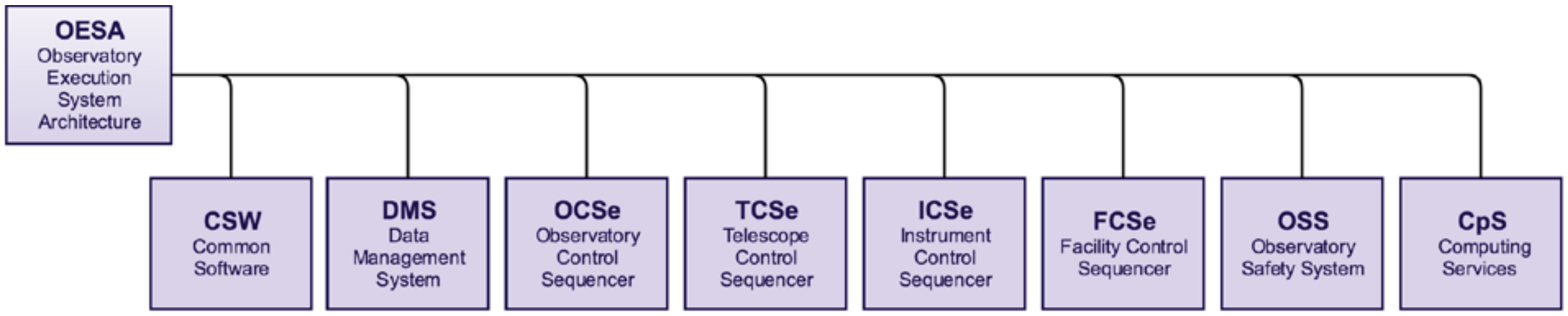

*Figure 104: OESA Product Breakdown Structure*

In addition to the aforementioned control sequencers, OESA functionalities require other software and hardware subsystems, as listed in the Product Breakdown Structure shown in *Figure 104*. These include:

- Common Software (CSW) has services and libraries, which are shared and reused among the many subsystems when interfacing with the control sequencers, as well as services used by the autonomous rules and engineering interfaces, to enable access to OCSe. In additional, CSW automatically logs information in the DMS and hosts the system status server, which is made available via remote access interfaces.
- Data Management System (DMS) allows for the formatting, storage, and retrieval of all the observatory's operational data, including science data, engineering logs, telemetry, status, etc.
- Observatory Safety System (OSS) is an independent hardware-based system, which monitors the observatory and uses safety logic to take automatic action to shutdown, override, and lockout observatory activities, as needed, to ensure the safety of personnel and equipment.
- Computing Services (CpS) provide the common infrastructure for the software system, including the computing facility, network, network services, and cabling at both the summit facility and the Waimea building. They also provide network services for inter-process communication, a domain name resolver, authentication services, messaging and alerts, etc.

Additional functionalities and attributes of the OESA products envisaged by the CFHT design team for the next design phase will be described in more detail in the following sections, according to their hierarchy within the OESA architecture.

### 4.5.2. Observatory Control Sequencer

The OCSe receives commands for telescope, instrument, and facility actions, via a TCP/IP socket. As *Figure 103* illustrates, commands are received either via an observing interface or an autonomous rules component. Commands received from either interface are received as human-readable command strings, with optional arguments. Commands from the observing interface either originate from engineering-style Graphical User Interfaces (GUIs) or from the command execution component within PESA. Each of the control systems has a similar multi-layered architecture (*Figure 105*).

During science operations, OCSe receives multiple commands from PESA. For instance, PESA provides the on-sky coordinates needed to point the telescope and enclosure, target coordinates by which to position the fibers, and configurations for the spectrographs. The OCSe organizes those commands, provides them to the relevant lower-level control sequencers (TCSe, FCSe, and ICSe) in the proper chronological order, and ensures that TCSe, FCSe, and ICSe collaborate harmoniously.

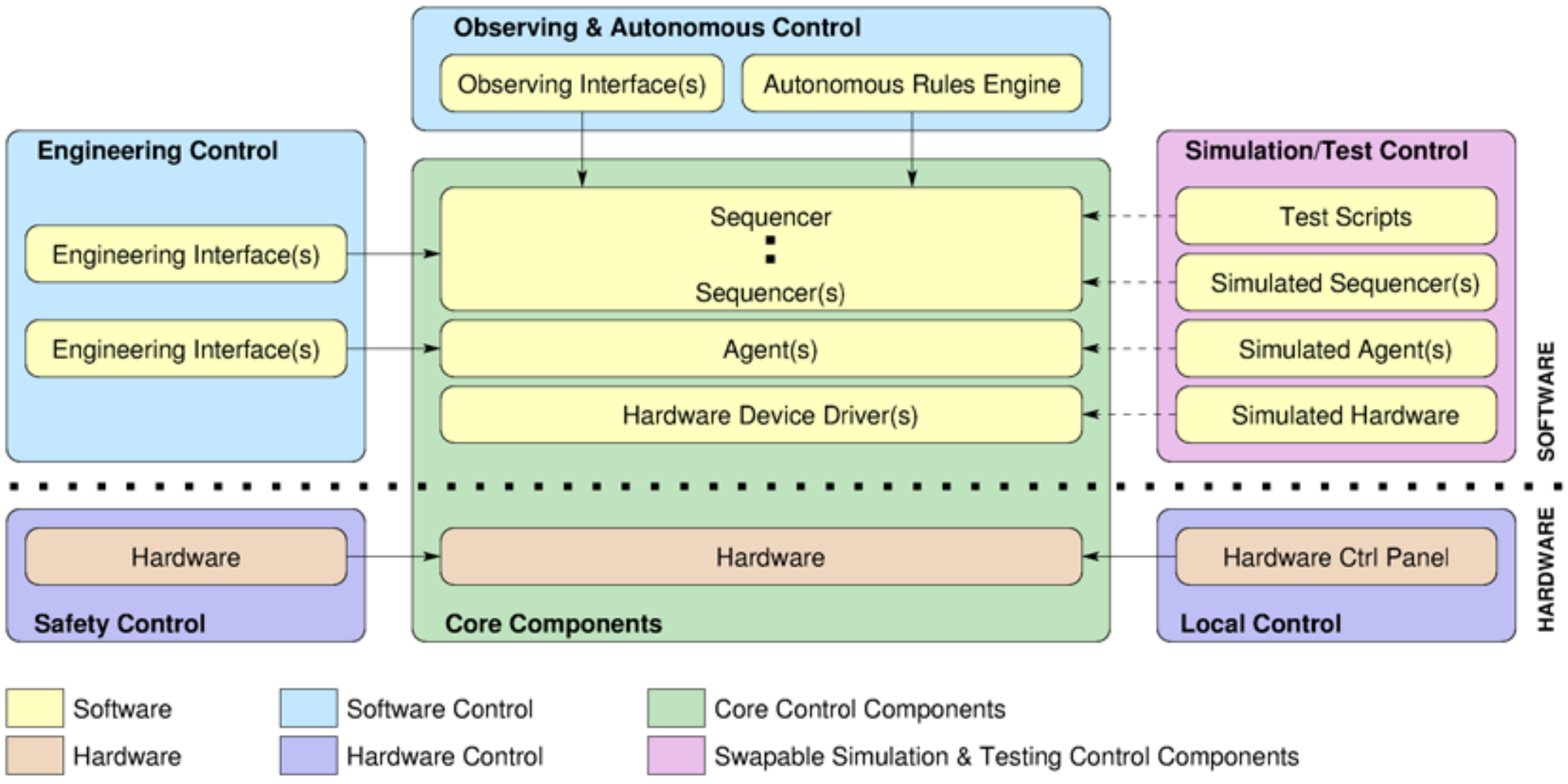

*Figure 105: Common control system architecture*

### 4.5.3. Telescope, Instrument, and Facility Control Sequencers

The TCSe, FCSe, and ICSe convert the information provided by the OCSe into parameters relevant to the subsystems they control. For instance, the TCSe computes the telescope's altitude and azimuth coordinates and the angular base and cap positions for the enclosure, along with the best path for reaching those coordinates, based on current pointing and other available system status information (e.g. current available range of the cable wraps). The ICSe computes the positions of each fiber positioner in the focal surface, along with the individual positioners' sequences of closed-loop motions, using feedback from the metrology system. TCSe, FCSe, and ICSe communicate with each other through the OCSe.

As shown in *Figure 103,* TCSe commands the execution of the observing set-up for the telescope, enclosure, InRo, PFHS, ADC setting, M1 shape, and TOFS. Once an observing field has been configured, the TCSe commences closed-loop guiding by controlling the MSTR, AGC, and InRo. These three subsystems are highly interdependent and their motions must be synchronized. For example, MSTR and InRo track the target field synchronously.

The ICSe controls the spectrographs, fiber positioning via the metrology camera, and the science calibration unit. These instruments are highly interdependent and it is therefore expedient to control and coordinate them all via the ICSe, instead of using different, separate control sequencers. For example, the 4,332 fibers are positioned in the target field by the PosS metrology system, which provides closed-loop feedback control to align the fibers with their targets iteratively during configuration, using back illumination of the fiber inputs, provided by the spectrographs (LMR and HR). The shutters in the wavelength arms of the LMR and HR spectrographs are closed during back illumination, to enable readout of the optical and infrared detectors.

The FCSe governs activities such as cooling control loops, monitoring air and hydraulics levels, and managing the dome vent positions, based on wind speed, direction, and anticipated precipitation conditions. However, these actions are typically not as closely coupled or as highly interdependent as they are for the ICSe and TCSe.

#### 4.5.4. Data Management System

The DMS encompasses the common infrastructure for the management of operational data across the observatory systems. It provides infrastructure that satisfies the data flow requirements of the data-intensive MSE subsystems. It supports, as far as possible, uniform interfaces for getting data into and out of the DMS.

The DMS comprises the Observatory Data Repository (ODR), which contains all science data (raw and reduced), all engineering data (logs and status server), and all environmental data (weather- and non-weather-related ambient conditions). The ODR is the main database that technical staff access when they need to troubleshoot, investigate systematic abnormalities, or establish statistics.

The status server is part of the DMS and provides real-time state and status information for the entire observatory system. The data structure is hierarchical: every (reporting) node has only one value at any moment in time, along with attributes such as a timestamp of the last update. The status server also provides notifications to event-driven subscriptions, which monitor changes in the state of selected nodes, and sends alerts to subscribers when predefined conditions occur.

#### 4.5.5. Common Software (CSW)

Generally, the CSW designed and developed for MSE follows the traditional waterfall development approach (requirements, conceptual design, preliminary design, final design, implementation, integration, and testing phases). Areas with high user involvement, such as graphical user interfaces (GUIs), follows a more iterative design cycle, such as the Agile software development methodology. To the extent that it is necessary, prototyping may be performed, although it is not used as a way to bypass the design process. The waterfall and Agile methodologies have been successfully used for large-scale software projects at CFHT, such as Queued Service Observing (QSO), implementation of remote observing capability, and the CFHT Status Server. Each of these projects was integrated with virtually no downtime and with very low subsequent maintenance overheads.

MSE imposes additional constraints on software engineering. Open source software is the baseline, in order to facilitate transition, despite inevitable changes to hardware and operating systems over time. Adopting simple communications protocols based on TCP/IP socket protocols allows for looser coupling between software components and greater implementation flexibility for individual components. CFHT software is generally reused and improved where appropriate, to reduce risk and to enable MSE to commence operations on a foundation of robust and proven software. Control logic is separated into a layer that can be triggered in multiple ways, to allow for the future replacement of GUI or logic layers with minimal disruption. The selection of standardized languages reflects the need to strike a balance between popularity, maturity, and on-going support—as opposed to simply applying the latest available programming frameworks and technologies—and to maintain the software system integrity (design, integration and troubleshooting) over the life of the Observatory.

#### 4.5.6. Computing Services

The CpS encompasses the shared physical computing infrastructure for the software systems within the MSE observatory. As a physical system, the CpS services both the control systems (OESA) and observing systems (PESA).The CpS is optimized for operations of the common software and the MSE database structure. It also includes the computer network and network configuration, network services (directory, authentication), computer rooms, file servers and backups, etc.

The most important attributes of CpS are its high levels of availability and redundancy, high performance, resilience to tampering, automatic and rapid self-configuration, centralized configuration, and easy administration.

The MSE’s CpS reuses as much of the current CFHT computing infrastructure as possible and incorporates the natural evolution of new technologies and best practices.

### 4.6. Program Execution System Architecture (PESA)

The Program Execution System Architecture (PESA) is responsible for planning, executing, reducing, and distributing science survey programs. PESA includes several sophisticated software tools, organized into an operational framework as the Phases described in Section 2.4.3. These are grouped into products as functional units, as shown in *Figure 106* and can be described as follows:

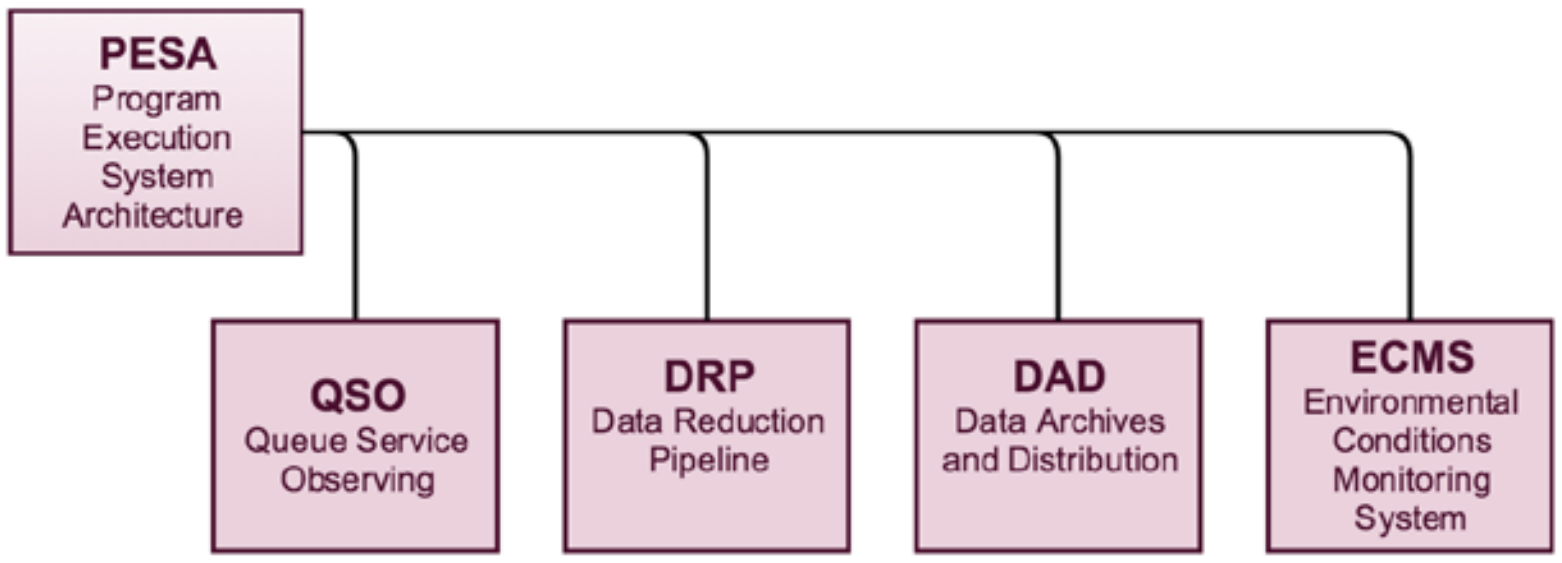

*Figure 106: PESA Product Breakdown Structure.*

- Queue Service Observing (QSO) provides the functionalities needed to prepare and schedule observations, including proposals, observation design, nighttime scheduling, validation, accounting, and statistics. QSO corresponds to Phases 1 through 3 of the science operations and provides MSE staff and scientific community with observing tools that are of high-level capabilities.
- Data Reduction Pipeline (DRP) operates during Phase 4, providing real-time feedback to the QSO system, by monitoring data quality and survey progress. This interaction enables the QSO to regenerate optimal observing schedules in real time.
- Data Archives and Distribution (DAD) represents Phase 5 of the operational framework. It accepts the science products generated by DRP in Phase 4 (or the enhanced reduction provided by the survey and science teams) for long-term storage and distribution. It provides the external MSE community with a database, which contrasts with the OESA DMS, which is intended for the internal use and contains supplemental operational data unconnected with science applications. DAD organizes the relevant metadata provided by DMS and the Environmental Conditions Monitoring Systems into the Science Products Archive (SPA). The SPA database also includes an interface, which enables MSE users to search and retrieve science data and metadata.
- Environmental Conditions Monitoring System (ECMS) monitors the observatory. It includes such instruments and sensors as anemometers, humidity and precipitation sensors, a Differential Image Motion Monitor (DIMM), a cloud/sky monitor, a water column density monitor, etc. ECMS does not include the standard sensing systems used for facility management and safety within the MSE observatory.

PESA has been developed at a feasibility level such that the core functions and requirements of the next design phase have been identified. The Project is currently assembling a collaborative team to plan this software effort.

The expected scope and concept of the QSO and DRP software modules, as defined by the MSE Operations Concept Document (OCD) are described here.

### 4.6.1. Queue-Scheduled Observing (QSO)

QSO is a complex suite of software that combines millions of targets used to dynamically generate the schedule of science observations defined by the survey teams according to their priorities, programmatic constraints, historical and current weather data, and the current status of the observatory's subsystems, Each science observation is defined by an Observing Matrix (OM), which contains the associated Observing Field (OF) and Observing Sequence (OS). The OF defines the allocation of fibers to targets, including the assignment of fibers to calibration stars and sky positions, and verifies that guide stars are available within the observing field. The OS also defines the number and duration of the exposures for each arm of each spectrograph. The Observing Matrix Generator (OMG) generates the optimal sequence of OMs to be observed, given all observing constraints and considerations. An OM is expected to be a very complex tool, merging targets from multiple OF surveys and using multiple exposure times on different arms and different spectrographs in the OS. For instance, both HR and LMR targets are included in every OF, since both types of fibers are always available concurrently. In addition, near-IR exposures are of limited duration because of the sky brightness (continuum and lines), therefore an OS combines multiple near-IR exposures, while longer single exposures progress in parallel at the shorter optical wavelengths.

Similar software tools under development will automate observing schedules at many upcoming observatories, including multi-object spectroscopic facilities. The publicly available software packages (e.g., Astropy's Astroplan, LSST's OpSim) provide valuable models, which will help bootstrap the development of the QSO tools for MSE.

### 4.6.2. Data Reduction, Validation, and Distribution

The Automatic Data Reduction Pipeline (ADRP) tool is triggered after every nighttime exposure. It processes the raw data (Level 0) to generate spectra in real time, for quick-look science analysis, schedule adjustments, and data validation. The data are designated as Level 1 products and immediately made available to the survey team and MSE staff. They are also used to determine the signal-to-noise ratio (SNR) of the spectra in real time and compare it with expectations, based on the estimates of the Exposure Time Calculator (ETC), and the actual recorded observing conditions. This analysis provides feedback to the OMG, allowing it to adjust the OS (exposure times and number of exposures) as required. Level 1 data products are also used to provide a validation and quality flag for each exposure. These are then used to update the target database, marking off those targets that have been validated or completed.

Level 1 data products are typically obtained without the full set of calibration exposures. For example, relevant daytime calibrations are typically obtained the day after the science targets have been observed. The ADRP is then triggered automatically again, once all the calibration exposures have been recorded, and is used to generate Level 2 data products. These are the main MSE data products, to which optimal data reduction methods are applied, using a basic set of measurements applied homogeneously across the spectra (e.g., line fluxes). Level 2 data products include stitched and stacked spectra reduced from multiple exposures of a given target, even if those exposures were spread over several nights.

Level 3 data products, which include spectra and catalogs specific to the survey's science programs, are the main data products provided by the survey team. Level 4 data products are value-added data products, provided by the science team or wider international community. Both Level 3 and Level 4 data are hosted by the same platform as Level 1 and 2 data products and are compatible in their format with other MSE data products. The definition and content of Level 2 and Level 3 data products may evolve, as the survey team and the MSE staff work together to improve all data products.

All data products include images and spectra, as well as metadata listing relevant information for the MSE staff and the survey team. Data products are archived in at least two databases, one of which is internal to the Observatory: the Observatory Data Repository (see Section 4.5.4), which contains Level 0–Level 2 data products, as well as all engineering and environmental metadata. It is primarily used by the MSE staff, to analyze, validate, and improve the data products. MSE provides a second database for the benefit of the external astronomical community. Access is restricted to authorized users. The Science Products Archive contains Level 0—Level 4 data products and a subset of the relevant observing metadata.

# 5. Moving Forward to the Preliminary Design Phase

The Conceptual Design Phase (CoDP) culminated in January 2018, with a system design review, to confirm whether the proposed MSE design met the science requirements. According to the review panel, chaired by Michael Strauss of Princeton University: "the bottom line is that this project is in very good shape, and at an appropriate level of maturity for the end of the Conceptual Design Phase. We have been very impressed by the level of sophistication that the MSE project team has brought to this project, and the tremendous amount of hard work that has been carried out thus far. This level of professionalism bodes well for the project as it enters the Preliminary Design Phase."

The following section describes the plan to move MSE through the Preliminary Design Phase (PDP) and will outline how we plan to achieve a formal partnership agreement for construction, including:

- Partnership model
- Advancement through science and engineering
- Governance of the PDP and beyond
- Execution plan for PDP completion
- On-going Project Office roles and responsibilities
- Establishment of the education and public outreach program

## 5.1. Partnership

Each participant's contribution to the design, construction, and operations of MSE is converted into a Beneficial Interest in MSE: i.e., a proportionate share in the scientific and technical usage and governance of the facility. The details of the partnership model that will govern MSE during its construction and operations will be finalized and approved by the MSE Board during the Preliminary Design Phase.

The science-driven survey model outlined in Section 2 has important implications for the MSE partnership model and for the scientific governance of the Observatory. Clearly, science leadership in MSE is ultimately contingent on a partner's ability to define, propose, lead, and use data from the science surveys on MSE, and influence the long-term scientific direction of the Observatory. As such, it is anticipated that each partner's share in survey leadership and governance will reflect their share of Beneficial Interest in MSE. Partners with larger beneficial interests are expect to define and lead more survey programs and have more influence over the scientific direction of MSE. MSE data will eventually be released to the international community, following a proprietary period set by the partnership.

It is anticipated that those communities who contribute more than a predetermined fraction of MSE costs during design, construction, and operations will become full partners in MSE. All scientists from one of MSE's full partners will be able to propose and lead surveys on MSE, and will have access to all MSE data. Limited-partnership categories (e.g., Associate Partners), who make smaller contributions, may be considered, at the discretion of the MSE Board. The board may decide to grant full partnership privileges to a limited number of scientists from a specific community, thereby allowing that smaller community to take visible and defining roles in the science leadership of MSE; alternatively, a larger community may be granted a limited partnership, which permits them data access, but does not allow them to take a role in survey definition.

### 5.2. Growth through Science

One of the review panel's strongest recommendations is that the project should develop a Design Reference Mission, a.k.a. the Design Reference Survey (DRS). The review panel stated that, for a survey-driven project like MSE, the DRS is an enormously useful tool for evaluating at every stage whether or not their science goals can be met by their system.

Soon after the design review, a call was issued to the astronomical community to join the MSE's international science team, with the purpose of "reaffirming" the science case, before starting the DRS process. Since then, the size of the science team has more than tripled: there are currently 327 members from thirty countries.[22] Nine different Science Working Groups (SWG) have been created, to focus on different areas of the MSE science case. Each SWG has two co-leads.

The SWGs are the primary groups responsible for the ongoing development of the MSE Science Case and the development of the DRS. A new version of the Detailed Science Case will be published at the start of 2019, and the DRS activities will start soon thereafter. The science team is the primary body to undertake all MSE's science development activities. The partnership recognizes that some scientists who are making significant contributions to this development activity may not be members of partner communities, once the MSE Observatory is operational. The partnership will therefore investigate procedures that would allow their involvement and collaboration during operations, in recognition of their significant contributions during the scientific development phases.

### 5.3. Progress through Engineering

The international design team and PO have completed the designs of many products for CoDP, under the MSE Production Breakdown Structure. For example, they have:

- Completed the conceptual designs of the enclosure and the telescope structure
- Selected a multiplexing configuration and generated functional positioner and metrology system and fiber transmission system designs
- Confirmed the designs of the telescope's optical system (segmented mirror and WFC/ADC)
- Produced LMR and HR spectrograph design concepts
- Established the observatory system execution architecture and a fiber testing facility to evaluate the performance of the fiber transmission system

The PO has also established the system performance budgets and produced the observatory operations concept and science calibration methodology, in order to set the requirements for the program execution system architecture.

Feasibility studies have been conducted for the subsystems that did not proceed through to the conceptual design phase. These studies have established the subsystems' functionalities and defined the requirements necessary to eliminate programmatic risks. The subsystems involved include the observatory building and facilities, and the telescope optical feedback system.

[22] As of September 2018. PhD astronomers from the international community are welcome to join the science team at any time.

The conceptual designs of the system and essential subsystem have been reviewed by independent external panels: their evaluations were favorable. Incorporating the reviews' recommendations, a PDP work plan has been established to advance the project through the next design phase.

### 5.4. Preliminary Design Phase Governance

PDP is scheduled to start in early 2019. The current PDP design team includes members of the Australian Astronomical Observatory, National Research Council (NRC) of Canada, National Astronomical Observatories of China (NAOC), Chinese Academy of Sciences, Centre National de la Recherche Scientifique (CNRS) of France, University of Hawaii, and the Indian Institute of Astrophysics (IIA). These six organizations were former participants in the CoDP.

Representatives from these organizations currently form the MSE Management Group (MG). In addition, representatives of the National Optical Astronomy Observatory (NOAO) in the US and Texas A&M University are also observers in the MG, and are exploring the possibility of joining as partners in the future.

The PDP's formal governance agreement is being developed as a Statement of Understanding (SoU) between CFHT Corp. and the MSE project participants. Once the SoU has been enacted, the MG will be replaced by the MSE Board, which will include representatives of all signatories. The SoU empowers the new MSE Board to set the direction of the project through the PDP and tasks the Board with defining the formal partnership agreement in subsequent phases, leading to science operations. The agreement will be developed within the framework and expectation of a growing partnership in the upcoming phases.

### 5.5. Preliminary Design Phase Execution Plan

As with the CoDP, execution of the PDP work will be organized by the Project Work Breakdown Structure (WBS)[23] and formalized by work package documents shared between the performing organizations and the Project Office (PO). Each work package document defines the scope of the work, the schedule, and the milestones, deliverables, designated resources, and corresponding work to be performed (quantified in hours), as well as the financial contributions from the performing organization and PO (such as procurements, travel costs, and reimbursements). The value of each work package is the sum of the labor costs[24] and the net financial contribution of the performing organization.

All PDP work packages will be authorized and approved by the MSE Board, based on recommendations from the PO. The values of the work packages will accumulate toward the final share of Beneficial Interest, according to the pending MSE partnership agreement. In the same way, the accumulated values of the CoDP contributions will be converted into corresponding Beneficial Interest.

---

[23] The Work Breakdown Structure is informed by the Product Breakdown Structure (Section 3). The latter identifies the products in terms of physical systems required to support and execute science observations. The former defines the work required in order to produce, qualify, deliver, integrate, and commission the individual systems in the PBS into a functioning observatory.

[24] Labor cost is currently calculated using hour estimates, at a pre-agreed rate of $110 USD per hour.

On the basis of the WBS, a PDP work plan has been developed, which summarizes the work envisaged and the corresponding resources required. For example, *Figure 107* shows examples of itemized Project Office labor costs for the PDP, over an anticipated duration of two years.

| WBS | Title | PDP estimate | Contingency | Total |
|---|---|---|---|---|
| MSE.PO.MGT.LOE | Management Level of Effort | $492,441 | $49,244 | $541,685 |
| MSE.PO.MGT.SOFT | Software Services | $18,750 | $1,500 | $20,250 |
| MSE.PO.MGT.HR | Human Resources | | | |
| MSE.PO.MGT.PuOu | Public Outreach | $395,058 | $39,506 | $434,564 |
| MSE.PO.MGT.BUS | Business Services | $255,087 | $25,508 | $280,595 |
| MSE.PO.MGT.MEE | Meetings and Review | $400,000 | $0 | $400,000 |
| MSE.PO.MGT.PER | Permitting | $325,000 | $74,750 | $399,750 |
| MSE.PO.MGT.SAF | Safety | | | |
| MSE.PO.MGT.PA | Product Assurance | $34,011 | $3,401 | $37,412 |
| MSE.PO.SCI.LOE | Science Level of Effort | $863,808 | $172,761 | $1,036,569 |
| MSE.PO.ENG.LOE | Engineering Level of Effort | $2,211,627 | $221,162 | $2,432,789 |
| MSE.PO.ENG.STA | Engineering Standards | | | |
| MSE.PO.ENG.SYS | System Design | | | |

*Figure 107: Examples from the PDP work plan—Project Office cost estimates*

Judging by the CoDP experience, the PDP will be challenging and exciting. Resource and budget pressures will be dynamic and sometimes arise at short notice. Progress on the project will therefore require on-going re-planning, anticipation, and adaptability, and participants will need to work with the MSE Board, to stay on schedule and manage risk exposure.

### 5.6. Project Office Roles and Responsibilities

As an international project with scientists and engineers from geographically scattered and culturally diverse teams from all over the world, the Project Office fills the vital role of organizer and coordinator of the PDP work. Working with the science team, the PO ensures engineering choices and design trades are well grounded in science requirements.

Globally, the PO manages the project from a system perspective, through standard systems engineering practices, and provides leadership in performance modeling and budgets, interface control, requirement management, configuration management, change control, requirement verification, etc. The PO also leads the planning of system integration, science commissioning, and observatory operations.

In addition to technical oversight, the PO manages the PDP costs and schedule, as well as the project risk register, with support from the MSE Board. In return, the PO supports the Board in the partnership building and promotion necessary to manage the construction partnership for the MSE Observatory. The major PO deliverables are the construction proposal, which describes the end-to-end project costs and schedule, and the engineering development plan (after the PDP), leading to science operations.

Beginning during the PDP, other essential PO activities will include:

- Coordinating the planning of a comprehensive education and outreach program, with inputs from national representatives of all participants

- Initiation of the construction permit application process following the University of Hawaii's Maunakea land use application[25] which is currently underway
- Introduction and enforcement of project-wide engineering standards, specifically with regard to health and safety, and quality assurance
- Organizing annual project-wide meetings for the design and science teams

### 5.7. Education and Public Outreach Program

In all its endeavors, MSE is committed to balancing social, environmental, and educational considerations in Hawaii and plans to extend this commitment to encompass all the MSE communities in the PDP. Recognizing that Education and Public Outreach (EPO) is an integral part of every next generation astronomical facility, the project plans to make an early start on building a cohesive program, which is meaningful, comprehensive, and matches the desires of the MSE participants.

MSE is thankful that it benefits from the goodwill generated by CFHT's forty-year presence on Maunakea and from a support staff who are engaged with and rooted in the Big Island community. Currently, CFHT's EPO program actively engages the Hawaii Islands communities by working with schools, community groups and workforce development programs, and by hosting numerous outreach events and activities across the Islands.

The Maunakea Observatories coordinate their efforts to nurture a lasting relationship with the community through a variety of means:

- Increasing community awareness of the existing astronomy outreach programs, and, in some cases, expanding into even larger programs
- Improving relations with government, through on-going dialog with legislators, the Governor's office, county officials, etc.
- Providing an infusion of new community outreach programs that invert the paradigm, bringing the community into the observatories, rather than relying mostly on observatory staff going out into the field (classrooms)
- Developing innovative educational, environmental, and cultural opportunities for the community, through unprecedented and unconditional philanthropy
- Focusing on a positive and inclusive future for Maunakea, with special emphasis on what the future of Hawaiʻi astronomy means to the *keiki* (children) of Hawaiʻi, who will, in a very real sense, "inherit" these remarkable facilities, as members of their technical, administrative, and scientific staffs
- Promoting community-based management of Maunakea, consistent with the wishes of the Office of Maunakea Management, Maunakea Management Board, Kahu Kū Mauna, Comprehensive Management Plan, etc.
- Establishing a lasting community-based vision for the future of Maunakea, for the next hundred years, which informs all aspects of policy and decision making about the summit of Maunakea

CFHT has taken a leadership role in many of these initiatives, facilitating the organization of:

---

[25] Also known as Master Lease

- Kamaʻāina Observatory Experience program, a free monthly program available to Hawaii residents, which includes presentations about the cultural importance and environmental sensitivity of the Maunakea summit, and a subsequent visit to the summit to see two telescopes up close. During the monthly online registration for this program, all 48 openings are generally filled within minutes—an indication of the popularity of the program and community interest in seeing Maunakea and the observatories.
- Maunakea Fund, a signature fund created within the Hawaii Community Foundation's portfolio, dedicated to sponsoring place-based STEM[26] education, and environmental, and cultural programs related to Maunakea. To date, the Maunakea Fund has provided $250,000 to sponsor a multitude of programs.
- Maunakea Scholars program, a partnership between the Maunakea Observatories, UH, and the State Department of Education, which provides mentored research opportunities for high school students across Hawaii. Students submit observing proposals, which are evaluated on their scientific merit and technical feasibility. Those awarded observing time are matched with one of the observatories on Maunakea. The program is conducted at ten public high schools and, for the first time, makes it possible for local high school students to use a Maunakea telescope to conduct their research. Reactions to this program have been overwhelmingly positive.
- Working closely with community leaders to establish the EnVision Maunakea (EM) program, which creates comfortable settings for small groups within the community to meet and discuss their visions for the future of Maunakea. In 2016–17, this program held 15 "listening sessions" or *ʻAha Kūkā* in different island communities. The common threads of these discussions were shared with the public in a final report—the intent being that the "soft voices" captured in the report would be heard and have an influence on upcoming decisions and policies pertaining to Maunakea. The program has been endorsed by the UH-Hilo Chancellor, who has given assurances that future policy on Maunakea will be critically informed by the results of EM. The Office of Maunakea Management reports to the UH-Hilo Chancellor, who is also responsible for leading the next Maunakea land-use application and will play a central role in determining its terms and conditions, as well as directing the Environmental Impact Statement process needed in the land use application.

These and many other education and outreach efforts, conducted on a sustained basis, will no doubt strengthen the relationship between the Hawaii community and the Maunakea Observatories.

Moreover, the CFHT–to–MSE transition marks the first large-scale redevelopment of an existing observatory into an entirely new facility. The transition will provide momentous EPO opportunities across the MSE partnership and form the core of the early EPO activities.

The MSE EPO will include a STEM-based program, which links the unique scientific and engineering capabilities enabled by the new observatory to educational and workforce development opportunities within the participant communities. As a survey facility, the MSE archive will potentially provide additional outreach opportunities for citizen science, whereby interested amateurs and the general public can participate in future discoveries.

---

[26] Promotion of science, technology, engineering, and math education.

The MG initiated the formation of the MSE EPO planning committee at the start of the PDP, with members from all participant communities engaged in education and public outreach. Leveraging the experience gained from CFHT's EPO efforts, the current CFHT outreach manager will co-chair the planning committee. It is expected that concepts tried and tested during CFHT's existing projects may have broader applicability to the MSE communities and will contribute to the EPO program. However, actual implementation will be tailored to the individual needs of each participant. Over the course of the PDP, the committee will consult a wide range of subject matter experts, scientists, and educators, in order to refine and target the program, to address the specific needs of individual communities.

By the end of PDP, a cohesive EPO will be available, along with its implementation plan, and staffing and budget needs. This will provide the means of cultivating community ownership and public interest in MSE. As MSE development advances into the construction phase, the appropriate plan and budget will be in place to support the EPO activities identified.

### 5.8. Final Words

Since the CoDP, the international astronomical community has categorically reaffirmed MSE as scientifically relevant, desirable, and important in the era of large imaging surveys and extra-large optical telescope projects. Building on a sound observatory design with enabling capabilities that are valued by the international astronomical community, as evidenced by recent US and European reports on wide field spectroscopy, MSE development will continue on a positive trajectory scientifically, technically, and programmatically—toward PDP and beyond.

---

[i] Bauman, S. E., G. Barrick, T. Benedict, et al. "Transforming the Canada France Hawaii Telescope (CFHT) into the Maunakea Spectroscopic Explorer (MSE): A conceptual observatory building and facilities design." Society of Photo-Optical Instrumentation Engineers (SPIE) Conference Series (2018): 10704, 107041E.
[ii] http://www.dynamicstructures.com/
[iii] Szeto, K., V. Konstantinos, H. Horia, et al. "Conceptual design study to determine optimal enclosure vent configuration for the Maunakea Spectroscopic Explorer (MSE)." *Ground-based and Airborne Telescopes V* (2014): 9145, 91450A.
[iv] Saunders, W. and P. R. Gillingham. "Optical designs for the Maunakea Spectroscopic Explorer Telescope." *Ground-based and Airborne Telescopes VI* (2016): 9906, 990638.
[v] https://www.idom.com/
[vi] Sheikh, D. A. "Improved silver mirror coating for ground and space-based astronomy." *Advances in Optical and Mechanical Technologies for Telescopes and Instrumentation II* (2016): 9912, 991239.
[vii] Schneider, T., T. Vucina, C. Ah Hee, et al. "The Gemini Observatory protected silver coating: Ten years in operation." *Ground-based and Airborne Telescopes VI* (2016): 9906, 990632.
[viii] http://www.symetrie.fr/en
[ix] http://www.lmtgtm.org
[x] Pazder, J., P. Fournier, R. Pawluczyk, and M. van Kooten. "The FRD and transmission of the 270-m GRACES optical fiber link and a high numerical aperture fiber for astronomy." *Advances in Optical and Mechanical Technologies for Telescopes and Instrumentation* (2014): 9151, 915124.
[xi] Brzeski, J., L. Gers, G. Smith, and N. Staszak. "Hermes: the engineering challenges." *Ground-based and Airborne Instrumentation for Astronomy IV* (2012): 8446, 84464N.
[xii] Monty, S., F. Jahandar, J. Lee, et al. "Automated testing of optical fibres: Towards the design of the Maunakea Spectroscopic Explorer Fibre Transmission System." *Ground-based and Airborne Instrumentation for Astronomy VII* (2018): 10702, 107027I.
[xiii] Smedley, S., G. Baker, R. Brown, et al. "Sphinx: A massively multiplexed fiber positioner for MSE." *Ground-based and Airborne Instrumentation for Astronomy VII* (2018): 10702, 107021M.

---

[xiv] Caillier, P., W. Saunders, W.; P.-H. Carton, et al. "Maunakea Spectroscopic Explorer Low Moderate Resolution Spectrograph Conceptual Design." *Ground-based and Airborne Instrumentation for Astronomy VII* (2018): 10702, 107028B.
[xv] Zhang, K.; J. R. Zheng, and W. Saunders. "High numerical aperture multimode fibers for prime focus use." *Advances in Optical and Mechanical Technologies for Telescopes and Instrumentation* II (2016): 9912, 99125J.
[xvi] http://www.specinst.com/
[xvii] Zhang, K., Z. Yifei, T. Zhen, et al. "Mauna Kea Spectroscopic Explorer (MSE): A preliminary design of multi-object high resolution spectrograph." *Ground-based and Airborne Instrumentation for Astronomy VII* (2018): 107027, 107027W.
[xviii] https://www.iof.fraunhofer.de/en.html
[xix] https://www.kosi.com/na_en/index.php
[xx] McConnachie, A., N. Flagey, et al. "The science calibration challenges of next generation highly multiplexed optical spectroscopy: The case of the Maunakea Spectroscopic Explorer." *Observatory Operations: Strategies, Processes, and Systems VII* (2018): 1070410.